%% file: thesis.tex
\documentstyle[12pt,epsfig,fleqn]{report}

\input{new}
\input{newlit}

\catcode`\"=\active \let"=\" \let\3=\ss

\begin{document}

\include{titd}

\pagenumbering{Roman}
\addcontentsline{toc}{chapter}{\protect\numberline{Contents}}
\tableofcontents
\newpage
\addcontentsline{toc}{chapter}{\protect\numberline{List of Figures}}
\listoffigures
\newpage
\addcontentsline{toc}{chapter}{\protect\numberline{List of Tables}}
\listoftables

\newpage
\pagenumbering{arabic}
\setcounter{page}{1}

\include{intro}

\input{mssm}

\input{decay}

\include{lep2}

\include{nlcprod}

\include{nlcmass}
\include{nlcslept}

\include{nlcstop}

\include{nlcsbottom}

\include{rar}

\include{conclusions}

\begin{appendix}
\include{appint}

\include{determin}
\include{wirk}

\end{appendix}

\addcontentsline{toc}{chapter}{\protect\numberline{Bibliography}}

\include{lit}
\include{leben}

\end{document}

%% file: new.tex
\newcommand {\acop} [2] {a^{\tilde{#1}}_{#2}}
\newcommand {\acopq} [2] {(a^{\tilde{#1}}_{#2})^2}
\newcommand {\alphas} {\alpha_s}
\newcommand {\app} [1] {Appendix~\ref{#1}}
\newcommand {\asbo} [1] {\overline{\tilde{b}_{#1}} }

\newcommand {\asfer} [1] {\overline{\tilde{f}_{#1}} }
\newcommand {\asnu} [1] {\overline{\tilde{\nu}_{#1} } }
\newcommand {\asta} [1] {\overline{\tilde{\tau}_{#1}} }

\newcommand {\asto} [1] {\overline{\tilde{t}_{#1}} }
\newcommand {\atan} { {\rm arctan} \hspace{0.5mm} }
\newcommand {\atsn} {\overline{\tilde{\nu}_{\tau}} }

\newcommand {\bcop} [2] {b^{\tilde{#1}}_{#2}}
\newcommand {\bcopq} [2] {(b^{\tilde{#1}}_{#2})^2}
\newcommand {\beq} {\begin{eqnarray}}

\newcommand {\chap} [1] {Chapter~\ref{#1}}
\newcommand {\chaps} [1] {Chapters~\ref{#1}}
\newcommand {\chim} [1] {\tilde{\chi}^{-}_{#1} }
\newcommand {\chipm} [1] {\tilde{\chi}^{\pm}_{#1} }
\newcommand {\chin} [1] {\tilde{\chi}^0_{#1} }
\newcommand {\chip} [1] {\tilde{\chi}^{+}_{#1} }
\newcommand {\col} [1] {O_{L_{#1}}'}
\newcommand {\colq} [1] {(O_{L_{#1}}')^2}
\newcommand {\cor} [1] {O_{R_{#1}}'}
\newcommand {\corq} [1] {(O_{R_{#1}}')^2}
\newcommand {\cosa} {\cos \alpha \hspace{0.5mm} }
\newcommand {\cosab} {\cos (\alpha + \beta) \hspace{0.5mm} }
\newcommand {\cosb} {\cos \theta_{\tilde{b}} \hspace{0.5mm} }
\newcommand {\cosbe} {\cos \beta \hspace{0.5mm} }

\newcommand {\cosbq} {\cos^2 \theta_{\tilde{b}} \hspace{0.5mm} }
\newcommand {\cosf} {\cos \theta_{\tilde{f}} \hspace{0.5mm} }
\newcommand {\cosfq} {\cos^2 \theta_{\tilde{f}} \hspace{0.5mm} }
\newcommand {\cosq} {\cos \theta_{\tilde{q}} \hspace{0.5mm} }
\newcommand {\cost} {\cos \theta_{\tilde{t}} \hspace{0.5mm} }
\newcommand {\costa} {\cos \theta_{\tilde{\tau}} \hspace{0.5mm} }
\newcommand {\costaq} {\cos^2 \theta_{\tilde{\tau}} \hspace{0.5mm} }
\newcommand {\costq} {\cos^2 \theta_{\tilde{t}} \hspace{0.5mm} }
\newcommand {\cosw} {\cos \theta_{W} \hspace{0.5mm} }
\newcommand {\coswq} {\cos^2 \theta_{W} \hspace{0.5mm} }
\newcommand {\coswv} {\cos^4 \theta_{W} \hspace{0.5mm} }
\newcommand {\coszb} {\cos 2 \theta_{\tilde{b}} \hspace{0.5mm} }
\newcommand {\coszbe} {\cos 2 \beta \hspace{0.5mm} }
\newcommand {\coszbeq} {\cos^2 2 \beta \hspace{0.5mm} }
\newcommand {\coszq} {\cos 2 \theta_{\tilde{q}} \hspace{0.5mm} }
\newcommand {\coszf} {\cos 2 \theta_{\tilde{f}} \hspace{0.5mm} }
\newcommand {\coszt} {\cos 2 \theta_{\tilde{t}} \hspace{0.5mm} }
\newcommand {\coszta} {\cos 2 \theta_{\tilde{\tau}} \hspace{0.5mm} }
\newcommand {\cotbe} {\cot \beta \hspace{0.5mm} }
\newcommand {\cotbeq} {\cot^2 \beta \hspace{0.5mm} }

\newcommand {\cottaq} {\cot^2 \theta_{\tilde{\tau}} \hspace{0.5mm} }
\newcommand {\cql} [1] {Q_{L_{#1}}'}
\newcommand {\cqlq} [1] {(Q_{L_{#1}}')^2}
\newcommand {\cqr} [1] {Q_{R_{#1}}'}
\newcommand {\cqrq} [1] {(Q_{R_{#1}}')^2}

\newcommand {\dreivi} {\textstyle \frac{3}{4}}
\newcommand {\dsum} [5] {m^2_{#1} + m^2_{#2} + m^2_{#3} + m^2_{#4}
                         - m^2_{#5} - s }

\newcommand {\ee} {e^+ e^- \to }
\newcommand {\eeq} {\end{eqnarray}}
\newcommand {\eindr} {\textstyle \frac{1}{3}}
\newcommand {\einha} {\textstyle \frac{1}{2}}
\newcommand {\einhag} {\frac{1}{2}}
\newcommand {\eint} [3] {\int\limits^{#1}_{#2}
                         \hspace{-2mm} d \, #3 \hspace{2mm}}
\newcommand {\einvi} {\textstyle \frac{1}{4}}

\newcommand {\eqn} [1] {Eq.~(\ref{#1})}
\newcommand {\eqns} [1] {Eqs.~(\ref{#1})}

\newcommand {\fbktOle} {\left(\acop{b}{11} \kte{1} \col{11}
                        + \bcop{b}{11} \lte{1} \cor{11} \right)}
\newcommand {\fbltOle} {\left(\acop{b}{11} \lte{1} \col{11}
                        + \bcop{b}{11} \kte{1} \cor{11} \right)}
\newcommand {\fbktQle} {\left(\acop{b}{11} \kte{1} \cql{11}
                        + \bcop{b}{11} \lte{1} \cqr{11} \right)}
\newcommand {\fbltQle} {\left(\acop{b}{11} \lte{1} \cql{11}
                        + \bcop{b}{11} \kte{1} \cqr{11} \right)}
\newcommand {\fig} [1] {Fig.~\ref{#1}}
\newcommand {\figcaption} [4]
            {\refstepcounter{figure}
             \label{#1}
             \addcontentsline{lof}{figure}
                {\protect\numberline{\protect\ref{#1}}{\ignorespaces #2}}
             {\small{\bf{Fig.~\ref{#1}:}}~#3}
             \vspace{#4mm}
            }

\newcommand {\glu} {\tilde{g}}
\newcommand {\gme} [1] {m_{#1} \Gamma_{#1} }
\newcommand {\gmz} [1] {m^2_{#1} \Gamma^2_{#1} }

\newcommand {\hbktOle} {\left(\bcop{b}{11} \kte{1} \col{11}
                        + \acop{b}{11} \lte{1} \cor{11} \right)}
\newcommand {\hbltOle} {\left(\bcop{b}{11} \lte{1} \col{11}
                        + \acop{b}{11} \kte{1} \cor{11} \right)}
\newcommand {\hbktQle} {\left(\bcop{b}{11} \kte{1} \cql{11}
                        + \acop{b}{11} \lte{1} \cqr{11} \right)}
\newcommand {\hbltQle} {\left(\bcop{b}{11} \lte{1} \cql{11}
                        + \acop{b}{11} \kte{1} \cqr{11} \right)}

\newcommand {\hp} {H^+}

\newcommand {\inwf} [3] { \lambda(#1,#2,#3) }

\newcommand {\jintz} [3] { J^{#2}_{#1}({#3})}

\newcommand {\kbe} [1] {k^{\tilde{b}}_{1 #1}}
\newcommand {\kbeq} [1] {k^{\tilde{b}^2}_{1 #1}}

\newcommand {\kcop} [2] {k^{\tilde{#1}}_{#2}}
\newcommand {\kcopq} [2] {(k^{\tilde{#1}}_{#2})^2}
\newcommand  {\kekzOleOlz}  {\left(\kte{1} \kte{2} \col{11} \col{12} +
                              \lte{1} \lte{2} \cor{11} \cor{12}\right) }
\newcommand  {\kekzOlzOre}  {\left(\kte{1} \kte{2} \col{12} \cor{11} +
                              \lte{1} \lte{2} \col{11} \cor{12}\right) }
\newcommand  {\kzleOleOlz}  {\left(\kte{2} \lte{1} \col{11} \col{12} +
                              \kte{1} \lte{2} \cor{11} \cor{12}\right) }
\newcommand  {\kzleOlzOre}  {\left(\kte{2} \lte{1} \col{12} \cor{11} +
                              \kte{1} \lte{2} \col{11} \cor{12}\right) }
\newcommand  {\kelzOleOlz}  {\left(\kte{1} \lte{2} \col{11} \col{12} +
                              \kte{2} \lte{1} \cor{11} \cor{12}\right) }
\newcommand  {\kelzOlzOre}  {\left(\kte{1} \lte{2} \col{12} \cor{11} +
                              \kte{2} \lte{1} \col{11} \cor{12}\right) }
\newcommand  {\kekzQleQlz}  {\left(\kte{1} \kte{2} \cql{11} \cql{12} +
                              \lte{1} \lte{2} \cqr{11} \cqr{12}\right) }
\newcommand  {\kekzQlzQre}  {\left(\kte{1} \kte{2} \cql{12} \cqr{11} +
                              \lte{1} \lte{2} \cql{11} \cqr{12}\right) }
\newcommand  {\kzleQleQlz}  {\left(\kte{2} \lte{1} \cql{11} \cql{12} +
                              \kte{1} \lte{2} \cqr{11} \cqr{12}\right) }
\newcommand  {\kzleQlzQre}  {\left(\kte{2} \lte{1} \cql{12} \cqr{11} +
                              \kte{1} \lte{2} \cql{11} \cqr{12}\right) }
\newcommand  {\kelzQleQlz}  {\left(\kte{1} \lte{2} \cql{11} \cql{12} +
                              \kte{2} \lte{1} \cqr{11} \cqr{12}\right) }
\newcommand  {\kelzQlzQre}  {\left(\kte{1} \lte{2} \cql{12} \cqr{11} +
                              \kte{2} \lte{1} \cql{11} \cqr{12}\right) }
\newcommand {\kleklzktektz} {\left( \ktsn{1} \ktsn{2} \kte{1} \kte{2}
                       + \ltsn{1} \ltsn{2} \lte{1}  \lte{2} \right) }
\newcommand {\kleklzkteltz} {\left( \ktsn{1} \ktsn{2} \kte{1} \lte{2}
                            + \kte{2} \ltsn{1} \ltsn{2} \lte{1} \right) }
\newcommand {\kleklzktzlte} {\left( \ktsn{1} \ktsn{2} \kte{2} \lte{1}
                            + \kte{1} \ltsn{1} \ltsn{2} \lte{2} \right) }
\newcommand {\kleklzlteltz} {\left( \ktsn{1} \ktsn{2} \lte{1} \lte{2}
                            + \kte{1} \kte{2} \ltsn{1} \ltsn{2} \right) }
\newcommand {\klektektzllz} {\left( \ktsn{1} \kte{1} \kte{2} \ltsn{2}
                            + \ktsn{2} \ltsn{1} \lte{1} \lte{2} \right) }
\newcommand {\klektellzltz} {\left( \ktsn{1} \kte{1} \ltsn{2} \lte{2}
                            + \ktsn{2} \kte{2} \ltsn{1} \lte{1} \right) }
\newcommand {\klektzllzlte} {\left( \ktsn{1} \kte{2} \ltsn{2} \lte{1}
                            + \ktsn{2} \kte{1} \ltsn{1} \lte{2} \right) }
\newcommand {\klellzlteltz} {\left( \ktsn{1} \ltsn{2} \lte{1} \lte{2}
                            + \ktsn{2} \kte{1} \kte{2} \ltsn{1} \right) }
\newcommand {\kkll} { \ktsn{1} \ltsn{1} \kte{1} \lte{1} }
\newcommand {\kllq} { \kte{1} \lte{1} \left(\ktsnq{1}+\ltsnq{1} \right)}
\newcommand {\klqktq} {\left( \ktsnq{1} \kteq{1}
                            + \ltsnq{1} \lteq{1} \right) }
\newcommand {\klqltq} {\left( \ktsnq{1} \lteq{1}
                            + \ltsnq{1} \kteq{1} \right) }
\newcommand {\ktlq} { \ktsn{1} \ltsn{1} \left(\kteq{1}+\lteq{1} \right)}
\newcommand {\ktsn} [1] {k^{\tilde{\nu}_{\tau}}_{#1}}
\newcommand {\ktsnq} [1] {(k^{\tilde{\nu}_{\tau}}_{#1})^2}
\newcommand {\kte} [1] {k^{\tilde{t}}_{1 #1}}
\newcommand {\kteq} [1] {(k^{\tilde{t}}_{1 #1})^2}

\def\lsim{\raise0.3ex\hbox{$\;<$\kern-0.75em\raise-1.1ex\hbox{$\sim\;$}}}
\def\gsim{\raise0.3ex\hbox{$\;>$\kern-0.75em\raise-1.1ex\hbox{$\sim\;$}}}
\newcommand {\lamh} [4] {\lambda^{#4}
                         (m^2_{#1},m^2_{#2},m^2_{#3})}
\newcommand {\lcop} [2] {l^{\tilde{#1}}_{#2}}
\newcommand {\lcopq} [2] {(l^{\tilde{#1}}_{#2})^2}
\newcommand {\lbe} [1] {l^{\tilde{b}}_{1 #1}}
\newcommand {\lbeq} [1] {l^{\tilde{b}^2}_{1 #1}}

\newcommand  {\lelzOleOlz}  {\left(\lte{1} \lte{2} \col{11} \col{12} +
                              \kte{1} \kte{2} \cor{11} \cor{12}\right) }
\newcommand  {\lelzOlzOre}  {\left(\lte{1} \lte{2} \col{12} \cor{11} +
                              \kte{1} \kte{2} \col{11} \cor{12}\right) }
\newcommand  {\lelzQleQlz}  {\left(\lte{1} \lte{2} \cql{11} \cql{12} +
                              \kte{1} \kte{2} \cqr{11} \cqr{12}\right) }
\newcommand  {\lelzQlzQre}  {\left(\lte{1} \lte{2} \cql{12} \cqr{11} +
                              \kte{1} \kte{2} \cql{11} \cqr{12}\right) }

\newcommand {\ltae} [1] {l^{\tilde{\tau}}_{1 #1}}
\newcommand {\ltaeq} [1] {(l^{\tilde{\tau}}_{1 #1})^2}
\newcommand {\ltaz} [1] {l^{\tilde{\tau}}_{2 #1}}

\newcommand {\ltsn} [1] {l^{\tilde{\nu}_{\tau}}_{#1}}
\newcommand {\ltsnq} [1] {(l^{\tilde{\nu}_{\tau}}_{#1})^2}
\newcommand {\lte} [1] {l^{\tilde{t}}_{1 #1}}
\newcommand {\lteq} [1] {(l^{\tilde{t}}_{1 #1})^2}

\newcommand{\Li}{\mbox{{\rm Li}$_2$}}

\newcommand {\ma} {m_{A^0} }
\newcommand {\maq} {m^2_{A^0} }
\newcommand {\mchin} [1] {m_{\chin{#1}}}
\newcommand {\mchinq} [1] { m^2_{\chin{#1}} }
\newcommand {\mchinv} [1] {m_{\tilde{\chi}^0_{#1}}^4 }
\newcommand {\mchip} [1] { m_{\tilde{\chi}^{+}_{#1}} }
\newcommand {\mchipq} [1] { m^2_{\tilde{\chi}^{+}_{#1}} }
\newcommand {\mchipm} [1] { m_{\tilde{\chi}^{\pm}_{#1}} }
\newcommand {\mchipmq} [1] { m^2_{\tilde{\chi}^{\pm}_{#1}} }
\newcommand {\mdq} {M^2_{\tilde{D}}}
\newcommand {\me} {E \hspace{-2.5mm} / \hspace{1.2mm}}
\newcommand{\mgev} [2] {m_{#1} = #2 \hspace{1mm} {\rm GeV} }
\newcommand {\mglu} {m_{\glu}}
\newcommand {\mgluq} {m^2_{\glu}}
\newcommand {\mh} {m_{h^0} }
\newcommand {\mhp} {m_{H^{\pm}} }

\newcommand {\mhpq} {m^2_{H^{\pm}} }
\newcommand {\mH} {m_{H^0} }

\newcommand {\mqq} {M^2_{\tilde{Q}}}
\newcommand {\msbot} [1] { m_{\tilde{b}_{#1}} }
\newcommand {\msbq} [1] { m^2_{\tilde{b}_{#1}} }
\newcommand {\mselec} [1] { m_{\tilde{e}_{#1}} }
\newcommand {\msfer} [1] { m_{\tilde{f}_{#1}} }
\newcommand {\msferp} [1] { m_{\tilde{f}_{#1}'} }
\newcommand {\msferq} [1] { m^2_{\tilde{f}_{#1}} }

\newcommand {\msnu} [1] {m_{\tilde{\nu}_{#1}} }

\newcommand {\msquq} [1] {m^2_{\tilde{q}_{#1}} }
\newcommand {\mstau} [1] { m_{\tilde{\tau}_{#1}} }
\newcommand {\mstaq} [1] { m^2_{\tilde{\tau}_{#1}} }
\newcommand {\mstop} [1] { m_{\tilde{t}_{#1}} }
\newcommand {\mstq} [1] { m^2_{\tilde{t}_{#1}} }
\newcommand {\mstv} [1] {m_{\tilde{t}_{#1}}^4 }

\newcommand {\mtau} {m_{\tau} }
\newcommand {\mtauq} {m^2_{\tau} }

\newcommand {\mtsn} {m_{\tilde{\nu}_{\tau}}  }
\newcommand {\mtsnq} {m^2_{\tilde{\nu}_{\tau}}  }
\newcommand {\muq} {M^2_{\tilde{U}}}
\newcommand {\mw} {m_{W}}
\newcommand {\mwq} {m^2_{W}}
\newcommand {\mzq} {m^2_{Z}}

\newcommand {\no} {\nonumber \\}
\newcommand {\nutau} {{\nu}_{\tau} }

\newcommand {\pbi} {p_{\sbo{i}}}
\newcommand {\plgl} { & \hspace{-1mm} =
                       \hspace{-1mm} & }
\newcommand {\plogl} [1] { &   & \hspace{#1 mm} }
\newcommand {\po} {p_{\chin{1}} }
\newcommand{\pxi }{ p_{\chipm{i}} }

\newcommand{\recht} {\begin{picture}(7,7)
                      \linethickness{1.7mm}
                      \put(2,1){\line(0,1){5}}
                      \thinlines
                     \end{picture}
                    } 
\newcommand{\rechtl} {\begin{picture}(7,7)
                      \put(2,1){\framebox(5,5){}}
                     \end{picture}
                    } 

\newcommand {\sbo} [1] {\tilde{b}_{#1}}
\newcommand {\sect} [1] {Section~\ref{#1}}
\newcommand {\sel} [1] {\tilde{e}_{#1} }
\newcommand {\sfer} [1] {\tilde{f}_{#1}}
\newcommand {\sferp} [1] {\tilde{f}_{#1}'}
\newcommand {\sina} {\sin \alpha \hspace{0.5mm} }
\newcommand {\sinab} {\sin (\alpha + \beta) \hspace{0.5mm} }
\newcommand {\sinb} {\sin \theta_{\tilde{b}} \hspace{0.5mm} }
\newcommand {\sinbe} {\sin \beta \hspace{0.5mm} }
\newcommand {\sinbeq} {\sin^2 \beta \hspace{0.5mm} }
\newcommand {\sinbq} {\sin^2 \theta_{\tilde{b}} \hspace{0.5mm} }
\newcommand {\sinf} {\sin \theta_{\tilde{f}} \hspace{0.5mm} }
\newcommand {\sinfq} {\sin^2 \theta_{\tilde{f}} \hspace{0.5mm} }
\newcommand {\sinq} {\sin \theta_{\tilde{q}} \hspace{0.5mm} }
\newcommand {\sint} {\sin \theta_{\tilde{t}} \hspace{0.5mm} }
\newcommand {\sinta} {\sin \theta_{\tilde{\tau}} \hspace{0.5mm} }
\newcommand {\sintaq} {\sin^2 \theta_{\tilde{\tau}} \hspace{0.5mm} }
\newcommand {\sintq} {\sin^2 \theta_{\tilde{t}} \hspace{0.5mm} }
\newcommand {\sinw} {\sin \theta_{W} \hspace{0.5mm} }
\newcommand {\sinwq} {\sin^2 \theta_{W} \hspace{0.5mm} }
\newcommand {\sinwv} {\sin^4 \theta_{W} \hspace{0.5mm} }
\newcommand {\sinzb} {\sin 2 \theta_{\tilde{b}} \hspace{0.5mm} }
\newcommand {\sinzbe} {\sin 2 \beta \hspace{0.5mm} }

\newcommand {\sinzf} {\sin 2 \theta_{\tilde{f}} \hspace{0.5mm} }
\newcommand {\sinzfq} {\sin^2 2 \theta_{\tilde{f}} \hspace{0.5mm} }
\newcommand {\sinzq} {\sin 2 \theta_{\tilde{q}} \hspace{0.5mm} }

\newcommand {\sinzt} {\sin 2 \theta_{\tilde{t}} \hspace{0.5mm} }
\newcommand {\sinzta} {\sin 2 \theta_{\tilde{\tau}} \hspace{0.5mm} }
\newcommand {\sinztq} {\sin^2 2 \theta_{\tilde{t}} \hspace{0.5mm} }
\newcommand {\smg} {SU(3) \times SU(2) \times U(1)}
\newcommand {\snu} [1] {\tilde{\nu}_{#1} }
\newcommand {\sta} [1] {\tilde{\tau}_{#1}}
\newcommand {\sthgbc} {\tilde{t}_1 \to H^+ + b + \chin{1} }
\newcommand {\stwbc} {\tilde{t}_1 \to W^+ + b + \chin{1} }
\newcommand {\sto} [1] {\tilde{t}_{#1}}

\newcommand {\squ} [1] {\tilde{q}_{#1}}

\newcommand {\tab} [1] {Tab.~\ref{#1}}
\newcommand {\tabcaption} [3]
            {\refstepcounter{table}
             \label{#1}
             \addcontentsline{lot}{table}
                {\protect\numberline{\protect\ref{#1}}{\ignorespaces #2}}
             {\small{\bf{Table~\ref{#1}:}}~#3}
             \vspace{3mm}
            }
\newcommand {\tanbe} {\tan \beta \hspace{0.5mm} }
\newcommand {\tanbeq} {\tan^2 \beta \hspace{0.5mm} }

\newcommand {\tant} {\tan \theta_{\tilde{t}} \hspace{0.5mm} }
\newcommand {\tanw} {\tan \theta_{W} \hspace{0.5mm} }
\newcommand {\tanwq} {\tan^2 \theta_{W} \hspace{0.5mm} }
\newcommand {\thline} {\\ \hline}
\newcommand {\tsn} {\tilde{\nu}_{\tau} }

\newcommand {\wzw} {\sqrt{2} \hspace{1mm} }

\newcommand {\zerfal} [3] { #1 \rightarrow #2 \hspace{0.5mm}
                            + \hspace{0.5mm} #3 }

\newcommand {\zerfallz} [3] {\Gamma(#1 \rightarrow #2 \hspace{0.5mm}
                            + \hspace{0.5mm} #3)}
\newcommand {\zwdr} {\frac{2}{3}}

%% file: newlit.tex
\newcommand {\apph} [3] {Astropart.~Phys.~{\bf #1}, #2 (19#3)}
\newcommand {\fsp} [3] {Fortschr.~Phy.~{\bf #1}, #2 (19#3)}
\newcommand {\ictp} [3] {ICTP series in Theo. Phys. {\bf #1}, #2 (19#3)}

\newcommand {\jetp} [3] {JETP Lett.~{\bf #1}, #2 (19#3)}

\newcommand {\npb} [3] {Nucl.~Phys.~{\bf B#1}, #2 (19#3)}
\newcommand {\pl} [3] {Phys.~Lett.~{\bf #1}, #2 (19#3)}
\newcommand {\plb} [3] {Phys.~Lett.~B {\bf #1}, #2 (19#3)}
\newcommand {\pr} [3] {Phys.~Rev.~{\bf #1}, #2 (19#3)}
\newcommand {\prd} [3] {Phys.~Rev.~D{\bf #1}, #2 (19#3)}
\newcommand {\prl} [3] {Phys.~Rev.~Lett. {\bf #1}, #2 (19#3)}

\newcommand {\ptp} [3] {Prog.~Theo.~Phys.~{\bf #1}, #2 (19#3)}

\newcommand {\rep} [3] {Phys.~Rep.~{\bf #1}, #2 (19#3)}

\newcommand {\snp} [3] {Sov.~J.~Nucl.~Phys.~{\bf #1}, #2 (19#3)}
\newcommand {\zfphc} [3] {Z.~Phys.~C--Particles and Fields 
                          {\bf #1}, #2 (19#3)}

\newcommand {\hepz} [2] {{\bf hep-ph/#1}, #2}

%% file: intro.tex
\chapter{Introduction}

Our understanding of the fundamental forces appearing in nature rests to a 
large extent on our increasing understanding of the underlying symmetries. 
Parallel to
this happened the development of quantum field theories for the
electromagnetic, weak and strong interactions
(see e.g. \cite{Itzykson80}). In these theories the
fundamental states are described by fields whose interactions obey the
underlying symmetries.
The theory which explains up to now all experimental facts is the so-called
Standard Model (SM) which is based on the gauge group $\smg$
(see e.g. \cite{Nachtmann86,Ross84,PDG96} and references therein).

Despite its enormous success the SM still leaves many questions open,
e.g.: Is it possible to explain the origin of the
various parameters of the model ?
Can gravity be included ? What is the origin of fermion masses ?
What is the origin of electroweak symmetry breaking ?
A first step to answering these questions could be the embedding of the
SM in a Grand Unified Theory (GUT) \cite{Ross84,Pati73}. In this 
case the SM appears as the low energy limit of a GUT with only one gauge
coupling. However, the simplest possibility based on $SU(5)$ has been ruled
out by experiment \cite{Amaldi91}.
Besides this, the SM and its simple GUT extensions exhibit the so-called
hierarchy problem \cite{Hierachy}. 
Here the question arises how one can stabilize the
masses of fundamental scalars in the range of the electroweak scale.
Radiative corrections push these masses up either to
the GUT- or the Planck-scale. The stabilization can be achieved through
cancellation between terms, each of the order $10^{15}$~GeV. This
seems to be rather unnatural, because these terms have to be adjusted up to
13 digits.

One possibility to solve this problem is supersymmetry (SUSY)\cite{Hierachy2}. 
SUSY is a symmetry, which relates bosons to fermions and vice
versa \cite{Ramond71,Wess74}.
Due to the fact that
fermionic loops enter with the opposite sign of bosonic loops, one
gets an automatic cancellation between different contributions of the
radiative corrections to scalar masses. Moreover,
repeated applications of supersymmetric transformations are equivalent
to a Poincar\'e transformation. Therefore, supersymmetry, if it is a local 
symmetry,
contains the graviton among its gauge fields and includes Einstein's
theory of gravity (see e.g. \cite{Bagger91}). For this reason local
supersymmetry is called supergravity (SUGRA). Beside these more theoretical
advantages it has been shown that within supersymmetric GUTs a unification
of gauge couplings consistent with experiments can be achieved
\cite{Amaldi91}.

Some other appealing features of SUSY are:
\begin{itemize}
 \item It can be shown that the SUSY algebra is the
  only nontrivial extension of the set of spacetime symmetries
  which forms one of the building blocks of relativistic quantum field
  theory \cite{Coleman67}.
 \item Electroweak symmetry breaking can be achieved due
   to radiative effects caused by the large top mass \cite{Ibanez82}.
 \item Some models contain a stable lightest supersymmetric particle
  which serves as a good
  candidate for dark matter \cite{Ellis84}.
\end{itemize}

Therefore, it is very interesting to study the phenomenological aspects
of supersymmetric theories. For technical details and how to construct 
a phenomenological acceptable theory we would like to refer the interested 
reader to the 
literature, e.g.
\cite{Bagger91,Haber84,Nath84,Gunion86a,Drees95b,Bailin95,Amundson96}.

The simplest supersymmetric extension of the SM is called
Minimal Supersymmetric Standard Model (MSSM). In this model every known SM
particle obtains a supersymmetric partner. One has to enlarge the Higgs sector
to obtain a consistent theory by adding an extra Higgs doublet of
opposite hypercharge.
The second Higgs doublet is needed to give mass to both up- and down-type quarks
and to cancel the triangle anomaly caused by the supersymmetric partners of
the Higgs-doublets. Due to the fact that no supersymmetric partner of the
SM particles has been found up to now, SUSY has to be broken. This can be
achieved by adding by hand soft SUSY breaking terms. These consist of
scalar and gaugino mass terms, as well as trilinear (A-terms) and bilinear
(B-terms) interactions \cite{Girardello82}. Such a model is the most
conservative approach to realistic model building, but the large parameter
space causes it to be rather unpredictable. However, 
there are many different ways
to construct a more fundamental theory which reduces the number of free
parameters. For example in minimal supergravity models there are
only five additional parameters which remain as free parameters (see e.g.
\cite{Drees95b,Bailin95,Amundson96}).

In this work we will use the MSSM as framework. In this model there exists
a conserved quantum number called R-parity $R = (-1)^{3 B + L + 2 S}$,
where $B$, $L$ and $S$ denote baryon number, lepton number and spin,
respectively.
This leads to two important phenomenological implications: Firstly, SUSY
particles can only be produced in pairs. Secondly, the lightest
supersymmetric particle (LSP) is stable.
In recent years it has turned out that third generation sfermions (the scalar
partners of the top- and bottom quark, the tau and the tau-neutrino) are
of special interest. This is mainly due to two reasons: Firstly, large
Yukawa couplings lead to a different phenomenology compared
to the sfermions of the first two generations. Secondly, if a SUSY GUT is
realized in nature, studies of renormalization group equations (RGEs)
\cite{Drees95b,Barger94a} indicate,
that third generation sfermions are in general lighter than the sfermions of the
first two generations (for the stop which is the scalar partner of the top
quark this has been pointed out in \cite{Ellis83}).
For these reasons it is possible that
one of them is the lightest charged SUSY particle and hence its appearance
could be the first experimental evidence of SUSY. In the following we will
study their phenomenology at $e^+ e^-$ colliders. However,
most of our results can also be applied to hadron colliders.

This work is organized in the following manner: In \chap{chap:mssm} we 
present the particle content of the MSSM. The relevant interaction
Lagrangian and the formulae for the various
decay widths of the sfermions are given in \chap{chap:decay}.
In \chap{chap:lep2} we discuss the phenomenology of sfermions at LEP2
($\sqrt{s} = 192$~GeV). In particular the influence of the Yukawa couplings
on the decay pattern will be discussed.
The production of sfermions at an $e^+ e^-$ Linear Collider (LC) with
an energy range between 500~GeV and 2~TeV is discussed in \chap{chap:nlcprod}.
In \chap{chap:nlcmass} we discuss our strategy for exploring the phenomenology
of sfermions with masses of several hundred GeV which will be presented in
\chaps{chap:nlcslept}, \ref{chap:nlcstop}, and \ref{chap:nlcsbottom}.
Here we put special emphasis on the decays into gauge and Higgs bosons.
Higher order decays of the light stop and the light sbottom are
discussed in \chap{chap:rardecay}.
In \chap{chap:con} we give a summary of the main topics. \app{appA} gives
the analytical solutions for the three body decay widths.
In \app{appB} we present the formulae for computation of the MSSM parameters
from sfermion masses and mixing angles. Finally, the formulae
for the production cross sections including ISR- and QCD-corrections
are given in \app{appC} .

%% file: mssm.tex
\chapter{The Minimal Supersymmetric Standard Model}
\label{chap:mssm}

\section{General Aspects}

The simplest linear extension of the Standard Model (SM) is the so-called
Minimal Supersymmetric Standard Model (MSSM). It has the same gauge
group as the SM: $SU(3)_C \times SU(2)_L \times U(1)_Y$ which breaks
spontaneously to $SU(3)_C \times U(1)_{em}$. This spontaneous symmetry
breaking can be achieved through radiative corrections due to the
large top mass \cite{Ibanez82}. In addition to the SM particles we
have a second Higgs doublet, spin $\einha$ partners for the Higgs fields
called higgsinos,  spin $\einha$ partners for the gauge bosons
called gauginos and two spin 0 partners for each fermion called sfermions
(one for each helicity state). In \tab{mssmtab1} all particles of the
MSSM are listed except the graviton and its partner, the gravitino.

In the following we will briefly discuss the 
mass eigenstates of this model.

\begin{table}[th]
\begin{tabular}{|c|l|c|l|} \hline
 \multicolumn{2}{|c|}{SM particles + additional Higgs}
     & \multicolumn{2}{c|}{SUSY particles} \thline
 $j = \einha$ & $q_{L_i} = \left( \begin{array}{c}
                                  u_{L_i} \\ d_{L_i} \end{array} \right)$,
                $u_{R_i}$, $d_{R_i}$
 & $j = 0$ & $\tilde{q}_{L_i} = \left( \begin{array}{c}
                  \tilde{u}_{L_i} \\ \tilde{d}_{L_i} \end{array}  \right)$,
                $\tilde{u}_{R_i}$, $\tilde{d}_{R_i}$   \\
           & $l_{L_i} = \left( \begin{array}{c}
                                  \nu_{L_i} \\ e_{L_i} \end{array} \right)$,
                $e_{R_i}$
 &         & $\tilde{l}_{L_i} = \left( \begin{array}{c}
                  \tilde{\nu}_{L_i} \\ \tilde{e}_{L_i} \end{array}  \right)$,
                $\tilde{e}_{R_i}$ \thline
 $j = 1$   & $g^a$  & $j=\einha$ & $\glu^a$  \\
           & $\gamma$, $Z$, $W^{\pm}$
           &  & $\chin{i} = a_{i} \tilde{\gamma}
                          + b_{i} \tilde{Z}
                          + c_{i} \tilde{H}^0_1
                          + d_{i} \tilde{H}^0_2 $ \\
 $j = 0$ &  $H_1 = \left( \begin{array}{c}
                                  H^0_1 \\ H^{-}_1 \end{array} \right)$,
            $H_2 = \left( \begin{array}{c}
                                  H^{+}_2 \\ H^{0}_2 \end{array} \right)$
           &  & $\chipm{k} = \alpha_{k} \tilde{W}^{\pm}
                          + \beta_{k} \tilde{H}^{\pm}_{2,1} $ \thline
\end{tabular} \\[0.3cm]
\tabcaption{mssmtab1}
    {Particle content of the Minimal Supersymmetric Standard Model.}
    {Particle content of the Minimal Supersymmetric Standard Model.}
\end{table} 

\section{Electroweak Gauge Bosons and Higgs Bosons}

As already mentioned there are two Higgs doublets $H_1$ and $H_2$
in the MSSM. The scalar potential
$V_{Higgs}$ reads at tree-level \cite{Gunion86a}
\beq
V_{Higgs} \plgl m^2_1 H^{\dagger}_1 H_1
               + m^2_2 H^{\dagger}_2 H_2
               - m^2_3 (H^{T}_{1i} \epsilon_{ij} H_{2j} + h.c. ) + \no
          \plogl{-2} + \frac{g^2}{2} \sum^3_{k=1}
                \left(H^{\dagger}_{1i} \frac{\sigma^k_{ij}}{2} H_{1j}
                     +H^{\dagger}_{2i} \frac{\sigma^k_{ij}}{2} H_{2j}
                \right)^2
            + \frac{g'}{4}
                \left(H^{\dagger}_{1} H_{1}
                     +H^{\dagger}_{2} H_{2} \right)^2 .
\eeq
The phases of the fields can be chosen such that $m^2_3 > 0$.
Note, that the quadric terms are determined through the gauge couplings
in contrast to the Standard Model or a general two Higgs doublet model
where these couplings are free parameters.

Due to spontaneous symmetry breaking of $SU(2)_L \times U(1)_Y$ one
gets for the vacuum expectation values (vevs) of the Higgs fields:
\beq
\langle H_1 \rangle =
   \left( \begin{array}{c} v_1 \\ 0 \end{array} \right) \hspace{5mm}
\langle H_2 \rangle =
  \left( \begin{array}{c} 0 \\ v_2 \end{array} \right).
\eeq
The diagonalization of the mass matrices for the charged and the neutral
Higgs fields leads to five physical Higgs bosons and
three Goldstone bosons. The physical Higgs bosons are two charged
Higgs bosons $H^{\pm}$, two scalar Higgs bosons $h^0,H^0$ and a
pseudoscalar Higgs boson $A^0$. The three Goldstone-bosons form
together with the $SU(2) \times U(1)$ gauge-bosons the photon and
the $Z^0$- and $W^{\pm}$-bosons. The masses of the gauge bosons
\beq
\mwq = \frac{g^2 v^2}{2}, \hspace{5mm} \mzq = \frac{(g'^2 + g^2) v^2}{2}
\eeq
fix the sum of the vevs $v^2 = v^2_1 + v^2_2$. The ratio
$\tanbe \equiv v_2 / v_1$ remains as a free parameter of the model.

At tree-level one gets the following mass relations:
\beq
\maq \plgl m^2_1 + m^2_2 \\
m^2_{h^0,H^0} \plgl \einha \left( \maq + \mzq \mp
           \sqrt{(\maq+\mzq)^2 - 4 \maq \mzq \coszbeq} \right) \\
\mhpq \plgl \maq + \mwq
\eeq
and thus
\beq
\mh \leq min(m_Z,\ma), \hspace{2mm}
\mH \geq m_Z, \hspace{2mm}
\mhp \geq m_W.
\eeq
It has been shown that one-loop corrections are important for the
Higgs-masses. In the effective potential approach \cite{Coleman73} the
potential $V_{Higgs}$ modifies to $V_{one-loop}$
\beq
V_{one-loop} = V_{Higgs} + \frac{1}{64 \pi^2}
STr {\cal{M}}^4 \left(\ln \frac{{\cal{M}}^2}{Q^2_0} - \frac{3}{2} \right).
\eeq
A total one-loop computation has been given in \cite{Chankowski94a,Eberl93}.
For the discussions in \chap{chap:nlcmass}-\ref{chap:nlcsbottom} the most
important corrections stemming from third generation quarks and squarks
\cite{Ellis91a,Brignole92a} have been included.

\section{Gauginos and Higgsinos}

\subsection{Charginos}

The mass eigenstates of the charged gauginos and higgsinos are called
charginos $\chipm{1,2}$. In the basis
\beq
\tilde{\psi}^{+}_j = \left( -i \lambda^{+}, \psi^1_{H_2} \right)
\hspace{5mm}
\tilde{\psi}^{-}_j = \left( -i \lambda^{-}, \psi^2_{H_1} \right)
\eeq
one finds the following mass term \cite{Bartl92a,Haber84a}
\beq
{\cal L}_{\mchipm{}} =
     - \einha \big( \tilde{\psi}^{+}_j, \tilde{\psi}^{-}_j \big)
         \left( \begin{array}{cc} 0 & X^T \\ X & 0 \end{array} \right)
         \left( \begin{array}{c}
             \tilde{\psi}^{+}_j \\ \tilde{\psi}^{-}_j \end{array} \right)
\eeq
with
\beq
X = \left( \begin{array}{cc} M &  \wzw m_W \sinbe \\
           \wzw m_W \cosbe & \mu \end{array} \right).
\eeq
The mass matrix $X$ can be diagonalized by two real, unitary matrices
$U$ and $V$
\beq
M_D = U^* X V^{-1}.
\eeq
The mass eigenstates are defined through:
\beq
\tilde{\chi}^{+}_{i} = V_{ij} \tilde{\psi}^{+}_{j}, \hspace{3mm}
\tilde{\chi}^{-}_{i} = U_{ij} \tilde{\psi}^{-}_{j} \hspace{3cm} i,j=1,2.
\eeq
 
For the eigenvalues $\mchipm{1,2}$ and the matrix elements
$U_{ij}$ and $V_{ij}$ one gets \cite{Bartl92a} for $\tanbe > 1$:
\beq
\mchipm{1,2} \plgl \einha \left(
         \sqrt{ \big(M-\mu\big)^2 + 2 m^2_W (1+\sinzbe)} \right .\no
     \plogl{1} \left. \mp
         \sqrt{ \big(M+\mu\big)^2+2m^2_W (1-\sinzbe)} \right)
  \label{char1}        \\
U_{22} \plgl - U_{11} = \frac{\epsilon_U}{\wzw}
         \sqrt{1 - \frac{M^2 - \mu^2 - 2 m^2_W \coszbe}{W} } \\
U_{12} \plgl  U_{21} = \frac{1}{\wzw}
         \sqrt{1 + \frac{M^2 - \mu^2 - 2 m^2_W \coszbe}{W} } \\
V_{11} \plgl  V_{22} = \frac{1}{\wzw}
         \sqrt{1 - \frac{M^2 - \mu^2 + 2 m^2_W \coszbe}{W} } \\
V_{21} \plgl - V_{12} = \frac{\epsilon_V}{\wzw}
         \sqrt{1 + \frac{M^2 - \mu^2 + 2 m^2_W \coszbe}{W} }
\label{eq:vij}
\eeq
with
\beq
 W = \sqrt{ (M^2+\mu^2+2 m^2_W)^2 - 4 (M \mu - m^2_W \sinzbe)^2 } \\
 \epsilon_U = \mathrm{sign} (M \cosbe + \mu \sinbe)
 \hspace{5mm} \mathrm{and} \hspace{5mm}
 \epsilon_V = \mathrm{sign} (M \sinbe + \mu \cosbe).
\eeq
For $\tanbe < 1$ one has to replace $U_{ij}$ by $\epsilon_U U_{ij}$
and $V_{ij}$ by $\epsilon_V V_{ij}$. In the chosen phase convention
the mass eigenvalue of the heavier chargino is always positive.
For $M \mu - m^2_W \sinzbe < 0$ ($ > 0$) the lighter mass eigenvalue is
positive (negative). The following
asymptotic behavior can be deduced from \eqns{char1}--(\ref{eq:vij}): 
for $M \gg |\mu|$
the lighter chargino is mainly a higgsino whereas the heavier chargino
is mainly gaugino--like. For $|\mu| \gg M$ the charginos interchange their
nature.

\subsection{Neutralinos}

Analogous to the chargino case a mixture takes place between neutral 
gauginos
and neutral higgsinos \cite{Haber84a,Bartl89a}. The mass eigenstates
are called neutralinos $\chin{j} \, (j=1,..,4)$. In the basis
\beq
\psi^0_j = \left(
              -i\lambda_\gamma, -i\lambda_Z,
               \psi^1_{H_1} \cosbe - \psi^{2}_{H_2} \sinbe,
               \psi^1_{H_1} \sinbe + \psi^{2}_{H_2} \cosbe \right)
\eeq
the mass matrix has the form \cite{Bartl89a}:
\beq
\hspace*{-10mm}
Y = \left( \begin{array}{cccc} M' \coswq + M \sinwq &
                ( M-M') \sinw \cosw
                & 0 & 0 \\  ( M-M') \sinw \cosw & M' \sinwq + M \coswq
                & m_Z & 0 \\ 0 & m_Z & \mu \sinzbe & - \mu \coszbe \\
                0 & 0 & - \mu \coszbe & -\mu \sinzbe  \end{array} \right)
   \nonumber
\eeq
\vspace{-6mm}
\beq
\eeq
This matrix can be diagonalized by a $4 \times 4$ unitary matrix N.
The neutralino mass eigenstates are given by:
\beq
\tilde{\chi}^0_i = N_{ij} \tilde{\psi}^0_j \hspace{3cm} i,j=1,..,4.
\label{neut1}
\eeq
Here $\tilde{\psi}^0 = (\psi^0_j ,\bar{\psi}^0_j)$ denotes a four component 
spinor whereas
the $\psi^0_j$ are two component Weyl spinors.
The ordering of the neutralino mass eigenstates is given by
$|\mchin{1}| < |\mchin{2}| < |\mchin{3}| < |\mchin{4}| $.
Under the assumption that the neutral gaugino--higgsino mixing contains
no source of $CP$ violation, the matrix N is real. The mass eigenvalues
can be either positive or negative. This sign is 
related to the $CP$ eigenvalue of the corresponding neutralino \cite{Bilenky84}.
In the following, the GUT relation
\beq
M' = 5/3 \tanwq M
\label{eq:gutone}
\eeq
is assumed for the $U(1)$ gaugino mass $M'$ and the $SU(2)$ gaugino 
mass $M$. For $M \gg |\mu|$
the lighter two neutralinos are mainly higgsinos and the heavier two
are mainly gauginos (where the heaviest is mainly the zino and the
second heaviest mainly the photino). For $|\mu| \gg M$ the lightest
neutralino is mainly the photino, the second one is mainly the zino and
the two heaviest states are mainly higgsinos.

\subsection{The Gluino}

For the gluino there is the following mass term
\beq
{\cal L}_{\mglu} = \einha \mglu \bar{\glu} \glu .
\eeq
In the framework of a GUT the gluino 
mass $\mglu$ is related to the $SU(2)$ gaugino mass $M$ through
\beq
M = \mglu \alpha / (\alphas \sinwq).
\label{eq:guttwo}
\eeq
where $\alpha$ and $\alpha_s$ are the electroweak and strong coupling
respectively.
It has been shown
that QCD corrections to this relation can be important \cite{Pierce94}.

\section{Leptons and Quarks}

The leptons and quarks have similar mass terms as in the SM:
\beq
{\cal L}_{f} = - v_1 \bar{E}_{L_a} h^e_{ab} E_{R_b}
        - v_1 \bar{D}_{L_a} h^d_{ab} D_{R_b}
        - v_2 \bar{U}_{L_a} h^d_{ab} U_{R_b} - h.c. \, .
\eeq
With the help of the unitary $3 \times 3$ matrices $V^l_{jk}$ one gets
the mass eigen states:
\beq
E^{phys}_{L_a,R_a} = V^{L,R}_{e,ab} E_{L_a,R_a}, \hspace{3mm}
D^{phys}_{L_a,R_a} = V^{L,R}_{d,ab} D_{L_a,R_a}, \hspace{3mm}
U^{phys}_{L_a,R_a} = V^{L,R}_{u,ab} U_{L_a,R_a}
\eeq
with $E^{phys} = (e,\mu,\tau)$, $D^{phys} = (d,s,b)$ and
$U^{phys} = (u,c,t)$. For the masses one finds
\beq
diag(m_e,m_{\mu},m_{\tau}) &=& v_1 V^L_e h^e V^{R^{\dagger}}_e, \hspace{3mm}
diag(m_d,m_s,m_b) = v_1 V^L_d h^d V^{R^{\dagger}}_d, \no
diag(m_u,m_c,m_t) &=& v_2 V^L_u h^u V^{R^{\dagger}}_u.
\eeq
Note, that there are no interactions between neutral gauge bosons
and fermions of different generations (GIM-mechanism \cite{Glashow70}).
However, there are interactions between the $W^{\pm}$ and quarks of
different generations. The strength of these interactions is related
to the so-called Cabbibo--Kobayashi--Maskawa (CKM) matrix $K_{CKM}$
\cite{Cabibbo63,Kobayashi73} which is given by 
$K_{CKM} \equiv (V^L_u)^{\dagger} V^L_d$.
The analogous matrix in the lepton sector does not exist because there are 
no right-handed neutrinos in the MSSM.

Note, that the masses of the u-quarks are proportional to
$v_2$ whereas the masses of the d-quarks and the leptons are
proportional to $v_1$.
Therefore, we have $h^t = h^b$ if $\tanbe = m_t / m_b$. In such a
case the $\tau$ Yukawa coupling is also of the order of the
top Yukawa coupling. This has
important consequences for the phenomenology of the supersymmetric
partners of the $\tau$~lepton and the b~quark in high $\tanbe$ scenarios
as will be shown in \chaps{chap:lep2}, \ref{chap:nlcsbottom} and
\ref{chap:nlcslept}.

\section{Sfermions}

As already mentioned every SM fermion has two spin 0 partners denoted by
$\sfer{L}$ and $\sfer{R}$, one for each helicity state. The obvious
exception are the neutrinos which are only left--handed
states in our model. In the following the scalar partners for the fermions
(quarks, leptons) will be called sfermions (squarks, sleptons). A similar 
notation will be used for the scalar partners of a specific fermion,
e.g. the scalar partners of the top-quarks
will be called stops. For the sfermions one gets the following mass terms:
\beq
{\cal L}_{\sfer{}} = - (\sfer{L}^*, \sfer{R}^* )
      \left( \begin{array}{cc}
        {\cal{M}}^2_{\cal{LL}} & {\cal{M}}^2_{\cal{LR}} \\
        {\cal{M}}^2_{\cal{RL}} & {\cal{M}}^2_{\cal{RR}} \end{array} \right)
      \left( \begin{array}{c}
        \sfer{L} \\ \sfer{R} \end{array} \right)
\eeq
with
\beq
 {\cal{M}}^2_{\cal{LL}} &=& M^2_F + v^2_i (h^f)^{\dagger} h^f
          + (T^3_I - e_f \sinwq ) \coszbe m^2_Z  {\bf 1}_3\no
 {\cal{M}}^2_{\cal{LR}} &=& {\cal{M}}^{2^{\dagger}}_{\cal{RL}}
                = v_i (A_f g^f - \mu h^f ) \no
 {\cal{M}}^2_{\cal{LL}} &=& M^2_{F'} + v^2_i (h^f)^{\dagger} h^f
          + e_f \sinwq \coszbe m^2_Z {\bf 1}_3 .
\eeq
Here $F$ denotes $Q$ in case of squarks and $L$ in case of sleptons,
$F'$ denotes $E,D,U$, and $f$ denotes $e,d,u$. $T^3_I$ is the third
component of the weak isospin and $e_q$ is the electric charge of the 
corresponding fermion.
In case of up-type squarks $v_i$ is $v_2$ whereas in
case of down-type squarks and charged sleptons $v_i = v_1$. $A_f$ is a
scalar quantity whereas $M^2_{F,F'}$ and $g^f$ are $3 \times 3$ matrices.
In principal the entries of these matrices are free quantities, but
especially for the first two generations they are very constrained
in order to avoid oversized Flavour Changing Neutral Currents (FCNCs) 
\cite{Wyler,HalGab}.
Therefore, we will assume that $M^2_F$ and $M^2_{F'}$ are diagonal
and that $g^f = h^f$. In this case one can perform a rotation in the sfermion
sector analogous to the fermion sector to decouple the
generations. The only exception will be the mixing between
the stops and the scalar left charm. This will be discussed in
section~\ref{sec:stchic}.

With the above assumptions the masses of the sneutrinos are given by
\beq
m^2_{\snu{}} = M^2_L + \einha \coszbe \mzq .
\label{eq:massneut}
\eeq
For the other sfermions one gets $2 \times 2$ mass matrices
of the following form \cite{Bartl94c}:
\beq
\hspace*{-10mm}
M^2_{\sfer{}} = \left( \begin{array}{cc}
    M^2_{ \tilde{F}} + (T^3_I - e_f \sinwq ) \coszbe \mzq + m^2_f &
    \hspace{-10mm} m_f \big( A_f - \mu \Theta(\beta) \big) \\
    m_f \big( A_f - \mu \Theta(\beta) \big) &
    \hspace{-10mm} M^2_{\tilde{F'}} +  e_f \sinwq \coszbe \mzq + m^2_f
                       \end{array} \right) \nonumber
\eeq
\vspace{-7mm}
\beq
\label{eq:sfmassmatrix}
\eeq
with
\beq
\Theta(\beta) = \left\{ 
\begin{array}{l}
 \cotbe \, \, \mathrm{for} \, \, T^3_I = \einha \\
 \tanbe \, \, \mathrm{for} \, \, T^3_I = - \einha 
\end{array} \nonumber \right.
\eeq
The mass eigenvalues are given by:
\beq
 m^2_{\sfer{1,2}} \hspace{-1mm}
        = \frac{1}{2} \big( M^2_{\sfer{LL}} + M^2_{\sfer{RR}} \big)
                  \mp \frac{1}{2}
                  \sqrt{ \big( M^2_{\sfer{LL}} - M^2_{\sfer{RR}} \big)^2
                    + 4 M^2_{\sfer{LR}} }
\eeq
with the eigenstates
\beq
 \left( \begin{array}{c} \sfer{1} \\ \sfer{2} \end{array} \right) &=&
 \left( \begin{array}{cc} \cosf & \sinf \\ -\sinf & \cosf \end{array}
 \right)
 \left( \begin{array}{c} \sfer{L} \\ \sfer{R} \end{array} \right) 
\label{sfer1} \\
 \cosf &=& \frac{- m_f \left( A_f - \mu \Theta(\beta) \right)}
                {\sqrt{ (M^2_{\sfer{LL}} - m^2_{\sfer{1}} )^2
                  + M^4_{\sfer{LR}} }}
\label{sfer2} \\
 \sinf &=& \frac{M^2_{\sfer{LL}} - m^2_{\sfer{1}} }
                {\sqrt{ (M^2_{\sfer{LL}} - m^2_{\sfer{1}} )^2
                  + M^4_{\sfer{LR}} }}.
\eeq
$\cosf$ 
is proportional to the fermion mass as can be read of \eqn{sfer2}. Therefore,
one can safely neglect the mixing for the first two generations but
not for the third generation. Especially for the stops one expects
a strong mixing and mass splitting between the heavier and the lighter
state due to the large top mass. For large $\tanbe$ ($\geq 10$) the
product $\mu \tanbe$ can be large leading to a strong mixing and
large mass splitting in the sbottom and in the stau sector too.
Note, that in this convention $\sinf$ is always positive because
\beq
M^2_{\sfer{LL}} - m^2_{\sfer{1}}
        = \frac{1}{2} \big( M^2_{\sfer{LL}} - M^2_{\sfer{RR}} \big)
                  + \frac{1}{2}
                  \sqrt{ \big( M^2_{\sfer{LL}} - M^2_{\sfer{RR}} \big)^2
                    + 4 M^2_{\sfer{LR}} } .
\eeq
A more detailed discussion on sfermion masses and mixing angles will be given
in \chap{chap:nlcmass}.

\section{Experimental bounds}
 
LEP1 has obtained a model independent experimental mass-bound on charged 
supersymmetric
particles which is $\tilde m \gsim 45$~GeV \cite{PDG96,Grivaz95}.
Stronger limits have been reported during the last two years from the
various LEP runs at higher energy \cite{LEP15,DELPHI96,OPAL96,ALEPH96,L396}. 
In the first half year LEP was running at $\sqrt{s} = 170-172$~GeV. There the
following limits have been obtained \cite{LEP170,Grivaz97}:
$\mstop{1} \gsim 73.3$~GeV, $\msbot{1} \gsim 73$~GeV, 
$\tilde{\mu}_R \gsim 59$~GeV, $\tilde{e}_R \gsim 58$~GeV, 
$\sta{R} \gsim{53}$~GeV, $m_{h^0} \gsim 69.5$~GeV, $\ma \gsim 62.5$~GeV, 
$m_{H^+} \gsim 51.5$~GeV,
$\mchip{1} \gsim 84.5 (67.6)$~GeV if $\mtsn = 1 (\sim 0.1)$~TeV.
Note, that these bounds depend on various assumptions.
The D0 experiment
at FNAL obtained additional mass limits for the stop \cite{Grivaz97,Abachi96a}
excluding the mass range 40~GeV $\lsim \mstop{1} \lsim 100$ GeV if the
mass difference $(\mstop{1} - \mchin{1}) \gsim 30$~GeV. The bound on the
gluino mass is $\mglu \gsim 175$~GeV \cite{Abe96a}.

%% file: decay.tex
\chapter{Decay Widths}
\label{chap:decay}

\section{General Remarks}

As already mentioned, in the MSSM exists a conserved quantum number
called R-parity leading to the existence of a lightest stable supersymmetric 
particle (LSP). 
We will assume that the lightest neutralino is the LSP
for cosmological reasons
(see e.g. \cite{Tata95} and references therein).

In general, sfermions decay according to $\tilde{f}_k \to
\tilde{\chi}^0_i + f$ and  $\tilde{f}_k \to\tilde{\chi}^{\pm}_j + f'$.
Normally, the flavour conserving decay into the lightest neutralino is always
possible except in case of the light stop because of the large $t$-quark mass.
Nevertheless, the decay of a light stop into the lighter chargino and a 
$b$-quark can still be accessible in this case. 
As we will show later there is a wide range of
SUSY parameters where all two body decays of the light stop are forbidden
at tree level. Therefore, the flavour changing decay into a $c$-quark
and the LSP becomes important \cite{Hikasa87}. Moreover, three body  decays
of the light stop can be competitive: $\sto{1} \to b + W^+ + \chin{1}, \,
b + H^+ + \chin{1}, \, b + \tilde{f}_{1,2} + \nu_f, \, 
b + f^+ + \tilde{\nu}_{f}$ where $f$ denotes $e,\mu,\tau$.

In the next section we collect the relevant terms of the interaction Lagrangian
which are needed for the computation of the decay widths. After this we
will list the formulae for the two body decays of the sfermions. In the
subsequent section we present the matrix elements, the Feynman diagrams
and the formulae for the above mentioned three body decays of the light stop.
For completeness we also present the formulae for the decay
$\sto{1} \to c + \chin{1}$ in our notation.

\section{The relevant parts of the Lagrangian}
\label{decaysect:lagrangian}

The part of the Lagrangian, which is needed for the calculation
of the subsequent decay widths, is given by:
\beq
{\cal L}_I \plgl g \sum_{ \stackrel{f=\nutau,t}{f' = \tau,b} }
   \left[ \bar{f} \left( \kcop{f'}{ij} P_L + \lcop{f'}{ij} P_R \right)
         \chip{j} \sfer{i}' +
         \bar{f'} \left( \kcop{f}{ij} P_L + \lcop{f}{ij} P_R \right)
         \chim{j} \sfer{i} + h.c. \right]  \no
 \plogl{-6} + \, g \sum_{f=\tau,\nutau,b,t} \left[ \bar{f}
         \left( b^{f}_{ki} P_L + a^{f}_{ki} P_R \right)
           {\tilde \chi_{i}^{0}} \tilde{f}_{k} + h.c. \right] \no
 \plogl{-6} - \, g \left[ W_{\mu }^{+} \overline{\chin{k}}
               \left( \col{kj} P_L + \cor{ki} P_R \right)
               \gamma^{\mu } \chip{j}  + h.c. \right] \no
 \plogl{-6} - \, g \left[ H^{+} \overline{\chin{k}}
               \left( \cqr{kj} P_L + \cql{ki} P_R \right)
               \chip{j}  + h.c. \right] \no
 \plogl{-6}  - \frac{g}{\wzw}
   \left[ W_{\mu }^{+} \bar{t} \gamma^{\mu } P_L b + h.c. \right] \no
 \plogl{-6} - \, g \left[ i W_{\mu }^{+} \left(
         \sum^2_{i,j=1,2} A^W_{\sto{i} \sbo{j}} \asto{i}
         \stackrel{\leftrightarrow }{ \partial_{\mu }} \sbo{j}
       + \sum^2_{i=1,2} A^W_{\snu{\tau} \sta{i}} \asnu{\tau}
         \stackrel{\leftrightarrow }{ \partial_{\mu }} \sta{i} \right)
          + h.c. \right] \no
 \plogl{-6} + \frac{g}{\wzw \mw}
    \left[ H^+ \bar{t} \left( m_b \tanbe P_R + m_t \cotbe P_L \right) b
       + h.c. \right] \no
 \plogl{-6} - \, g \left[ H^{+} \left(
         \sum^2_{i,j=1,2} C^H_{\sto{i} \sbo{j}} \asto{i} \, \sbo{j}
       + \sum^2_{i=1,2} C^H_{\snu{\tau} \sta{i}} \asnu{\tau} \, \sta{i}
         \right)  + h.c. \right] \no
 \plogl{-6} - \frac{i\,g}{2 \cosw} Z_{\mu}
   \bigg[ \cost \sint
   \left( \asto{1} \stackrel{\leftrightarrow }{ \partial_{\mu }} \sto{2}
   -  \asto{2} \stackrel{\leftrightarrow }{ \partial_{\mu }} \sto{1} \right)
   \no
  \plogl{25}
   - \sum_{f=\tau,b} \cosf \sinf
   \left( \asfer{1} \stackrel{\leftrightarrow }{ \partial_{\mu }} \sfer{2}
   - \asfer{2} \stackrel{\leftrightarrow }{ \partial_{\mu }} \sfer{1} \right)
      \bigg] \no
 \plogl{-6} - \, g \, h^0 \left( \sum_{f=\tau,b,t}
        B^{\sfer{}}_{h^0} \sfer{1} \asfer{2}  + h.c. \right)
  - \, g \, H^0 \left( \sum_{f=\tau,b,t}
        B^{\sfer{}}_{H^0} \sfer{1} \asfer{2}  + h.c. \right) \no
 \plogl{-6} + \, i \, g \, A^0 \left[ \sum_{f=\tau,b,t}
    B^{\sfer{}}_{A^0} \left(
    \sfer{1} \asfer{2} - \sfer{2} \asfer{1} \right) \right] \no
 \plogl{-6} - \wzw g_s T^a_{jk} \bigg[
   \sum_{q=b,t} \left( \bar{q}_j
    (\cosq P_R - \sinq P_L) \glu_a \tilde{q}_1^k \right. \no
 \plogl{25} \left. - \bar{q}_j (\sinq P_R + \cosq P_L) \glu_a \tilde{q}_2^k
     \right) + h.c. \bigg]
\label{eq:interaction}
\eeq
where $P_{R,L}= ( 1 \pm \gamma_5 ) / 2 $. The $T^a_{jk}$ are the
generators of $SU(3)$.
In the following we will use the abbreviation ${\cal{R}}^{\sfer{}}$ for
the mixing matrix (\eqn{sfer1}):
\beq
{\cal{R}}^{\sfer{}} =
 \left( \begin{array}{cc} \cosf & \sinf \\ -\sinf & \cosf \end{array}
 \right) .
\eeq
The Yukawa couplings of the sfermions are given by:
\beq
Y_{\tau} = \frac{m_{\tau}}{\wzw \mw \cosbe}, \hspace{5mm}
Y_{b} = \frac{m_{b}}{\wzw \mw \cosbe}, \hspace{5mm}
Y_{t} = \frac{m_{t}}{\wzw \mw \sinbe}.
\eeq
The $\squ{i}$-$q'$-$\chipm{j}$ couplings read
\beq
\lcop{q}{ij} = {\cal{R}}^{\squ{}}_{in} {\cal{O}}^{q}_{jn}, \hspace{5mm}
\kcop{q}{ij} = {\cal{R}}^{\squ{}}_{i1} {\cal{O}}^{q'}_{j2}
\label{eq:kte}
\eeq
with
\beq
 {\cal{O}}^{t}_{j} = \left( \begin{array}{c} -V_{j1}
                      \\ Y_t V_{j2} \end{array} \right), \hspace{5mm}
 {\cal{O}}^{b}_{j} = \left( \begin{array}{c} -U_{j1}
                      \\ Y_b U_{j2} \end{array} \right) .
\eeq
In case of sleptons we have:
\beq
\lcop{\tau}{ij} = {\cal{R}}^{\sta{}}_{in} {\cal{O}}^{\tau}_{jn}, \hspace{5mm}
\kcop{\tau}{ij} = 0, \hspace{5mm}
\lcop{\nu}{j} = - V_{j1}, \hspace{5mm} \kcop{\nu}{j} = Y_{\tau} U_{j2} .
\eeq
with
\beq
 {\cal{O}}^{\tau}_{j} = \left( \begin{array}{c} -U_{j1}
                      \\ Y_{\tau} U_{j2} \end{array} \right) .
\eeq
The $\sfer{i}$-$f$-$\chin{k}$ couplings are given by
\beq
\acop{f}{ik} = {\cal{R}}^{\sfer{}}_{in} {\cal{A}}^{f}_{kn}, \hspace{5mm}
\bcop{f}{ik} = {\cal{R}}^{\sfer{}}_{in} {\cal{B}}^{f}_{kn}
\eeq
with
\beq
 {\cal{A}}^{f}_{k} = \left( \begin{array}{c} f^f_{Lk}
                      \\ h^f_{Rk} \end{array} \right), \hspace{5mm}
 {\cal{B}}^{f}_{k} = \left( \begin{array}{c} h^f_{Lk}
                      \\ f^f_{Rk} \end{array} \right),
\eeq
and
\beq
\begin{array}{l}
h^t_{Lk} =  Y_t \big( \sinbe N^*_{k3} - \cosbe N^*_{k4} \big) \\
f^t_{Lk} = \frac{-2\wzw}{3} \sinw N_{k1} - \wzw \big( \einha -
         \frac{2}{3} \sinwq \big) \frac{N_{k2}}{\cosw} \\
h^t_{Rk} =  Y_t \big( \sinbe N_{k3} - \cosbe N_{k4} \big) \\
f^t_{Rk} =  \frac{-2\wzw}{3} \sinw \big( \tanw N^*_{k2} - N^*_{k1} \big)
\end{array}
\label{cop:ftl}
\eeq
\beq
\begin{array}{l}
h^b_{Lk} = - Y_b \big( \cosbe N^*_{k3} + \sinbe N^*_{k4} \big) \\
f^b_{Lk} = \frac{\wzw}{3} \sinw N_{k1} + \wzw \big( \einha -
         \frac{1}{3} \sinwq \big) \frac{N_{k2}}{\cosw} \\
h^b_{Rk} = - Y_b \big( \cosbe N_{k3} + \sinbe N_{k4} \big) \\
f^b_{Rk} =  \frac{\wzw}{3} \sinw \big( \tanw N^*_{k2} - N^*_{k1} \big)
\end{array}
\eeq
\beq
\begin{array}{l}
h^{\tau}_{Lk} = - Y_{\tau} \big( \cosbe N^*_{k3} + \sinbe N^*_{k4} \big) \\
f^{\tau}_{Lk} = \wzw \sinw N_{k1} + \wzw \big( \einha -
               \sinwq \big) \frac{N_{k2}}{\cosw} \\
h^{\tau}_{Rk} = - Y_{\tau} \big( \cosbe N_{k3} + \sinbe N_{k4} \big) \\
f^{\tau}_{Rk} =  \wzw \sinw \big( \tanw N^*_{k2} - N^*_{k1} \big)
\end{array}
\eeq
In case of $\snu{\tau}$ we get
$\acop{\nutau}{k} = - \frac{N_{k2}}{\wzw \cosw}$ and $\bcop{\nutau}{k} = 0$.
The couplings $\asto{i}$-$\sbo{j}$-$W^{+}$ read
\beq
A^W_{\sto{i}\sbo{j}} = (A^W_{\sbo{i}\sto{j}})^T
= \frac{1}{\wzw} \left( \begin{array}{rr}
 \cosb \cost & - \sinb \cost \\
 -\cosb \sint & \sinb \sint \end{array} \right)
\label{eq:coupstoWsbo}
\eeq
The couplings $\sta{i}$-$\snu{\tau}$-$W^{+}$ are given by
\beq
A^W_{\sta{1}\snu{\tau}} = \frac{\costa}{\wzw}, \hspace{5mm}
A^W_{\sta{2}\snu{\tau}} = - \frac{\sinta}{\wzw} .
\eeq
The couplings between light sfermion, heavy sfermion and neutral Higgs
read:
\beq
\hspace{-10mm}
\begin{array}{l}
B^{\sfer{}}_{h^0} = -
    \frac{m_{Z} \sinzf}{2 \, \cosw} ( \einha + 2 e_f \sinwq )
      \sinab - \frac{m_{f} \coszf}{2 \, m_{W}\cosbe}
       ( \mu \cosa + A_{f} \sina ) \\
B^{\sfer{}}_{H^0} =
     \frac{m_{Z} \sinzf}{2 \, \cosw} ( \einha + 2 e_f \sinwq )
      \cosab - \frac{m_{f} \coszf}{2 \, m_{W} \cosbe}
         ( \mu \sina - A_{f} \cosa ) \\
B^{\sfer{}}_{A^0} =
      - \frac{m_{f}}{2 \, m_{W} } ( A_{f} \tanbe +\mu )
\end{array}
\label{eq:coupHsf1sf2a}
\eeq
if $f = b,\tau$ and in case of $f = t$ we get
\beq
\hspace{-10mm}
\begin{array}{l}
B^{\sto{}}_{h^0} = - 
    \frac{m_{Z} \sinzt}{2 \, \cosw} ( \frac{4}{3} \sinwq - \frac{1}{2} )
              \sinab  + \frac{m_{t} \coszt}{2 \, m_{W} \sinbe}
        ( \mu \sina + A_{t} \cosa )  \\
B^{\sto{}}_{H^0} =  
   \frac{m_{Z} \sinzt}{2 \, \cosw} ( \frac{4}{3} \sinwq - \frac{1}{2} )
              \cosab  - \frac{m_{t} \coszt}{2 \, m_{W} \sinbe}
             ( \mu \cosa - A_{t} \sina )   \\
B^{\sto{}}_{A^0} = 
     - \frac{m_{t}}{2 \, m_{W} } ( A_{t} \cotbe +\mu )
\end{array}
\label{eq:couphsto2sto1}
\eeq
The couplings $\asto{i}$-$\sbo{j}$-$H^{+}$ are given by
\beq \hspace*{-6mm}
C^H_{\sto{i}\sbo{j}} \plgl (C^H_{\sbo{i}\sto{j}})^T \no
 \plogl{-12} =
\frac{1}{\wzw \mw} {\cal{R}}^{\sto{}}
\left( \begin{array}{cc}
   m^2_b \tanbe + m^2_t \cotbe - \mwq \sinzbe & m_b (A_b \tanbe + \mu) \\
   m_t (A_t \cotbe + \mu)  & 2 m_b m_t / \sinzbe
     \end{array}  \right)
\left(  {\cal{R}}^{\sbo{}} \right)^{\dagger} \no
\label{eq:coupstoHsbo}
\eeq
and the couplings $\sta{i}$-$\snu{\tau}$-$H^{+}$ are given by
\beq
C^H_{\sta{i}\snu{\tau}} = 
\frac{1}{\wzw \mw} {\cal{R}}^{\sta{}}
\left( \begin{array}{c}
   m^2_{\tau} \tanbe - \mwq \sinzbe \\ m_{\tau} (A_{\tau} \tanbe + \mu)
     \end{array}  \right)
\eeq
The $W^+$-$\chim{j}$-$\chin{k}$ couplings read:
\beq
 \col{kj} \plgl \frac{V_{j2}}{\wzw}
    \left( \sinbe N_{k3} - \cosbe N_{k4} \right)
  + V_{j1} \left( \sinw N_{k1} + \cosw N_{k2} \right) \\
 \cor{kj} \plgl \frac{U_{j2}}{\wzw}
    \left( \cosbe N_{k3} + \sinbe N_{k4} \right)
  + U_{j1} \left( \sinw N_{k1} + \cosw N_{k2} \right)
\eeq
The $H^+$-$\chim{j}$-$\chin{k}$ couplings are given by:
\beq
 \cql{kj} \plgl \cosbe \bigg[ V_{j1}
    \left( \cosbe N_{k4} - \sinbe N_{k3} \right) \no
 \plogl{10} + \frac{V_{j2}}{\wzw} \left( 2 \, \sinw N_{k1}
      + \left( \cosw  - \sinw \tanw \right) N_{k2} \right) \bigg] \\
 \cqr{kj} \plgl \sinbe \bigg[ U_{j1}
    \left( \cosbe N_{k3} + \sinbe N_{k4} \right) \no
 \plogl{12} - \frac{U_{j2}}{\wzw} \left( 2 \, \sinw N_{k1}
      + \left( \cosw  - \sinw \tanw \right) N_{k2} \right) \bigg]
\eeq

\section{Two body decays}
\label{decaysect:twobody}

The decay widths for the electroweak two body decays of the sfermions
$\sfer{i}$ are given by ($\sfer{i}$
denotes $\sto{1,2}$, $\sbo{1,2}$, $\sta{1,2}$ or $\snu{\tau}$):
\beq
\zerfallz{\sfer{i}}{f}{\chin{k}} \plgl
 \frac{ g^2 \lamh{\sfer{i}}{f}{\chin{k}}{\einha} }
      {16 \pi  m^3_{\sfer{i}} } \no
   \plogl{-15} * \left[ \Big( \acopq{f}{ik} + \bcopq{f}{ik} \Big)
        \Big( \msferq{i} - m^2_{f} - \mchinq{k} \Big)
    - 4 \acop{f}{ik} \bcop{f}{ik} m_f \mchin{k} \right]
\label{decaygl1}
\eeq
\beq
\zerfallz{\sfer{i}}{f'}{\chipm{k}} \plgl
 \frac{ g^2 \lamh{\sfer{i}}{f'}{\chipm{k}}{\einha} }
      {16 \pi  m^3_{\sfer{i}} } \no
  \plogl{-15}
     *  \left[ \Big( \kcopq{f}{ik} + \lcopq{f}{ik}
        \Big) \Big( \msferq{i} - m^2_{f'} - \mchipmq{k} \Big)
  - 4 \kcop{f}{ik} \lcop{f}{ik} m_{f'} \mchipm{k} \right]
\label{decaygl2}
\eeq
\beq
\zerfallz{\sfer{i}}{W^{\pm}}{\sferp{j}} =
   \frac{g^2 (A^W_{\sfer{i}\sferp{j}})^2
         \lamh{\sfer{i}}{W}{\msferp{j}}{\frac{3}{2}}}
        {16 \pi m^2_{W} m^3_{\sfer{i}} }
\label{decaygl3}
\eeq
\beq
\zerfallz{\sfer{i}}{H^{\pm}}{\sferp{j}} \plgl
   \frac{g^2 (C^H_{\sfer{i}\sferp{j}})^2
       \lamh{\sfer{i}}{H}{\sferp{j}}{\frac{1}{2}}}
        {16 \pi m^3_{\sfer{i}} }
\label{decaygl4}
\eeq
\beq
\zerfallz{\sfer{2}}{Z}{\sfer{1}} \plgl
   \frac{g^2 \sinzfq \lamh{\sfer{2}}{Z}{\sfer{1}}{\frac{3}{2}}}
        {256 \pi m^2_{W} m^3_{\sfer{2}}}
\label{decaygl5}
\eeq
\beq
\zerfallz{\sfer{2}}{H}{\sfer{1}} \plgl
   \frac{g^2 (B^{\sfer{}}_H)^2 \lamh{\sfer{2}}{H}{\sfer{1}}{\frac{1}{2}}}
        {16 \pi  m^3_{\sfer{2}}}
\label{decaygl6}
\eeq
with $\inwf{x}{y}{z} = x^2 + y^2 + z^2 - 2 x y - 2 x z - 2 y z$.
In \eqn{decaygl6} H denotes $h^0$, $H^0$ or $A^0$.
Squarks can also decay into a gluino and a quark:
\beq
\zerfallz{\squ{1}}{\glu}{q} \plgl
   \frac{2 \alphas \lamh{\squ{1}}{\glu}{q}{\frac{1}{2}}}
        {3 \, m^3_{\tilde{q}_1}}
       \left(\msquq{1} - \mgluq - m^2_q + 2 \sinzq \mglu m_q \right) \no
\label{decaygl7}
\eeq
\beq
\zerfallz{\squ{2}}{\glu}{q} \plgl
   \frac{2 \alphas \lamh{\squ{2}}{\glu}{q}{\frac{1}{2}}}
        {3 \, m^3_{\tilde{q}_2}}
       \left(\msquq{2} - \mgluq - m^2_q - 2 \sinzq \mglu m_q \right) \no
\label{decaygl8}
\eeq

%

\section{Three Body Decays of the Light Stop}
\label{sec:threebody}

\subsection{The Decay of the Light Stop into a W-Boson, a Bottom Quark and
         the Lightest Neutralino}

In \fig{fig:stbWchi} we show the Feynman diagrams for the decay $\stwbc$.
The matrix element for this decay is given by:
\beq
\hspace*{-7mm}
T_{fi} \plgl  - \frac{g^2}{\sqrt{2}} \sum^2_{i=1} A^W_{\sto{1} \sbo{i}}
     \frac{(p_{\tilde t} + \pbi )^{\mu }}
     {\pbi^{2} - \msbq{i} - i \msbot{i} \Gamma_{\tilde{b}_i} } \bar u(p_b)
     \left[ \bcop{b}{i1} P_L + \acop{b}{i1} P_R \right] v (\po )
     \epsilon_{\mu} (p_W ) \no
 \plogl{-5} + \, g^2  \sum^2_{i=1} \bar u(p_b)
    \left[ \lcop{t}{1i} P_R + \kcop{t}{1i} P_L \right]
   \frac{ \not \! \pxi - \mchipm{i} }
  { \pxi^{2} -\mchipmq{i}  - i \mchipm{i} \Gamma_{\tilde{\chi}^{\pm}_i}} \no
 \plogl{20} *  \left[ \col{1i} P_L + \cor{1i} P_R  \right]
    \gamma^{\mu} v(\po ) \epsilon_{\mu} (p_W) \no
 \plogl{-5} - \frac{g^2}{\sqrt{2}} \bar u (p_b ) \gamma^{\mu } P_L
   \frac{ \not \! p_t +m_t }{ p_{t}^{2} -m_{t}^{2}  - i m_t \Gamma_{t}}
   \left[ \bcop{t}{11} P_L + \acop{t}{11} P_R \right] v (\po )
   \epsilon_{\mu } (p_W )
\label{eq:TfistbWchi}
\end{eqnarray}

\noindent
The decay width is given by
\beq
 \Gamma(\stwbc) \plgl \no
 \plogl{-39} = \frac{\alpha^2}{16 \, \pi m^3_{\sto{1}} \sinwv}
        \eint{(\mstop{1}-m_W)^2}{(m_b + \mchin{1})^2}{s}
   \left( F_{\chip{} \chip{}} +
   F_{\chip{} t} +
   F_{\chip{} \sbo{}} +
   F_{t t} +
   F_{t \sbo{}} +
   F_{\sbo{} \sbo{}} \right). \no
\eeq
The explicit expressions for the $F_{ij}$ is given are \app{appA}.

\begin{figure}
 \unitlength 1mm
 \begin{picture}(135,175)
   \put(43,151){\large $\tilde{t}_1$}
   \put(106,142){\large $\tilde{\chi}^0_1$}
   \put(65,161){\large $W^+$}
   \put(75,151){\large $\tilde{b}_{1,2}$}
   \put(106,154){\large $b$}
   \put(43,93){\large $\tilde{t}_1$}
   \put(108,95){\large $W^+$}
   \put(107,82){\large $\tilde{\chi}^0_1$}
   \put(75,93){\large $\tilde{\chi}^-_{1,2}$}
   \put(66,98){\large $b$}
   \put(43,33){\large $\tilde{t}_1$}
   \put(65,42){\large $\tilde{\chi}^0_1$}
   \put(108,36){\large $W^+$}
   \put(77,32){\large $t$}
   \put(108,19){\large $b$}
\put(30,10){\psfig{file=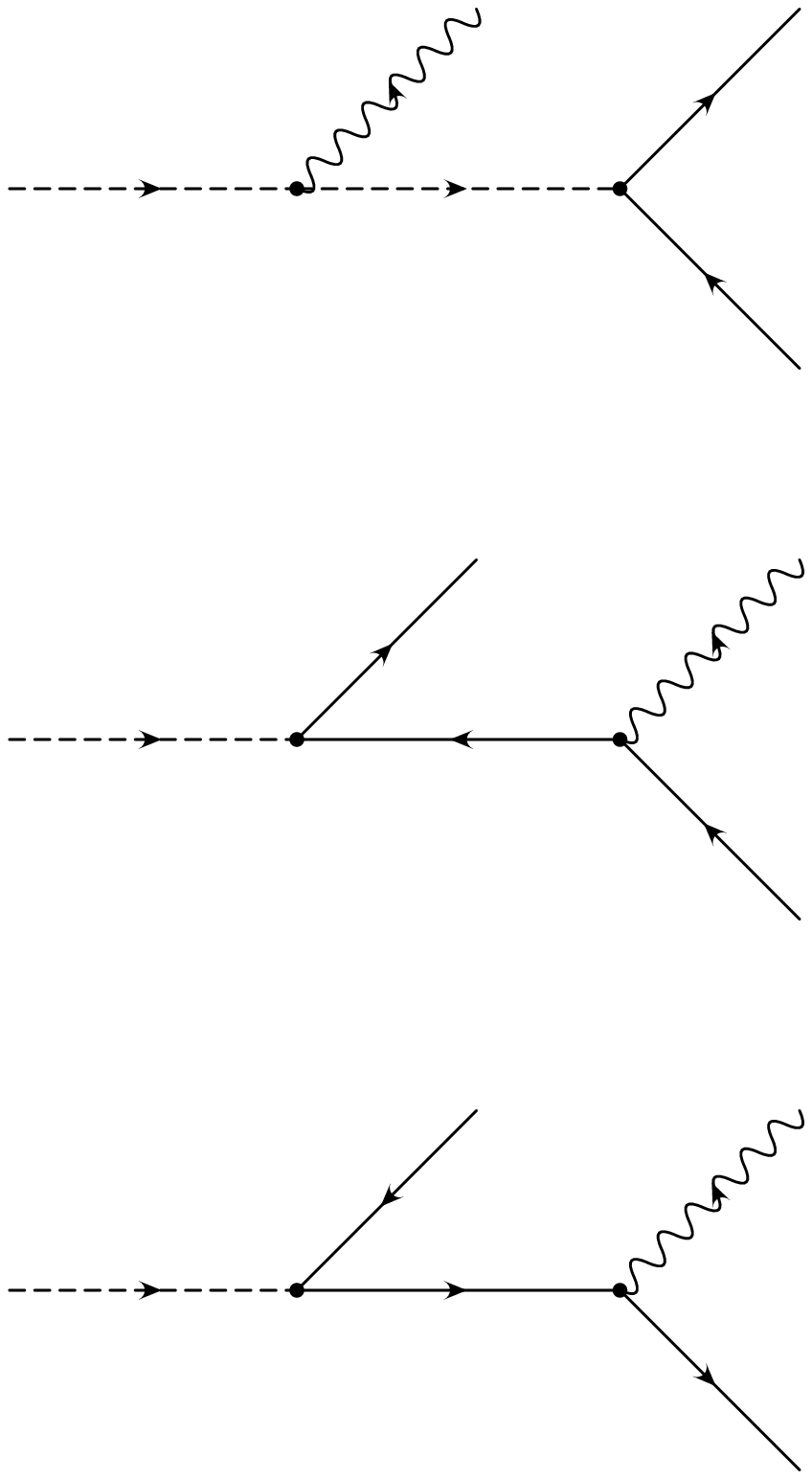,height=16cm}}
\end{picture}
\figcaption{fig:stbWchi}{Feynman  diagrams for the decay
         $\protect\sto{1} \to b \, W \, \protect\chin{1}$}
 {Feynman  diagrams for the decay $\stwbc$.
 The arrow of a fermionic line defines a fermion flow
 and is not necessarily identical with the momentum flow used in our
 calculations.}{0}
\end{figure}

\subsection{The Decay of the Light Stop into a charged Higgs Boson,
        a Bottom Quark and the Lightest Neutralino}

In \fig{fig:stbHgchi} we show the Feynman diagrams for the decay $\sthgbc$.
The matrix element for this decay is given by:
\beq
T_{fi} \plgl  - g^2 \sum^2_{i=1} C^H_{\sto{1} \sbo{i}}
     \frac{ \bar u(p_b)
     \left[ \bcop{b}{i1} P_L + \acop{b}{i1} P_R \right] v (\po )}
     {\pbi^{2} - \msbq{i} - i \msbot{i} \Gamma_{\tilde{b}_i} } \no
 \plogl{-10} - \, g^2  \sum^2_{i=1}
    \frac{ \bar u(p_b)
    \left[ \lcop{t}{1i} P_R + \kcop{t}{1i} P_L \right]
          [\not \! \pxi - \mchipm{i} ]
  \left[ \cql{1i} P_L + \cqr{1i} P_R  \right] v(\po )}
  { \pxi^{2} -\mchipmq{i}  - i \mchipm{i} \Gamma_{\tilde{\chi}^{\pm}_i}} \no
 \plogl{-10} + \frac{g^2} {\sqrt{2} \mw }
    \frac{ \bar u (p_b )
   \left[ m_b \tanbe P_L + m_t \cotbe P_R \right]
        [ \not \! p_t +m_t ]
   \left[ \bcop{t}{11} P_L + \acop{t}{11} P_R \right] v (\po ) }
       {p_{t}^{2} -m_{t}^{2}  - i m_t \Gamma_{t}} \no
\label{eq:TfistbHchi}
\end{eqnarray}
The decay width is given by
\beq
 \Gamma(\sthgbc) \plgl \no
 \plogl{-43} = \frac{\alpha^2}{16 \, \pi m^3_{\sto{1}} \sinwv}
        \eint{(\mstop{1}-\mhp)^2}{(m_b + \mchin{1})^2}{s}
   \left( G_{\chip{} \chip{}} +
   G_{\chip{} t} +
   G_{\chip{} \sbo{}} +
   G_{t t} +
   G_{t \sbo{}} +
   G_{\sbo{} \sbo{}} \right) \no
\eeq
The explicit expressions for the $G_{ij}$ are given in \app{appA}.

\begin{figure}
 \unitlength 1mm
 \begin{picture}(135,175)
   \put(43,153){\large $\tilde{t}_1$}
   \put(106,144){\large $\tilde{\chi}^0_1$}
   \put(65,161){\large $H^+$}
   \put(76,153){\large $\tilde{b}_{1,2}$}
   \put(106,155){\large $b$}
   \put(43,93){\large $\tilde{t}_1$}
   \put(108,98){\large $H^+$}
   \put(107,83){\large $\tilde{\chi}^0_1$}
   \put(75,94){\large $\tilde{\chi}^-_{1,2}$}
   \put(68,101){\large $b$}
   \put(43,33){\large $\tilde{t}_1$}
   \put(65,42){\large $\tilde{\chi}^0_1$}
   \put(108,37){\large $H^+$}
   \put(77,32){\large $t$}
   \put(108,20){\large $b$}
\put(30,10){\psfig{file=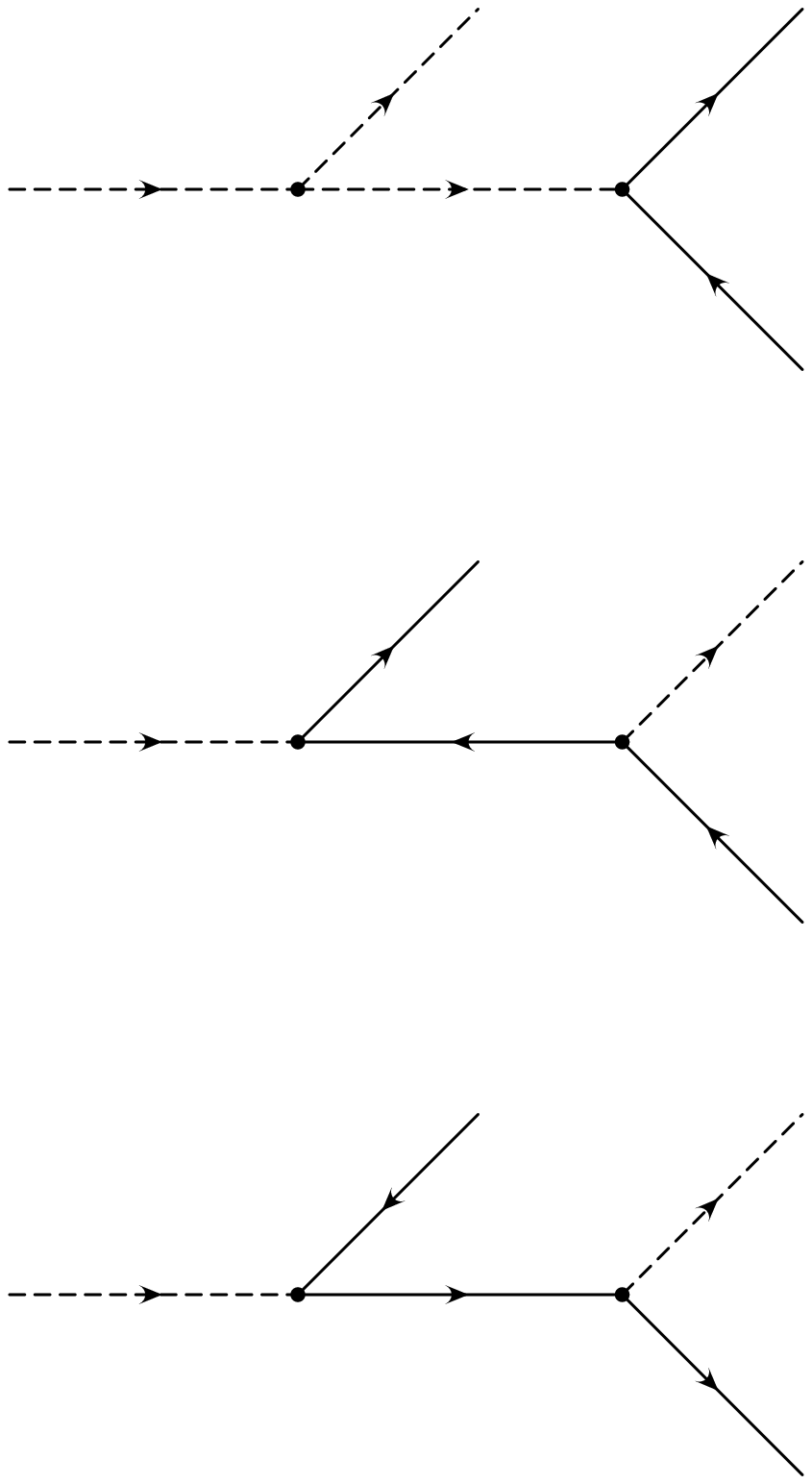,height=16cm}}
\end{picture}
\figcaption{fig:stbHgchi}{Feynman  diagrams for the decay
         $\protect\sto{1} \to b \, \protect \hp \, \protect\chin{1}$}
 {Feynman  diagrams for the decay $\sthgbc$.
 The arrow of a fermionic line defines a fermion flow
 and is not necessarily identical with the momentum flow used in our
 calculations.}{0}
\end{figure}

\subsection{The Decay of the Light Stop into a Bottom Quark, a Slepton and
         a Lepton}

Here we have the following possibilities:
\begin{itemize}
 \item $\sto{1} \to b + \tilde{\nu}_e + e^+,
                \,  b + \tilde{\nu}_\mu + \mu^+$
 \item $\sto{1} \to b + \tilde{e}^+_L + \nu_{e},
         \, b + \tilde{\mu}^+_L + \nu_{\mu}$
 \item $\sto{1} \to b + \tsn + \tau^+$
 \item $\sto{1} \to b + \sta{1,2}^+ + \nu_{\tau}$
\end{itemize}

\newpage
\noindent
Note, that the decays into $\tilde{e}_R$ and $\tilde{\mu}_R$ are negligible
because the couplings to the charginos are proportional to $m_e / \mw$
and $\mu / \mw$ respectively.
In the case of decays into sneutrinos and leptons the matrix elements have
the generic form:
\beq
T_{fi} \plgl  g^2  \sum^2_{i=1} 
 \frac{ \bar u(p_b)
    \left[ \lcop{t}{1i} P_R + \kcop{t}{1i} P_L \right]
        \left[ \not \! \pxi - \mchipm{i} \right]
 \left[ \lcop{f}{ki} P_R + \kcop{f}{ki} P_L  \right] v(p_{f'} )}
  { \pxi^{2} -\mchipmq{i}  - i \mchipm{i} \Gamma_{\tilde{\chi}^{\pm}_i}}, \no
\label{eq:Tfistbsneut}
\end{eqnarray}
whereas for the decays into sleptons and neutrinos we get:
\beq
T_{fi} \plgl  g^2  \sum^2_{i=1} 
 \frac{ \bar u(p_b)
    \left[ \lcop{t}{1i} P_R + \kcop{t}{1i} P_L \right]
        \left[ \not \! \pxi - \mchipm{i} \right]
 \left[ \lcop{f}{ki} P_L \right] v(p_{f'} )}
  { \pxi^{2} -\mchipmq{i}  - i \mchipm{i} \Gamma_{\tilde{\chi}^{\pm}_i}}, \no
\label{eq:Tfistbslept}
\eeq
In both cases the decay width is given by
\beq
 \Gamma(\sto{1} \to b + \tilde{l} + l') \plgl \no
 \plogl{-35} = \frac{\alpha^2}{16 \, \pi m^3_{\sto{1}} \sinwv}
        \eint{(\mstop{1}-m_b)^2}{(m_{l'} + m_{\tilde{l}})^2}{s}
        W_{l' \tilde{l}}(s)
        \sum^3_{i=1} \left( \sum^5_{j=1} c_{ij} s^{(j-4)} \right) D_i (s)
\eeq
with
\beq
 D_{1,2} (s) \plgl
  \frac{1}{(s-\mchipmq{1,2})^2 + \mchipmq{1,2} \Gamma^2_{\chipm{1,2}}} \\
 D_3 (s) \plgl Re \left(
  \frac{1}{(s-\mchipmq{1} + i \mchipm{1} \Gamma_{\chipm{1}})
            (s-\mchipmq{2} - i \mchipm{2} \Gamma_{\chipm{2}}) }  \right).
\eeq
The explicit expressions for the $W_{l' \tilde{l}}$ and $c_{ij}$ are given in 
\app{appA}.

\section{The Decay of the light Stop into a Charm-quark and a
         Neutralino}
\label{sec:stchic}

For completeness we will also rewrite the results of
\cite{Hikasa87} for the two body decay $\sto{1} \to c + \chin{1}$.
They found that the decay is
dominated by top--charm squark mixing, which is induced at one loop level.
In the limit $m_c \rightarrow 0$ only the left charm squark
contributes to this mixing.
In the basis of \eqn{sfer1} the respective 
$\tilde t_1 - \tilde t_2 - \tilde c_L$ mixing matrix is given by
\beq
\hspace*{-10mm}
{\cal M}^{2}_{ \tilde t_1\tilde t_2  \tilde c_L } \plgl \no
\plogl{-10} \left(
\begin{array}{ccc}
m^{2}_{ \tilde t_1 } & 0 & \Delta_L \cos \theta_t + \Delta_R
\sin \theta_t \\
0 & m^{2}_{ \tilde t_2 } & - \Delta_L \sin \theta_t + \Delta_R \cos
\theta_t \\
\Delta_{L}^{\ast } \cos \theta_t + \Delta_{R}^{\ast} \sin \theta_t &
-\Delta_{L}^{\ast} \sin \theta_t + \Delta_{R}^{\ast} \cos \theta_t &
m^{2}_{ \tilde c_L}
\label{fcnmat}
\end{array} \right) \no
\eeq
$\Delta_L$ ($\Delta_R$) are $\tilde t_L - \tilde c_L $ 
($\tilde t_R - \tilde c_L $) mixing terms with
\begin{eqnarray}
\Delta_L & = & - \frac{g^2}{16 \pi^2} \ln \left(\frac{M_X^2}{m_W^2} \right)
\frac{K^{\ast}_{tb} K_{cb} m_b^2 }{2 m_W^2 \cos^2 \beta }
( M_Q^2 + M_D^2 + M_{H_1}^{2} + A_b^2 )
\label{deltal} \\[0.2cm]
\Delta_R & = & \frac{g^2}{16 \pi^2} \ln \left(\frac{M_X^2}{m_W^2}
\right)
\frac{K^{\ast}_{tb} K_{cb} m_b^2 }{2 m_W^2 \cos^2 \beta } m_t A_b
\label{deltar}
\end{eqnarray}
where $M_X$ is a high scale which we assume to be the Planck mass to get a
maximal mixing. $ M_Q ,\, M_D $ and
$ M_{H_1} $ are soft SUSY breaking squark and Higgs mass terms
and $K_{tb} $ and $  K_{cb} $ are the respective elements of the
CKM matrix.

One gets \eqn{deltal} and (\ref{deltar}) as one step solutions in
$\ln(M^2_P / M^2_W)$ of the renormalization group equation in the
framework of supergravity theories.  Note, that one should stay away
from $A_b = 0$ because in this case higher order terms in $\ln(M^2_P / M^2_W)$
become important for $\Delta_R$. Note, also that in this
approximation $M_D$, $M_Q$ and $M_{H_1}$  can be evaluated at any scale
because the induced error is of higher order. Therefore, the expressions
should be treated as rough estimates giving the order of magnitude for
the mixing. For that reason we will use the 
formula for the decay width in \eqn{eq:stchic} mainly as a check if this
decay clearly dominates or if it is negligible.

In the following
$\epsilon $ gives the size of the charm squark component of the lighter
stop, which we calculated numerically.
Therefore, in this decay mode the charm-squark component
of the lighter stop couples with the charm quark and the LSP $\tilde
\chi_1^0 $ and the width is given by
\begin{equation}
\label{eq:stchic}
\Gamma (\tilde t_1 \rightarrow c \tilde \chi_1^0 ) =
\frac{g^2}{16 \pi} \epsilon^2 |f^{c}_{L1}|^2
m_{\tilde t_1 } \left( 1- \frac{m_{\tilde \chi_1^0}^{2}}{m_{\tilde t_1 }^{2}}
\right)^2
\end{equation}
where $f^c_{L1} = f^t_{L1}$ as given in \eqn{cop:ftl}.
Before discussing our results in detail we give the conditions 
leading to a large $\epsilon$:
($i$) $\mstop{1}$ and $m_{\tilde c_L}$
have almost the same size, $(ii)$ $\tanbe$ becomes large ($\cosbe$
small) which will enhance $\Delta_L$ and $\Delta_R$ $(iii)$
$\tant \sim \Delta_L / \Delta_R$
which will maximize the ${\cal M}^2_{13}$ and ${\cal M}^2_{31}$
components of the mixing matrix
${\cal M}^2_{\sto{1} \sto{2} \tilde{c}_L}$ (\eqn{fcnmat})
and $(iv)$ the parameters $M_D,\,M_Q,\,M_{H_1}$ and $A_b$ entering
$\Delta_L$ and $\Delta_R$ are big.

%% file: lep2.tex
\chapter{Numerical results for LEP2}
\label{chap:lep2}

\section{Light Stop}
\label{sect:stoplep2}

\begin{minipage}{72mm}
The total cross sections for the process $\ee \sto{1} \asto{1}$ at
$\sqrt{s} = 200$~GeV are shown in \fig{fig:pstoplep} as a function of
$|\cos\theta_{\sto{}}|$ for several mass values of $\sto{1}$. Here ISR- and
SUSY-QCD-corrections are included \cite{Drees90,Eberl96}. For completeness the
formulae are given in \app{appC}. The main corrections are 
due to ISR- and gluonic QCD-corrections \cite{Bartl97a}. For
$\mstop{1} = 80$~GeV the cross section reaches 0.41~pb.
Therefore, one can expect $\sim$ 60 to 123 $\sto{1} \asto{1}$ events assuming
an integrated luminosity of 300 $\rm pb^{-1}$. Moreover, the cross section
shows a clear dependence on the mixing angle for $\mstop{1} \lsim 80$~GeV and
$|\cos\theta_{\sto{}}| \gsim 0.6$. In this region cross section measurements
should therefore allow to determine the stop mixing angle once $\mstop{1}$
is known.

\hspace*{4mm}
Assuming $\mstop{1} < m_{{\tilde l},{\tilde \nu}}$ the main decay modes are
$\sto{1} \to c \, \chin{1}$ and $\sto{1} \to b \, \chip{1}$. As 
\end{minipage}
\hspace{4mm}
\begin{minipage}{73mm}
{\setlength{\unitlength}{1mm}
\begin{picture}(73,77)
\put(0,1){\mbox{\psfig{figure=pstoplep.eps,height=6.9cm}}}
\put(3,71.5){\makebox(0,0)[bl]{{\small $\sigma$~[pb]}}}
\put(69.5, 0.3){\makebox(0,0)[br]{{\small $|\cos \theta_{\sto{}}|$}}}
\end{picture}}
\figcaption{fig:pstoplep}
           {Production cross sections for a light stop at LEP2}
           {Total cross section for
            $e^+ e^- \to \sto{1} \bar{\sto{1}}$ in pb at
            $\sqrt{s} = 200$~GeV as a function of $|\cos\theta_{\sto{}}|$
            for $\sto{1}$ masses 50, 60, 70, 80 and 90~GeV. ISR- and
            QCD-corrections are included.
           }{7}
\end{minipage}

\noindent
already
mentioned in \chap{chap:decay} the first decay is a FCNC decay occurring at
one loop level whereas the second one occurs
at tree level. Therefore, $\sto{1} \to b \, \chip{1}$ dominates
with practically 100~\% branching ratio if it is kinematically allowed.
As $\chip{1}$ further decays into $\chin{1} l^+ \nu_l$ or
$\chin{1} q \bar{q}'$ the signature is two  acoplanar $b$ jets accompanied
\noindent
\begin{minipage}[b]{150mm}
{\setlength{\unitlength}{1mm}
\begin{picture}(150,77)                        
\put(0,75){{\small \bf a)}}
\put(0,2.5){\mbox{\epsfig{figure=sto15lep.eps,height=6.9cm}}}
\put(70.0,1){\makebox(0,0)[br]{{\small $\mu$~[GeV]}}}
\put(5,72){\makebox(0,0)[bl]{{\small $M$~[GeV]}}}
\put(12,58.5){{\small $m_{\tilde t_1} < m_{\tilde \chi^0_1}$}}
\put(50,58.5){{\small $m_{\tilde t_1} < m_{\tilde \chi^0_1}$}}
\put(20,35){{\small a}}
\put(60.5,40){{\small a}}
\put(32.7,24){{\small b}}
\put(63.5,33.7){{\small b}}
\put(66,34.1){\vector(0,-1){4}}
\put(20.7,25.8){{\small c}}
\put(22.5,26.2){\vector(1,-1){3.9}}
\put(60.5,32.3){{\small c}}
\put(63.0,33.1){\vector(0,-1){3.7}}
\put(78.5,2.5){\mbox{\epsfig{figure=sto40lep.eps,height=6.9cm}}}
\put(78.5,75){{\small \bf b)}}
\put(147.5,1){\makebox(0,0)[br]{{\small $\mu$~[GeV]}}}
\put(82.5,72){\makebox(0,0)[bl]{{\small $M$~[GeV]}}}
\put(89.5,58.5){{\small $m_{\tilde t_1} < m_{\tilde \chi^0_1}$}}
\put(125.5,58.5){{\small $m_{\tilde t_1} < m_{\tilde \chi^0_1}$}}
\put(97.5,36.5){{\small a}}
\put(132.5,36.5){{\small a}}
\put(92.5,31.4){{\small b}}
\put(95.5,31.7){\vector(0,-1){3.7}}
\put(137.5,31.4){{\small b}}
\put(140.5,31.7){\vector(0,-1){3.7}}
\put(97.5,23.2){{\small c}}
\put(100,23.0){\vector(0,1){4}}
\put(132.5,23.5){{\small c}}
\put(135,23.3){\vector(0,1){4}}
\end{picture}}
\figcaption{fig:mmustolep}
           {Parameter domains in the $(M,\mu)$ plane for
            $\protect \sto{1}$ decays at LEP2}
           {Kinematically allowed parameter domains in the $(M,\,\mu)$
            plane for $\mgev{\tilde t_1}{90}$, a) $\tan \beta = 1.5$ and  b)
            $\tan \beta = 40$ for the decays:
            a) $\tilde t_1 \to c \, \tilde \chi^0_1$,
            b) $\tilde t_1 \to c \, \tilde \chi^0_2$,
            c) $\tilde t_1 \to b \, \tilde \chi^+_1$.
            The dark grey area is excluded by LEP1 and the bright grey area
            is excluded by LEP1.5 ($\sqrt{s} = 170$~GeV, $\msnu{e} = 90$~GeV).
           }{7}
\end{minipage}

\noindent
by two charged leptons
+ large missing energy ($\me$), or a single charged 
lepton
+ jets + $\me$, or jets + $\me$. Here $b$ tagging techniques can be
used to extract 
the signal. Moreover, in this case it is most likely that
$\chip{1}$ will be observed first  and its decay properties can be
used for identifying $\sto{1}$. If $\mstop{1} < m_b + \mchip{1}$ the
decay $\sto{1} \to c \, \chin{1}$ has practically 100~\% branching ratio.
The signature is then two acoplanar jets + $\me$. Generally, in this case
the invisibly energy will be larger than in the case of
$\sto{1} \to b \, \chip{1}$.
In \fig{fig:mmustolep}~a (b) we show the domains of the stop decay modes in
the ($M,\mu$) plane for $\mgev{\sto{1}}{90}$ and $\tanbe = 1.5$~(40).
Note, that there is a small strip where the decay $\sto{1} \to c \, \chin{2}$
is also possible.

If the lifetime of $\sto{1}$ is longer than the typical hadronization time of
${\cal O} (10^{-23} s)$, i.e. $\Gamma_{\sto{1}} \lsim 0.2$~GeV,
it will first hadronize into a colourless
($\sto{1} \bar{q}$) or ($\sto{1} q q$) bound state before decaying.
The process of $\sto{1}$ fragmentation was discussed in detail in
\cite{Beenaker95b}. Fast moving stops first radiate off gluons at small
angles. This process can be treated perturbatively. After that the
non-perturbative hadronization phase follows leading to
$(\sto{1} \bar{q})$ and $(\sto{1} q q)$ hadrons. If the velocity
$\beta_{\sto{}}$ is $\sim 1/2$ the energy loss of the stop due to gluon
radiation and due to hadronization is of comparable size. Near the
threshold, the gluon emission is suppressed by $\beta^4_{\sto{}}$.

Hadronization is generally expected in case of $\sto{1} \to c \, \chin{1}$,
$\sto{1} \to b \, l^+ \, \tilde{\nu}_l$ and
$\sto{1} \to b \, \nu \, \tilde{l}^+$
since these decays involve the electroweak coupling twice
(see \fig{fig:stoplep15g3}). However,
also in case of $\sto{1} \to b \, \chip{1}$ this can happen as
illustrated in \fig{fig:stolep15d2} and \ref{fig:stolep40d2}.

\begin{table}[ht]
\begin{tabular}{|c|c|c|c|c|c|c|c|}
\hline 
Scenario & $\mu$ & $M_2$ & $\tanbe$ & $\mchin{1}$ & $\mchin{2}$ &
$\mchipm{1}$ & $\mchipm{2}$ \thline \hline
pt1 & -500.0 &  70.0 & 1.5 & 37.3 & 80.7 & 80.0 & 511.2 \thline
pt2 & -138.0 &  55.0 & 1.5 & 31.5 & 82.1 & 80.0 & 168.7 \thline
pt3 & -64.9  & 300.0 & 1.5 & 63.5 & 88.6 & 80.0 & 317.2 \thline
pt4 &  230.6 & 120.0 & 1.5 & 43.3 & 89.7 & 80.0 & 272.0 \thline
pt5 & -500.0 &  81.4 & 40  & 40.4 & 80.0 & 80.0 & 512.9 \thline
pt6 & -138.0 & 110.0 & 40  & 48.4 & 82.4 & 80.0 & 193.8 \thline
pt7 & -84.8  & 300.0 & 40  & 64.7 & 98.0 & 80.0 & 321.9 \thline
pt8 &  133.9 & 120.0 & 40  & 49.4 & 85.8 & 80.0 & 196.8 \thline
pt9 & -250.0 &  75.0 & 1.5 & 41.1 & 93.0  & 91.6  & 269.3 \thline
pt10 & -67.0 &  90.0 & 1.5 & 50.5 & 64.2  & 91.5  & 130.4 \thline
\end{tabular} \\[0.3cm]
\tabcaption{tab:scenlep2}{Parameters for different scenarios at LEP2}
{Parameters used in the the different scenarios for LEP2.
In addition, we give the respective values for the masses (in GeV) of the two
lighter neutralinos and the charginos. The other parameters are given in
the text.
}
\end{table}

\noindent
\begin{minipage}[t]{72mm}
{\setlength{\unitlength}{1mm}
\begin{picture}(72,77)
\put(3,1){\mbox{\psfig{figure=sto15lepg2.eps,height=6.9cm}}}
\put(3,70.0){\makebox(0,0)[bl]{{\small $\Gamma_{\sto{1}}$~[GeV]}}}
\put(69.5,1.0){\makebox(0,0)[br]{{\small $\cos \theta_{\sto{}}$}}}
\end{picture}}
\figcaption{fig:stolep15d2}
           {Total decay widths of a light stop at LEP2 for $\tanbe =1.5$}
           {Total decay width of the light stop as a function of $\cost$
            for the scenarios pt1 (solid line), pt2 (dashed line), pt3
            (dashed dotted line)  and pt4 (long dashed line) of
            \tab{tab:scenlep2}. ($\tanbe = 1.5$)}{4}
\end{minipage}
\hspace{3mm}
\noindent
\begin{minipage}[t]{72mm}
{\setlength{\unitlength}{1mm}
\begin{picture}(72,77)
\put(3,1){\mbox{\psfig{figure=sto40lepg2.eps,height=6.9cm}}}
\put(3,70.0){\makebox(0,0)[bl]{{\small $\Gamma_{\sto{1}}$~[GeV]}}}
\put(69.5,1.0){\makebox(0,0)[br]{{\small $\cos \theta_{\sto{}}$}}}
\end{picture}}
\figcaption{fig:stolep40d2}
           {Total decay widths of a light stop at LEP2 for $\tanbe =40$}
           {Total decay width of the light stop as a function of $\cost$
            for the scenarios pt5 (solid line), pt6 (dashed line), pt7
            (dashed dotted line)  and pt8 (long dashed line) of
            \tab{tab:scenlep2}. ($\tanbe = 40$)}{4}
\end{minipage}

\newpage

\noindent 
In \fig{fig:stolep15d2} (\ref{fig:stolep40d2}) we show the width of
$\sto{1} \to b \, \chip{1}$ as a function of $\cost$ for $\mgev{\sto{1}}{90}$,
$\mchip{1} \simeq 80$~GeV, $\tanbe = 1.5$ (40), and four 
different sets of $M$ and $\mu$ as given in \tab{tab:scenlep2}. 
In addition we present there the masses of the two lighter neutralinos and of 
both charginos.
In case of pt1 (pt5) the chargino is mainly gaugino-like ($M \ll \mu$),
in case of pt3 (pt7) the chargino is mainly higgsino-like ($M \gg \mu$),
whereas in the other cases we have a strongly mixed chargino ($M \sim \mu$).
In the last case we have taken two different points in the ($M,\mu$) plane
with a different sign for $\mu$.

For small $\tanbe$ and gaugino-like charginos (pt1 in \fig{fig:stolep15d2})
the $\sto{1}-b-\chip{1}$
interaction is dominated by $V_{11} \cost$. The deviation of the decay width
from the $\costq$ shape is due to constructive and destructive interference
with the term proportional to the top Yukawa coupling ($Y_t V_{12} \sint$)
which becomes especially important for $\sto{1} \sim \sto{R}$ and
increases with decreasing $|\mu|$.
In the case of pt3 this part of the coupling leads to a $\sintq$ shape.
If the gaugino- and higgsino-components of the chargino are of comparable
size (pt2 and pt4) a more complicated interplay of the gaugino and higgsino
couplings give rise to an intricate dependence on the mixing angle and
a large asymmetry in the sign of $\mu$.

In \fig{fig:stolep40d2} we show the total decay widths for the large $\tanbe$
(=40) scenarios pt5, pt6, pt7 and pt8 of \tab{tab:scenlep2}.
Here the term $Y_b U_{12} \cost$
becomes important if the chargino has a sizable higgsino-component.
This leads to an enhancement of the decay width and to a shift of
the maxima compared to the case $\tanbe = 1.5$ (\fig{fig:stolep15d2}).

Let us now turn to the case 
that three body decays of $\sto{1}$ into sleptons are
kinematically possible and that $\mchip{1} > \mstop{1}$.
In this case the decay chain will be either
$\sto{1} \to b \, l^+ \, \tilde{\nu}_l \to b \, l^+ \, \nu_l \, \chin{1}$
or
$\sto{1} \to b \, \nu_l \, \tilde{l}^+_{1,2}
         \to b \, l^+ \, \nu_l \, \chin{1}$.
These decays compete with the decay $\sto{1} \to c \, \chin{1}$. The
signature is 2 $b$ jets + 2 charged leptons + $\me$ or
1 $b$ jet + 1 $c$ jet + 1 charged lepton + $\me$ or 2 $c$ jets + $\me$.
If the three body decays dominate, one may ask if there is a chance to
distinguish between the different sleptons. It should be no problem to
identify the generation by identifying the nature of  the charged lepton.
To answer the question whether the final state occurs through sneutrinos and/or 
charged sleptons requires most likely the study of the
angular distribution of the $b$-jets, the lepton and the angle between the
jet and the lepton.

In the following examples we will assume that the parameters
$M_{\tilde E}$ and $M_{\tilde L}$ are the same for every generation.
For small $\tanbe$ this is justified by RGE studies. The modifications for
a high $\tanbe$ scenario will be discussed later. Before going into
detail we want to remark that: (i) The charged sleptons are expected to 
be heavier than the sneutrinos because of the  $D$-term contributions.
Therefore, the discovery
of those decay modes is not only an evidence for a light stop but at the same
time most likely the first evidence of light sneutrinos. 
The sneutrino decays invisibly into $\nu_l \, \chin{1}$. Therefore, it is 
difficult to detect. (ii) Even for small $\tanbe$ the influence of the
$\tau$ Yukawa coupling and stau mixing
can lead to sizable effects. (iii) In case that the 
\noindent
\begin{minipage}[t]{149mm}
{\setlength{\unitlength}{1mm}
\begin{picture}(155,76)                        
\put(3,4){\mbox{\epsfig{figure=sto15lepbr3p9.eps,height=6.6cm}}}
\put(0,72){{\small \bf a)}}
\put(5,71){\makebox(0,0)[bl]{{\small $BR(\sto{1})$}}}
\put(69.5,1){\makebox(0,0)[br]{{\small $\cost$}}}
\put(80.5,4){\mbox{\epsfig{figure=sto15lepbr3p10.eps,height=6.6cm}}}
\put(77.5,72){{\small \bf b)}}
\put(82.5,71){\makebox(0,0)[bl]{{\small $BR(\sto{1})$}}}
\put(147,1){\makebox(0,0)[br]{{\small $\cost$}}}
\end{picture}}
\figcaption{fig:stoplep15br3}
           {Branching ratios for stop decays into sleptons at LEP2}           
           {Branching ratios for stop decays into sleptons for the scenarios
            pt9 (a) and pt10 (b) of \tab{tab:scenlep2}.
            The other parameters are given in the text. The curves correspond
            to the following decays:
            \rechtl \hspace{1mm}$\sto{1} \to b \, \nutau \, \sta{1}$,
            $ \bullet \hspace{1mm} \sto{1} \to b \, e^+ \, \tilde{\nu}_e
                      \hspace{1mm} (b \, \mu^+ \, \tilde{\nu}_\mu)$,
            and \recht $\sto{1} \to b \, \tau^+ \, \tsn$.
            }{7}
\end{minipage}

\noindent
chargino is mainly
higgsino-like there is only a small mass difference between the chargino and
the lightest neutralino. Therefore, the charged particles will most likely
have too little energy to pass the experimental cuts \cite{Giudice96}. 
We will therefore
discuss examples where the lighter chargino is either mainly a gaugino
(pt9 of \tab{tab:scenlep2}) or where it is highly mixed 
(pt10 of \tab{tab:scenlep2}).
(iv) The existence of this decay and the fact that the stop should be
heavier than the lightest neutralino implies an upper bound on the
lighter chargino which is approximately given by
$\mchip{1} \lsim 2 \, \mstop{1}$. Here it is assumed that $\chin{1}$
is the LSP and that the GUT relation \ref{eq:gutone} between the gaugino 
masses $M'$ and $M$ is valid.

In \fig{fig:stoplep15br3} we show the branching ratios for
the three body decays for the scenarios pt9 and pt10 of \tab{tab:scenlep2}
and for $\mstop{1}= 80$~GeV.
Both cases are low $\tanbe$ scenarios ($\tanbe = 1.5$).
We have used the following (flavor independent) parameters in the slepton
sector: $M_{\tilde E} = 59.8$~GeV, $M_{\tilde L} = 72.1$~GeV and
$A_{\tau} = 500$~GeV (774.5GeV) in case of scenario pt9 (pt10). They lead
to the following physical quantities: $m_{\tilde{\nu}_i} = 60$~GeV 
(i = $e,\mu,\tau$), 
$m_{\tilde{e}_L} = 77.9$~GeV, $m_{\tilde{e}_R} = 65.7$~GeV,
$\mstau{1} = 58.4$, $\mstau{2} = 83.5$ and $\costa = -0.5$. Therefore, the
decays into all sneutrinos and the lighter stau are possible.
Note, that also the decays into $\tilde{e}_R$ and $\tilde{\mu}_R$ are
kinematically allowed. However,
as already mentioned in \chap{chap:decay}, these decays are negligible because
the coupling of $\chip{i}$-$\tilde{e}_R$-$\nu_e$
($\chip{i}$-$\tilde{\mu}_R$-$\nu_\mu$) is proportional $m_e$ ($m_\mu$).
We expect that the numbers of produced electrons and muons are nearly
equal and that they are smaller than the number of produced $\tau$ leptons.

\noindent
\begin{minipage}[t]{73mm}
{\setlength{\unitlength}{1mm}
\begin{picture}(80,75)
\put(-1.0,-1.7){\mbox{\psfig{figure=sto15lepg3.eps,width=7.2cm,height=7.6cm}}}
\put(0,71){\makebox(0,0)[bl]{{\small $\Gamma_{\sto{1}}$~[GeV]}}}
\put(73.5, 0.8){\makebox(0,0)[br]{{\small $\cost$}}}
\end{picture}}
\figcaption{fig:stoplep15g3}
           {Total decay widths of a light stop at LEP2 (three body decays)}
           {Total decay widths of a light stop for the scenarios pt9
            (full line) and pt10 (dashed line) as given in \tab{tab:scenlep2}.
            The other parameters are given in the text.
           }{6}
\end{minipage}
\hspace{4mm}
\begin{minipage}[t]{73mm}
{\setlength{\unitlength}{1mm}
\begin{picture}(80,75)
\put(-1,3.2){\mbox{\psfig{figure=sto15lepbr3p9c5.eps,height=6.6cm,width=7.2cm}}}
\put(2,71){\makebox(0,0)[bl]{{\small $BR(\sto{1} \to b \, \nutau \, \sta{1})$}}}
\put(73.5, 0.8){\makebox(0,0)[br]{{\small $\cost$}}}
\end{picture}}
\figcaption{fig:sto15lepbr3p9c5}
           {Branching ratios of the stop decay into stau at LEP2}
           {Branching ratio for the decay
            $\sto{1} \to b \, \nutau \, \sta{1}$ as a function of $\cost$ for
            $M = 75$~GeV, $\mu = -250$~GeV and $\tanbe = 1.5$
            (pt9 of \tab{tab:scenlep2}). The masses of the sleptons are:
            $m_{\tilde{\nu}_i} = 60$~GeV (i = $e,\mu,\tau$),
            $\mstau{1} = 58.4$ and $\costa =$ 0 (long dashed line),
            $-0.24$ (short long dashed line), 0.39 (short dashed line), $-0.5$
            (full line) and $-0.68$ (middle long dashed line).
            The other parameters are given in the text.
           }{6}
\end{minipage}

\noindent

In \fig{fig:stoplep15br3}a the situation for scenario pt9 is shown.
The decays into sneutrinos dominate over
the decays into $\sta{1}$ except in the region $0 \lsim \cost \lsim 0.25$.
This behaviour can be understood by combining the following facts: 
i) The sleptons couple mainly to $\chip{1}$ because $\tanbe$ is small. 
ii) $\costa = -0.5$ and therefore $\sta{1} \simeq \sta{R}$.
iii) The $\sto{1}$-$\chip{1,2}$-$b$ couplings are dominated by $\lte{1,12}$.
iv) Near $\cost = 0.25$ the coupling $\lte{1}$ vanishes whereas $\lte{2}$
is near its maximum. Therefore, the exchange of the heavier chargino
becomes important. Note, that $\chip{2}$ couples stronger to $\sta{1}$
than to $\tsn$ in this example. Moreover, the interference
terms have opposite signs: in case of the stau (sneutrino) it is positive 
(negative)
for $\cost \lsim 0.25$ and negative (positive) for $\cost \gsim 0.25$.

Let us now turn to scenario pt10 where we have two main differences
compared to the above example: (i) the heavier chargino has rather
large gaugino components ($U_{12} = 0.997$ and $V_{12} = -0.651$).
(ii) The mass difference between the charginos is much smaller compared to the 
previous example (see \tab{tab:scenlep2}). The decay into the stau proceeds
mainly through $\chipm{2}$ exchange whereas the decays into the 
sneutrinos proceed mainly through $\chip{1}$ exchange. Looking
at the $\sto{1}$-$\chip{1,2}$-$b$ couplings one sees that
$|\lte{1}| < |\lte{2}|$ ($|\lte{1}| > |\lte{2}|$) if $\cost \lsim 0.1$
($\cost \gsim 0.1$). $\lte{1}$ ($\lte{2}$) vanishes near $ \cost = - 0.8$
($\cost =0.8$). Moreover, the interference term is negative for all
decay modes if $|\cost| \lsim 0.8$. The combination of all facts leads to
the pronounced maximum of $BR(\sto{1} \to b \, \nutau \, \sta{1})$ near
$\cost = -0.7$. Because of the large asymmetry it should be possible to 
determine 
the sign of $\cost$ if $BR(\sto{1} \to b \, \nutau \, \sta{1})$ can be
measured.

In \fig{fig:stoplep15g3} we show the total decay width for both examples.
One can easily see that the decay width has a minimum near the above
mentioned $\cost$ values where $\lte{1}$ vanishes. In case of scenario pt9 
(pt10) the
decay width varies over 2 (1) orders of magnitude. This difference in the
order is due to the different mass of the heavier chargino. In both cases
the stop will hadronize before decaying.

The branching ratios depend not only on the kinematics but also on the stau 
mixing angle.
To get a feeling for this
dependence let us first look at the possible range for $\costa$. As
can be seen from the formulae given in \app{appB} we can fix $\mu$ and $\tanbe$,
vary $A_\tau$ and calculate $\costa$. Keeping $\mstau{1}$ and
$\mtsn$ fixed one can first calculate $\mstau{2}$ (see \eqn{eq:determl2}) and
then $M_{\tilde E}$ (\eqn{eq:determl1}) to get a feeling if the parameters
are reasonable. One gets $0.39 > \costa > -0.68$, 77.9~GeV $< \mstau{2} <$
91.1~GeV and 64.3~GeV $< M_{\tilde E} <$ 79.9~GeV if $A_\tau$ is varied
between $- 1$~TeV and 1 TeV in
scenario pt9. In \fig{fig:sto15lepbr3p9c5} we show the branching ratio
for the decay into a stau for the following 
($A_\tau$, $M_{\tilde E}$, $\costa$, $\mstau{2}$) sets:
(-1000, 55.97, 0.39, 80.8), (-375, 51.69, 0, 77.9), (0, 53.27, -0.24, 78.9),
(500, 59.8, -0.5, 83.5) and (1000, 70.04, -0.68, 91.1). 
The masses are given in GeV. A large ratio
$A_\tau / M_{\tilde E}$ implies the possible danger of a charge breaking
minimum. However, to our knowledge there is no sufficient way to
treat this problem. Therefore, we will leave it aside for the moment. For a 
discussion of this topic see e.g. \cite{Gunion88a} and references therein. 
One can easily see that the importance of the decay 
$\sto{1} \to b \, \nutau \, \sta{1}$ 
grows with $|\costa|$ because this
implies a stronger coupling to the lighter chargino. We have checked
that the dependence on $\costa$ is similar in scenario pt10.
An interesting detail is that even for $\costa = 0$ there is small
range of $\cost$ where  $BR(\sto{1} \to b \, \nutau \, \sta{1})$ is of $O(0.1)$
because of the Yukawa coupling to the heavier chargino.

In the case that a large $\tanbe$ scenario is realized in nature, it is
expected that the lighter stau is the lightest slepton. In the energy range
of LEP2 the most likely scenario would be that the stau is lighter than
the stop and the stop lighter than the other sleptons. This
can be seen in the following way: let us first assume
$M_{\tilde E} = M_{\tilde L}$ for simplicity. The mass of the lighter
stau is then given by:
\beq
\hspace*{-9mm}
\mstaq{1} \plogl{-10} = 
           M^2_{\tilde L} + m^2_\tau - \einvi \mzq \coszbe
 - \sqrt{\einvi m^4_Z \cos^2 2 \beta (1/2 - 2 \sinw)^2
        + (m_\tau \mu \tanbe)^2} \no
  &\sim& M^2_{\tilde L} + m^2_\tau - \einvi \mzq \coszbe
 - m_\tau |\mu| \tanbe.
\eeq
The last relation holds because $1/2 - 2 \sinw \sim 0.04$.
The mass of the tau sneutrino is given by
\beq
\mtsnq = M^2_{\tilde L} + \einha \mzq \coszbe.
\eeq
Therefore, $\sta{1}$ is lighter than $\tsn$ if
\beq
\label{eq:lepsmassdiff}
 - m_\tau |\mu| \tanbe
< \textstyle \frac{3}{4} \mzq \coszbe < 0
\eeq
There is only a small parameter range in the large $\tanbe$ regime where 
$\chip{1}$ is mainly a higgsino and at the same time 
$\mchin{1} < min(\mstau{1},mtsn)$.
Therefore, $|\mu|$ is larger than $\mchipm{1}$
implying that relation \ref{eq:lepsmassdiff} will be fulfilled in most cases.
Now we should remember that RGE studies
\cite{Drees95b,Barger94a} indicate
that $M_{\tilde E}$ and $M_{\tilde L}$ are smaller for the third generation
than for the other two and  that $M_{\tilde E} < M_{\tilde L}$. For these
reasons even the tau sneutrino is expected to be heavier than the lighter
stau in such a 
scenario. This finishes our chain of hints for the above assumption.
Therefore, the two competing modes are
$\sto{1} \to c \, \chin{1}$ and
$\sto{1} \to b \, \nutau \, \sta{1} \to b \, \nutau \, \tau \, \chin{1}$.
As mentioned in \sect{sec:stchic} we use the formula for the decay-width
$\Gamma(\sto{1} \to c \, \chin{1})$ only as indication of the order of 
magnitude. Therefore, we show no figures for the large $\tanbe$ scenarios.

Let us shortly comment on the  possibility of 
$\mchin{1} < min(\mstau{1},\mtsn)$.
This, for example is realized in gauge mediated SUSY models 
(see e.g. \cite{GMSB1}
and references therein) if $\tanbe$ is large. In this class of models, the
gravitino $\tilde{G}$
is the LSP and the lighter stau is the next heavier SUSY particle.
In such a scenario the lighter stop can decay in the following way:
$\sto{1} \to  b \, \nutau \, \sta{1} \to b \, \nutau \, \tau \, \tilde{G}$.
This leads to the same signature as in the previous case.

Monte Carlo studies for $\sto{1}$ pair production have been performed
within the CERN-LEP2 Workshop 1995 \cite{Giudice96}. They have mainly
concentrated on the decay $\sto{1} \to c \, \chin{1}$ since in this
case the $\sto{1}$ would most likely be the first SUSY particle to be
discovered. For the simulation of $\sto{1}$ hadronization different
approaches have been used. The conclusions have been a $5 \sigma$
discovery reach for $\mstop{1} \simeq 75$ to 90 GeV and a 95\% confidence
level exclusion of $\mstop{1} = 84$ to 92 GeV at $\sqrt{s} = 190$~GeV,
for ${\cal L} = 300 {\rm pb^{-1}}$ depending on $\cost$ and $\mchin{1}$.
In case of $\sto{1} \to b \, \chip{1}$ the experimental reach for
$\mstop{1}$ is $\sim 85$~GeV. Here $b$-tagging is important for identifying
the stop. For the case of dominance of the three body decays the limits on
$\mstop{1}$ given in \cite{OPAL96} are about 5~GeV lower than in the case
where the main decay mode is $\sto{1} \to c \, \chin{1}$.

\section{Light Sbottom}
\label{sec:lep2sbottom}

A considerable $\sbo{L}$-$\sbo{R}$ mixing is possible if $\tanbe$
is large (see \eqn{sfer2}). In this case $\sbo{1}$ can be rather
light. Therefore, it is interesting to discuss the phenomenology of $\sbo{1}$
at LEP2.

The total cross sections of $\sbo{1}$ pair production at $\sqrt{s} =200$~GeV
are shown in \fig{fig:psbotlep} as a function of $|\cosb|$ for \hfill several 
\hfill mass \hfill
\hfill values \hfill of \hfill $\sbo{1}$ \hfill including \hfill ISR- \hfill and

\noindent
\begin{minipage}[t]{72mm}  
{\setlength{\unitlength}{1mm}
\begin{picture}(75,75)
\put(0,1){\mbox{\psfig{figure=psbotlep.eps,height=6.9cm}}}
\put(3,71.5){\makebox(0,0)[bl]{{\small $\sigma$~[pb]}}}
\put(69.5, 0.3){\makebox(0,0)[br]{{\small $|\cos \theta_{\sbo{}}|$}}}
\end{picture}}
\figcaption{fig:psbotlep}
           {Production cross sections for a light sbottom at LEP2}
           {Total cross section for
            $e^+ e^- \to \sbo{1} \asbo{1}$ in pb at
            $\sqrt{s} = 200$~GeV as a function of $|\cos\theta_{\sbo{}}|$
            for $\sbo{1}$ masses 50, 60, 70, 80 and 90~GeV. ISR- and
            QCD-corrections are included.
           }{7}
\end{minipage}
\hspace{4mm}
\noindent
\begin{minipage}[t]{72mm}
{\setlength{\unitlength}{1mm}
\begin{picture}(75,75)                        
\put(0.0,1.5){\mbox{\epsfig{figure=sbo40lep.eps,height=6.9cm}}}
\put(69.5,1){\makebox(0,0)[br]{{\small $\mu$~[GeV]}}}
\put(0.0,71.5){\makebox(0,0)[bl]{{\small $M$~[GeV]}}}
\put(12.0,58.5){{\small $m_{\tilde b_1} < m_{\tilde \chi^0_1}$}}
\put(47.0,58.5){{\small $m_{\tilde b_1} < m_{\tilde \chi^0_1}$}}
\put(20.0,35.0){{\small a}}
\put(57.0,35.0){{\small a}}
\put(20.0,22.0){{\small b}}
\put(22.5,22.5){\vector(0,1){4}}
\put(57.0,21.8){{\small b}}
\put(59.5,21.8){\vector(0,1){4}}
\put(27.2,35.5){{\small c}}
\put(27.2,35.0){\vector(1,0){4.0}}
\put(48.0,35.5){{\small c}}
\put(50.0,35.0){\vector(-1,0){3.7}}
\end{picture}}
\figcaption{fig:mmusbolep}
           {Parameter domains in the $(M,\mu)$ plane for
            $\protect \sbo{1}$ decays at LEP2}
           {Kinematically allowed parameter domains in the $(M,\,\mu)$
            plane for $\mgev{\tilde b_1}{90}$ and $\tan \beta = 40$ for
            the decays: a) $\tilde b_1 \to b \, \tilde \chi^0_1$,
            b) $\tilde b_1 \to b \, \tilde \chi^0_2$ and
            c) $\tilde b_1 \to c \, \tilde \chi^-_1$.
            The dark grey area is excluded by LEP1 and the bright grey area
            is excluded by LEP1.5 ($\sqrt{s} = 170$~GeV, $\msnu{e} = 90$~GeV).
           }{7}
\end{minipage}

\noindent
QCD-corrections. If
$|\cosb| \gsim 0.6$ the dependence on the mixing angle is even more pronounced
than in the case of $\sto{1}$ production. The production cross section of
$\ee \sbo{1} \asbo{1}$ is smaller than in the case of $\sto{1}$
production by a factor $\sim 5/4$ to 5. For
a sbottom mass of 80~GeV
the cross section varies between 0.05 and 0.34 pb. Therefore, 15 to 100
events per year are expected for an integrated luminosity of
$300 \, \, {\rm pb^{-1}}$.

The main decay modes of $\sbo{1}$ are $\sbo{1} \to b \, \chin{1}$ and
$\sbo{1} \to b \, \chin{2}$ if $\msbot{1} < \mglu$. The domains of the
$\sbo{1}$ decays in the ($M,\mu$) plane are shown in \fig{fig:mmusbolep}
for $\msbot{1} = 90$~GeV and $\tanbe = 40$.
If only $\sbo{1} \to b \, \chin{1}$ is kinematically allowed, the
signature is two $b$-jets plus $\me$. In order to distinguish the sbottom from
a stop $b$-tagging can be  necessary because the decay 
$\sto{1} \to c \, \chin{1}$ 
leads to a similar signature. If in addition, the decay into
the second lightest neutralino is possible, there can be additional
charged leptons and/or jets stemming from
$\chin{2} \to \chin{1} \, q \, \bar{q}$
and $\chin{2} \to \chin{1} \, l^+ \, l^-$. In any case $b$-tagging will 
enhance the signal. In principal the 
flavour changing decay
$\sbo{1} \to c \, \chim{1}$ is also possible 
(region b and c in \fig{fig:mmusbolep}).
Assuming that the mixing between
different squark generations is of the same order as in the quark sector,
it turns 
\noindent
\begin{minipage}[t]{72mm}  
{\setlength{\unitlength}{1mm}
\begin{picture}(75,77)
\put(3,4){\mbox{\epsfig{figure=sbo40lepg.eps,height=6.6cm}}}
\put(70.5,1){\makebox(0,0)[br]{{\small $\cosb$}}}
\put(2,73){\makebox(0,0)[bl]{{\small $\Gamma_{\sbo{1}}$~[GeV]}}}
\end{picture}}
\figcaption{fig:sbolep40g}
           {Decay widths of a light sbottom at LEP2 for $\tanbe =40$}
           {Decay width of the light sbottom as a function of $\cosb$
            for the scenarios pt5 (solid line), pt6 (dashed line), and pt7
            (dashed dotted line) of \tab{tab:scenlep2}.}{7}
\end{minipage}
\hspace{5mm}
\noindent
\begin{minipage}[t]{72mm}
{\setlength{\unitlength}{1mm}
\begin{picture}(75,77)                        
\put(3,4){\mbox{\epsfig{figure=sbo40lepbr.eps,height=6.6cm}}}
\put(70.5,1){\makebox(0,0)[br]{{\small $\cosb$}}}
\put(2,72){\makebox(0,0)[bl]{{\small $BR(\sbo{1} \to \chin{1} + b)$}}}
\end{picture}}
\figcaption{fig:sbolep40br}
           {Branching ratios for $\protect\sbo{1} \to b \, \protect\chin{1}$
            at LEP2 for $\tanbe =40$}
           {Branching ratio for $\sbo{1} \to b \, \chin{1}$ as a function
            of $\cosb$ for the scenarios pt5 (solid line), pt6 (dashed line)
            and pt7 (dashed dotted line) of \tab{tab:scenlep2}.}{7}
\end{minipage}

\noindent
out that this decay mode is negligible. Nevertheless, this decay can 
be important if one studies the flavour structure of the squark
sector. Moreover, this mode gains 
some importance for larger $\msbot{1}$ as will be shown in 
\sect{sect:rarsbot}.

In \fig{fig:sbolep40g} we show the total decay width of $\sbo{1}$ for
the scenarios pt5, pt6 and pt7 of \tab{tab:scenlep2} and for 
$\msbot{1} = 90$~GeV. Scenario pt8 gives nearly
the same result as scenario pt6 for equal $|\cosb|$.
Here we show only
the dependence on negative values of $\cosb$ because for large $\tanbe$ the 
sign of $\cosb$ is determined by the sign of $\mu$ 
(\eqn{sfer2}) if one wants to avoid unnaturally large $A_b$.
As can be seen, the total decay width is smaller than 0.2~GeV if
the neutralinos are mainly gaugino-like (pt5). Therefore, hadronization becomes
important. Its effects are the same as in the case of $\sto{1}$ because
QCD is flavour blind. If $\chin{1}$ and $\chin{2}$ have a sizable 
higgsino-component
(pt6 and pt7) the total decay widths increase  because of the large
bottom Yukawa coupling although the neutralino masses
are larger when compared to the previous case.

The branching ratio for $\sbo{1} \to b \, \chin{1}$ is shown in
\fig{fig:sbolep40br} as a function of $\cosb$ for the above examples.
If the neutralinos are mainly gauginos the branching ratio depends strongly
on $\cosb$ (pt5). This is because $\sbo{L}$ has a strong coupling
to the zino components whereas $\sbo{R}$ couples
only to the bino components. Due to the assumption
$M' = 5/3 \tanwq M$ the lightest neutralino is 
mainly a bino and the
second lightest neutralino is mainly a zino for the parameter choices of
pt5. Evidently, 
\noindent
\begin{minipage}[t]{72mm}  
{\setlength{\unitlength}{1mm}
\begin{picture}(75,75)
\put(0,1){\mbox{\psfig{figure=pstaulep.eps,height=6.9cm}}}
\put(3,71.5){\makebox(0,0)[bl]{{\small $\sigma$~[pb]}}}
\put(69.5, 0.3){\makebox(0,0)[br]{{\small $|\cos \theta_{\sta{}}|$}}}
\end{picture}}
\figcaption{fig:pstaulep}
           {Production cross sections for a light stau at LEP2}
           {Total cross section for
            $e^+ e^- \to \sta{1} \asta{1}$ in pb at
            $\sqrt{s} = 200$~GeV as a function of $|\cos\theta_{\sta{}}|$
            for $\sta{1}$ masses 50, 60, 70, 80 and 90~GeV. ISR-corrections
            are included.
           }{7}
\end{minipage}
\hspace{5mm}
\noindent
\begin{minipage}[t]{72mm}
{\setlength{\unitlength}{1mm}
\begin{picture}(75,75)                        
\put(1.0,2){\mbox{\epsfig{figure=sta40lep.eps,height=6.9cm}}}
\put(70.5,1){\makebox(0,0)[br]{{\small $\mu$~[GeV]}}}
\put(0.0,72){\makebox(0,0)[bl]{{\small $M$~[GeV]}}}
\put(12.0,58.5){{\small $m_{\tilde \tau_1} < m_{\tilde \chi^0_1}$}}
\put(49.0,58.5){{\small $m_{\tilde \tau_1} < m_{\tilde \chi^0_1}$}}
\put(20.0,35.5){{\small a}}
\put(57.0,35.5){{\small a}}
\put(20.0,22.5){{\small b}}
\put(22.5,23.0){\vector(0,1){4.5}}
\put(57.0,23.0){{\small b}}
\put(59.5,23.0){\vector(0,1){4.5}}
\put(28.0,36.5){{\small c}}
\put(28.0,36.0){\vector(1,0){4.5}}
\put(49.0,36.5){{\small c}}
\put(51.0,36.0){\vector(-1,0){4.2}}
\end{picture}}
\figcaption{fig:mmustalep}
           {Parameter domains in the $(M,\mu)$ plane for
            $\protect \sta{1}$ decays at LEP2}
           {Kinematically allowed parameter domains in the $(M,\,\mu)$
            $\mgev{\sta{1}}{90}$ and $\tan \beta = 40$ for the decays:
            a) $\tilde \tau_1 \to \tau \, \tilde \chi^0_1$,
            b) $\tilde \tau_1 \to \tau \, \tilde \chi^0_2$ and
            c) $\tilde \tau_1 \to \nu_\tau \, \tilde \chi^-_1$.
            The dark grey area is excluded by LEP1 and the bright grey area
            is excluded by LEP1.5 ($\sqrt{s} = 170$~GeV, $\msnu{e} = 90$~GeV).
           }{7}
\end{minipage}

\noindent
the dependence on the mixing angle will decrease if the mass
of the second lightest 
neutralino increases.
The increase of $BR(\sbo{1} \to b \, \chin{1})$ in case of the mixed scenario
pt6 is an effect of the larger higgsino components of $\chin{1,2}$.

DELPHI has studied $\sbo{1}$ search for the case that only the decay
$\sbo{1} \to b \, \chin{1}$ is possible \cite{Giudice96}. Their conclusion
was that the discovery potential for $\sbo{1}$ is similar to
$\sto{1} \to c \, \chin{1}$ ($\msbot{1} \lsim 75$ to 90 GeV depending on
$\cosb$ and $\mchin{1}$).

\section{Light Stau}

There are two reasons that at least one of the staus could be within the mass
range which will be explored by LEP2: Firstly, the soft SUSY breaking mass terms
for sleptons are expected to be smaller than the corresponding ones for
squarks. Secondly, as in the sbottom sector one expects a large
$\sta{L}$-$\sta{R}$ mixing for large $\tanbe$.

The cross sections for stau pair production are plotted in \fig{fig:pstaulep}
as a function of $|\costa|$ for $\sqrt{s} = 200$~GeV. Here we have included
ISR-corrections. As can be seen, the
dependence on the mixing angle is much weaker for stau production than
in case of squark production. The cross section is of ${\cal O}(0.15)$~pb
for $\mstau{1} = 80$~GeV 
\noindent
\begin{minipage}[t]{150mm}
{\setlength{\unitlength}{1mm}
\begin{picture}(155,130)
\put(-4,-16){\mbox{\epsfig{figure=sta40lepbr.eps,height=16.0cm,width=15.6cm}}}
\put(0,127){{\small \bf a)}}
\put(4,126){\makebox(0,0)[bl]{{\small $BR(\sta{1})$}}}
\put(77,128){{\small \bf b)}}
\put(81,127){\makebox(0,0)[bl]{{\small $BR(\sta{1})$}}}
\put(0,62.5){{\small \bf c)}}
\put(4,61.5){\makebox(0,0)[bl]{{\small $BR(\sta{1})$}}}
\put(77,64){{\small \bf d)}}
\put(81,63){\makebox(0,0)[bl]{{\small $BR(\sta{1})$}}}
\put(71,1){\makebox(0,0)[br]{{\small $\costa$}}}
\put(71,65){\makebox(0,0)[br]{{\small $\costa$}}}
\put(149,1){\makebox(0,0)[br]{{\small $\costa$}}}
\put(149,65){\makebox(0,0)[br]{{\small $\costa$}}}
\end{picture}}
\figcaption{fig:stalep40d}
           {Branching ratios for $\protect\sta{1}$ decays at LEP2 for
             $\protect \tanbe =40$}
           {Branching ratios for the decays of the lighter stau as a function of
            $\costa$ for the scenarios pt5 (a), pt6 (b), pt7 (c) and pt8 (d)
            of  \tab{tab:scenlep2}, $\mstau{1} = 90$~GeV.
            The graphs correspond to the following decays:
            $ \circ \hspace{1mm} \zerfal{\sta{1}}{\chin{1}}{\tau}$,
            \rechtl \hspace{1mm}$ \zerfal{\sta{1}}{\chin{2}}{\tau}$,
            and \recht $\zerfal{\sta{1}}{\chim{1}}{\nutau}$.
           }{7}
\end{minipage}

\noindent
corresponding to $\sim 45$ events for an integrated
luminosity of ${\cal L} = 300$ ${\rm pb^{-1}}$.

We get the classical SUSY signature consisting of 2 $\tau$-leptons + $\me$
if only the decay $\sta{1} \to \tau \, \chin{1}$ is kinematically allowed.
The decays $\sta{1} \to \tau \, \chin{2}$ and/or
$\sta{1} \to \nutau \, \chim{1}$ 
lead to additional jets and/or leptons. The parameter 
domains of the various
stau decays in the $(M, \mu)$ plane are shown in \fig{fig:mmustalep} for
$\mstau{1} = 90$~GeV and $\tanbe = 40$.

In \fig{fig:stalep40d} we show the branching ratios of $\sta{1}$ decays into
$\tau \, \chin{1}$, $\tau \, \chin{2}$, and $\nutau \, \chim{1}$ as a
function of $\costa$ for $\mstau{1} = 90$~GeV and $\tanbe = 40$ for
the scenarios pt5, pt6, pt7, and pt8 of \tab{tab:scenlep2}. As in the case of
$\sbo{1}$ decays we have taken the same sign for $\costa$ and $\mu$ in order to
avoid unnatural large values for $A_{\tau}$. 
The branching ratio $BR(\sta{1} \to \tau \, \chin{1})$ is always larger 
than 50\% and it
is nearly 100\% 
\noindent
\begin{minipage}[t]{72mm}
{\setlength{\unitlength}{1mm}
\begin{picture}(75,71)
\put(0.5,2.0){\mbox{\psfig{figure=psneulep.eps,height=6.8cm}}}
\put(0,71.0){\makebox(0,0)[bl]{{\small $\sigma$~[pb]}}}
\put(70.5,0.7){\makebox(0,0)[br]{{\small $m_{\tilde{\nu}_{\tau}}~[GeV]$}}}
\end{picture}}
\figcaption{fig:psneulep}
           {Production cross sections for $\protect\tsn$ at LEP2}
           {Total cross section for  $e^+ e^- \to \tsn \bar{\tsn}$ in pb at
            $\sqrt{s} = 200$~GeV as a function of $\mtsn$. ISR-corrections
            are included.
           }{7}
\end{minipage}
\hspace{5mm}
\noindent
\begin{minipage}[t]{72mm}
{\setlength{\unitlength}{1mm}
\begin{picture}(75,76)                        
\put(1,2){\mbox{\epsfig{figure=sne15lep.eps,height=6.9cm}}}
\put(70.5,1){\makebox(0,0)[br]{{\small $\mu$~[GeV]}}}
\put(2,71.5){\makebox(0,0)[bl]{{\small $M$~[GeV]}}}
\put(12,59.5){{\small $m_{\tilde \nu_{\tau}} < m_{\tilde \chi^0_1}$}}
\put(50,59.5){{\small $m_{\tilde \nu_{\tau}} < m_{\tilde \chi^0_1}$}}
\put(20,35){{\small a}}
\put(60.5,40){{\small a}}
\put(29.0,25.0){{\small b}}
\put(29.5,24.6){\vector(1,0){4.8}}
\put(60.5,32){{\small b}}
\put(63.0,33.2){\vector(0,-1){3.8}}
\put(24.5,25.5){{\small c}}
\put(26.5,25.4){\vector(0,-1){3.3}}
\put(51.9,39.1){{\small c}}
\put(51,40.0){\vector(0,-1){4}}
\end{picture}}
\figcaption{fig:mmutsnlep}
           {Parameter domains in the $(M,\mu)$ plane for
            $\protect \tsn$ decays at LEP2}
           {Kinematically allowed parameter domains in the $(M,\,\mu)$
            $\mgev{\tilde \nu_{\tau}}{90}$ and $\tan \beta = 1.5$ for the
            decays: a) $\tilde \nu_{\tau} \to \nu_{\tau} \, \tilde \chi^0_1$,
            b) $\tilde \nu_{\tau} \to \nu_{\tau} \, \tilde \chi^0_2$ and
            c) $\tilde \nu_{\tau} \to \tau \, \tilde \chi^+_1$.
            The dark grey area is excluded by LEP1 and the bright grey area
            is excluded by LEP1.5 
            ($\sqrt{s} = 170$~GeV, $m_{\snu{e}} = 90$~GeV).
           }{7}
\end{minipage}

\noindent
for $\costa \sim 0$ if the neutralinos
and the lighter chargino are mainly gauginos (pt5). 
In contrast, $BR(\sta{1} \to \nutau \, \chim{1})$ is larger for smaller $\costa$
if the lighter chargino is mainly a higgsino
(\fig{fig:stalep40d}c). In the case that the gaugino- and the
higgsino-content of the chargino are of the same order there is less
dependence of the branching ratio on $\costa$ (\fig{fig:stalep40d}b and d).

OPAL has studied stau search at LEP2 for the case
$\mstau{R} \ll \mstau{L}$ \cite{Giudice96}. At $\sqrt{s} = 190$~GeV and
for ${\cal L} = 300$ ${\rm pb^{-1}}$ they obtained a $5 \sigma$ detectability
for $\mstau{R} \simeq$ 70 to 80 GeV for neutralino masses in the
range between 20 and 72 GeV. However, they have not considered the interesting
case of $\sta{R}$-$\sta{L}$ mixing.

\section{Tau Sneutrino}

In \fig{fig:psneulep} we show the production cross section
$\ee \tsn \atsn$ as a function of $\mtsn$ for $\sqrt{s} = 200$~GeV
(ISR corrections are included). For $\mtsn = 80$ (50)~GeV the cross section
is 0.094 (0.35) pb leading to 28 (105) events per year for an integrated
luminosity of 300 ${\rm pb^{-1}}$.

\noindent
\begin{minipage}[t]{149mm}
{\setlength{\unitlength}{1mm}
\begin{picture}(155,74)                        
\put(-3,2){\mbox{\epsfig{figure=sne15lepbr.eps,height=6.6cm,width=15.4cm}}}
\put(0,69){{\small \bf a)}}
\put(5,68){\makebox(0,0)[bl]{{\small $BR(\tilde \nutau )$}}}
\put(71.5,1){\makebox(0,0)[br]{{\small $\mu$~[GeV]}}}
\put(74,69){{\small \bf b)}}
\put(79,68){\makebox(0,0)[bl]{{\small $BR(\tilde \nutau )$}}}
\put(147.0,1){\makebox(0,0)[br]{{\small $M$~[GeV]}}}
\end{picture}}
\figcaption{fig:snelep15d}
           {Branching ratios for $\protect\tsn$ decays at LEP2 for
             $\protect \tanbe =1.5$}
           {Branching ratios for the decays of the tau sneutrino for
            $\tanbe = 1.5$ and $\mtsn = 90$~GeV. In Fig.~a we show the
            branching ratios as a function of $\mu$ for $M = 70$~GeV.
            In Fig.~b we show the branching ratios as a function of $M$ for
            $\mu = -64.9$~GeV. The curves correspond to the following
            transitions:
            $\circ \hspace{1mm} \tilde \nutau \to \nutau \, \tilde \chi^0_1$,
            \rechtl $\tilde \nutau \to \nutau \, \tilde \chi^0_2$ and
            \recht $\tilde \nutau \to \tau \, \tilde \chi^+_1$.
            The grey area is excluded by LEP1.5
            ($\sqrt{s} = 170$~GeV, $m_{\snu{e}} = 90$~GeV).
           }{7}
\end{minipage}

In the case that only the invisible decay $\tsn \to \nutau \, \chin{1}$ is
possible the signature is a single photon + $\me$, the photon coming from the
incoming electron or positron. 
The main background is single photon
production stemming from $\ee \nu \bar{\nu} \gamma$. The probability for 
this process is several magnitudes higher than that for sneutrino 
production. Therefore, it seems not to be clear if the signal for sneutrino 
production can be extracted in this case. If $\tsn$ 
decays into $\nutau \, \chin{2}$ or $\tau \, \chip{1}$ the signature is
charged leptons and/or jets + $\me$,
a single $\tau$ + leptons and/or jets+ $\me$, or
2 $\tau$ + leptons and/or jets. Note, that these signatures 
are similar to
that of neutralino production. However, a study of the differential cross 
sections should clarify the situation, because sneutrinos have a
$\sin^2 \theta$ angular distribution whereas neutralinos have a
$E^2 (1+\cos^2 \theta) + \mchinq{i} \sin^2 \theta$
angular distribution due to their different spins.
Here $\cos \theta$ is the 
angle between the $e^-$ beam and the outgoing
sneutrino or neutralino. In the case that
$\tsn \to \tau \, \chip{1}$ is allowed, the breaking of lepton universality
gives a clear distinction. As an example the domains of the various decays in the
$(M, \mu)$ plane are shown in \fig{fig:mmutsnlep} for $\mtsn = 90$~GeV and
$\tanbe = 1.5$.

In \fig{fig:snelep15d} we show the branching ratios of the various sneutrino
decays for $\mtsn = 90$~GeV and $\tanbe = 1.5$. 
In \fig{fig:snelep15d}a we show the branching
ratios as a function of $\mu$ for $M = 70.0$~GeV. Here the invisible decay
into $\nutau \, \chin{1}$ is the most important one. However for
$\mu \lsim -350$~GeV the sum of the cascade decays is at least as important as
the invisible decay leading to $\sim 50\%$ one-sided events. In the case
that $M \sim |\mu|$ the branching ratios depend strongly on $\mu$. With
increasing $|\mu|$ the number of one-sided events increases until
$\mu \lsim -100$~GeV where $\tsn$ decays completely invisible.
In \fig{fig:snelep15d}b
the branching ratios are shown as a function of $M$ for $\mu = -64.9$~GeV.
In the range 70~GeV $\lsim M \lsim 200$~GeV the sneutrino mainly decays 
invisibly.
In the other parameter range the decay into $\tau \, \chip{1}$
becomes important and 
even dominant while the decay into $\chin{2}$ can be 
important for $M \lsim 150$~GeV. Especially for
$M \gg |\mu|$ the main signatures are one or two $\tau$ + additional leptons
and/or jets + $\me$.

%% file: nlcprod.tex
\chapter{Production of sfermions at a Linear Collider}
\label{chap:nlcprod}

In this chapter we systematically present the production cross sections
of the various sfermions. The formulae for these processes including ISR-
and SUSY-QCD corrections are given in \app{appC} \cite{Drees90,Eberl96}. 
In the following examples these corrections will be included.
A Linear Collider will most likely offer the possibility of polarized beams.
As will be shown, beam polarization is an important tool for the
determination of $\cosfq$ from the measurement of the production cross sections. 
Conservative assumptions for the expected
luminosities are: 10~fb$^{-1}$ for $\sqrt{s} = 0.5$~TeV, 100~fb$^{-1}$ for 
$\sqrt{s} = 1$~TeV and 2~TeV \cite{Schulte96}.

In \fig{fig:psneutnlc2} the total cross section for $\ee \tsn \atsn$ is shown
as a function of $\mtsn$ for a) $\sqrt{s} = 500$~GeV and b) $\sqrt{s} = 2$~TeV,
for unpolarized, left polarized and right polarized
$e^-$ beams. The ratio $\sigma_L : \sigma_U : \sigma_R$ is given by
$2 L^2_e : (L^2_e + R^2_e) : 2 R^2_e \simeq 1.16 : 1 : 0.84$.
Here $L,U$ and $R$ denote 
left-polarized, unpolarized and right-polarized respectively, 
$L_e = -1/2 + \sinwq$ and $R_e = \sinwq$. 
This ratio is independent of the 
underlying parameters at tree level (and even when ISR corrections are 
included), because the $\tsn$ is a pure left state in the MSSM. Therefore,
a deviation from this ratio is a clear
hint for the existence of a "right" sneutrino and/or a mixing with other
sneutrinos. 

In \fig{fig:pstau11nlc2}a the total cross section $\ee \sta{1} \asta{1}$ is
shown as a function of $\costa$ for $\sqrt{s} = 500$~GeV and several 
stau masses. For unpolarized beams and $\mstau{1} \gsim 150$~GeV the dependence 
on the mixing angle is rather weak.
The dependence is much stronger if a polarized
electron beam is available as can be seen in \fig{fig:pstau11nlc2}b. It is 
important to note that this dependence is opposite for left and right 
polarization. In particular the quantity
\beq
\sigma(\sta{1})_{LR} \equiv \sigma(e^+ e^-_L \to  \sta{1} \asta{1})
                  -  \sigma(e^+ e^-_R \to  \sta{1} \asta{1})
\eeq
\noindent
\begin{minipage}[t]{150mm}
{\setlength{\unitlength}{1mm}
\begin{picture}(150,60)                        
\put(0,1.5){\mbox{\epsfig{figure=psneutnlc2.eps,height=5.7cm,width=15.4cm}}}
\put(3,60){{\small \bf a)}}
\put(23,50){{\small $\sigma_L$}}
\put(15,49){{\small $\sigma_U$}}
\put(15,37){{\small $\sigma_R$}}
\put(7,59){\makebox(0,0)[bl]{{\small $\sigma(\tsn \atsn)$~[fb]}}}
\put(76.5,0){\makebox(0,0)[br]{{\small $\mtsn$~[GeV]}}}
\put(79,60){{\small \bf b)}}
\put(98,50){{\small $\sigma_L$}}
\put(91,48){{\small $\sigma_U$}}
\put(88.5,38){{\small $\sigma_R$}}
\put(83,59){\makebox(0,0)[bl]{{\small $\sigma(\tsn \atsn)$~[fb]}}}
\put(149.5,0){\makebox(0,0)[br]{{\small $\mtsn$~[GeV]}}}
\end{picture}}
\figcaption{fig:psneutnlc2}
           {Total cross section 
            $\sigma(e^+ e^- \to \protect \tsn \protect \atsn)$}
           {Total cross section 
            $\sigma(e^+ e^- \to \protect \tsn \protect \atsn)$
            in $fb$ as a function of $\mtsn$ for
            unpolarized $e^-$ ($\sigma_U$, full line), left polarized $e^-$ 
            ($\sigma_L$, dashed line),
            right polarized $e^-$ ($\sigma_R$, dashed dotted line),
            a) $\sqrt{s} = 500$~GeV; b) $\sqrt{s} = 2$~TeV. ISR-corrections
            are included.
      }{7}
\end{minipage}

\noindent
\begin{minipage}[t]{150mm}
{\setlength{\unitlength}{1mm}
\begin{picture}(150,60)                        
\put(-6,0){\mbox{\epsfig{figure=pstau11nlc2.eps,height=5.9cm,width=15.9cm}}}
\put(-3,59){{\small \bf a)}}
\put(2,58.5){\makebox(0,0)[bl]{{\small $\sigma(\sta{1} \asta{1})$~[fb]}}}
\put(73,1){\makebox(0,0)[br]{{\small $\costa$}}}
\put(35,48){{\small 100}}
\put(35,40){{\small 125}}
\put(35,32){{\small 150}}
\put(35,24){{\small 175}}
\put(35,16){{\small 200}}
\put(35,10){{\small 225}}
\put(75,59){{\small \bf b)}}
\put(80,58.5){\makebox(0,0)[bl]{{\small $\sigma(\sta{1} \asta{1})$~[fb]}}}
\put(150.5,1){\makebox(0,0)[br]{{\small $\costa$}}}
\end{picture}}
\figcaption{fig:pstau11nlc2}
           {Total cross section 
            $\sigma(e^+ e^- \to \protect \sta{1} \protect \asta{1})$}
           {Total cross section 
            $\sigma(e^+ e^- \to \protect \sta{1} \protect \asta{1})$
            in $fb$ as a function of $\costa$ for
            $\sqrt{s} = 500$~GeV; a) 
            unpolarized $e^-$ and various $\mstau{1}$: 100, 125, 150, 175,
            200, and 225 GeV; b) 
            unpolarized $e^-$ (full line), left polarized $e^-$ (dashed line),
            right polarized $e^-$ (dashed dotted line), for
            $\mstau{1} = 150$~GeV. ISR-corrections are included.
      }{7}
\end{minipage}

\noindent
shows a strong dependence on $\costa$ (see also \cite{Nojiri95}). 
Here $e^-_L$ ($e^-_R$) denotes a left-polarized 
(right-polarized) electron.
Note, that the sign of $\costa$ can not be 
determined, because the cross
section depends on $\costaq$ (see \eqn{sigtree} and (\ref{acoup})).

In \fig{fig:pstau22nlc2}a the total cross section 
$\sigma(\ee \sta{2} \asta{2})$ is
shown as a function of $\costa$ for $\sqrt{s} = 2$~TeV and several 
stau masses. Concerning the dependence on the mixing angle the situation 
is obviously similar to the case of $\sta{1}$ pair production: there 
is hardly any
dependence on $\costa$ for $\mstau{2} \gsim 700$~GeV and unpolarized electrons.
\noindent
\begin{minipage}[t]{150mm}
{\setlength{\unitlength}{1mm}
\begin{picture}(150,70)                        
\put(-6,1.7){\mbox{\epsfig{figure=pstau22nlc2.eps,height=5.7cm,width=15.9cm}}}
\put(-3,61){{\small \bf a)}}
\put(2,60.5){\makebox(0,0)[bl]{{\small $\sigma(\sta{2} \asta{2})$~[fb]}}}
\put(73,1){\makebox(0,0)[br]{{\small $\costa$}}}
\put(45,51.5){\makebox(0,0)[br]{{\small 400}}}
\put(45,42.5){\makebox(0,0)[br]{{\small 500}}}
\put(45,33.5){\makebox(0,0)[br]{{\small 600}}}
\put(45,25){\makebox(0,0)[br]{{\small 700}}}
\put(45,17){\makebox(0,0)[br]{{\small 800}}}
\put(45,10.5){\makebox(0,0)[br]{{\small 900}}}
\put(75,61){{\small \bf b)}}
\put(80,60.5){\makebox(0,0)[bl]{{\small $\sigma(\sta{2} \asta{2})$~[fb]}}}
\put(150.5,1){\makebox(0,0)[br]{{\small $\costa$}}}
\end{picture}}
\figcaption{fig:pstau22nlc2}
           {Total cross section 
            $\sigma(e^+ e^- \to \protect \sta{2} \protect \asta{2})$}
           {Total cross section 
            $\sigma(e^+ e^- \to \protect \sta{2} \protect \asta{2})$
            in $fb$ as a function of $\costa$ for
            $\sqrt{s} = 2$~TeV; a) 
            unpolarized $e^-$ and various $\mstau{2}$: 400, 500, 600, 700,
            800, and 900 GeV; b) 
            unpolarized $e^-$ (full line), left polarized $e^-$ (dashed line),
            right polarized $e^-$ (dashed dotted line), for 
            $\mstau{2} = 700$~GeV. ISR-corrections are included.
      }{7}
\end{minipage}

\noindent
\begin{minipage}[t]{150mm}
{\setlength{\unitlength}{1mm}
\begin{picture}(150,60)                        
\put(-6,0){\mbox{\epsfig{figure=pstop11nlc2.eps,height=5.9cm,width=15.9cm}}}
\put(-3,59){{\small \bf a)}}
\put(2,58.5){\makebox(0,0)[bl]{{\small $\sigma(\sto{1} \asto{1})$~[fb]}}}
\put(73,1){\makebox(0,0)[br]{{\small $\cost$}}}
\put(35,36){{\small 100}}
\put(35,31){{\small 125}}
\put(35,25){{\small 150}}
\put(35,20){{\small 175}}
\put(35,15){{\small 200}}
\put(35,10){{\small 225}}
\put(75,59){{\small \bf b)}}
\put(80,58.5){\makebox(0,0)[bl]{{\small $\sigma(\sto{1} \asto{1})$~[fb]}}}
\put(150.5,1){\makebox(0,0)[br]{{\small $\cost$}}}
\end{picture}}
\figcaption{fig:pstop11nlc2}
           {Total cross section 
            $\sigma(e^+ e^- \to \protect \sto{1} \protect \asto{1})$}
           {Total cross section 
            $\sigma(e^+ e^- \to \protect \sto{1} \protect \asto{1})$
            in $fb$ as a function of $\cost$ for
            $\sqrt{s} = 500$~GeV; a) 
            unpolarized $e^-$ and various $\mstop{1}$: 100, 125, 150, 175,
            200, and 225 GeV; b) 
            unpolarized $e^-$ (full line), left polarized $e^-$ (dashed line),
            right polarized $e^-$ (dashed dotted line), for
            $\mstop{1} = 150$~GeV.
            ISR- and SUSY-QCD corrections are included ($\mglu = 825$~GeV,
            $\mstop{2} = 400$~GeV).
      }{7}
\end{minipage}

\noindent
Again the dependence is much stronger if a polarized
electron beam can be used as 
demonstrated in \fig{fig:pstau22nlc2}b.
In principal also the processes $\ee \sta{1} \asta{2} , \, \asta{1} \sta{2}$ 
are possible. It turns out that the corresponding cross sections are 
rather small. We get for 
\noindent
\begin{minipage}{69mm}
example $\sigma = 1.3$~fb for $\sqrt{s} = 1$~TeV, 
$\mstau{1} = 250$~GeV, $\mstau{2} = 400$~GeV and 
$\costa = 1 / \wzw$. Therefore, we do not show the cross section in an extra 
figure. However, one 
can take as a guideline 
$\sigma(\ee \sta{1} \asta{2} + \asta{1} \sta{2}) \simeq 
\sigma(\ee \sto{1} \asto{2} + \asto{1} \sto{2}) / 3 $
for the 
results presented in \fig{fig:pstop12nlc2}.

\hspace*{4mm}
In \fig{fig:pstop11nlc2}a the total cross section for the process
$\ee \sto{1} \asto{1}$ at $\sqrt{s} = 500$~GeV is 
shown for various stop 
masses. For the 
calculation of the SUSY-QCD corrections we have assumed $\mstop{2} = 400$~GeV.
The cross section can reach 220~fb for $\mstop{1} = 100$~GeV 
and $\cost = 1$. There is a strong dependence on $\cost$ if $\cost \gsim 0.6$. 
As already mentioned in the case of stau production the dependence 
on $\cost$ is stronger if polarized 
\end{minipage}
\hspace{4mm}
\noindent
\begin{minipage}{74mm}
{\setlength{\unitlength}{1mm}
\begin{picture}(74,60)                        
\put(-2,3){\mbox{\epsfig{figure=pstop12nlc2.eps,height=5.3cm,width=7.4cm}}}
\put(2,58.5){\makebox(0,0)[bl]{{\small $\sigma(\sto{1} \asto{2})$~[fb]}}}
\put(73,1){\makebox(0,0)[br]{{\small $\cost$}}}
\put(35,47){{\small 400}}
\put(50,37){{\small 500}}
\put(50,22){{\small 600}}
\put(50,12){{\small 700}}
\end{picture}}
\figcaption{fig:pstop12nlc2}
           {Total cross section $e^+ e^- \to \protect \sto{1} \protect \asto{2}$
            }
           {Total cross section $e^+ e^- \to \protect \sto{1} \protect \asto{2}$
            in $fb$ as a function of $\cost$ for
            $\sqrt{s} = 1$~TeV, $\mstop{1} = 250$~GeV and various 
            $\sto{2}$ masses: 400~GeV, 500~GeV, 600~GeV and 700~GeV.
            ISR- and SUSY-QCD corrections are included ($\mglu = 755$~GeV).
      }{6}
\end{minipage}

\noindent
$e^-$-beams are used. This can be seen in \fig{fig:pstop11nlc2}b 
where the cross section is shown for various beam polarizations. The quantity
$\sigma_{LR}(\sto{1})$ varies from
$-117$~fb ($\cost = 0$) to 250~fb ($\cost = 1$). 

In \fig{fig:pstop12nlc2} the cross section 
$\ee \sto{1} \asto{2} + \asto{1} \sto{2}$ is shown as a function of $\cost$ for
$\sqrt{s} = 1$~TeV, $\mstop{1} = 250$~GeV and 
various $\sto{2}$ masses. For the calculation 
of the SUSY-QCD corrections we have taken $\mglu = 755$~GeV. For
this process only the $Z$-exchange contributes at tree-level. The cross section
has its maximum at $\cost = 1 / \wzw \simeq 0.71$ because the
$Z \, \sto{1} \, \sto{2}$ coupling is proportional to $\sinzt$. Note, that this 
cross section is rather small even at its maximum compared to 
$\sto{1} \asto{1}$
and $\sto{2} \asto{2}$ production: $\sigma(\ee \sto{1} \asto{1}) = 29.3$~fb,
$\sigma(\ee \sto{1} \asto{2} + \asto{1} \sto{2}) = 4.8$~fb
and $\sigma(\ee \sto{2} \asto{2}) = 8.9$~fb for $\sqrt{s} = 1$~TeV, 
$\mstop{1} = 250$~GeV, $\mstop{2} = 400$~GeV and 
$\cost = 1 / \wzw$. Therefore, this process is mainly of interest for the case
that only $\sto{1} \asto{1}$ can be produced at a given energy but not 
$\sto{2} \asto{2}$. Assuming an integrated luminosity of 50~fb$^{-1}$
and that $\sim 100$ stop pairs are needed for a successful detection, one
can probe $\sto{2}$ masses up to $\sim 550$~GeV in our example if the mixing is
maximal. Beam polarization changes the cross section in the following way:
$\sigma(e^+ e^-_L) : \sigma(e^+ e^-_U) : \sigma(e^+ e^-_R)
 = (2 L^2_e) : (L^2_e + R^2_e) : (2 R^2_e)$ as in the case of sneutrino 
production. 

In \fig{fig:pstop22nlc2}a the total cross section 
$\sigma(\ee \sto{2} \asto{2})$ is
shown as a function of $\cost$ for $\sqrt{s} = 2$~TeV. For the calculation of
the SUSY-QCD corrections we have assumed $\mstop{1} = 300$~GeV and 
$\mglu = 693$~GeV. The total cross section goes up to 
$\sim 14$~fb 
\noindent
\begin{minipage}[t]{150mm}
{\setlength{\unitlength}{1mm}
\begin{picture}(150,60)                        
\put(-6,1){\mbox{\epsfig{figure=pstop22nlc2.eps,height=5.7cm,width=15.9cm}}}
\put(-3,58){{\small \bf a)}}
\put(2,57.5){\makebox(0,0)[bl]{{\small $\sigma(\sto{2} \asto{2})$~[fb]}}}
\put(73,1){\makebox(0,0)[br]{{\small $\cost$}}}
\put(35,47){{\small 400}}
\put(35,39){{\small 500}}
\put(35,31){{\small 600}}
\put(35,23.5){{\small 700}}
\put(35,16.5){{\small 800}}
\put(35,11){{\small 900}}
\put(75,59){{\small \bf b)}}
\put(80,58.5){\makebox(0,0)[bl]{{\small $\sigma(\sto{2} \asto{2})$~[fb]}}}
\put(150.5,1){\makebox(0,0)[br]{{\small $\cost$}}}
\end{picture}}
\figcaption{fig:pstop22nlc2}
           {Total cross section 
            $\sigma(e^+ e^- \to \protect \sto{2} \protect \asto{2})$}
           {Total cross section 
            $\sigma(e^+ e^- \to \protect \sto{2} \protect \asto{2})$
            in $fb$ as a function of $\cost$ for
            $\sqrt{s} = 2$~TeV; a) 
            unpolarized $e^-$ and various $\sto{2}$ masses
            (in GeV): 400, 500, 600, 700, 800, and 900; b) 
            unpolarized $e^-$ (full line), left polarized $e^-$ (dashed line),
            right polarized $e^-$ (dashed dotted line), for
            $\mstop{2} = 700$~GeV.
            ISR- and SUSY-QCD corrections are included
            ($\mstop{1} = 300$~GeV and $\mglu = 693$~GeV).
      }{7}
\end{minipage}

\noindent
for $\mstop{2} = 400$~GeV and
$\cost = 0$. Even for $\mstop{2} = 900$~GeV it can reach 1.7~fb if $\cost = 0$.
This corresponds to 
340 produced $\sto{2}$ pairs for an integrated luminosity 
of 
200~fb$^{-1}$. As in the case of $\sto{1} \asto{1}$ production the
quantity $\sigma_{LR}(\sto{2})$ 
shows the strongest dependence on $\cost$ as
can be seen in \fig{fig:pstop22nlc2}b. Here the production cross section is
shown for $\mstop{2} = 700$~GeV and polarized $e^-$ beams.  

In \fig{fig:psbot11nlc2}a we present the total cross section for the process
$\ee \sbo{1} \asbo{1}$ for $\sqrt{s} = 500$~GeV and various sbottom
masses. For the SUSY-QCD corrections we have taken
$\msbot{2} = 400$~GeV and $\mglu = 825$~GeV.
The cross section can reach 155~fb for $\msbot{1} = 100$~GeV 
and $\cosb = 1$. The dependence on $\cosb$ is similar to the $\sto{1}$ and
$\sta{1}$ cases: i) there is hardly any dependence for $\cosb \lsim 0.6$ and ii)
there is a clear dependence for polarized beams (\fig{fig:psbot11nlc2}b). 

In \fig{fig:psbot22nlc2}a the total cross section for $\ee \sbo{2} \asbo{2}$ is
shown as a function of $\cosb$ for $\sqrt{s} = 2$~TeV. For the calculation of
the SUSY-QCD corrections we have taken $\msbot{1} = 300$~GeV and 
$\mglu = 693$~GeV. The cross section can reach up to $\sim 9.5$~fb if 
$\sbo{2} = \sbo{R}$ and $\mstop{2} = 400$~GeV.
Even for $\msbot{2} = 900$~GeV it can reach 0.9~fb corresponding to 180
produced $\sbo{2}$ pairs for an integrated luminosity of 
200~fb$^{-1}$. As in the previous cases the
quantity $\sigma_{LR}(\sbo{2})$ shows the strongest dependence on $\cosb$ as
can be seen in \fig{fig:psbot22nlc2}b. Here the production cross section is
shown for $\msbot{2} = 700$~GeV and polarized $e^-$ beams.  

As in the case of the stops the processes 
$\ee \sbo{1} \asbo{2} , \, \asbo{1} \sbo{2}$ are possible. The results are 
very similar to the stop case (\fig{fig:pstop12nlc2}). The only differences
come from SUSY-QCD corrections which are of $O(0.1 \%)$. 

\noindent
\begin{minipage}[t]{150mm}
{\setlength{\unitlength}{1mm}
\begin{picture}(150,60)                        
\put(-6,0){\mbox{\epsfig{figure=psbot11nlc2.eps,height=5.9cm,width=15.9cm}}}
\put(-3,59){{\small \bf a)}}
\put(2,58.5){\makebox(0,0)[bl]{{\small $\sigma(\sbo{1} \asbo{1})$~[fb]}}}
\put(73,1){\makebox(0,0)[br]{{\small $\cosb$}}}
\put(60,51){{\small 100}}
\put(63,45){{\small 125}}
\put(63,36){{\small 150}}
\put(63,29){{\small 175}}
\put(63,20.5){{\small 200}}
\put(63,13){{\small 225}}
\put(75,59){{\small \bf b)}}
\put(80,58.5){\makebox(0,0)[bl]{{\small $\sigma(\sbo{1} \asbo{1})$~[fb]}}}
\put(150.5,1){\makebox(0,0)[br]{{\small $\cosb$}}}
\end{picture}}
\figcaption{fig:psbot11nlc2}
           {Total cross section 
            $\sigma(e^+ e^- \to \protect \sbo{1} \protect \asbo{1})$}
           {Total cross section 
            $\sigma(e^+ e^- \to \protect \sbo{1} \protect \asbo{1})$
            in $fb$ as a function of $\cosb$ for
            $\sqrt{s} = 500$~GeV; a) 
            unpolarized $e^-$ and various $\msbot{1}$: 100, 125, 150, 175,
            200, and 225 GeV; b) 
            unpolarized $e^-$ (full line), left polarized $e^-$ (dashed line),
            right polarized $e^-$ (dashed dotted line), for 
            $\msbot{1} = 150$~GeV.
            ISR- and SUSY-QCD corrections are included ($\mglu = 825$~GeV,
            $\msbot{2} = 400$~GeV).
      }{5}
\end{minipage}

\noindent
\begin{minipage}[t]{150mm}
{\setlength{\unitlength}{1mm}
\begin{picture}(150,65)                        
\put(-6,1){\mbox{\epsfig{figure=psbot22nlc2.eps,height=5.7cm,width=15.9cm}}}
\put(-3,59){{\small \bf a)}}
\put(2,58.5){\makebox(0,0)[bl]{{\small $\sigma(\sbo{2} \asbo{2})$~[fb]}}}
\put(73,0.5){\makebox(0,0)[br]{{\small $\cosb$}}}
\put(25,49.5){{\small 400}}
\put(25,41.5){{\small 500}}
\put(25,33.5){{\small 600}}
\put(25,25.5){{\small 700}}
\put(25,18){{\small 800}}
\put(25,12.5){{\small 900}}
\put(75,59){{\small \bf b)}}
\put(80,58.5){\makebox(0,0)[bl]{{\small $\sigma(\sbo{2} \asbo{2})$~[fb]}}}
\put(150.5,0.5){\makebox(0,0)[br]{{\small $\cosb$}}}
\end{picture}}
\figcaption{fig:psbot22nlc2}
           {Total cross section 
            $\sigma(e^+ e^- \to \protect \sbo{2} \protect \asbo{2})$}
           {Total cross section
            $\sigma(e^+ e^- \to \protect \sbo{2} \protect \asbo{2})$
            in $fb$ as a function of $\cosb$ for
            $\sqrt{s} = 2$~TeV; a) 
            unpolarized $e^-$ and various $\msbot{2}$: 400~GeV, 500~GeV,
            600~GeV, 700~GeV, 800~GeV and 900~GeV; b) 
            unpolarized $e^-$ (full line), left polarized $e^-$ (dashed line),
            right polarized $e^-$ (dashed dotted line), for
            $\msbot{2} = 700$~GeV.
            ISR- and SUSY-QCD corrections are included
            ($\msbot{1} = 300$~GeV and $\mglu = 693$~GeV).
      }{5}
\end{minipage}

%% file: nlcmass.tex
\chapter{Masses and mixing angles}
\label{chap:nlcmass}

\section{Outlook on the following chapters}

In the following chapters we discuss the phenomenology of the various 
sfermions at colliders where sfermion masses can be probed up to the TeV range,
for example the LHC or a future Linear Collider. Sfermions in that mass range
cannot only decay into fermions but also in another sfermion plus either a
gauge boson or a Higgs boson. We have seen in \chap{chap:lep2} that the mixing
between left and right sfermions strongly influences the fermionic decay modes.
As we will demonstrate in the next chapters this is also true for the bosonic
decay modes.

As it is shown in \app{appB} there
are relations between sfermion masses and mixing angles. 
Therefore, we will change our strategy and use the soft SUSY breaking parameters
as input for the discussion of the decays instead of the physical quantities.
Let us shortly the review underlying parameters for the masses and mixing 
angles:
\begin{itemize}
\item sleptons: $M_E, M_L, A_\tau, \tanbe$ and $\mu$
\item squarks: $M_D, M_Q, M_U, A_b, A_t, \tanbe$ and $\mu$
\item gluino: $\mglu$
\item charginos and neutralinos: $M', M, \mu$ and $\tanbe$
\item Higgs bosons: $\ma$ and $\tanbe$. The squark parameters also enter the
      mass matrix due to radiative corrections.
\end{itemize}
Even for a detailed discussion it is not necessary to study the dependence
on all 14 parameters. Therefore, we fix $M = 350$~GeV and use
the GUT relations to compute $M'$ and $\mglu$ as stated in \chap{chap:mssm}.
We choose $\ma = 150$~GeV so that the decays into Higgs bosons are kinematically
possible. We further assume $A_b = A_t = A_\tau$. Moreover, we have chosen
three ($M_D,M_Q,M_U;M_E,M_L$) sets: (700,700,700;700,700), (980,350,700;350,700)
and (500,500,700;980,700). At first glance one may ask how strongly the 
squark parameters influence the slepton phenomenology due to the radiative
correction on the Higgs masses and the mixing angle. Here we want to note first,
that also $A_b$, $A_t$, $\mu$ and $\tanbe$ enter these corrections so that 
there are seven
input parameters for the four physical quantities $m_{h^0},m_{H^0}$, 
$\cosa$, and $m_{H^+}$. Therefore, the resulting Higgs masses 
and Higgs mixing angle can be seen as typical examples independent of the
specific input parameters if one studies the phenomenology of the sleptons.
We have explicitly checked, that a variation of $M_D, M_Q$ and/or $M_U$ does
not change the general features of slepton decays into Higgs bosons.

In the remaining sections of this chapter we present the results for the various
masses and mixing angles as needed for the following chapters. In the next
chapter we start with the discussion of slepton phenomenology. Here the 
situation is easier, because there are fewer parameters compared to the
squark sector. Afterwards the results for stops and sbottoms will be presented.
As implicitly mentioned above we study the cases 
$M_E < M_L$, $M_E = M_L$, $M_D = M_Q = M_U$, $M_U < M_Q < M_D$ and
$M_D = M_U < M_Q$. Here results from the study of 
renormalization group equations
in supergravity models \cite{Drees95b,Barger94a} and models 
with gauge mediated SUSY
breaking \cite{GMSB1} have served as guidelines for the ordering of the masses.
The only exception is the case $M_E > M_L$. It has been included to study the 
phenomenology if SUSY is realized and broken in a different way than 
proposed in those models. Examples are models with
additional gauge bosons \cite{Kolda96}.

\section{Masses and mixing angles of staus and sbottoms}
\label{sect:msbot}

We treat the masses and the mixing angles of staus and sbottoms together because
their mass matrices have the same structure including their $\tanbe$ dependence
(see \eqn{eq:sfmassmatrix}).

In \fig{fig:mstaunlc2mu} we show $\mstau{1}$, $\mstau{2}$, and $\costa$ 
as function of 
$\mu$. We fix $M_L = 700$~GeV and choose three different $M_E$ values:
700~GeV (full line), 350~GeV (dashed line), and 980~GeV (dashed dotted line).
In \fig{fig:mstaunlc2mu}a $\mstau{1,2}$ are shown for
$\tanbe = 1.5$ and $A_\tau = 500$~GeV. Clearly the masses are mainly determined
by $M_E$ and $M_L$ in this case, because the off-diagonal element 
$m_\tau (A_\tau - \mu \tanbe)$ is rather small compared to $M^2_{E,L}$. 
Nevertheless this mixing term can lead to a strong mixing for $M_E \sim M_L$
as can be seen in \fig{fig:mstaunlc2mu}b. The reason for this at first glance
surprising effect is that $(M^2_{\sta{11}} - \mstaq{1})^2$ is of the same order
as $m^2_\tau (A_\tau - \mu \tanbe)^2$ (see \eqn{sfer2}). Moreover, there is
a rather strong dependence of $\costa$ on $\mu$ if $\mu \sim A_\tau \cotbe$ and
$M_E \sim M_L$. In the case $M_E = 350$~GeV (980 GeV) $\sta{1}$
is nearly $\sta{R}$ ($\sta{L}$).

\noindent
\begin{minipage}[t]{150mm}
{\setlength{\unitlength}{1mm}
\begin{picture}(150,125)                        
\put(-4,-8){\mbox{\epsfig{figure=mstaunlc2mu.eps,height=14.0cm,width=15.6cm}}}
\put(0,122){{\small \bf a)}}
\put(5,121){\makebox(0,0)[bl]{{$\mstau{1},\mstau{2}$ \small [GeV]}}}
\put(15,113){\makebox(0,0)[bl]{{\small $\tanbe = 1.5$}}}
\put(80,120.5){{\small \bf b)}}
\put(85,119.5){\makebox(0,0)[bl]{{$\costa$}}}
\put(95,113){\makebox(0,0)[bl]{{\small $\tanbe = 1.5$}}}
\put(0,59){{\small \bf c)}}
\put(5,58){\makebox(0,0)[bl]{{$\mstau{1},\mstau{2}$ \small [GeV]}}}
\put(15,50){\makebox(0,0)[bl]{{\small $\tanbe = 40$}}}
\put(80,58.5){{\small \bf d)}}
\put(85,57.5){\makebox(0,0)[bl]{{$\costa$}}}
\put(95,50){\makebox(0,0)[bl]{{\small $\tanbe = 40$}}}
\put(71.5,3){\makebox(0,0)[br]{{$\mu$ \small [GeV]}}}
\put(71.5,64.5){\makebox(0,0)[br]{{$\mu$ \small [GeV]}}}
\put(149.5,3){\makebox(0,0)[br]{{$\mu$ \small [GeV]}}}
\put(149.5,64.5){\makebox(0,0)[br]{{$\mu$ \small [GeV]}}}
\end{picture}}
\figcaption{fig:mstaunlc2mu}
           {Stau masses and mixing angle as a function
            of $\mu$}
           {Stau masses (a,c) and mixing angle (b,d) as a function of 
            $\mu$ for $A_{\tau} = 500$~GeV, $\tanbe$ = 1.5 (a,b) and 
            $\tanbe$ = 40 (c,d). 
            The graphs correspond to the following parameter sets
            ($M_E,M_L$) (in GeV): (700,700) full line, (350,700)
            dashed line, and (980,700) dashed-dotted line.}{7}
\end{minipage}

In \fig{fig:mstaunlc2mu}c and \ref{fig:mstaunlc2mu}d the $\mu$ dependence is
shown for $\tanbe = 40$. The dependence on $\mu$ is the stronger the closer
$M_E$ and $M_L$ are because then the mass splitting is 
mainly due to the off diagonal element. Concerning the mixing angle we
have again a strong mixing 
for the case $M_E = M_L$. As in the case of small $\tanbe$ the lighter mass 
eigenstate is 
nearly a right state (left state) if $M_E = 350$~GeV (980 GeV).
We have found that the masses are
nearly independent of $A_\tau$. The same is true for the mixing angle except
the case where $A_\tau \sim \mu \tanbe$ and and at the same time $M_E \sim M_L$.

In \fig{fig:msbotnlc2mu} we show the masses and mixing angle of the sbottoms.
The differences to the stau case are only due to the fact that 
$m_b \simeq 3 \, m_\tau$. For small $\tanbe$ the situation is nearly the same
except that for the case $M_D = M_Q$ there is a slight 
\noindent
\begin{minipage}[t]{150mm}
{\setlength{\unitlength}{1mm}
\begin{picture}(150,125)                        
\put(-3,-8){\mbox{\epsfig{figure=msbotnlc2mu.eps,height=14.0cm,width=15.4cm}}}
\put(0,122){{\small \bf a)}}
\put(5,121){\makebox(0,0)[bl]{{$\msbot{1},\msbot{2}$ \small [GeV]}}}
\put(15,112){\makebox(0,0)[bl]{{\small $\tanbe = 1.5$}}}
\put(80,122){{\small \bf b)}}
\put(85,121){\makebox(0,0)[bl]{{$\cosb$}}}
\put(95,113){\makebox(0,0)[bl]{{\small $\tanbe = 1.5$}}}
\put(0,60){{\small \bf c)}}
\put(5,59){\makebox(0,0)[bl]{{$\msbot{1},\msbot{2}$ \small [GeV]}}}
\put(15,49){\makebox(0,0)[bl]{{\small $\tanbe = 40$}}}
\put(80,60){{\small \bf d)}}
\put(85,59){\makebox(0,0)[bl]{{$\cosb$}}}
\put(95,50){\makebox(0,0)[bl]{{\small $\tanbe = 40$}}}
\put(72.5,2){\makebox(0,0)[br]{{$\mu$ \small [GeV]}}}
\put(72.5,63.5){\makebox(0,0)[br]{{$\mu$ \small [GeV]}}}
\put(149.5,1.5){\makebox(0,0)[br]{{$\mu$ \small [GeV]}}}
\put(149.5,63){\makebox(0,0)[br]{{$\mu$ \small [GeV]}}}
\end{picture}}
\figcaption{fig:msbotnlc2mu}
           {Sbottom masses and mixing angle as a function of $\mu$}
           {Sbottom masses (a,c) and mixing angle (b,d) as a function of 
            $\mu$ for $A_b = 500$~GeV and $\tanbe$ = 1.5 (a,b) 
            ($\tanbe$ = 40 (c,d)).
            The graphs correspond to the following parameter sets
            ($M_D,M_Q$) (in GeV): (700,700) full line, (980,700)
            dashed line, and (500,700) dashed-dotted line.}{5}
\end{minipage}

\noindent
visible dependence of
$\mu$ on the masses. In case of large $\tanbe$ the dependence on $\mu$ is 
much more pronounced than in the stau case.

\section{Masses and mixing angles of the stops}

The stops differ in two ways from the staus and sbottoms: Firstly, $m_t$ is much
larger than  $m_b$ or $m_\tau$. Secondly, the off diagonal term in the mass 
matrix is
given by $m_t(A_t - \mu \cotbe)$. The combination of these two facts leads to a 
strong dependence of the masses on $\mu$ for the case 
$M_Q = M_U$ and small $\tanbe$
as can be seen in \fig{fig:mstopnlc2mu}. This is a case of maximal mixing 
($\cost \simeq \pm 1/\sqrt{2}$) except the range where $\mu \simeq A_t \tanbe$.
Clearly the dependence of the masses on $\mu$ becomes weaker
for larger 
\noindent
\begin{minipage}[t]{150mm}
{\setlength{\unitlength}{1mm}
\begin{picture}(150,65)                        
\put(-4,-1){\mbox{\epsfig{figure=mstopnlc2mu.eps,height=6.5cm,width=15.8cm}}}
\put(0,62.5){{\small \bf a)}}
\put(5,62){\makebox(0,0)[bl]{{$\mstop{1}, \mstop{2}$ \small [GeV]}}}
\put(80,61){{\small \bf b)}}
\put(85,60.5){\makebox(0,0)[bl]{{$\cost$}}}
\put(73.5,0){\makebox(0,0)[br]{{$\mu$ \small [GeV]}}}
\put(149.5,1){\makebox(0,0)[br]{{$\mu$ \small [GeV]}}}
\end{picture}}
\figcaption{fig:mstopnlc2mu}
           {Stop masses and mixing angle as a function of $\mu$ 
            for $\protect \tanbe = 1.5$}
           {Stop masses (a) and mixing angle (b) as a function of 
            $\mu$ for $A_t = 500$~GeV and $\tanbe$ = 1.5.
            The graphs correspond to the following parameter sets
            ($M_Q,M_U$) (in GeV): (700,700) full line, (700,350)
            dashed line, and (700,500) dashed-dotted line.}{7}
\end{minipage}

\noindent
\begin{minipage}[t]{150mm}
{\setlength{\unitlength}{1mm}
\begin{picture}(150,65)                        
\put(-4,-2){\mbox{\epsfig{figure=mstopnlc2A.eps,height=6.5cm,width=15.8cm}}}
\put(0,61.5){{\small \bf a)}}
\put(5,61){\makebox(0,0)[bl]{{$\mstop{1}, \mstop{2}$ \small [GeV]}}}
\put(80,60){{\small \bf b)}}
\put(85,59.5){\makebox(0,0)[bl]{{$\cost$}}}
\put(73.5,0){\makebox(0,0)[br]{{$A_t$ \small [GeV]}}}
\put(149.5,0){\makebox(0,0)[br]{{$A_t$ \small [GeV]}}}
\end{picture}}
\figcaption{fig:mstopnlc2A}
           {Stop masses and mixing angle as a function of $A_t$ 
            for $\protect \tanbe = 1.5$}
           {Stop masses (a) and mixing angle (b) as a function of 
            $A_t$ for $\mu = -500$~GeV and $\tanbe$ = 1.5.
            The graphs correspond to the following parameter sets
            ($M_Q,M_U$) (in GeV): (700,700) full line, (700,350)
            dashed line, and (700,500) dashed-dotted line.}{5}
\end{minipage}

\noindent
$|M_U^2 - M^2_Q|$. In this case $\cost$ shows a stronger dependence on $\mu$ 
than in the previous case.
Note, that the masses and $\cost$ depend hardly 
on $\mu$ for large $\tanbe$.

The masses show also a strong dependence on $A_t$ as it is
demonstrated in \fig{fig:mstopnlc2A} for the different ($M_Q,M_U$) sets and 
small $\tanbe$. This is also true for $\cost$ except in the case 
$M_U \simeq M_Q$. There is still a strong 
dependence on $A_t$ if $\tanbe$ increases. The only difference is 
that the point where the stops are pure left or right eigenstates 
is shifted towards $A_t = 0$.

\noindent
\begin{minipage}[t]{150mm}
{\setlength{\unitlength}{1mm}
\begin{picture}(150,123)                        
\put(-4,-10){\mbox{\epsfig{figure=mhiggsnlc2mu.eps,height=14.0cm,width=15.6cm}}}
\put(0,120){{\small \bf a)}}
\put(5,119){\makebox(0,0)[bl]{{$m_{h^0,H^0,H^\pm}$ \small [GeV]}}}
\put(13,100){\makebox(0,0)[bl]{{\small $\tanbe = 1.5$}}}
\put(35,112.5){\makebox(0,0)[bl]{{$m_{H^0}$}}}
\put(35,105){\makebox(0,0)[bl]{{$m_{H^\pm}$}}}
\put(35,77){\makebox(0,0)[bl]{{$m_{h^0}$}}}
\put(80,118){{\small \bf b)}}
\put(85,117){\makebox(0,0)[bl]{{$\sina$}}}
\put(94,110){\makebox(0,0)[bl]{{\small $\tanbe = 1.5$}}}
\put(0,59){{\small \bf c)}}
\put(5,58){\makebox(0,0)[bl]{{$m_{h^0,H^0,H^\pm}$ \small [GeV]}}}
\put(13,45){\makebox(0,0)[bl]{{\small $\tanbe = 40$}}}
\put(35,51.5){\makebox(0,0)[bl]{{$m_{H^\pm}$}}}
\put(35,30){\makebox(0,0)[bl]{{$m_{H^0}$}}}
\put(35,20){\makebox(0,0)[bl]{{$m_{h^0}$}}}
\put(80,56){{\small \bf d)}}
\put(85,55){\makebox(0,0)[bl]{{$\sina$}}}
\put(100,48){\makebox(0,0)[bl]{{\small $\tanbe = 40$}}}
\put(70.5,1.5){\makebox(0,0)[br]{{$\mu$ \small [GeV]}}}
\put(70.5,62.5){\makebox(0,0)[br]{{$\mu$ \small [GeV]}}}
\put(148.5,2.5){\makebox(0,0)[br]{{$\mu$ \small [GeV]}}}
\put(148.5,64){\makebox(0,0)[br]{{$\mu$ \small [GeV]}}}
\end{picture}}
\figcaption{fig:mhiggsnlc2mu}
           {Higgs masses and mixing angle as a function of $\mu$}
           {Higgs masses and mixing angle as a function of 
            $\mu$ for $A_t = A_b = 500$~GeV, $m_{A^0}$ = 150 GeV, 
            a) and c) $\tanbe$ = 1.5, b) and d) $\tanbe = 40$.
            In (a,c) we show $m_{h^0}$, $m_{H^0}$ and $m_{H^\pm}$, 
            and in (b,d) $\sina$.
            Radiative corrections are included,
            the graphs correspond to the following ($M_D,M_Q,M_U$) 
            sets (in GeV): (700,700,700) full line, (980,700,350)
            dashed line, and (500,700,500) dashed-dotted line.}{5}
\end{minipage}

\section{Higgs masses and mixing angle}

Last but not least we present the Higgs masses and the Higgs mixing angle. In
\fig{fig:mhiggsnlc2mu} the $\mu$ dependence is shown. $m_{H^+}$ is nearly
independent of the ratio of the scalar squark masses as can be seen in 
\fig{fig:mhiggsnlc2mu}a and c. In the small $\tanbe$ case its mass varies 
between 160 and 166~GeV and is therefore somewhat smaller than its tree level
value (= 170~GeV).
For large $\tanbe$ the corrections are positive and more important. In this case
the mass can be as high as 183~GeV. The same features hold if
one varies $A_t$ for fixed $\mu$ (\fig{fig:mhiggsnlc2A}a and c).

\noindent
\begin{minipage}[t]{150mm}
{\setlength{\unitlength}{1mm}
\begin{picture}(150,123)                        
\put(-4,-10){\mbox{\epsfig{figure=mhiggsnlc2A.eps,height=14.0cm,width=15.6cm}}}
\put(0,120){{\small \bf a)}}
\put(5,119){\makebox(0,0)[bl]{{$m_{h^0,H^0,H^\pm}$ \small [GeV]}}}
\put(10,108){\makebox(0,0)[bl]{{\small $\tanbe = 1.5$}}}
\put(35,113.5){\makebox(0,0)[bl]{{$m_{H^0}$}}}
\put(35,103){\makebox(0,0)[bl]{{$m_{H^\pm}$}}}
\put(35,76.5){\makebox(0,0)[bl]{{$m_{h^0}$}}}
\put(80,117){{\small \bf b)}}
\put(85,116){\makebox(0,0)[bl]{{$\sina$}}}
\put(94,81){\makebox(0,0)[bl]{{\small $\tanbe = 1.5$}}}
\put(0,58){{\small \bf c)}}
\put(5,57){\makebox(0,0)[bl]{{$m_{h^0,H^0,H^\pm}$ \small [GeV]}}}
\put(11,47){\makebox(0,0)[bl]{{\small $\tanbe = 40$}}}
\put(35,51.5){\makebox(0,0)[bl]{{$m_{H^\pm}$}}}
\put(35,27){\makebox(0,0)[bl]{{$m_{H^0}$}}}
\put(28,15){\makebox(0,0)[bl]{{$m_{h^0}$}}}
\put(80,56){{\small \bf d)}}
\put(85,55){\makebox(0,0)[bl]{{$\sina$}}}
\put(94,48){\makebox(0,0)[bl]{{\small $\tanbe = 40$}}}
\put(70.5,1){\makebox(0,0)[br]{{$A_t (= A_b)$ \small [GeV]}}}
\put(70.5,63){\makebox(0,0)[br]{{$A_t (= A_b)$ \small [GeV]}}}
\put(148.5,1.5){\makebox(0,0)[br]{{$A_t (= A_b)$ \small [GeV]}}}
\put(148.5,65){\makebox(0,0)[br]{{$A_t (= A_b)$ \small [GeV]}}}
\end{picture}}
\figcaption{fig:mhiggsnlc2A}
           {Higgs masses and mixing angle as a function of $A_t = A_b$}
           {Higgs masses as a function of 
            $A_t = A_b$ for $\mu = - 500$~GeV, $m_{A^0}$ = 150 GeV, 
            a) and c) $\tanbe$ = 1.5, b) and d) $\tanbe = 40$.
            In (a,c) we show $m_{h^0}$, $m_{H^0}$ and $m_{H^\pm}$, 
            and in (b,d) $\sina$.
            Radiative corrections are included,
            the graphs correspond to the following ($M_D,M_Q,M_U$) 
            sets (in GeV): (700,700,700) full line, (980,700,350)
            dashed line, and (500,700,500) dashed-dotted line.}{7}
\end{minipage}

The masses and the mixing angle of the neutral scalar Higgs bosons depend on the
ratio of the scalar squark masses. One finds that with 
increasing $\tanbe$
$m_{h^0}$ increases whereas $m_{H^0}$ decreases (\fig{fig:mhiggsnlc2mu}c 
and d). $m_{H^0}$ also decreases with
increasing $|\mu|$. Note, that $|\sina|$ is rather small for large $\tanbe$ 
implying $\cosa \simeq 1$.
These general features hold also
if one studies the $A_t$ dependence except for the fact
that $m_{H^0}$ is increasing
with increasing $|A_t|$ for small $\tanbe$ (\fig{fig:mhiggsnlc2A}a and c).

%% file: nlcslept.tex
\chapter{Numerical results for sleptons}
\label{chap:nlcslept}

\section{Decays of $\protect \tsn$}
\label{sect:nlcsneut}

As mentioned in the previous chapter we start our discussion  with the 
phenomenology of the tau sneutrino because here the fewest parameters enter.
In \fig{fig:brsneutnlc2mu} branching ratios for $\tsn$ decays are shown as a 
function of $\mu$ for $M_L = 700$~GeV, $M = 350$~GeV, a) $\tanbe = 1.5$ and
b) $\tanbe = 40$. Moreover, we have chosen $A_\tau = 500$~GeV and 
$M_E \geq M_L$ to ensure that the decays $\tsn \to W^+ \, \sta{1}$ and
$\tsn \to H^+ \, \sta{1}$ are kinematically forbidden. The sneutrino mass
is 689.9~GeV (697,0~GeV) if $\tanbe = 1.5$ (40). In the case of $\tanbe = 1.5$
the decays into $\tau \, \chip{1}$ and $\nutau \, \chin{2}$ 
($\tau \, \chip{2}$ and $\nutau \, \chin{4}$) are the most important ones
if $M < |\mu|$ ($M > |\mu|$). The reason is that the $\tsn$ couples
mainly to the zino and to the wino in such a scenario. The more $\tanbe$ 
increases the more the coupling to the charged higgsino becomes important
as can be seen in \fig{fig:brsneutnlc2mu}b. It is interesting to note that
in these examples the branching ratio for the invisible mode 
$\tsn \to \nutau \, \chin{1}$ is always smaller than 20\%, leading
to much fewer one side events compared to the examples we have studied for
LEP2 (see \fig{fig:snelep15d} and the corresponding discussion).

In \fig{fig:brsneutnlc2muA} we study examples where the decays 
$\tsn \to W^+ \, \sta{1}$ and $\tsn \to H^+ \, \sta{1}$ are kinematically
allowed. We have fixed 
$M_E = 350$~GeV, $\tanbe = 40$, and the other parameters as above.
In \fig{fig:brsneutnlc2muA}a the branching ratios are shown as a function of
$\mu$ for $A_\tau = 500$~GeV. The decay $\tsn \to W^+ \, \sta{1}$ can reach
$ \sim 34\%$ for $|\mu| \sim 1$~TeV. At first glance this is a surprising fact,
because the lighter stau is mainly a right state and therefore the coupling
to the $W$-boson is rather small ($\costa \simeq 2 \cdot 10^{-4} * \mu$ 
in our example).
To understand this, let us compare this decay with $\tsn \to \tau \, \chip{1}$
which is the most important one for $|\mu| = 1$~TeV: 
\beq
 \frac{ \Gamma ( \tsn \to W^+ \, \sta{1} ) }
      { \Gamma ( \tsn \to \tau \, \chip{1} ) } \simeq
 \frac{\costaq \lamh{\mtsnq}{\mwq}{\mstaq{1}}{\frac{3}{2}} }
      {V^2_{11} \, \mwq \, (\mtsnq - \mchipq{1})^2}
 = \frac{\costaq}{V^2_{11}} C_1
\eeq
\noindent
\begin{minipage}[t]{150mm}
{\setlength{\unitlength}{1mm}
\begin{picture}(150,65)                        
\put(-6,0){\mbox{\epsfig{figure=brsneutnlc2a.eps,height=6.0cm,width=16.0cm}}}
\put(-1,60){{\small \bf a)}}
\put(3,59){\makebox(0,0)[bl]{{\small $BR(\tsn )$}}}
\put(71.5,0){\makebox(0,0)[br]{{\small $\mu$~[GeV]}}}
\put(76,60){{\small \bf b)}}
\put(80,59){\makebox(0,0)[bl]{{\small $BR(\tsn )$}}}
\put(149.5,0){\makebox(0,0)[br]{{\small $\mu$~[GeV]}}}
\end{picture}}
\figcaption{fig:brsneutnlc2mu}
           {Branching ratios for $\protect \tsn $ decays as a function 
            of $\mu$}
           {Branching ratios for $\tsn$ decays as a function of 
            $\mu$ for $M_L$ = 700~GeV, $M$ = 350~GeV,
            a) $\tanbe = 1.5$ and b) $\tanbe = 40$.
           The curves correspond to the transitions:
           $\circ \hspace{1mm} \tsn \to \nutau \, \chin{1}$,
           \rechtl \hspace{1mm} $\tsn \to \nutau \, \chin{2}$,
           $\triangle \hspace{1mm} \tsn \to \nutau \, \chin{3}$,
           $\diamondsuit \hspace{1mm} \tsn \to \nutau \, \chin{4}$,
           \recht $\tsn \to \tau \, \chip{1}$, and
           $\bullet \hspace{1mm} \tsn \to \tau \, \chip{2}$.
           The grey area
           will be covered by LEP2 ($\mchipm{1} \leq 95$~GeV).
      }{7}
\end{minipage}

\noindent
\begin{minipage}[t]{150mm}
{\setlength{\unitlength}{1mm}
\begin{picture}(150,60)                        
\put(-6,0){\mbox{\epsfig{figure=brsneutnlc2b.eps,height=6.0cm,width=16.0cm}}}
\put(-1,60){{\small \bf a)}}
\put(3,59){\makebox(0,0)[bl]{{\small $BR(\tsn )$}}}
\put(71.5,0){\makebox(0,0)[br]{{\small $\mu$~[GeV]}}}
\put(76,60){{\small \bf b)}}
\put(80,59){\makebox(0,0)[bl]{{\small $BR(\tsn )$}}}
\put(149.5,0){\makebox(0,0)[br]{{\small $A_\tau$~[GeV]}}}
\end{picture}}
\figcaption{fig:brsneutnlc2muA}
           {Branching ratios for $\protect \tsn $ decays as a function 
            of $\mu$}
           {Branching ratios for $\tsn$ decays as a function 
            a) of $\mu$ for $A_\tau$ = 500~GeV, and b) of $A_\tau$ for
            $\mu = -500$~GeV. The other parameters are $\tanbe = 40$,
            $M$ = 350~GeV, $M_L$ = 700~GeV and $M_E = 350$~GeV.
           The curves correspond to the transitions:
           $\circ \hspace{1mm} \tsn \to \nutau \, \chin{1}$,
           \rechtl \hspace{1mm} $\tsn \to \nutau \, \chin{2}$,
           $\triangle \hspace{1mm} \tsn \to \nutau \, \chin{3}$,
           $\diamondsuit \hspace{1mm} \tsn \to \nutau \, \chin{4}$,
           \recht $\tsn \to \tau \, \chip{1}$,
           $\bullet \hspace{1mm} \tsn \to \tau \, \chip{2}$,
           $\star \hspace{1mm}\tsn \to W^+ \, \sta{1}$, and
           $\Join \tsn \to H^+ \, \sta{1}$.
           The grey area
           will be covered by LEP2 ($\mchipm{1} \leq 95$~GeV).
      }{7}
\end{minipage}

\noindent
Note that $C_1$ is $\sim 50$ for $\mchip{1} \simeq M = 350$~GeV which is caused
by the factor $\inwf{\mtsnq}{\mwq}{\mstaq{1}} / \mwq$. This factor
arises from the derivative in the corresponding part of the interaction
Lagrangian (see \eqn{eq:interaction}).
Here $\lambda$ is the kinematic function $\lambda(x,y,z) = (x-y-z)^2 - 4 y z$.

Let us now compare the decay into a $W^+$ with the decay into a $H^+$.
For $|A_\tau| \gsim 50$~GeV and $\costaq < 1/2$ the ratio of the two decay 
widths
is approximately given by
\beq
\hspace*{-3mm}
 \frac{ \Gamma ( \tsn \to W^+ \, \sta{1} ) }
      { \Gamma ( \tsn \to H^+ \, \sta{1} ) }& \simeq &
 \frac{ \lamh{\mtsnq}{\mwq}{\mstaq{1}}{\frac{1}{2}} }
      { \lamh{\mtsnq}{m^2_{H^+}}{\mstaq{1}}{\frac{1}{2}} }
 \frac{ \cottaq \inwf{\mtsnq}{\mwq}{\mstaq{1}} }
      { m^2_\tau \, A^2_\tau \tanbeq } \no
 & = &
 C_2  \frac{ \cottaq \inwf{\mtsnq}{\mwq}{\mstaq{1}} }
      { m^2_\tau \, A^2_\tau \tanbeq } 
 \simeq C_2 \frac{ \mu^2 \inwf{\mtsnq}{\mwq}{\mstaq{1}} }
                 { A^2_\tau (M^2_L - \mstaq{1})^2 }
\label{eq:ratiogam1}
\eeq
where $C_2 \simeq 1.07$ for the examples shown in \fig{fig:brsneutnlc2muA}.
Especially the last relation of \eqn{eq:ratiogam1} explains the relative
importance of $\tsn \to W^+ \, \sta{1}$ to $\tsn \to H^+ \, \sta{1}$ in
\fig{fig:brsneutnlc2muA}.
Note, that in \fig{fig:brsneutnlc2muA}b all partial decay widths are nearly 
independent of $A_\tau$ except the one into $H^+$.

Let us now turn to the question, which cascade decays are induced by these
decay modes.
For most cases one has to keep in mind that a right stau decays mainly into
a tau-lepton and the lightest neutralino (see \fig{fig:brstau1nlc2mu}d of the 
next section). Therefore, the decay 
chains are:
\beq
\tsn \to W^+ \, \sta{1} \to W^+ \, \tau \, \chin{1} \to
\left\{ \begin{array}{lr} q \, q' \, \tau \, \chin{1} & (\sim 70 \%) \\
                         l^+ \, \tau \, \nu_l \, \chin{1} & (\sim 30 \%)
        \end{array} \right. \\
\tsn \to H^+ \, \sta{1} \to H^+  \, \tau \, \chin{1} \to
\left\{ \begin{array}{lr}
               \tau^+  \tau \, \nutau \, \chin{1} & (\sim 90 \%) \\
               c \, \bar{s} \, \tau \, \chin{1} & (\sim 10 \%)
        \end{array} \right. 
\eeq
For the discussion of $H^+$
decays in this mass range see for example \cite{Sopczak93}.
As can be seen in \fig{fig:brstau1nlc2mu}d the lighter stau can also decay
into charginos and the other neutralinos if $|\mu| \lsim M$ leading to
additional quarks and/or leptons for these final states.

\section{Decays of $\protect \sta{1}$}

In \fig{fig:brstau1nlc2mu} we show branching ratios for $\sta{1}$ decays as a 
function of $\mu$ for various scenarios. We have seen in \chap{chap:nlcmass}
that the nature of the staus is mainly determined by $M_E$, $M_L$ and the
product $\mu \tanbe$. For the following discussion we have fixed 
$A_\tau = 500$~GeV. In \fig{fig:brstau1nlc2mu}a a
maximal mixing scenario for $\tanbe = 1.5$ is realized except a small range
near $\mu \simeq 333$~GeV where $\sta{1} \simeq \sta{R}$. We have chosen
$M_L = M_E = 700$~GeV. For $|\mu| \gsim 200$~GeV the decay into 
$\tau \, \chin{1}$ dominates. Note, that the decay into $\nutau \, \chim{1}$
has nearly the same branching ratio for a wide range of $\mu$. For large
$\tanbe$ this decay becomes more important (\fig{fig:brstau1nlc2mu}b). 
This results from the increasing Yukawa coupling and the
interferences between the gaugino and the higgsino components of the chargino 
in the corresponding decay widths.

\noindent
\begin{minipage}[t]{150mm}
{\setlength{\unitlength}{1mm}
\begin{picture}(150,173)                        
\put(-6,-22){\mbox{\epsfig{figure=brstau1nlc2mu.eps,
                           height=21.5cm,width=15.6cm}}}
\put(-2,171){{\small \bf a)}}
\put(3,170){\makebox(0,0)[bl]{{$BR(\sta{1})$ }}}
\put(75,171){{\small \bf b)}}
\put(80,170){\makebox(0,0)[bl]{{$BR(\sta{1})$}}}
\put(-2,113){{\small \bf c)}}
\put(2,112){\makebox(0,0)[bl]{{$BR(\sta{1})$}}}
\put(75,113){{\small \bf d)}}
\put(80,112){\makebox(0,0)[bl]{{$BR(\sta{1})$}}}
\put(-2,56){{\small \bf e)}}
\put(2,55){\makebox(0,0)[bl]{{$BR(\sta{1})$}}}
\put(75,58){{\small \bf f)}}
\put(80,57){\makebox(0,0)[bl]{{$BR(\sta{1})$}}}
\put(67.5,1){\makebox(0,0)[br]{{$\mu$ \small [GeV]}}}
\put(69,58){\makebox(0,0)[br]{{$\mu$ \small [GeV]}}}
\put(69.5,115){\makebox(0,0)[br]{{$\mu$ \small [GeV]}}}
\put(147.5,1){\makebox(0,0)[br]{{$\mu$ \small [GeV]}}}
\put(147,58){\makebox(0,0)[br]{{$\mu$ \small [GeV]}}}
\put(147.5,115){\makebox(0,0)[br]{{$\mu$ \small [GeV]}}}
\end{picture}}
\figcaption{fig:brstau1nlc2mu}
           {Branching ratios for $\protect \sta{1}$ decays as a
            function of $\mu$}
           {Branching ratios for $\sta{1}$ decays as a function of 
            $\mu$ for $A_\tau$ = 500~GeV, $M$ = 350~GeV, $M_L$ = 700~GeV,
            in a) and b) $M_E$ = 700~GeV, 
            in c) and d) $M_E$ = 350~GeV, and
            in e) and f) $M_E$ = 980~GeV. 
           In a), c) and e) $\tanbe = 1.5$ and in b), d) and f)
           $\tanbe$ = 40.
           The curves correspond to the following transitions:
           $\circ \hspace{1mm} \sta{1} \to \tau \, \chin{1}$,
           \rechtl \hspace{1mm} $\sta{1} \to \tau \, \chin{2}$,
           $\triangle \hspace{1mm} \sta{1} \to \tau \, \chin{3}$,
           $\diamondsuit \hspace{1mm} \sta{1} \to \tau \, \chin{4}$,
           \recht $\sta{1} \to \nutau \, \chim{1}$, and
           $\bullet \hspace{1mm} \sta{1} \to \nutau \, \chim{2}$. 
           The grey area
           will be covered by LEP2 ($\mchipm{1} \leq 95$~GeV).
      }{5}
\end{minipage}

\newpage

In \fig{fig:brstau1nlc2mu}c ($\tanbe = 1.5$) and \fig{fig:brstau1nlc2mu}d
($\tanbe = 40$) we show the situation $\sta{1} \simeq \sta{R}$. We have
taken $M_E = 350$~GeV and the other parameters as above. 
The coupling to the bino is the dominating one for small $\tanbe$. Therefore,
the decay into $\chin{1}$ ($\chin{3}$) is the most important one 
for $|\mu| \gsim 200$~GeV
($|\mu| \lsim 200$~GeV). Note, that for 200~GeV $\lsim \mu \lsim 350$~GeV
also the decay into $\chin{2}$ gains some importance. The results
for small $\tanbe$ are
nearly independent of the flavour. Therefore, the branching ratios for
$\sel{R}$- and $\tilde \mu_R$-decays are nearly the same (see e.g.
\cite{Bartl92b}).
The differences between small and large $\tanbe$ scenarios 
(\fig{fig:brstau1nlc2mu}c and d) are again due to the different Yukawa 
couplings.

Let us now turn to the case $\sta{1} \simeq \sta{L}$ 
(\fig{fig:brstau1nlc2mu}e and \ref{fig:brstau1nlc2mu}f). Here we have
chosen $M_E = 980$~GeV and the other parameters as above. 
The couplings to the zino and the wino dominate if $\tanbe = 1.5$. The most 
important decays are
those into charginos followed by the decays into $\chin{2}$ ($M \lsim |\mu|$)
and $\chin{4}$ ($M \gsim |\mu|$). Similar as in the above example the 
couplings to the higgsino 
components become important for $\tanbe = 40$ and $|\mu| \lsim M$ 
(\fig{fig:brstau1nlc2mu}f).

\section{Decays of $\protect \sta{2}$}

In \fig{fig:brstau2anlc2mu} the branching ratios for $\sta{2}$ decays are
shown as a function of $\mu$ for $M_E = M_L = 700$~GeV, $A_\tau = 500$~GeV,
and $M = 350$~GeV. In \fig{fig:brstau2anlc2mu}a we have chosen $\tanbe = 1.5$. 
Comparing \fig{fig:brstau2anlc2mu}a with \fig{fig:brstau1nlc2mu}a one 
notices that the branching ratio for the various decays of $\sta{1}$ and
$\sta{2}$
are nearly the same for a wide range of $\mu$. Here it is likely that very 
accurate 
measurements will be needed to separate the signals of the staus and to 
determine their properties. Only in the range where
$\mu \simeq A_\tau \tanbe$ does the heavier stau have completely different 
properties because the mass eigenstates are nearly identical with the
weak eigenstates.

Let us now see how the phenomenology differs by changing the ratio $M_L : M_E$.
In \fig{fig:brstau2bnlc2mu} the branching ratios are shown 
as function of $\mu$ for $M_L = 700$~GeV, $M_E = 350$~GeV,  $m_{A^0} = 150$~GeV 
and $\tanbe = 40$. As a qualitatively new feature the decays into the Z boson 
and neutral Higgs bosons are possible.
In \fig{fig:brstau2bnlc2mu}a the branching ratios for the decays 
into fermions are shown. For $|\mu| > M$ the decay into $\nutau \, \chim{1}$
dominates followed by $\tau \, \chin{2}$ and $\tau \, \chin{1}$.
The decays into $\chim{2}, \, \chin{3,4}$ become kinematically
possible if $|\mu| \lsim M$.  Especially the decay into the heavier chargino 
gets important in
that range. The importance of these decays is expected because $\sta{2}$ is
mainly a left state.
In \fig{fig:brstau2bnlc2mu}b the branching ratios for the decays into bosons 
are shown.
Their sum varies between $\sim 10$\%
and $\sim 40$\%. The decays into $Z \, \sta{1}$ and $h^0 \, \sta{1}$ are
important for large $\mu$ where the branching ratios go up to $\sim 18\%$ and
$\sim 14\%$ respectively. 
Their varying importance can be understood by having a look on the 
\noindent
\begin{minipage}[t]{150mm}
{\setlength{\unitlength}{1mm}
\begin{picture}(150,63)                        
\put(-6,-1){\mbox{\epsfig{figure=brstau2anlc2mu.eps,height=6.0cm,width=16.0cm}}}
\put(-1,59){{\small \bf a)}}
\put(4,58){\makebox(0,0)[bl]{{\small $BR(\sta{2} )$}}}
\put(73.5,0){\makebox(0,0)[br]{{\small $\mu$~[GeV]}}}
\put(77,59){{\small \bf b)}}
\put(82,58){\makebox(0,0)[bl]{{\small $BR(\sta{2} )$}}}
\put(149.5,0){\makebox(0,0)[br]{{\small $\mu$~[GeV]}}}
\end{picture}}
\figcaption{fig:brstau2anlc2mu}
           {Branching ratios for $\protect \sta{2} $ decays as a
            function of $\mu$}
           {Branching ratios for $\sta{2}$ decays as a function 
            of $\mu$ for $A_\tau$ = 500~GeV, $M = 350$~GeV,
            $m_{A^0} = 150$~GeV, $M_L = M_E = 700$~GeV,
            a) $\tanbe = 1.5$ and b) $\tanbe = 40$.
           The curves correspond to the transitions:
           $\circ \hspace{1mm} \sta{2} \to \tau \, \chin{1}$,
           \rechtl \hspace{1mm} $\sta{2} \to \tau \, \chin{2}$,
           $\triangle \hspace{1mm} \sta{2} \to \tau \, \chin{3}$,
           $\diamondsuit \hspace{1mm} \sta{2} \to \tau \, \chin{4}$,
           \recht $\sta{2} \to \nutau \, \chim{1}$, and
           $\bullet \hspace{1mm} \sta{2} \to \nutau \, \chim{2}$.
           The grey area
           will be covered by LEP2 ($\mchipm{1} \leq 95$~GeV).
      }{7}
\end{minipage}

\noindent
\begin{minipage}[t]{150mm}
{\setlength{\unitlength}{1mm}
\begin{picture}(150,63)
\put(-6,-6.0){\mbox{\epsfig{figure=brstau2bnlc2mu.eps,
                           height=6.9cm,width=15.8cm}}}
\put(-1,57){{\small \bf a)}}
\put(3,56){\makebox(0,0)[bl]{{$BR(\sta{2})$}}}
\put(76,57){{\small \bf b)}}
\put(82,56){\makebox(0,0)[bl]{{$BR(\sta{2})$}}}
\put(70.5,0.5){\makebox(0,0)[br]{{$\mu$ \small [GeV]}}}
\put(149.5,0.5){\makebox(0,0)[br]{{$\mu$ \small [GeV]}}}
\end{picture}}
\figcaption{fig:brstau2bnlc2mu}
           {Branching ratios for $\protect \sta{2} $ decays as a
            function of $\mu$ for $\protect \tanbe = 40$}
           {Branching ratios for $\sta{2}$ decays as a function of 
            $\mu$ for $\tanbe = 40$, $A_\tau$ = 500~GeV, $M$ = 350~GeV,
            $m_{A^0} = 150$~GeV, $M_L$ = 700~GeV, and $M_E$ = 350~GeV.
           The curves in a) correspond to:
           $\circ \hspace{1mm} \sta{2} \to \tau \, \chin{1}$,
           \rechtl \hspace{1mm} $\sta{2} \to \tau \, \chin{2}$,
           $\triangle \hspace{1mm} \sta{2} \to \tau \, \chin{3}$,
           $\diamondsuit \hspace{1mm} \sta{2} \to \tau \, \chin{4}$,
           \recht $\sta{2} \to \nutau \, \chim{1}$, and
           $\bullet \hspace{1mm} \sta{2} \to \nutau \, \chim{2}$.
           The curves in b) correspond to:
           $\circ \hspace{1mm} \sta{2} \to Z^0 \, \sta{1}$,
           \rechtl \hspace{1mm} $\sta{2} \to h^0 \, \sta{1}$,
           $\triangle \hspace{1mm} \sta{2} \to H^0 \, \sta{1}$, and
           $\diamondsuit \hspace{1mm} \sta{2} \to A^0 \, \sta{1}$.
           The grey area
           will be covered by LEP2 ($\mchipm{1} \leq 95$~GeV).
      }{7}
\end{minipage}

\noindent
\begin{minipage}[t]{150mm}
{\setlength{\unitlength}{1mm}
\begin{picture}(150,62)
\put(-6,-6){\mbox{\epsfig{figure=brstau2nlc2A.eps,
                           height=6.9cm,width=15.9cm}}}
\put(-1,57.5){{\small \bf a)}}
\put(4,56.5){\makebox(0,0)[bl]{{$BR(\sta{2})$}}}
\put(76,58){{\small \bf b)}}
\put(82,57){\makebox(0,0)[bl]{{$BR(\sta{2})$}}}
\put(70.5,1.5){\makebox(0,0)[br]{{$A_\tau$ \small [GeV]}}}
\put(149.5,0.5){\makebox(0,0)[br]{{$A_\tau$ \small [GeV]}}}
\end{picture}}
\figcaption{fig:brstau2nlc2A}
           {Branching ratios for $\protect \sta{2} $ decays as a
            function of $A_\tau$ for $\protect \tanbe = 40$}
           {Branching ratios for $\sta{2}$ decays as a function of 
            $A_\tau$ for $\tanbe = 40$, $\mu = -500$~GeV, $M$ = 350~GeV,
            $m_{A^0} = 150$~GeV, $M_L$ = 700~GeV, and $M_E$ = 350~GeV.
           The curves in a) correspond to:
           $\circ \hspace{1mm} \sta{2} \to \tau \, \chin{1}$,
           \rechtl \hspace{1mm} $\sta{2} \to \tau \, \chin{2}$,
           $\triangle \hspace{1mm} \sta{2} \to \tau \, \chin{3}$,
           $\diamondsuit \hspace{1mm} \sta{2} \to \tau \, \chin{4}$,
           \recht $\sta{2} \to \nutau \, \chim{1}$, and
           $\bullet \hspace{1mm} \sta{2} \to \nutau \, \chim{2}$.
           The curves in b) correspond to:
           $\circ \hspace{1mm} \sta{2} \to Z^0 \, \sta{1}$,
           \rechtl \hspace{1mm} $\sta{2} \to h^0 \, \sta{1}$,
           $\triangle \hspace{1mm} \sta{2} \to H^0 \, \sta{1}$,
           $\diamondsuit \hspace{1mm} \sta{2} \to A^0 \, \sta{1}$,
           \recht $\sta{2} \to W^- \, \tsn$,  and
           $\bullet \hspace{1mm} \sta{2} \to H^- \, \tsn$.
      }{7}
\end{minipage}

\noindent
$\mu$ dependence
of the couplings (see also \eqn{eq:coupHsf1sf2a} and \eqn{decaygl5}):
\beq
\begin{array}{ll}
Z \sta{2} \sta{1} \sim \sinzta  &
h^0 \sta{2} \sta{1} \sim \coszta (A_\tau \sina + \mu \cosa) \\
A^0 \sta{2} \sta{1} \sim (A_\tau \tanbe + \mu) &
H^0 \sta{2} \sta{1} \sim \coszta (-A_\tau \cosa + \mu \sina)
\end{array}
\label{eq:coupneutrstau}
\eeq
Here we have only considered the most
important parts of the couplings to the Higgs bosons.
From \eqn{eq:coupneutrstau} the ''parabolic'' form of the branching ratios into 
$Z \, \sta{1}$ and $h^0 \, \sta{1}$ follows ($\cosa \simeq \pm 1$)
whereas the branching ratios for the decays into $H^0 \, \sta{1}$ and
$A^0 \, \sta{1}$ show only a mild dependence on $\mu$. 
Note, that the appearance of $\sinzta$ in the $Z \, \sta{2} \, \sta{1}$
coupling corresponds to the fact that the $Z \, \tau \, \tau$ coupling
preserves the chirality of the $\tau$. In the same way $\coszta$
corresponds to the chirality change of the $\tau$-lepton in the Higgs tau 
coupling.
In \fig{fig:brstau2nlc2A}b we show the branching ratios as a function of 
$A_\tau$ for $\mu = -500$~GeV and the other
parameters as above. Here the decays into $H^0 \, \sta{1}$ and $A^0 \, \sta{1}$
become important for large $|A_\tau|$. 
Note that $BR(\sta{2} \to h^0 \, \sta{1})$ is very small for large negative 
$A_\tau$ whereas at the same time $BR(\sta{2} \to H^0 \, \sta{1})$ increases.
This can be understood by having a look at \fig{fig:mhiggsnlc2A}d 
where we can see that $\sina \simeq -0.4$ for $A_\tau \lsim -850$~GeV.
Now the signs in front of $A_\tau$ in the couplings come into the
play (\eqn{eq:coupneutrstau}) leading to a reduction of the coupling in the 
case of $h^0$ and an increase in case of $H^0$.  
Note, that the partial widths for the decays into charginos and neutralinos
are nearly independent 
of $A_\tau$. The variation of the corresponding 
branching ratios
in \fig{fig:brstau2nlc2A}a is mainly due to the variation of the partial widths
for the decays into bosons.

\noindent
\begin{minipage}[t]{150mm}
{\setlength{\unitlength}{1mm}
\begin{picture}(150,63)
\put(-6,-5.0){\mbox{\epsfig{figure=brstau2cnlc2mu.eps,
                           height=6.9cm,width=15.8cm}}}
\put(-1,58.5){{\small \bf a)}}
\put(4,57.5){\makebox(0,0)[bl]{{$BR(\sta{2})$}}}
\put(76,59.0){{\small \bf b)}}
\put(82,58.0){\makebox(0,0)[bl]{{$BR(\sta{2})$}}}
\put(70.5,0.5){\makebox(0,0)[br]{{$\mu$ \small [GeV]}}}
\put(149.5,1){\makebox(0,0)[br]{{$\mu$ \small [GeV]}}}
\end{picture}}
\figcaption{fig:brstau2cnlc2mu}
           {Branching ratios for $\protect \sta{2} $ decays as a
            function of $\mu$ for $\protect \tanbe = 40$}
           {Branching ratios for $\sta{2}$ decays as a function of 
            $\mu$ for $\tanbe = 40$, $A_\tau$ = 500~GeV, $M$ = 350~GeV,
            $m_{A^0} = 150$~GeV, $M_L$ = 700~GeV, and $M_E$ = 980~GeV.
           The curves in a) correspond to:
           $\circ \hspace{1mm} \sta{2} \to \tau \, \chin{1}$,
           \rechtl \hspace{1mm} $\sta{2} \to \tau \, \chin{2}$,
           $\triangle \hspace{1mm} \sta{2} \to \tau \, \chin{3}$,
           $\diamondsuit \hspace{1mm} \sta{2} \to \tau \, \chin{4}$,
           \recht $\sta{2} \to \nutau \, \chim{1}$, and
           $\bullet \hspace{1mm} \sta{2} \to \nutau \, \chim{2}$.
           The curves in b) correspond to:
           $\circ \hspace{1mm} \sta{2} \to Z^0 \, \sta{1}$,
           \rechtl \hspace{1mm} $\sta{2} \to h^0 \, \sta{1}$,
           $\triangle \hspace{1mm} \sta{2} \to H^0 \, \sta{1}$,
           $\diamondsuit \hspace{1mm} \sta{2} \to A^0 \, \sta{1}$,
           \recht $\sta{2} \to W^- \, \tsn$,  and
           $\bullet \hspace{1mm} \sta{2} \to H^- \, \tsn$.
           The grey area
           will be covered by LEP2 ($\mchipm{1} \leq 95$~GeV).
      }{7}
\end{minipage}

Let us now turn to the question, which decay chains are induced by 
these decay modes:
\beq
\sta{2} &\to& Z \, \sta{1} \to Z \, \tau \, \chin{1} \to
\left\{ \begin{array}{lr} q \, \bar{q} \, \tau \, \chin{1} & (\sim 70 \%) \\
                         l^+ \, l^- \, \tau \, \chin{1} & (\sim 10 \%)    \\
                         \nu_l \, \bar{\nu}_l \, \tau \, \chin{1} & (\sim 20 \%)
        \end{array} \right. \\
\sta{2} &\to& h^0 \, \sta{1} \to h^0 \, \tau \, \chin{1} \to
\left\{ \begin{array}{lr}
    \left. \begin{array}{lr}
         b \, \bar{b} \, \tau \, \chin{1} & (\sim 96 \%) \\
         \tau^+ \, \tau^- \, \tau \, \chin{1} & (\sim 4 \%) 
        \end{array}  \right\} \mathrm{if} \, \, |\sina| \gsim 0.05 \\
    \left. \begin{array}{lr}
         W \, W^* \, \tau \, \chin{1} & (\sim 90 \%) \\
         Z \, Z^* \, \tau \, \chin{1} & (\sim 10 \%) 
        \end{array}  \right\} \mathrm{if} \, \, |\sina| \lsim 0.01 
   \end{array}  \right. \\
\sta{2} &\to& H^0 \, \sta{1} \to H^0 \, \tau \, \chin{1} \to
\left\{ \begin{array}{lr}
            b \, \bar{b} \, \tau \, \chin{1} & (\sim 96 \%) \\
         \tau^+ \, \tau^- \, \tau \, \chin{1} & (\sim 4 \%) 
        \end{array} \right.  \\
\sta{2} &\to& A^0 \, \sta{1} \to A^0 \, \tau \, \chin{1} \to
\left\{ \begin{array}{lr}
            b \, \bar{b} \, \tau \, \chin{1} & (\sim 96 \%) \\
         \tau^+ \, \tau^- \, \tau \, \chin{1} & (\sim 4 \%) 
        \end{array}\right.  
\eeq
As mentioned above  the lighter stau can also decay
into charginos and the other neutralinos if $|\mu| \lsim M$
(\fig{fig:brstau1nlc2mu}d) leading to
additional quarks and/or leptons for these final states.

In \fig{fig:brstau2cnlc2mu} the branching ratios are shown 
for the case $M_E > M_L$ as a function of $\mu$. Here we fix $M_E = 980$~GeV, 
$M_L = 700$~GeV, $A_\tau = 500$~GeV and the other parameters as above. 
Therefore, $\sta{2}$ is nearly a right state and the decay into
the lightest neutralino dominates except for a small range where $\chin{3}$ is
mainly a bino (\fig{fig:brstau2cnlc2mu}a). 
The decays into higgsino-like particles also have sizable branching ratios
due to the Yukawa couplings.
Note, that in this case the influence of the Yukawa couplings  is much
stronger than in the case of $M_E < M_L$ because now they only compete with
the gauge coupling to the bino. 
As a new feature the decays into $W^- \, \tsn$ and $H^- \, \tsn$ are possible
(\fig{fig:brstau2cnlc2mu}b). Especially the 
decay into $W^- \, \tsn$ becomes important for large $|\mu|$ because $|\sinta|$
becomes larger. The sum of the
branching ratios into bosons varies between $\sim 6\%$ ($|\mu| < 150$~GeV)
and $\sim 54\%$ ($|\mu| = 1$~TeV). Beside the decay into $W^+$, also the
decays into $Z \, \sta{1}$ and $h^0 \, \sta{1}$ get sizable branching
ratios for large
$\mu$.

%% file: nlcstop.tex
\chapter{Numerical results for $\protect \sto{1,2}$}
\label{chap:nlcstop}

\section{Decays of $\protect \sto{1}$}
\label{sect:stopnlc}

In \chap{chap:lep2} we have seen that in the mass range of LEP2 the lighter
stop has only one two body decay mode at tree level, namely 
$\sto{1} \to b \, \chip{1}$. For larger stop masses the decays into 
$t \, \chin{i}$ are possible if $\mstop{1} \gsim min(M/2,|\mu|) + m_t$. In
\fig{fig:brstop1anlc2mu}a branching ratios are shown as a function of $\mu$ for
$M_D = M_Q = M_U = 700$~GeV, $A_b = A_t = 500$~GeV, $M = 350$~GeV, 
$\tanbe = 1.5$ and $\ma = 150$~GeV. For $|\mu| \gsim \mstop{1}$ only the decays
into $\chip{1}$ and $\chin{1,2}$ are kinematically allowed, where the decay into
$\chip{1}$ is dominating. Note, that the kinks near $\mu = 750$~GeV are
due to the fact that $\sto{1} = \sto{L}$ for this specific value. Moreover, the
sign of $\cost$ changes at this point leading to an interchange between
positive and negative interferences of the gaugino and higgsino couplings.
For $|\mu| \lsim \mstop{1}$ the other charginos and neutralinos also enter the
game. Here complicated decay chains are expected. In this parameter range the 
different branching ratios depend strongly on $\mu$ as a result of various 
interferences between  the gaugino and higgsino couplings. 
Moreover, kinematics changes from
both sides: light stop and charginos/neutralinos. In addition 
the nature of the charginos and neutralinos depends strongly on the sign 
of $\mu$ for low $\tanbe$.

In \fig{fig:brstop1anlc2mu}b the branching ratios are shown for $\tanbe = 40$
and the other parameters as above. Here $\mstop{1}$ and $\cost$ are nearly
independent of $\mu$. For the decays into charginos also the bottom Yukawa
coupling becomes important leading to an enhancement of the sum of both
partial decay widths. The branching ratio for $\sto{1} \to b \, \chip{1}$
($b \, \chip{2}$) can reach 60\% (55\%). For large $\mu$ the decays into
$t \, \chin{1,2}$ have branching ratios of $\sim 20\%$. Note, that for 
$|\mu| \sim 350$~GeV the decay into $t \, \chin{4}$ can reach 25\% being the
dominant one among the
decays into neutralinos. As a new feature compared to the previous 
case, the decay $\sto{1} \to W^+ \, \sbo{1}$ becomes kinematically possible
for large $\mu$ (\fig{fig:msbotnlc2mu}). The branching ratio
can reach 20\% because there is a strong mixing in the stop sector as well 
as in the sbottom sector. The $\sbo{1}$ decays further into $t \chim{1}$
($\sim 30\%$), $b \, \chin{1}$ ($\sim 25\%$) and $b \, \chin{2}$ ($\sim 45\%$)
(see \fig{fig:brsbot1anlc2mu}b for details).

\noindent
\begin{minipage}[t]{150mm}
{\setlength{\unitlength}{1mm}
\begin{picture}(150,130)                        
\put(-7,-6){\mbox{\epsfig{figure=brstop1anlc2mu.eps,
                           height=14.0cm,width=15.8cm}}}
\put(-4,127.5){{\small \bf a)}}
\put(1,126.5){\makebox(0,0)[bl]{{$BR(\sto{1})$}}}
\put(74,127.5){{\small \bf b)}}
\put(79,126.5){\makebox(0,0)[bl]{{$BR(\sto{1})$}}}
\put(-4,63){{\small \bf c)}}
\put(2,62){\makebox(0,0)[bl]{{$BR(\sto{1})$}}}
\put(74,63){{\small \bf d)}}
\put(79,62){\makebox(0,0)[bl]{{$BR(\sto{1})$}}}
\put(69.5,1.5){\makebox(0,0)[br]{{$\mu$ \small [GeV]}}}
\put(69.5,65.2){\makebox(0,0)[br]{{$\mu$ \small [GeV]}}}
\put(149,1){\makebox(0,0)[br]{{$\mu$ \small [GeV]}}}
\put(149,65.5){\makebox(0,0)[br]{{$\mu$ \small [GeV]}}}
\end{picture}}
\figcaption{fig:brstop1anlc2mu}
           {Branching ratios for $\protect \sto{1}$ decays as a
            function of $\mu$}
           {Branching ratios for $\sto{1}$ decays as a function of 
      $\mu$ for $A_t = A_b$ = 500~GeV, $M$ = 350~GeV, $M_Q$ = 700~GeV,
            in a) and b) $M_U = M_D$ = 700~GeV, 
            and in c) and d) $M_U$ = 350~GeV, $M_D$ = 980~GeV.
           In a), and c) $\tanbe = 1.5$ and in b), and d)
           $\tanbe$ = 40.
           The curves correspond to the following transitions:
           $\circ \hspace{1mm} \sto{1} \to t \, \chin{1}$,
           \rechtl \hspace{1mm} $\sto{1} \to t \, \chin{2}$,
           $\triangle \hspace{1mm} \sto{1} \to t \, \chin{3}$,
           $\diamondsuit \hspace{1mm} \sto{1} \to t \, \chin{4}$,
           \recht $\sto{1} \to b \, \chip{1}$, and
           $\bullet \hspace{1mm} \sto{1} \to b \, \chip{2}$. In addition
           $\star$ in b) denotes the transition
           $\sto{1} \to W^+ \, \sbo{1}$, and in c) $\star$ the
           transition $\sto{1} \to W^+ \, b \, \chin{1}$ whereas
           $\Join$ denotes $\sto{1} \to c \, \chin{1}$. 
           The grey area
           will be covered by LEP2 ($\mchipm{1} \leq 95$~GeV).
      }{7}
\end{minipage}

In \fig{fig:brstop1anlc2mu}c we show the branching ratios for $M_D = 980$~GeV,
$M_U = 350$~GeV, $\tanbe = 1.5$ and the other parameters as above. 
With this parameter choice $\sto{1}$ is mainly $\sto{R}$.
For $\mu \lsim -300$~GeV only higher order decays are possible. 
The dominant decay mode is
$\sto{1} \to b \, W^+ \, \chin{1}$ wherever it is kinematically possible.
The decay mode $\sto{1} \to c \, \chin{1}$ is dominant if all three 
body decay modes are kinematically forbidden. 
A more detailed discussion of these decays will be given in 
\chap{chap:rardecay} to which we refer for further details. 
As can be seen in \fig{fig:gstop1nlc2mu}a the total decay width can be 
\begin{minipage}[t]{150mm}
{\setlength{\unitlength}{1mm}
\begin{picture}(150,58)                        
\put(-5,-4){\mbox{\epsfig{figure=gstop1nlc2mu.eps,height=6.2cm,width=15.8cm}}}
\put(-3,56){{\small \bf a)}}
\put(3,55){\makebox(0,0)[bl]{{$\log_{10}(\Gamma_i(\sto{1}))$ \small [GeV]}}}
\put(70.5,1){\makebox(0,0)[br]{{$\mu$ \small [GeV]}}}
\put(76,55){{\small \bf b)}}
\put(82,54){\makebox(0,0)[bl]
               {{$\log_{10}(\Gamma_i(\sto{1}))$ \small [GeV]}}}
\put(149,1){\makebox(0,0)[br]{{$\mu$ \small [GeV]}}}
\end{picture}}
\figcaption{fig:gstop1nlc2mu}
           {Total decay widths of $\protect \sto{1}$ as a function
            of $\mu$}
           {Total decay widths of $\sto{1}$ as function of 
            $\mu$ for $A_t=A_b=500$~GeV, $m_{A^0}$ = 150 GeV, 
            $M = 350$~GeV, a) $\tanbe = 1.5$ and b) $\tanbe = 40$.
            The graphs correspond to the following ($M_D,M_Q,M_U$) sets
            (in GeV): full line (700,700,700), dashed line 
            (980,700,350), and dashed-dotted line (500,700,500).
            }{7}
\end{minipage}

\noindent
much smaller
than 0.2~GeV for this parameter range. Therefore, hadronization effects
will become important. 
The remaining part of \fig{fig:brstop1anlc2mu} can be understood by kinematics 
and the fact that a right stop couples only to the bino and and the higgsinos.
This leads to the dominance of 
$\sto{1} \to t \, \chin{1}$ for $\mu \gsim 500$~GeV, because there the lightest
neutralino is mainly a bino and the lighter chargino is mainly a wino. 
\noindent
For $|\mu| \lsim 500$~GeV the decay into 
$b \, \chip{1}$ dominates due to the strong coupling to the charged 
higgsino and as a result of kinematics. The situation is similar for large
$\tanbe$  except that now tree level decays are kinematically allowed over the 
whole $\mu$ range considered (\fig{fig:brstop1anlc2mu}d). 

Let us have a short look on the decay widths shown in \fig{fig:gstop1nlc2mu}.
In general the decay width is too large for hadronization effects to become
important. The obvious exception is that only higher order decays are allowed
as mentioned above (\fig{fig:gstop1nlc2mu}a). For $M_U = 350$~GeV, 
$|\mu| \gsim 500$~GeV and
$\tanbe = 40$ (dashed line of \fig{fig:gstop1nlc2mu}b) the decay width is in
the order of 0.1~GeV. Here remains of a hadronization process could be
visible although tree level two body decays are allowed and 
$\mstop{1} \simeq 360$~GeV. 

In \fig{fig:brstop1anlc2A}a we study the dependence of the branching ratios on
$A_t (=A_b)$ for $M_D = M_Q = M_U = 700$~GeV, $\mu = -500$~GeV
and $\tanbe = 1.5$. With this the chargino/neutralino sector is completely fixed
and only the properties of the stop are varied. At 
$A_t = \mu \cotbe = 1000/3$~GeV the lighter stop is a pure left state and here
also the sign of $\cost$ changes. The stop mass and the absolute value of $\cost$
are symmetric with 
respect to this point (\fig{fig:mstopnlc2A}) but not the branching
ratios. Therefore, it should be possible to
determine the sign of $\cost$ and $A_t$ in such a scenario from the branching 
ratios once $\mu$ and $\tanbe$ are known. The assumption $A_t = A_b$ is not
restrictive in this case, because $A_b$ does not enter in the two body 
\noindent
\begin{minipage}[t]{150mm}
{\setlength{\unitlength}{1mm}
\begin{picture}(150,130)                        
\put(-7,-6){\mbox{\epsfig{figure=brstop1anlc2A.eps,
                           height=14.0cm,width=15.8cm}}}
\put(-4,127.5){{\small \bf a)}}
\put(1,126.5){\makebox(0,0)[bl]{{$BR(\sto{1})$}}}
\put(74,127.5){{\small \bf b)}}
\put(79,126.5){\makebox(0,0)[bl]{{$BR(\sto{1})$}}}
\put(-4,63){{\small \bf c)}}
\put(2,62){\makebox(0,0)[bl]{{$BR(\sto{1})$}}}
\put(74,63){{\small \bf d)}}
\put(79,62){\makebox(0,0)[bl]{{$BR(\sto{1})$}}}
\put(69.5,1.5){\makebox(0,0)[br]{{$A_t (= A_b)$ \small [GeV]}}}
\put(69.5,65.5){\makebox(0,0)[br]{{$A_t (= A_b)$ \small [GeV]}}}
\put(148.5,1){\makebox(0,0)[br]{{$A_t (= A_b)$ \small [GeV]}}}
\put(148.5,65.5){\makebox(0,0)[br]{{$A_t (= A_b)$ \small [GeV]}}}
\end{picture}}
\figcaption{fig:brstop1anlc2A}
           {Branching ratios for $\protect \sto{1}$ decays as a
            function of $A_t (= A_b)$}
           {Branching ratios for $\sto{1}$ decays as a function of 
   $A_t (= A_b)$ for $\mu = -500$~GeV, $M$ = 350~GeV, $M_Q$ = 700~GeV,
            in a) and b) $M_U = M_D$ = 700~GeV, and
            in c) and d) $M_U$ = 350~GeV, $M_D = 980$~GeV.
           In a), and c) $\tanbe = 1.5$ and in b), and d) $\tanbe$ = 40.
           The curves in a), and b) correspond to the transitions:
           $\circ \hspace{1mm} \sto{1} \to t \, \chin{1}$,
           \rechtl \hspace{1mm} $\sto{1} \to t \, \chin{2}$,
           $\triangle \hspace{1mm} \sto{1} \to t \, \chin{3}$,
           $\diamondsuit \hspace{1mm} \sto{1} \to t \, \chin{4}$,
           \recht $\sto{1} \to b \, \chip{1}$, 
           $\bullet \hspace{1mm} \sto{1} \to b \, \chip{2}$, and
           $\star \sto{1} \to W^+ \sbo{1}$.
           The curves in c) and d) correspond to the transitions: 
           $\circ \hspace{1mm} \sto{1} \to t \, \chin{1}$,
           \recht $\sto{1} \to b \, \chip{1}$, 
           $\Join \sto{1} \to c \, \chin{1}$, and 
           $\star \sto{1} \to W^+ \, b \, \chin{1}$.
      }{7}
\end{minipage}

\noindent
decay widths at tree level. Note, that the variation of 
the branching ratios with $A_t$ starting from
$A_t = \mu \cotbe$ is mainly determined by kinematics and hardly by the 
couplings except near $A_t = \mu \cotbe$.
The change in the ordering of the branching
ratios caused be the change of the sign of $\cost$ is even more pronounced for 
large $\tanbe$ as can be seen in \fig{fig:brstop1anlc2A}b. 

The case $M_U \ll M_Q$ is shown in \fig{fig:brstop1anlc2A}c and d. If $\tanbe$
is small (\fig{fig:brstop1anlc2A}c) then there is hardly any dependence on 
$A_t$. As expected by the fact that the light stop is mainly a right state the 
decay into $t \, \chin{1}$ dominates if kinematically allowed. In the range
of $A_t$ where only higher order decays are possible again the decay into 
$b \, W^+ \, \chin{1}$ dominates. For $A_t \gsim 800$~GeV only the
decay into $c \, \chin{1}$ is possible. For large $\tanbe$ 
(\fig{fig:brstop1anlc2A}d) the situation changes for the two body decays at tree
level. Here the bottom Yukawa coupling becomes important for the decay into
$b \, \chip{1}$ leading to somewhat stronger dependence on $A_t$.
 
\section{Decays of $\protect \sto{2}$}

Before discussing the different scenarios for $\sto{2}$ decays in
detail we want to note that there is a wide parameter range where the decays 
into bosons dominate.
In particular this holds for large values of $\mu$ and/or $A_t$.

In \fig{fig:brstop2anlc2mu}a and b the branching ratios are shown as a function
of $\mu$ for  $M_D = M_Q = M_U$ = 700~GeV, $A_t = A_b = 500$~GeV, 
$\tanbe = 1.5$, and $m_{A^0} = 150$~GeV. For positive $\mu$ only decays into 
fermions are kinematically allowed. Here the decay into $b \, \chip{1}$
dominates with branching ratios up to $\sim 60\%$ followed by the decays into 
$t \, \chin{2}$ and $t \, \chin{3}$. For negative $\mu$ the decays into bosons
become kinematically accessible (\fig{fig:brstop2anlc2mu}b). Before discussing 
them we want to note that in the range -500~GeV $< \mu < -300$~GeV the decay 
into $t \, \chin{4}$ is the most important one followed by $b \, \chip{2}$. 
This remarkable feature is again due to the positive interference between 
gaugino and higgsino components of $\chip{2}$ and $\chin{4}$.
For the discussion of the bosonic decays it is useful to 
consider the most important parts of the relevant couplings:
\beq
\begin{array}{ll}
Z^0 \, \sto{2} \, \sto{1} \sim \sinzt &
h^0 \, \sto{2} \, \sto{1} \sim \coszt (\mu \sina + A_t \cosa) \\
A^0 \, \sto{2} \, \sto{1} \sim (A_t \cotbe + \mu) &
H^0 \, \sto{2} \, \sto{1} \sim \coszt (\mu \cosa - A_t \sina)
\end{array}
\label{eq:coupst2st1}
\eeq
For $\mu \lsim -600$~GeV the decay into $Z \, \sto{1}$ dominates for two 
reasons: firstly, similar to the case $\tsn \to W^+ \, \sta{1}$ 
(\sect{sect:nlcsneut}),  there is a large factor
$\lambda(\mstq{2},\mstq{1},\mzq) / \mwq$. Secondly, the coupling
is proportional to $\sinzt$ which is near its maximum.
The branching ratio for $\sto{2} \to A^0 \, \sto{1}$ increases for large 
$| \mu |$ because of kinematics and because the coupling is proportional to
$(A_t \cotbe + \mu)$. The occurrence of decays into both sbottoms is typical for
scenarios where $\tanbe$ is small and $M_D \simeq M_Q$ because here 
$\msbot{1} \simeq \msbot{2}$ and there is a strong mixing in the sbottom
sector as explained in
\chap{chap:nlcmass} (see especially \fig{fig:msbotnlc2mu}). 
In this scenario the properties of the sbottoms are
rather similar. Therefore, it could be difficult to distinguish between
these two decay modes. The decays 
$\sto{2} \to \sto{1} \, h^0, \, \sto{1} \, H^0$ are also possible. 
However, $\coszt$ is rather small and therefore their branching ratios are 
so tiny that we won't show them. 

In \fig{fig:brstop2anlc2mu}c and d we present the scenario where 
$M_U = 350$~GeV,
$M_Q = 700$~GeV, and $M_D = 980$~GeV and the other parameters as above. Now 
decays into bosons are 
\noindent
\begin{minipage}[t]{150mm}
{\setlength{\unitlength}{1mm}
\begin{picture}(150,130)
\put(-7,-6){\mbox{\epsfig{figure=brstop2anlc2mu.eps,
                           height=13.8cm,width=15.8cm}}}
\put(-4,126){{\small \bf a)}}
\put(1,125){\makebox(0,0)[bl]{{$BR(\sto{2})$}}}
\put(75,127){{\small \bf b)}}
\put(80,126){\makebox(0,0)[bl]{{$BR(\sto{2})$}}}
\put(-4,62){{\small \bf c)}}
\put(2,61){\makebox(0,0)[bl]{{$BR(\sto{2})$}}}
\put(75,62){{\small \bf d)}}
\put(80,61){\makebox(0,0)[bl]{{$BR(\sto{2})$}}}
\put(71.5,1.5){\makebox(0,0)[br]{{$\mu$ \small [GeV]}}}
\put(71.5,65){\makebox(0,0)[br]{{$\mu$ \small [GeV]}}}
\put(148.5,1.5){\makebox(0,0)[br]{{$\mu$ \small [GeV]}}}
\put(148.5,64.5){\makebox(0,0)[br]{{$\mu$ \small [GeV]}}}
\end{picture}}
\figcaption{fig:brstop2anlc2mu}
           {Branching ratios for $\protect \sto{2}$ decays as a function
            of $\mu$ for $\protect \tanbe = 1.5$}
           {Branching ratios for $\sto{2}$ decays as a function of 
          $\mu$ for $\tanbe = 1.5$, $M$ = 350~GeV, 
            $m_{A^0} = 150$~GeV, $A_t = A_b = 500$~GeV, $M_Q$ = 700~GeV,
        a) and b) $M_U = M_D$ = 700~GeV, c) and d) $M_U$ = 350~GeV,
        $M_D$ = 980 GeV. 
           The curves in a), and c) correspond to:
           $\circ \hspace{1mm} \sto{2} \to t \, \chin{1}$,
           \rechtl \hspace{1mm} $\sto{2} \to t \, \chin{2}$,
           $\triangle \hspace{1mm} \sto{2} \to t \, \chin{3}$,
           $\diamondsuit \hspace{1mm} \sto{2} \to t \, \chin{4}$,
           \recht $\sto{2} \to b \, \chip{1}$, and
           $\bullet \hspace{1mm} \sto{2} \to b \, \chip{2}$.
           The curves in b), and d) correspond to:
           $\circ \hspace{1mm} \sto{2} \to Z^0 \, \sto{1}$,
           \rechtl \hspace{1mm} $\sto{2} \to h^0 \, \sto{1}$,
           $\triangle \hspace{1mm} \sto{2} \to H^0 \, \sto{1}$,
           $\diamondsuit \hspace{1mm} \sto{2} \to A^0 \, \sto{1}$,
           \recht $\sto{2} \to W^+ \, \sbo{1}$, and
           $\star \hspace{1mm} \sto{2} \to W^+ \, \sbo{2}$.
           The grey area
           will be covered by LEP2 ($\mchipm{1} \leq 95$~GeV).
      }{7}
\end{minipage}

\noindent
allowed for the whole $\mu$ range. Note, that the decays
into sbottoms are not shown
because they are kinematically 
highly suppressed (therefore, they are not shown). 
For $|\mu| \lsim 350$~GeV the 
decays into fermions 
are kinematically favoured. Here the decays into neutralinos dominate. 
For large positive $\mu$ the decays into $H^0 \, \sto{1}$ and 
$A^0 \, \sto{1}$ dominate whereas the branching ratios for 
$\sto{2} \to Z^0 \, \sto{1}, \, h^0 \, \sto{1}$ are very small, because 
$|\cost|$ is rather small ($\lsim 0.1$) and 
$\sina \simeq -0.75$. Note, that the sign in front of $A_t$ in 
\eqn{eq:coupst2st1} leads to an increase of the $H^0 \, \sto{1}$ channel and
to a decrease in the $h^0 \, \sto{1}$ channel. On the other side for large 
negative $\mu$ we find $\cost \simeq -0.4$ and $-0.9 \lsim \cosa \lsim -0.8$. 
This and the kinematics lead to the dominance of $\sto{2} \to Z \, \sto{1}$. 
For $\mu \lsim -800$~GeV this branching ratio shows a slight decrease although
the corresponding decay width is growing. The reason is that the decay width
for $\sto{2} \to A^0 \, \sto{1}$ increases much stronger. Note that in this
parameter are hardly any decays into $H^0$.
The kink in $BR(\sto{2} \to h^0 \sto{1})$ arises because 
the coupling has a maximum near this point and because of the growth of the 
above mentioned decay widths. We want to stress
that for large $|\mu|$ more than 80\% of 
the $\sto{2}$ decay into bosons.

In \fig{fig:brstop2bnlc2mu} the situation is shown for $\tanbe = 40$. In the
case $M_D = M_Q = M_U$ (\fig{fig:brstop2bnlc2mu}a and b) 
we encounter the following differences compared to $\tanbe = 1.5$. Among the
decays into fermions the decay into $b \, \chip{1}$ is the most important one
for the whole $\mu$ range (in fact this is independent of the ratio
$M_D : M_Q : M_U$ in the considered examples). 
For large $|\mu|$ the decays into $W^+ \, \sbo{1}$
and $H^+ \, \sbo{1}$ are the dominant ones. This is partly due to the fact that 
$\sbo{1}$ becomes relatively light for these values of $\mu$. 
In case of $H^+ \, \sbo{1}$ also the
term $(- \sinb \sint m_b A_b \tanbe + \cosb \cost m_t \mu)/\mw$ becomes large
(see also \eqn{eq:coupstoHsbo}). Moreover, the decays into $\sbo{2}$ are 
kinematically forbidden and also the decay $\sto{2} \to Z \, \sto{1}$ is 
suppressed by kinematics. 
 
Similar to the small $\tanbe$ scenarios the situation changes in two important
ways, if one turns to the case $M_U = 350$~GeV and $M_D = 980$~GeV
(\fig{fig:brstop2bnlc2mu}c and d). Firstly, the decays into $\sbo{1}$ are at
least highly suppressed by kinematics if not forbidden. Secondly, the decays 
into $h^0 \, \sto{1}$ and $H^0 \, \sto{1}$ get sizable branching ratios. 
Besides 
this, there are some differences stemming mainly from the fact that 
$\mstop{1}$, $\mstop{2}$, $\cost$, 
and $\cosa$ are nearly independent of $\mu$ for 
large $\tanbe$. The decay into $A^0 \, \sto{1}$ becomes the most important one 
for large $|\mu|$ followed by $\sto{2} \to H^0 \, \sto{1}$.
The branching ratio for the decay into $Z \, \sto{1}$ never exceeds
13\% because $\cost$ is small ($\simeq - 0.2$ in this example).

In the case $M_U = M_D = 500$~GeV and $M_Q = 700$~GeV the decays into 
$\sbo{1}$ become important as can be seen in \fig{fig:brstop2bnlc2mu}f. 
Especially for large $|\mu|$ they are the dominant decays and they are in 
general more important than the decays into $\sto{1}$ because 
$\msbot{1} < \mstop{1}$. For small $|\mu|$ the decay into $W^+ \, \sbo{1}$
becomes less important because 
$|\cosb| \lsim 0.1$ in this range. Note, that 
$BR(\sto{2} \to H^+ \, \sbo{1})$ is always larger than 8\% and can
even reach 22\%.

Let us now turn to the $A_t$ dependence for
fixed $\mu$. In \fig{fig:brstop2bnlc2A}a and b the branching ratios are shown
for $M_D = M_Q = M_U = 700$~GeV and $\tanbe = 40$. Similar to the case of the
light stop the branching ratios for the decays into fermions are sensitive to
the sign of $\cost$ which changes at $A_t = \mu \cotbe$. 
This does not happen in the case of the decays into 
bosons as their decay widths are mainly determined by 
$\costq$ (see \sect{decaysect:lagrangian} and \ref{decaysect:twobody})
except $\sto{2} \to H^+ \, \sbo{1}$. In this case the most important 
contribution
\noindent
\begin{minipage}[t]{150mm}
{\setlength{\unitlength}{1mm}
\begin{picture}(150,173)                        
\put(-7,-21){\mbox{\epsfig{figure=brstop2bnlc2mu.eps,
                           height=21.3cm,width=15.8cm}}}
\put(-4,170){{\ \bf a)}}
\put(2,169){\makebox(0,0)[bl]{{$BR(\sto{2})$}}}
\put(74,171){{\ \bf b)}}
\put(80,170){\makebox(0,0)[bl]{{$BR(\sto{2})$}}}
\put(-4,113){{\ \bf c)}}
\put(2,112){\makebox(0,0)[bl]{{$BR(\sto{2})$}}}
\put(74,113){{\ \bf d)}}
\put(80,112){\makebox(0,0)[bl]{{$BR(\sto{2})$}}}
\put(-4,57.5){{\ \bf e)}}
\put(2,56.5){\makebox(0,0)[bl]{{$BR(\sto{2})$}}}
\put(74,57.5){{\ \bf f)}}
\put(80,56.5){\makebox(0,0)[bl]{{$BR(\sto{2})$}}}
\put(68.5,2){\makebox(0,0)[br]{{$\mu$ \small [GeV]}}}
\put(68.5,59){\makebox(0,0)[br]{{$\mu$ \small [GeV]}}}
\put(68.5,115){\makebox(0,0)[br]{{$\mu$ \small [GeV]}}}
\put(148,2){\makebox(0,0)[br]{{$\mu$ \small [GeV]}}}
\put(148,59){\makebox(0,0)[br]{{$\mu$ \small [GeV]}}}
\put(148,115){\makebox(0,0)[br]{{$\mu$ \small [GeV]}}}
\end{picture}}
\figcaption{fig:brstop2bnlc2mu}
           {Branching ratios for $\protect \sto{2}$ decays as a function
            of $\mu$ for $\protect \tanbe = 40$}
           {Branching ratios for $\sto{2}$ decays as a function of 
          $\mu$ for $\tanbe = 40$, $A_t = A_b = 500$~GeV, $M$ = 350~GeV, 
            $m_{A^0} = 150$~GeV, $M_Q$ = 700~GeV,
        a) and b) $M_U = M_D$ = 700~GeV, c) and d) $M_U$ = 350~GeV,
        $M_D$ = 980 GeV, e) and f) $M_U = M_D$ = 500 GeV.
           The curves in a), c) and e) correspond to:
           $\circ \hspace{1mm} \sto{2} \to t \, \chin{1}$,
           \rechtl \hspace{1mm} $\sto{2} \to t \, \chin{2}$,
           $\triangle \hspace{1mm} \sto{2} \to t \, \chin{3}$,
           $\diamondsuit \hspace{1mm} \sto{2} \to t \, \chin{4}$,
           \recht $\sto{2} \to b \, \chip{1}$, and
           $\bullet \hspace{1mm} \sto{2} \to b \, \chip{2}$.
           The curves in b), d) and f) correspond to:
           $\circ \hspace{1mm} \sto{2} \to Z^0 \, \sto{1}$,
           \rechtl \hspace{1mm} $\sto{2} \to h^0 \, \sto{1}$,
           $\triangle \hspace{1mm} \sto{2} \to H^0 \, \sto{1}$,
           $\diamondsuit \hspace{1mm} \sto{2} \to A^0 \, \sto{1}$,
           \recht $\sto{2} \to W^+ \, \sbo{1}$, and 
           $\bullet \hspace{1mm} \sto{2} \to H^+ \, \sbo{1}$.
           The grey area
           will be covered by LEP2 ($\mchipm{1} \leq 95$~GeV).
      }{0}
\end{minipage}


\noindent
\begin{minipage}[t]{150mm}
{\setlength{\unitlength}{1mm}
\begin{picture}(150,132)                        
\put(-6,-6){\mbox{\epsfig{figure=brstop2bnlc2A.eps,
                           height=14.0cm,width=15.7cm}}}
\put(-4,127.5){{\small \bf a)}}
\put(1,126.5){\makebox(0,0)[bl]{{$BR(\sto{2})$}}}
\put(76,125.5){{\small \bf b)}}
\put(81,124.5){\makebox(0,0)[bl]{{$BR(\sto{2})$}}}
\put(-4,62){{\small \bf c)}}
\put(2,61){\makebox(0,0)[bl]{{$BR(\sto{2})$}}}
\put(76,62){{\small \bf d)}}
\put(82,61){\makebox(0,0)[bl]{{$BR(\sto{2})$}}}
\put(69.5,2.5){\makebox(0,0)[br]{{$A_t (=A_b)$ \small [GeV]}}}
\put(69.5,65.5){\makebox(0,0)[br]{{$A_t (=A_b)$ \small [GeV]}}}
\put(148.5,2.0){\makebox(0,0)[br]{{$A_t (=A_b)$ \small [GeV]}}}
\put(148.5,67.5){\makebox(0,0)[br]{{$A_t (=A_b)$ \small [GeV]}}}
\end{picture}}
\figcaption{fig:brstop2bnlc2A}
           {Branching ratios for $\protect \sto{2} $ decays as a
            function of $A_t$ for $\protect \tanbe = 40$}
           {Branching ratios for $\sto{2}$ decays as a function of 
   $A_t (=A_b)$ for $\tanbe = 40$, $\mu = - 500$~GeV, $M$ = 350~GeV, 
            $m_{A^0} = 100$~GeV, $M_Q$ = 700~GeV,
        a) and b) $M_U = M_D$ = 700~GeV, c) and d) $M_U = M_D$ = 500 GeV.
           The curves in a) and c) correspond to:
           $\circ \hspace{1mm} \sto{2} \to t \, \chin{1}$,
           \rechtl \hspace{1mm} $\sto{2} \to t \, \chin{2}$,
           $\triangle \hspace{1mm} \sto{2} \to t \, \chin{3}$,
           $\diamondsuit \hspace{1mm} \sto{2} \to t \, \chin{4}$,
           \recht $\sto{2} \to b \, \chip{1}$, and
           $\bullet \hspace{1mm} \sto{2} \to b \, \chip{2}$.
           The curves in b) and d) correspond to:
           $\circ \hspace{1mm} \sto{2} \to Z^0 \, \sto{1}$,
           \rechtl \hspace{1mm} $\sto{2} \to h^0 \, \sto{1}$,
           $\triangle \hspace{1mm} \sto{2} \to H^0 \, \sto{1}$,
           $\diamondsuit \hspace{1mm} \sto{2} \to A^0 \, \sto{1}$,
           \recht $\sto{2} \to W^+ \, \sbo{1}$, and
           $\bullet \hspace{1mm} \sto{2} \to H^+ \, \sbo{1}$.
      }{7}
\end{minipage}

\noindent
to the coupling is 
$(- \sinb \sint m_b A_b \tanbe + \cosb \cost m_t \mu)/\mw$ as mentioned above.
Here the sign of $A_b$ is correlated to the 
sign of $\cost$ due to the 
assumption $A_b = A_t$. The assumption $A_b = - A_t$ would lead to a
reduction of this branching ratio. This is even true if one changes 
simultaneous the sign
of $\mu$ as then, in general, $\cosb$ also changes its sign. Note, that all
partial decay widths are growing with $|A_t|$ because of 
kinematics. This 
increase is stronger for the decays into bosons than for the decays into 
fermions 
except for the decay into $A^0 \, \sto{1}$. In case of the vector bosons
\noindent
\begin{minipage}[t]{150mm}
{\setlength{\unitlength}{1mm}
\begin{picture}(150,133)                        
\put(-7,-6){\mbox{\epsfig{figure=brstop2bnlc2tb.eps,
                           height=14.0cm,width=15.8cm}}}
\put(-4,127.5){{\small \bf a)}}
\put(1,126.5){\makebox(0,0)[bl]{{$BR(\sto{2})$}}}
\put(75,127.5){{\small \bf b)}}
\put(80,126.5){\makebox(0,0)[bl]{{$BR(\sto{2})$}}}
\put(-4,63){{\small \bf c)}}
\put(2,62){\makebox(0,0)[bl]{{$BR(\sto{2})$}}}
\put(75,63){{\small \bf d)}}
\put(80,62){\makebox(0,0)[bl]{{$BR(\sto{2})$}}}
\put(71.5,1.0){\makebox(0,0)[br]{{$\tanbe$}}}
\put(71.5,66){\makebox(0,0)[br]{{$\tanbe$}}}
\put(149.5,1.7){\makebox(0,0)[br]{{$\tanbe$}}}
\put(149.5,66){\makebox(0,0)[br]{{$\tanbe$}}}
\end{picture}}
\figcaption{fig:brstop2bnlc2tb}
           {Branching ratios for $\protect \sto{2}$ decays as a function
            of $\protect \tanbe$ for $A_t = 500$~GeV}
           {Branching ratios for $\sto{2}$ decays as a function of 
          $\tanbe$ for $A_t = A_b = 500$~GeV, $\mu = -500$~GeV, $M$ = 350~GeV, 
            $m_{A^0} = 100$~GeV, $M_Q$ = 700~GeV,
          a) and b) $M_U = M_D$ = 700~GeV, c) and d) $M_U = M_D$ = 500 GeV.
           The curves in a) and c) correspond to:
           $\circ \hspace{1mm} \sto{2} \to t \, \chin{1}$,
           \rechtl \hspace{1mm} $\sto{2} \to t \, \chin{2}$,
           $\triangle \hspace{1mm} \sto{2} \to t \, \chin{3}$,
           $\diamondsuit \hspace{1mm} \sto{2} \to t \, \chin{4}$,
           \recht $\sto{2} \to b \, \chip{1}$, and
           $\bullet \hspace{1mm} \sto{2} \to b \, \chip{2}$.
           The curves in b) and d) correspond to:
           $\circ \hspace{1mm} \sto{2} \to Z^0 \, \sto{1}$,
           \rechtl \hspace{1mm} $\sto{2} \to h^0 \, \sto{1}$,
           $\triangle \hspace{1mm} \sto{2} \to H^0 \, \sto{1}$,
           $\diamondsuit \hspace{1mm} \sto{2} \to A^0 \, \sto{1}$,
           \recht $\sto{2} \to W^+ \, \sbo{1}$, and 
           $\star \hspace{1mm} \sto{2} \to W^+ \, \sbo{2}$,
           $\bullet \hspace{1mm} \sto{2} \to H^+ \, \sbo{1}$.
      }{7}
\end{minipage}

\noindent
again the extra factor $\lambda(\mstq{2},m^2_V,\msferq{1})/m^2_V$ is the reason,
whereas in case of $H^+$ it is the term proportional to $A_b$ in the coupling.

Similar to the case of $\sto{1}$, the situation changes if $M_U \ll M_Q$.
This can be seen in \fig{fig:brstop2bnlc2A}c and d where we have taken
$M_U = M_D = 500$~GeV and the other parameters as above. The main reason is
that $\cost$ changes smoothly its sign at $A_t = \mu \cotbe$ 
($\sto{2} \simeq \sto{L}$). Therefore, the asymmetry of the fermionic decays
with respect to this point is less pronounced. The measurements of these
decay modes could nevertheless give a clear hint for the relative sign between
$A_t$ and $\mu$. The decrease in the $W^+ \, \sbo{1}$ channel is due to the
decrease of $\sint$ for large $|\mu|$ which, on the contrary, leads to the 
increase
of the $Z^0 \, \sto{1}$ mode. Note, that $BR(\sto{2} \to H^0 \, \sto{1})$ 
nearly vanishes near $\mu = 1$~TeV.

In \fig{fig:brstop2bnlc2tb} we study the $\tanbe$ dependence for 
$A_t = A_b = 500$~GeV, and $\mu = -500$~GeV. Again we compare the case
$M_D = M_Q = M_U = 700$~GeV with $M_D = M_U = 500$~GeV and $M_Q = 700$~GeV.
In \fig{fig:brstop2bnlc2tb}a and c the fermionic decay modes are shown. As
a general feature the decays into gaugino-like particles increase with 
$\tanbe$ whereas the decays into higgsino-like particles decrease for 
$\tanbe \lsim 10$. The main reason is that for small $\tanbe$ the mixture 
between gauginos and higgsinos is smaller than for large $\tanbe$. The
only exception is $\sto{2} \to t \, \chin{1}$ for $M_U = 500$~GeV because
here the $\sto{L}$ component of $\sto{2}$ increases with $\tanbe$.
Near $\tanbe = 2.72$ the two heaviest neutralinos have the same 
mass and change their (higgsino) nature (for a discussion of such degenerate
points of the neutralino mass matrix see for example \cite{Bartl89a}).
For large $\tanbe$ the branching ratios decrease except for
$BR(\sto{2} \to b \, \chip{2})$ which grows because of the bottom Yukawa
coupling. The decay into $Z^0 \, \sto{1}$ in both cases decreases 
with $\tanbe$ because of phase space as can be seen in \fig{fig:brstop2bnlc2tb}b
and d. In case of $M_U = 500$~GeV the decrease of $\sinztq$ also leads to a 
stronger decrease in this branching ratio. Note, that with increasing $\tanbe$
$\cosa$ and $m_{h^0}$ also grow whereas $m_{H^0}$ decreases leading to the
shown dependence of the branching ratios. $BR(\sto{2} \to A^0 \, \sto{1})$
grows with $\tanbe$ for small $\tanbe$. Here, the decrease of phase space is
compensated be the increase of $(A_t \cotbe + \mu)$. The decays into sbottoms
are mainly influenced by kinematics and the fact that $\cosbq$ grows with
$\tanbe$. It is important to note that various branching ratios have minima and 
maxima for $\tanbe \lsim 25$ making an easy interpolation between 
small and large $\tanbe$ scenarios impossible.

%% file: nlcsbottom.tex
\chapter{Numerical results for $\protect \sbo{1,2}$}
\label{chap:nlcsbottom}

\section{Decays of $\protect \sbo{1}$}

Many of the properties of the sbottoms can be understood from the discussions
in the previous chapters. Therefore, we concentrate in this chapter mainly on 
the decays into bosons. In
\fig{fig:brsbot1anlc2mu}a we show the branching ratios as a
function of $\mu$ for $M_D = M_Q = M_U = 700$~GeV and $\tanbe = 1.5$. As
explained in \sect{sect:msbot}, $\cosb$ is $\sim 0.7$ for most of the 
$\mu$ range although $\tanbe$ is small. This leads to the remarkable result 
that the decays into $t \, \chim{1}, \, t \, \chim{2}$ 
dominate in the whole $\mu$
range except near $\mu = A_b \cotbe$ where $\cosb$ vanishes. Note, that
the dominance of $\sbo{1} \to t \, \chim{2}$ ($t \, \chim{1}$) for 
$-450$~GeV $\lsim \mu \lsim -350$~GeV ($|\mu| \lsim 350$~GeV) is caused by the
top Yukawa coupling. A second consequence of $\cosb \simeq 0.7$ is
the appearance of $\sbo{1} \to W^- \, \sto{1}$ with a branching ratio up
to 30\%.
In \fig{fig:brsbot1anlc2mu}b the branching ratios are shown for 
$\tanbe = 40$. Here
the influence of the bottom Yukawa coupling leads to an enhancement of the
decays into neutralinos. Moreover, there is less
phase space compared to the small $\tanbe$ case.

In \fig{fig:brsbot1anlc2mu}c the situation is shown for $M_D = 980$~GeV,
$M_Q = 700$~GeV, $M_U = 350$~GeV, and $\tanbe = 1.5$. 
Here we do not show the branching ratios for the decays
into $b \, \chin{1}, b \, \chin{3}$ which turn out to be rather small.
For
$\mu \lsim -350$~GeV the decay into $W^- \, \sto{1}$ dominates whereas
for $\mu \gsim 350$~GeV the decay into $H^- \, \sto{1}$ is the dominating one.
For $|\mu| \lsim 350$~GeV 
$\sbo{1} \to t \, \chim{1}$ is the most important decay.
Qualitatively this can be understood by having a look on the most
relevant parts of the couplings (see also \eqn{eq:coupstoWsbo} and 
(\ref{eq:coupstoHsbo})):
\beq
\begin{array}{l}
W^{\pm}  \sbo{1} \sto{1}  \sim  \cost \cosb
                           \sim (A_t - \mu \cotbe)  (A_b - \mu \tanbe) \\
H^{\pm} \sbo{1} \sto{1}  \sim  \sint \cosb m_t (\mu + A_t \cotbe) 
\end{array}
\label{eq:charHstosbo}
\eeq
Here the relative sign between $A_b$ and $\mu$ is important. 
Now $\cost$ varies between -0.4 ($\mu = -1$~TeV) and 0.1 ($\mu = 1$~TeV) and 
$|\cosb| \simeq 1$ leading to the dominance of 
\noindent
\begin{minipage}[t]{150mm}
{\setlength{\unitlength}{1mm}
\begin{picture}(150,130)                        
\put(-6,-7){\mbox{\epsfig{figure=brsbot1anlc2mu.eps,
                           height=14.0cm,width=15.8cm}}}
\put(-1,126){{\small \bf a)}}
\put(3,125){\makebox(0,0)[bl]{{$BR(\sbo{1})$}}}
\put(76,126){{\small \bf b)}}
\put(81,125){\makebox(0,0)[bl]{{$BR(\sbo{1})$}}}
\put(-1,62){{\small \bf c)}}
\put(3,61){\makebox(0,0)[bl]{{$BR(\sbo{1})$}}}
\put(76,62){{\small \bf d)}}
\put(82,61){\makebox(0,0)[bl]{{$BR(\sbo{1})$}}}
\put(70.5,0.8){\makebox(0,0)[br]{{$\mu$ \small [GeV]}}}
\put(70.5,66.2){\makebox(0,0)[br]{{$\mu$ \small [GeV]}}}
\put(149,1){\makebox(0,0)[br]{{$\mu$ \small [GeV]}}}
\put(149,65.2){\makebox(0,0)[br]{{$\mu$ \small [GeV]}}}
\end{picture}}
\figcaption{fig:brsbot1anlc2mu}
           {Branching ratios for $\protect \sbo{1}$ decays as a
            function of $\mu$}
           {Branching ratios for $\sbo{1}$ decays as a function of 
      $\mu$ for $A_t = A_b$ = 500~GeV, $M$ = 350~GeV, $M_Q$ = 700~GeV,
            in a) and b) $M_U = M_D$ = 700~GeV,  and
            in c) and d) $M_U$ = 350~GeV, $M_D$ = 980~GeV.
           In a), and c) $\tanbe = 1.5$ and in b), and d)
           $\tanbe$ = 40.
           The curves correspond to the following transitions:
           $\circ \hspace{1mm} \sbo{1} \to b \, \chin{1}$,
           \rechtl \hspace{1mm} $\sbo{1} \to b \, \chin{2}$,
           $\triangle \hspace{1mm} \sbo{1} \to b \, \chin{3}$,
           $\diamondsuit \hspace{1mm} \sbo{1} \to b \, \chin{4}$,
           \recht $\sbo{1} \to t \, \chim{1}$,
           $\bullet \hspace{1mm} \sbo{1} \to t \, \chim{2}$,
           $\star \hspace{1mm} \sbo{1} \to W^- \, \sto{1}$, and
           $\Join \sbo{1} \to H^- \, \sto{1}$. 
           The grey area
           will be covered by LEP2 ($\mchipm{1} \leq 95$~GeV).
      }{7}
\end{minipage}

\noindent
$\sbo{1} \to W^- \, \sto{1}$ ($\sbo{1} \to H^- \, \sto{1}$) 
for negative (positive) $\mu$. The sum of 
both branching ratios constitutes at least 
20\% ($\mu \simeq 100$~GeV) and can reach up to 85\% 
($|\mu| \simeq 1$~TeV). Concerning the importance of the decays
into charginos for $|\mu| \lsim 500$~GeV we would like to note, that this is 
again due to the top Yukawa coupling.
In \fig{fig:brsbot1anlc2mu}d the branching ratios are shown for $\tanbe = 40$
and the other parameters as above. Concerning the decays into fermions, the
situation is similar to the case $M_D = M_Q = M_U$, the main differences 
arise due to kinematics since $\msbot{1}$ is now, in general, larger.
Concerning 
\noindent
\begin{minipage}[t]{150mm}
{\setlength{\unitlength}{1mm}
\begin{picture}(150,63)                        
\put(-6,0){\mbox{\epsfig{figure=brsbot1nlc2A.eps,height=6.0cm,width=15.8cm}}}
\put(-1,60){{\small \bf a)}}
\put(3,59){\makebox(0,0)[bl]{{\small $BR(\sbo{1} )$}}}
\put(70.5,1.5){\makebox(0,0)[br]{{\small $A_b (=A_t)$~[GeV]}}}
\put(75,60){{\small \bf b)}}
\put(79,59){\makebox(0,0)[bl]{{\small $BR(\sbo{1} )$}}}
\put(148,1.5){\makebox(0,0)[br]{{\small $A_b (=A_t)$~[GeV]}}}
\end{picture}}
\figcaption{fig:brsbot1anlc2A}
           {Branching ratios for $\protect \sbo{1}$ decays as a
            function of $A_b (= A_t)$}
           {Branching ratios for $\sbo{1}$ decays as a function of 
   $A_b (= A_t)$ for $\mu = -500$~GeV, $M$ = 350~GeV, $M_Q$ = 700~GeV,
            $M_U$ = 350~GeV, $M_D = 980$~GeV, a) $\tanbe = 1.5$ and b)
           $\tanbe$ = 40.
           The curves correspond to the transitions:
           $\circ \hspace{1mm} \sbo{1} \to b \, \chin{1}$,
           \rechtl \hspace{1mm} $\sbo{1} \to b \, \chin{2}$,
           $\triangle \hspace{1mm} \sbo{1} \to b \, \chin{3}$,
           $\diamondsuit \hspace{1mm} \sbo{1} \to b \, \chin{4}$,
           \recht $\sbo{1} \to t \, \chim{1}$,
           $\bullet \hspace{1mm} \sbo{1} \to t \, \chim{2}$,
           $\star \hspace{1mm} \sbo{1} \to W^- \, \sto{1}$, and
           $\Join \sbo{1} \to H^- \, \sto{1}$. 
      }{5}
\end{minipage}

\noindent
\begin{minipage}{75mm}
{\setlength{\unitlength}{1mm}
\begin{picture}(80,72)
\put(-1,0.8){\mbox{\psfig{figure=brsbot1bnlc2mu.eps,height=6.6cm,width=7.4cm}}}
\put(1,66.5){\makebox(0,0)[bl]{{\small $BR(\sbo{1})$}}}
\put(74.5,1){\makebox(0,0)[br]{{\small $\mu$~[GeV]}}}
\end{picture}}
\figcaption{fig:brsbot1bnlc2mu}
           {Branching ratios for $\protect \sbo{1}$ decays as a
            function of $\mu$}
           {Branching ratios for $\sbo{1}$ decays as a function of 
      $\mu$ for $A_t = A_b$ = 500~GeV, $M$ = 350~GeV, $M_Q$ = 700~GeV,
           $M_U = M_D$ = 500~GeV,  and
           $\tanbe$ = 40.
           The curves correspond to the following transitions:
           $\circ \hspace{1mm} \sbo{1} \to b \, \chin{1}$,
           \rechtl \hspace{1mm} $\sbo{1} \to b \, \chin{2}$,
           $\triangle \hspace{1mm} \sbo{1} \to b \, \chin{3}$,
           $\diamondsuit \hspace{1mm} \sbo{1} \to b \, \chin{4}$, and
           \recht $\sbo{1} \to t \, \chim{1}$.
           The grey area
           will be covered by LEP2 ($\mchipm{1} \leq 95$~GeV).
      }{0}
\end{minipage}
\hspace{4mm}
\begin{minipage}{70mm}
the decay into $H^- \, \sto{1}$ we would like to note, that the term
$\cost \sinb m_b (\mu + A_b \tanbe)$ 
is still suppressed because 
$\cost \simeq -0.21$ and $|\cosb| \gsim 0.9$. This leads 
to the dominance of the above mentioned term in \eqn{eq:charHstosbo}.
For the same reason, $BR(\sbo{1} \to H^- \, \sto{1})$ is larger than
$BR(\sbo{1} \to W^- \, \sto{1})$ for $|\mu| \gsim 500$~GeV.

\hspace*{3mm}
Note, that a larger $|A_t|$ leads to an enhancement of the
$W^- \, \sto{1}$ mode. This is demonstrated in \fig{fig:brsbot1anlc2A}b where
we show the branching ratios as a function of $A_t$ for $\mu = -500$~GeV and
the other parameters as above. Note, that $BR(\sbo{1} \to W^- \, \sto{1})$ and
$BR(\sbo{1} \to H^- \, \sto{1})$ are of the same size if $|A_t| \simeq |\mu|$.
This is a general result for large $\tanbe$ scenarios if $M_U \ll M_Q \ll M_D$.
In \fig{fig:brsbot1anlc2A}a the branching ratios are shown 
for $\tanbe = 1.5$. The differences between positive and 
negative $A_t$ are
due the same reasons as between negative and positive $\mu$.
\end{minipage}

\noindent
\begin{minipage}[t]{150mm}
{\setlength{\unitlength}{1mm}
\begin{picture}(150,63)                        
\put(-6,-4){\mbox{\epsfig{figure=brsbot2nlc2muA.eps,height=7.0cm,width=15.8cm}}}
\put(-1,60){{\small \bf a)}}
\put(3,59){\makebox(0,0)[bl]{{\small $BR(\sbo{2} )$}}}
\put(70.5,1){\makebox(0,0)[br]{{$\mu$ \small [GeV]}}}
\put(75,60){{\small \bf b)}}
\put(79,59){\makebox(0,0)[bl]{{\small $BR(\sbo{2} )$}}}
\put(149,1){\makebox(0,0)[br]{{$A_b (=A_t)$ \small [GeV]}}}
\end{picture}}
\figcaption{fig:brsbot2anlc2mu}
           {Branching ratios for $\protect \sbo{2}$ decays as a function
            of $\mu$ and $A_b$ for $\protect \tanbe = 1.5$}
           {Branching ratios for $\sbo{2}$ decays a) as a function of 
            $\mu$ for $A_t = A_b = 500$~GeV and b) as a function of 
            $A_b (=A_t)$ for $\mu = -500$~GeV. The other parameters are 
            $\tanbe = 1.5$, $M$ = 350~GeV,  $\ma = 150$~GeV, $M_Q$ = 700~GeV,
            and $M_U = M_D$ = 500~GeV.
           The curves correspond to:
           \rechtl \hspace{1mm} $\sbo{2} \to b \, \chin{2}$,
           $\diamondsuit \hspace{1mm} \sbo{2} \to b \, \chin{4}$,
           \recht $\sbo{2} \to t \, \chim{1}$,
           $\bullet \hspace{1mm} \sbo{2} \to t \, \chim{2}$,
           $\star \sbo{2} \to W^- \, \sto{1}$, and
           $\Join \hspace{1mm} \sbo{2} \to H^- \, \sto{1}$.
           The grey area in a)
           will be covered by LEP2 ($\mchipm{1} \leq 95$~GeV).
      }{7}
\end{minipage}

RGE studies favour the case 
$M_D \simeq M_U \ll M_Q$ if $\tanbe$ is large.
In such a case the decays into bosons will be kinematically forbidden and
one has only decays into fermions as in the case $M_D \simeq M_U \simeq M_Q$.
An example is presented in \fig{fig:brsbot1bnlc2mu} where we have chosen
$M_U = M_D$ = 500~GeV, and $M_Q = 700$~GeV. 

\section{Decays of $\protect \sbo{2}$}

In the case that $\tanbe$ is small it turns out that the decays of $\sbo{2}$ 
into bosons are negligible. The exceptions are the decays 
into $W^- \, \sto{1}, \, H^- \sto{1}$
if $M_U \simeq M_D \ll M_Q$ or in other words if $\sbo{2} \simeq \sbo{L}$
and $\msbot{2} > \mstop{1} + \mw, \,  \mstop{1} + m_{H^{\pm}}$. 
A typical example is shown in \fig{fig:brsbot2anlc2mu}
where the branching ratios are shown as a function of $\mu$ and $A_b$ for
$M_D = M_U = 500$~GeV, $M_Q = 700$~GeV and $\tanbe = 1.5$. Here, we do not 
show the decays into $b \, \chin{1,3}; \, Z^0,h^0,H^0,A^0 \, \sbo{1}$
because the maximum of their branching ratios is less than 2\%. Even their sum 
is always less than 5\%.
For the importance of the decays into $W^- \, \sto{1}$ and $H^- \, \sto{1}$
similar arguments hold as in the $\sbo{1}$ case if one replaces
$\cosb$ by $-\sinb$ and $\sinb$ by $\cosb$. 
In the case $M_D \gg M_Q$ also $BR(\sbo{2} \to W^- \, \sto{1})$ and
$BR(\sbo{2} \to H^- \, \sto{1})$ are small (1-5\%). This can also be seen in
\fig{fig:brsbot2bnlc2tb} where we show the branching ratios as a function
of $\tanbe$ for $A_b = A_t = 500$~GeV and $\mu = -500$~GeV. Similar as in
the case of $\sto{2}$ it can be misleading if one simply interpolates the
branching ratios between
small and large $\tanbe$ scenarios 
because of the minima and maxima  in between. Note, that the smallness
\noindent
\begin{minipage}[t]{150mm}
{\setlength{\unitlength}{1mm}
\begin{picture}(150,169)                        
\put(-7,-21.2){\mbox{\epsfig{figure=brsbot2bnlc2tb.eps,
                           height=21.0cm,width=16.0cm}}}
\put(-1,165.5){{\small \bf a)}}
\put(3,164.5){\makebox(0,0)[bl]{{$BR(\sbo{2})$ }}}
\put(76,166){{\small \bf b)}}
\put(82,165){\makebox(0,0)[bl]{{$BR(\sbo{2})$}}}
\put(-1,111.5){{\small \bf c)}}
\put(3,110.5){\makebox(0,0)[bl]{{$BR(\sbo{2})$}}}
\put(76,112.0){{\small \bf d)}}
\put(82,111.0){\makebox(0,0)[bl]{{$BR(\sbo{2})$}}}
\put(-1,56.0){{\small \bf e)}}
\put(3,55.0){\makebox(0,0)[bl]{{$BR(\sbo{2})$}}}
\put(77,54.5){{\small \bf f)}}
\put(81,53.5){\makebox(0,0)[bl]{{$BR(\sbo{2})$}}}
\put(71,1.1){\makebox(0,0)[br]{{$\tanbe$}}}
\put(70.5,56.5){\makebox(0,0)[br]{{$\tanbe$}}}
\put(68.5,112){\makebox(0,0)[br]{{$\tanbe$}}}
\put(149,0.5){\makebox(0,0)[br]{{$\tanbe$}}}
\put(150.5,56.5){\makebox(0,0)[br]{{$\tanbe$}}}
\put(149.5,112){\makebox(0,0)[br]{{$\tanbe$}}}
\end{picture}}
\figcaption{fig:brsbot2bnlc2tb}
           {Branching ratios for $\protect \sbo{2} $ decays as a
            function of $\protect \tanbe$ for $A_b = 500$~GeV}
           {Branching ratios for $\sbo{2}$ decays as a function of 
            $\tanbe$ for $\mu = -500$~GeV, $A_t = A_b$ = 500~GeV,
            $M$ = 350~GeV, $\ma = 150$~GeV, $M_Q$ = 700~GeV,
        a) and b) $M_U = M_D$ = 700~GeV, c) and d) $M_U$ = 350~GeV,
        $M_D$ = 980 GeV, e) and f) $M_U = M_D$ = 500 GeV.
           The curves in a), c) and e) correspond to:
           $\circ \hspace{1mm} \sbo{2} \to b \, \chin{1}$,
           \rechtl \hspace{1mm} $\sbo{2} \to b \, \chin{2}$,
           $\triangle \hspace{1mm} \sbo{2} \to b \, \chin{3}$,
           $\diamondsuit \hspace{1mm} \sbo{2} \to b \, \chin{4}$,
           \recht $\sbo{2} \to t \, \chim{1}$,
           $\bullet \hspace{1mm} \sbo{2} \to t \, \chim{2}$, and
           $\star \hspace{1mm} \sbo{2} \to b \, \glu$.
           The curves in b), d) and f) correspond to:
           $\circ \hspace{1mm} \sbo{2} \to Z^0 \, \sbo{1}$,
           \rechtl \hspace{1mm} $\sbo{2} \to h^0 \, \sbo{1}$,
           $\triangle \hspace{1mm} \sbo{2} \to H^0 \, \sbo{1}$,
           $\diamondsuit \hspace{1mm} \sbo{2} \to A^0 \, \sbo{1}$,
           \recht $\sbo{2} \to W^- \, \sto{1}$, 
           $\star \hspace{1mm} \sbo{2} \to W^- \, \sto{2}$,
           $\bullet \hspace{1mm} \sbo{2} \to H^- \, \sto{1}$ and
           $ \Join \hspace{1mm}\sbo{2} \to H^- \, \sto{2}$.
      }{4}
\end{minipage}

\noindent
of the 
branching ratios into bosonic states for $\tanbe = 40$ in 
\fig{fig:brsbot2bnlc2tb}b and d is
due to our choices of $A_b, A_t$ and $\mu$. The larger these parameters are 
the larger the branching ratios of the bosonic decay modes.

This is demonstrated in \fig{fig:brsbot2bnlc2mu} where the branching ratios
are shown as a function of $\mu$ for $\tanbe = 40$. It is interesting to note
that the sum of the branching ratios into bosonic modes is greater than 80\%
if $max(M_D,M_Q) \lsim |\mu|$. To explain their relative 
importance
it is again useful to rewrite the most important parts of the couplings:
(see also \eqns{eq:coupstoWsbo}, (\ref{eq:coupHsf1sf2a}), (\ref{eq:coupstoHsbo})
and (\ref{decaygl5})):
\beq
\begin{array}{ll}
Z \sbo{2} \sbo{1} \sim \sinzb  &
h^0 \sbo{2} \sbo{1} \sim \coszb (A_b \sina + \mu \cosa) \\
A^0 \sbo{2} \sbo{1} \sim (A_b \tanbe + \mu) &
H^0 \sbo{2} \sbo{1} \sim \coszb (-A_b \cosa + \mu \sina) \\
W^{\pm}  \sbo{2} \sto{1}  \sim  \cost \sinb &
W^{\pm}  \sbo{2} \sto{2}  \sim  \sint \sinb \\
\multicolumn{2}{l}{
H^{\pm} \sbo{2} \sto{1}  \sim  \sint \sinb m_t (\mu + A_t \cotbe)
                               - \cost \cosb m_b (\mu + A_b \tanbe)} \\
\multicolumn{2}{l}{
H^{\pm} \sbo{2} \sto{2}  \sim  \cost \sinb m_t (\mu + A_t \cotbe)
                               + \sint \cosb m_b (\mu + A_b \tanbe)}
\end{array}
\label{eq:coupsbot2}
\eeq
Let us start with the case $M_D = M_Q = M_U$ shown in \fig{fig:brsbot2bnlc2mu}a
and b. The dominance of the decays into gauge bosons is again due to the 
factor $\lambda / m^2_V$ and because of the strong mixing in the sbottom and
in the stop sector. The ordering of these two modes is caused by 
$\msbot{1} < \mstop{1}$. The decrease of $BR(\sbo{2} \to A^0 \, \sbo{1})$ occurs
because $\Gamma(\sbo{2} \to A^0 \, \sbo{1})$ 
grows less with $\mu$ compared to the
widths for the decays into gauge bosons.

In the case $M_D = 980$~GeV, $M_Q = 700$~GeV and $M_U = 350$~GeV all possible
decays into gauge and Higgs bosons are kinematically allowed 
(\fig{fig:brsbot2bnlc2tb}c and d).
For large $|\mu|$ the decay $\sbo{2} \to W^- \, \sto{2}$ is the most important 
one
followed by $\sbo{2} \to Z \, \sbo{1}$ and $\sbo{2} \to h^0 \, \sbo{1}$. This
can be understood by noticing that $|\cost| \lsim 0.3$, $|\cosb| \gsim 0.9$, 
and $\cosa \simeq 1$. This leads to the remarkable result that 
$BR(\sbo{2} \to W^- \, \sto{2}) > BR(\sbo{2} \to W^- \, \sto{1})$ and
$BR(\sbo{2} \to H^- \, \sto{2}) > BR(\sbo{2} \to H^- \, \sto{1})$ because
$\sto{2}$ is mainly a left state. 
Note, that the $H^{\pm} \squ{L} \squ{R}'$ coupling is in general larger than the
$H^{\pm} \squ{R} \squ{R}'$ coupling and the $H^{\pm} \squ{L} \squ{L}'$ 
coupling. As in the case of the neutral Higgs this corresponds to the chirality
change in the $H^{\pm} \bar{q} q'$ coupling. The decay
$\sbo{2} \to Z \sbo{1}$ is less important than 
$\sbo{2} \to W^- \, \sto{2}$ because there is only a small mixing in the
sbottom sector. Especially in the range $|\mu| \lsim M$, the smallness of 
$\sinb$ leads to the suppression of both branching ratios. In this
range the most important bosonic decay modes
are $\sbo{2} \to H^- \, \sto{2}, \, H^0 \sbo{1}, \, A^0 \, \sbo{1}$. Here
their decay widths depend hardly on $\mu$. The sum of their
branching ratios does not exceed 12\% and therefore they are less important
than the fermionic decay modes. Here $\sbo{2} \to t \, \chim{1}$
is the most important decay because of the large top Yukawa coupling.
Finally, we would like to note that the decay $\sbo{2} \to b \, \glu$ also
is possible in this example. 
However, its branching ratio is always smaller than 2\% and 
thus it is now shown.

In the case $M_D = M_U = 500$~GeV and $M_Q = 700$~GeV 
$\sbo{2} \to Z \, \sbo{1}$ is the most important decay for large $|\mu|$. 
The mixing in the sbottom sector is larger than in the
\noindent
\begin{minipage}[t]{150mm}
{\setlength{\unitlength}{1mm}
\begin{picture}(150,171)                        
\put(-7,-21.5){\mbox{\epsfig{figure=brsbot2bnlc2mu.eps,
                           height=21.1cm,width=16.0cm}}}
\put(-2,168){{\ \bf a)}}
\put(3,167){\makebox(0,0)[bl]{{$BR(\sbo{2})$}}}
\put(76,166.5){{\ \bf b)}}
\put(82,165.5){\makebox(0,0)[bl]{{$BR(\sbo{2})$}}}
\put(-2,112.5){{\ \bf c)}}
\put(3,111.5){\makebox(0,0)[bl]{{$BR(\sbo{2})$}}}
\put(76,111.5){{\ \bf d)}}
\put(82,110.5){\makebox(0,0)[bl]{{$BR(\sbo{2})$}}}
\put(-2,56){{\ \bf e)}}
\put(3,55){\makebox(0,0)[bl]{{$BR(\sbo{2})$}}}
\put(76,55){{\ \bf f)}}
\put(82,54){\makebox(0,0)[bl]{{$BR(\sbo{2})$}}}
\put(70.5,1){\makebox(0,0)[br]{{$\mu$ \small [GeV]}}}
\put(70.5,57){\makebox(0,0)[br]{{$\mu$ \small [GeV]}}}
\put(70.5,113){\makebox(0,0)[br]{{$\mu$ \small [GeV]}}}
\put(150,1){\makebox(0,0)[br]{{$\mu$ \small [GeV]}}}
\put(150,58){\makebox(0,0)[br]{{$\mu$ \small [GeV]}}}
\put(149,113){\makebox(0,0)[br]{{$\mu$ \small [GeV]}}}
\end{picture}}
\figcaption{fig:brsbot2bnlc2mu}
           {Branching ratios for $\protect \sbo{2}$ decays as a function
            of $\mu$ for $\protect \tanbe = 40$}
           {Branching ratios for $\sbo{2}$ decays as a function of 
          $\mu$ for $\tanbe = 40$, $A_t = A_b = 500$~GeV, $M$ = 350~GeV, 
            $\ma = 150$~GeV, $M_Q$ = 700~GeV,
        a) and b) $M_U = M_D$ = 700~GeV, c) and d) $M_U$ = 350~GeV,
        $M_D$ = 980 GeV, e) and f) $M_U = M_D$ = 500 GeV.
           The curves in a), c) and e) correspond to:
           $\circ \hspace{1mm} \sbo{2} \to b \, \chin{1}$,
           \rechtl \hspace{1mm} $\sbo{2} \to b \, \chin{2}$,
           $\triangle \hspace{1mm} \sbo{2} \to b \, \chin{3}$,
           $\diamondsuit \hspace{1mm} \sbo{2} \to b \, \chin{4}$,
           \recht $\sbo{2} \to t \, \chim{1}$, and
           $\bullet \hspace{1mm} \sbo{2} \to t \, \chim{2}$.
           The curves in b), d) and f) correspond to:
           $\circ \hspace{1mm} \sbo{2} \to Z^0 \, \sbo{1}$,
           \rechtl \hspace{1mm} $\sbo{2} \to h^0 \, \sbo{1}$,
           $\triangle \hspace{1mm} \sbo{2} \to H^0 \, \sbo{1}$,
           $\diamondsuit \hspace{1mm} \sbo{2} \to A^0 \, \sbo{1}$,
           \recht $\sbo{2} \to W^- \, \sto{1}$,
           $\star \hspace{1mm} \sbo{2} \to W^- \, \sto{2}$,
           $\bullet \hspace{1mm} \sbo{2} \to H^- \, \sto{1}$, and
           $\Join \hspace{1mm} \sbo{2} \to H^- \, \sto{2}$.
           The grey area
           will be covered by LEP2 ($\mchipm{1} \leq 95$~GeV).
      }{4}
\end{minipage}

\noindent
\begin{minipage}[t]{150mm}
{\setlength{\unitlength}{1mm}
\begin{picture}(150,130)                        
\put(-6,-7){\mbox{\epsfig{figure=brsbot2bnlc2A.eps,
                           height=14.0cm,width=15.8cm}}}
\put(-2,126){{\small \bf a)}}
\put(3,125){\makebox(0,0)[bl]{{$BR(\sbo{2})$ }}}
\put(76,126){{\small \bf b)}}
\put(81,125){\makebox(0,0)[bl]{{$BR(\sbo{2})$}}}
\put(-2,61){{\small \bf c)}}
\put(3,60){\makebox(0,0)[bl]{{$BR(\sbo{2})$}}}
\put(76,61){{\small \bf d)}}
\put(81,60){\makebox(0,0)[bl]{{$BR(\sbo{2})$}}}
\put(70.5,1){\makebox(0,0)[br]{{$A_b$ \small [GeV]}}}
\put(70.5,65.5){\makebox(0,0)[br]{{$A_b$ \small [GeV]}}}
\put(149.5,1){\makebox(0,0)[br]{{$A_b (=A_t)$ \small [GeV]}}}
\put(149.5,65.5){\makebox(0,0)[br]{{$A_b (=A_t)$ \small [GeV]}}}
\end{picture}}
\figcaption{fig:brsbot2bnlc2A}
           {Branching ratios for $\protect \sbo{2} $ decays as a
            function of $A_b$ for $\protect \tanbe = 40$}
           {Branching ratios for $\sbo{2}$ decays as a function of 
            $A_b (=A_t)$ for $\tanbe = 40$, $\mu = -500$~GeV,
            $M$ = 350~GeV, $\ma = 150$~GeV, $M_Q$ = 700~GeV,
            a) and b) $M_U$ = 350~GeV,
            $M_D$ = 980 GeV, c) and d) $M_U = M_D$ = 500 GeV.
           The curves in a), and c) correspond to:
           $\circ \hspace{1mm} \sbo{2} \to b \, \chin{1}$,
           \rechtl \hspace{1mm} $\sbo{2} \to b \, \chin{2}$,
           $\triangle \hspace{1mm} \sbo{2} \to b \, \chin{3}$,
           $\diamondsuit \hspace{1mm} \sbo{2} \to b \, \chin{4}$,
           \recht $\sbo{2} \to t \, \chim{1}$,
           $\bullet \hspace{1mm} \sbo{2} \to t \, \chim{2}$, and
           $\star \hspace{1mm} \sbo{2} \to b \, \glu$.
           The curves in b), and d) correspond to:
           $\circ \hspace{1mm} \sbo{2} \to Z^0 \, \sbo{1}$,
           \rechtl \hspace{1mm} $\sbo{2} \to h^0 \, \sbo{1}$,
           $\triangle \hspace{1mm} \sbo{2} \to H^0 \, \sbo{1}$,
           $\diamondsuit \hspace{1mm} \sbo{2} \to A^0 \, \sbo{1}$,
           \recht $\sbo{2} \to W^- \, \sto{1}$, 
           $\star \hspace{1mm} \sbo{2} \to W^- \, \sto{2}$,
           $\bullet \hspace{1mm} \sbo{2} \to H^- \, \sto{1}$ and
           $ \Join \hspace{1mm}\sbo{2} \to H^- \, \sto{2}$.
      }{7}
\end{minipage}

\noindent
previous case. Moreover, the decays into
$\sto{2}$ are kinematically not possible. In this example $\sto{1}$ 
and $\sbo{1}$ are mainly right states leading to
$BR(\sbo{2} \to H^- \, \sto{1}) > BR(\sbo{2} \to W^- \, \sto{1})$ for large
$|\mu|$ although $\mw < m_{H^\pm}$. $\Gamma(\sbo{2} \to H^- \, \sto{1})$ 
vanishes for small $\mu$ because of kinematics 
(see also \chap{chap:nlcmass}). 

Let us have a short look on the fermionic decay modes. In general the most
important ones among them are those into a chargino independent of 
$\cosb$. Whenever the decays into neutralinos are more important it
turns out that they are higgsino like. In each case the (relative) dominance
is due to large Yukawa couplings.

Let us finally discuss the $A_b$ dependence for  the cases $M_D < M_Q$ and 
$M_D > M_Q$ \fig{fig:brsbot2bnlc2A}. Here $\msbot{1}$,
$\msbot{2}$, and $\cosb$ are nearly
independent of $A_b$ because $\tanbe = 40$. This implies that the 
widths for the decays into charginos, into neutralinos, and into 
$Z \, \sbo{1}$ also are nearly independent of $A_b$. We start with the case
$M_U < M_Q < M_D$ shown in \fig{fig:brsbot2bnlc2A}a and b. 
The ordering of the various
modes can be understood by noting that i) $\cost \simeq 0$ for small $\mu$,
and for large $\mu$ $|\cost| \simeq 1/2$, ii) $m_{h^0}$ and $m_{H^{\pm}}$ 
grow with $|A_b|$ whereas $m_{H^0}$ decreases, iii) $\sina \simeq -0.45$ for
$A_b \simeq -1$~TeV and $\sina \simeq 0.05$ for $A_b \simeq 1$~TeV. The first 
point
leads to a decrease of $BR(\sbo{2} \to W^- \, \sto{2})$ for large $|A_b|$ and
at the same time to an increase in the $BR(\sbo{2} \to W^- \, \sto{1})$.
The second observation implies a slower increase of 
$\Gamma(\sbo{2} \to H^- \, \sto{2})$ compared to 
$\Gamma(\sbo{2} \to A^0 \, \sbo{1})$ 
which grows as $A^2_b \tanbeq$ leading to the decrease in the
$BR(\sbo{2} \to H^- \, \sto{2})$  for $|A_b| \simeq 1$~TeV. 
The third observation 
explains the dependence of $BR(\sbo{2} \to h^0 \, \sbo{1})$ and
$BR(\sbo{2} \to H^0 \, \sbo{1})$ on the sign of $A_b$. Similar arguments hold
for the case $M_D = M_U < M_Q$ shown in \fig{fig:brsbot2bnlc2A}c and d.

%% file: rar.tex
\chapter{Higher order decays of $\protect \sto{1}$ and $\protect \sbo{1}$}
\label{chap:rardecay}

\section{Three body decays of $\protect \sto{1}$}

In \sect{sect:stoplep2} scenarios have been presented where the main decays
of $\sto{1}$
are three body decays into sleptons. For larger stop masses also the decays
into $b \, W^+ \, \chin{1}$ and $b \, H^+ \, \chin{1}$ become possible as we
have seen in \sect{sect:stopnlc}. There only scenarios have been considered
where decays into sleptons are kinematically forbidden. In this section
we are going to compare both possibilities.

For fixing the parameters we have chosen the following
procedure: additional to $\tanbe$ and $\mu$ we have used within the stop
sector $\mstop{1}$ and $\cost$ as input parameters.
For the sbottom (stau) sector we have fixed
$M_Q, M_D$ and $A_b$ ($M_E, M_L, A_\tau$) as input parameters.
We have used this mixed set of parameters in
order to avoid unnatural parameters in the sbottom (stau) sector. Moreover,
we have assumed for simplicity that the soft SUSY breaking parameters are
equal for all generations. Note, that
because of $SU(2)$ invariance $M_Q$ also appears in the stop mass matrix
(see \eqn{eq:sfmassmatrix} and also \app{appB}). 
After some algebraic manipulation of the formulae in \app{appB} it can be seen 
that by variation of $\mu$ or
$\tanbe$ for fixed $\mstop{1}$ and $\cost$ one also varies $A_t$ and $M_U$.
Therefore, the mass of the heavier stop can be calculated
from this set of input parameters:
\begin{equation}
\mstq{2} = 
     \frac{2 M^2_{Q} 
           +2 \mzq \coszbe \left(\frac{1}{2} - \frac{2}{3} \sinwq \right)
           +2 m^2_t
           - \mstq{1} (1 + \coszt ) }
           { 1 - \coszt }
\end{equation}
In the sbottom (stau) sector obviously the physical quantities $\msbot{1}$,
$\msbot{2}$,and $\cosb$ 
($\mstau{1}, \mstau{2}$, $\costa$) change with $\mu$ and $\tanbe$.

A typical example is given in \fig{fig:brst3cosa} where we show branching
ratios as a function of $\cost$. We have restricted the $\cost$ range in such
a way that $|A_t| \lsim 1$~TeV to avoid color/charge breaking minima. The
parameters and physical quantities are given in \tab{tab:brst3cosa}.
The slepton parameters have been chosen in such a way that 
\noindent
\begin{minipage}[t]{150mm}
{\setlength{\unitlength}{1mm}
\begin{picture}(150,65)
\put(-4,-4){\mbox{\epsfig{figure=brst3cosa.eps,height=7.0cm,width=15.6cm}}}
\put(-1,61){{\small \bf a)}}
\put(4,60){\makebox(0,0)[bl]{{\small $BR(\sto{1} )$}}}
\put(70.5,1){\makebox(0,0)[br]{{\small $\cost$}}}
\put(75,60.5){{\small \bf b)}}
\put(80,59.5){\makebox(0,0)[bl]{{\small $BR(\sto{1} )$}}}
\put(148.5,1.5){\makebox(0,0)[br]{{\small $\cost$}}}
\end{picture}}
\figcaption{fig:brst3cosa}
           {Branching ratios for three body decays of 
            $\protect \sto{1}$ as a function of $\protect \cost$}
           {Branching ratios for $\sto{1}$ decays as a function of $\cost$
            for $\mstop{1} = 250$~GeV, $\tanbe = 2$, $\mu = 530$~GeV, and
            $M = 270$~GeV. The other parameters are given in 
            \tab{tab:brst3cosa}.
           The curves in a) correspond to the transitions:
           $\circ \hspace{1mm} \sto{1} \to b \, W^+ \, \chin{1}$,
           $\triangle \hspace{1mm} \sto{1} \to c \chin{1}$, 
           \recht $(\sto{1} \to b \, e^+ \, \tilde{\nu}_e)$
                + $(\sto{1} \to b \, \nu_e \, \tilde{e}^+_L)$, and
           $\bullet \hspace{1mm} (\sto{1} \to b \, \tau^+ \, \tsn)$
                + $(\sto{1} \to b \, \nutau \, \sta{1})$
                + $(\sto{1} \to b \, \nutau \, \sta{2})$.
           The curves in b) correspond to the transitions:
           $\circ \hspace{1mm} \sto{1} \to b \, \nu_e \, \tilde{e}^+_L$, 
           \rechtl $\sto{1} \to b \, \nutau \, \sta{1}$,
           $\triangle \hspace{1mm} \sto{1} \to b \, \nutau \, \sta{2}$,
           \recht $\sto{1} \to b \, e^+ \, \tilde{\nu}_e$, and
           $\bullet \hspace{1mm} \sto{1} \to b \, \tau^+ \, \tsn$.
      }{7}
\end{minipage}

\noindent
\begin{minipage}{75mm}
\begin{tabular}{|c|c|c||c|c|c|}
\hline 
$\tanbe$ & $\mu$ & $M$ & $\mchin{1}$ & $\mchip{1}$ & $\mchip{2}$ \\ \hline
2 & 530 & 270 &  130 & 250 & 551 \\ \hline \hline
$M_D$ & $M_Q$ & $A_b$  & $\msbot{1}$ & $\msbot{2}$ & $\cosb$ \\ \hline
370 & 340 & 150 & 342 & 372 & 0.98 \\ \hline \hline
$M_E$ & $M_L$ & $A_\tau$  & $\mstau{1}$ & $\mstau{2}$ & $\costa$ \\ \hline
210 & 210 & 150 & 209 & 217 & 0.68 \\ \hline \hline
 & & $\mstop{1}$ &
            $\mselec{L}$ & $\mselec{R}$ & $\mtsn$ \\ \hline
 & & 250 & 213 & 212 & 204 \\ \hline
\end{tabular}
\\[2mm]
\tabcaption{tab:brst3cosa}
           {Parameters and physical quantities used in 
            {\protect \fig{fig:brst3cosa}}}
           {Parameters and physical quantities used in \fig{fig:brst3cosa}.
            All masses are given in GeV.
           }
\end{minipage}
\hspace*{4mm}
\begin{minipage}{70mm}
the sum of
the final state particles are $215 \pm 5$~GeV leading to comparable
kinematics for each decay mode.
In \fig{fig:brst3cosa}a we present 
$BR(\sto{1} \to b \, W^+ \, \chin{1})$,
$BR(\sto{1} \to c \, \chin{1})$,
$BR(\sto{1} \to b \, e^+ \, \tilde{\nu}_e$) + 
$BR(\sto{1} \to b \, \nu_e \, \tilde{e}^+_L)$, and
$BR(\sto{1} \to b \, \tau^+ \, \tsn)$ +
$BR(\sto{1} \to b \, \nutau \, \sta{1})$ +
$BR(\sto{1} \to b \, \nutau \, \sta{2})$. 
Here we have not included the
possibility $BR(\sto{1} \to b \, H^+ \, \chin{1})$ because it was not possible
to find a $\ma$ which simultaneously allowed this decay and fulfilled the
condition $m_{h^0} \gsim 70$~GeV. However, we will 
\end{minipage} \\[1mm]
discuss this 
decay later on.
We have added those branching ratios
for the decays into sleptons that give the same final state after
the sleptons have decayed. For example:
\beq
\sto{1} \to b \, \nutau \, \sta{1} \,
          \to \, b \, \tau \, \nutau \, \chin{1}; \hspace{5mm}
\sto{1} \to b \, \tau \, \tsn \,
          \to \, b \, \tau \, \nutau \, \chin{1}
\eeq
Note, that the requirement $\mstop{1} - m_b < \mchip{1}$
implies that the sleptons can only decay into the corresponding lepton plus
the lightest neutralino except a small parameter 
\noindent
\begin{minipage}[t]{150mm}
{\setlength{\unitlength}{1mm}
\begin{picture}(150,65)                        
\put(-4,-3.5){\mbox{\epsfig{figure=brst3tana.eps,height=7.0cm,width=15.6cm}}}
\put(-1,61.5){{\small \bf a)}}
\put(4,60.5){\makebox(0,0)[bl]{{\small $BR(\sto{1} )$}}}
\put(70.5,1){\makebox(0,0)[br]{{\small $\tanbe$}}}
\put(75,61.5){{\small \bf b)}}
\put(80,60.5){\makebox(0,0)[bl]{{\small $BR(\sto{1} )$}}}
\put(148.5,1.0){\makebox(0,0)[br]{{\small $\tanbe$}}}
\end{picture}}
\figcaption{fig:brst3tana}
           {Branching ratios for three body decays of 
            $\protect \sto{1}$ as a function of $\protect \tanbe$}
           {Branching ratios for $\sto{1}$ decays as a function of $\tanbe$
            for $\mstop{1} = 250$~GeV, $\cost = 0.6$, $\mu = 530$~GeV,
            $M = 270$~GeV. The other parameters are given 
            in \tab{tab:brst3cosa}.
           The curves in a) correspond to the transitions:
           $\circ \hspace{1mm} \sto{1} \to b \, W^+ \, \chin{1}$,
           $\triangle \hspace{1mm} \sto{1} \to c \chin{1}$, 
           \recht $(\sto{1} \to b \, e^+ \, \tilde{\nu}_e)$
                + $(\sto{1} \to b \, \nu_e \, \tilde{e}^+_L)$, and
           $\bullet \hspace{1mm} (\sto{1} \to b \, \tau^+ \, \tsn)$
                + $(\sto{1} \to b \, \nutau \, \sta{1})$
                + $(\sto{1} \to b \, \nutau \, \sta{2})$.
           The curves in b) correspond to the transitions:
           $\circ \hspace{1mm} \sto{1} \to b \, \nu_e \, \tilde{e}^+_L$, 
           \rechtl $\sto{1} \to b \, \nutau \, \sta{1}$,
           $\triangle \hspace{1mm} \sto{1} \to b \, \nutau \, \sta{2}$,
           \recht $\sto{1} \to b \, e^+ \, \tilde{\nu}_e$, and
           $\bullet \hspace{1mm} \sto{1} \to b \, \tau^+ \, \tsn$.
      }{7}
\end{minipage}

\noindent
region where
the decay into $\chin{2}$ is possible. However, there this decay will be
negligible due to kinematics. The branching ratios for 
decays into
$\tilde{\mu}_{L}, \, \tilde{\nu}_{\mu}$ are not shown because they are the
same as
in the case of $\tilde{e}_{L}, \, \tilde{\nu}_{e}$ up to very tiny mass
effects. The sum of the branching ratios
for the decays into $\sta{1}$ and $\sta{2}$ also has nearly the same size 
because of the small $\tanbe$. The
$BR(\sto{1} \to c \, \chin{1})$ is $O(10^{-4})$ 
independent of $\cost$ and therefore is
negligible. Near $\cost = -0.3$ the decay into $b \, W^+ \, \chin{1}$
has a branching ratio of $\sim 100\%$. Here $\lte{1}$ vanishes leading to the
reduction of the decays into sleptons (see also \sect{sect:stoplep2}).

In \fig{fig:brst3cosa}b the branching ratios for the decays into the
different sleptons are shown. As already mentioned in \sect{sect:stoplep2}
the sleptons couple mainly to the gaugino components of $\chip{1}$ if
$\tanbe$ is small. Therefore, the branching ratios for decays into the staus 
are reduced because they are strongly mixed.
However, the sum of both branching ratios is
nearly the same as $BR(\sto{1} \to b \, \nu_e \, \tilde{e}^+_L)$.
The decays into sneutrinos are preferred by kinematics. Moreover, the
matrix elements (\eqn{eq:Tfistbsneut} and (\ref{eq:Tfistbslept}))
for the decays into charged and neutral sleptons have
a different structure in the limit $m_b, m_l \to 0$ leading to an
additional difference:
\beq
T_{fi}(\sto{1} \to b \, l^+ \, \tilde{\nu}) & \sim &  
         \mchipm{i} \bar u(p_b) P_R v(p_l) \\
T_{fi}(\sto{1} \to b \, \nu_l \, \tilde{l}_k) & \sim &  
         \bar u(p_b) P_R  \not \! \pxi v(p_{\nu_l})
\eeq
\noindent
\begin{minipage}[t]{150mm}
{\setlength{\unitlength}{1mm}
\begin{picture}(150,65)                        
\put(-4,-3.5){\mbox{\epsfig{figure=brst3cospmu.eps,height=7.0cm,width=15.6cm}}}
\put(-1,61){{\small \bf a)}}
\put(4,60){\makebox(0,0)[bl]{{\small $BR(\sto{1} )$}}}
\put(70.5,1){\makebox(0,0)[br]{{\small $\cost$}}}
\put(75,60){{\small \bf b)}}
\put(80,59){\makebox(0,0)[bl]{{\small $BR(\sto{1} )$}}}
\put(148.5,2.0){\makebox(0,0)[br]{{\small $\cost$}}}
\end{picture}}
\figcaption{fig:brst3cospmu}
           {Branching ratios for three body decays of 
            $\protect \sto{1}$ as a function of $\protect \cost$}
           {Branching ratios for $\sto{1}$ decays as a function of $\cost$
            for $\mstop{1} = 350$~GeV, $\tanbe = 2$, $\mu = 750$~GeV,
            $M = 380$~GeV, and $\ma = 110$~GeV. The other parameters are given 
           in the text.
           The curves in a) correspond to the transitions:
           $\circ \hspace{1mm} \sto{1} \to b \, W^+ \, \chin{1}$,
           \rechtl $\sto{1} \to b \,H^+ \,  \chin{1}$, 
           $\triangle \hspace{1mm} \sto{1} \to c \chin{1}$, 
           \recht $(\sto{1} \to b \, e^+ \, \tilde{\nu}_e)$
                + $(\sto{1} \to b \, \nu_e \, \tilde{e}^+_L)$, and
           $\bullet \hspace{1mm} (\sto{1} \to b \, \tau^+ \, \tsn)$
                + $(\sto{1} \to b \, \nutau \, \sta{1})$
                + $(\sto{1} \to b \, \nutau \, \sta{2})$.
           The curves in b) correspond to the transitions:
           $\circ \hspace{1mm} \sto{1} \to b \, \nu_e \, \tilde{e}^+_L$, 
           \rechtl $\sto{1} \to b \, \nutau \, \sta{1}$,
           $\triangle \hspace{1mm} \sto{1} \to b \, \nutau \, \sta{2}$,
           \recht $\sto{1} \to b \, e^+ \, \tilde{\nu}_e$, and
           $\bullet \hspace{1mm} \sto{1} \to b \, \tau^+ \, \tsn$.
           In the grey area $m_{h^0}$ is smaller than 70~GeV.
      }{7}
\end{minipage}

\noindent
\begin{minipage}{8.0cm}
\begin{tabular}{|c|c|c||c|c|c|}
\hline 
$\tanbe$ & $\mu$ & $M$ & $\mchin{1}$ & $\mchip{1}$ & $\mchip{2}$ \\ \hline
2 & 750 & 380 &  186 & 366 & 766 \\ \hline \hline
$M_D$ & $M_Q$ & $A_b$  & $\msbot{1}$ & $\msbot{2}$ & $\cosb$ \\ \hline
550 & 500 & 400 & 502 & 551 & 0.99 \\ \hline \hline
$M_E$ & $M_L$ & $A_\tau$  & $\mstau{1}$ & $\mstau{2}$ & $\costa$ \\ \hline
275 & 275 & 400 & 274 & 281 & 0.69 \\ \hline \hline
$m_{H^+}$ & $\ma$ & $\mstop{1}$ &
            $\mselec{L}$ & $\mselec{R}$ & $\mtsn$ \\ \hline
$130 \pm 1$ & 110 & 350 & 278 & 277 & 270 \\ \hline
\end{tabular}
\\[2mm]
\tabcaption{tab:brst3cospmu}
           {Parameters and physical quantities used in 
            {\protect \fig{fig:brst3cospmu}}}
           {Parameters and physical quantities used in \fig{fig:brst3cospmu}.
            All masses are given in GeV.
           }
\end{minipage}
\hspace*{4mm}
\begin{minipage}{65mm}
This leads to different decay widths even in the case where one assumes
equal masses for all sleptons.
The decay $\sto{1} \to b \, W^+ \, \chin{1}$ is dominated by the $t$-quark
contribution 
followed by the chargino contributions. In many cases
the interference term between $t$ and $\chip{1,2}$ is 
more important than
the $\chip{1,2}$ part. Moreover, we have found that the contribution 
from the exchange of the sbottoms are in general negligible.
\end{minipage} \\[1mm]

In \fig{fig:brst3tana} we show the branching ratios as a function of $\tanbe$
for $\cost = 0.6$ and the other parameters as above. For small $\tanbe$ the
decay into $\sto{1} \to b \, W^+ \, \chin{1}$ is the most 
important one.
The branching ratios 
for the decays into sleptons are reduced in the range 
$\tanbe \lsim 5$ because the 
gaugino component of $\chip{1}$ decreases and its mass increases.
For $\tanbe \gsim 10$ the decays into the
$b \, \tau \, \me$ final state becomes
more important because of the growing $\tau$ Yukawa coupling and because of
kinematics ($\mstau{1}$ decreases if $\tanbe$ increases and the other
parameters are 
\noindent
\begin{minipage}[t]{150mm}
{\setlength{\unitlength}{1mm}
\begin{picture}(150,65)                        
\put(-4,-5){\mbox{\epsfig{figure=brst3tanb.eps,height=7.0cm,width=15.6cm}}}
\put(-1,60.5){{\small \bf a)}}
\put(4,59.5){\makebox(0,0)[bl]{{\small $BR(\sto{1} )$}}}
\put(70.5,1){\makebox(0,0)[br]{{\small $\tanbe$}}}
\put(75,59.5){{\small \bf b)}}
\put(80,58.5){\makebox(0,0)[bl]{{\small $BR(\sto{1} )$}}}
\put(148.5,2.0){\makebox(0,0)[br]{{\small $\tanbe$}}}
\end{picture}}
\figcaption{fig:brst3tanb}
           {Branching ratios for three body decays of 
            $\protect \sto{1}$ as a function of $\protect \tanbe$}
           {Branching ratios for $\sto{1}$ decays as a function of $\tanbe$
            for $\mstop{1} = 350$~GeV, $\cost = 0.7$, $\mu = 750$~GeV,
            $M = 380$~GeV and $\ma = 110$~GeV. The other parameters are given 
           in \tab{tab:brst3cospmu}.
           The curves in a) correspond to the transitions:
           $\circ \hspace{1mm} \sto{1} \to b \, W^+ \, \chin{1}$,
           \rechtl $\sto{1} \to b \,H^+ \,  \chin{1}$, 
           $\triangle \hspace{1mm} \sto{1} \to c \chin{1}$, 
           \recht $(\sto{1} \to b \, e^+ \, \tilde{\nu}_e)$
                + $(\sto{1} \to b \, \nu_e \, \tilde{e}^+_L)$, and
           $\bullet \hspace{1mm} (\sto{1} \to b \, \tau^+ \, \tsn)$
                + $(\sto{1} \to b \, \nutau \, \sta{1})$
                + $(\sto{1} \to b \, \nutau \, \sta{2})$.
           The curves in b) correspond to the transitions:
           $\circ \hspace{1mm} \sto{1} \to b \, \nu_e \, \tilde{e}^+_L$, 
           \rechtl $\sto{1} \to b \, \nutau \, \sta{1}$,
           $\triangle \hspace{1mm} \sto{1} \to b \, \nutau \, \sta{2}$,
           \recht $\sto{1} \to b \, e^+ \, \tilde{\nu}_e$, and
           $\bullet \hspace{1mm} \sto{1} \to b \, \tau^+ \, \tsn$.
      }{7}
\end{minipage}

\noindent
fixed. See also the discussion at the end of
\sect{sect:stoplep2}). Here $\sto{1} \to b \, \nutau \, \sta{1}$
is the most important contribution as can be seen in \fig{fig:brst3tana}b.
For large 
$\tanbe$ the decay into $c \, \chin{1}$ also gains some importance
because the width is proportional to the bottom
Yukawa coupling in the approximation used (\eqn{deltal} and (\ref{deltar})).

From the requirement that no two body decays are allowed at
tree level follows that $\mchip{1} > \mstop{1} - m_b$. Therefore, one expects
an increase of $BR(\sto{1} \to b \, W^+ \, \chin{1})$ if $\mstop{1}$ increases, 
because
the decay into $b \, W^+ \, \chin{1}$ is dominated by the $t$ exchange
whereas for the decays into sleptons the $\chipm{1}$ contribution is
the dominating one. This is demonstrated in \fig{fig:brst3cospmu} where we
have fixed $\mstop{1} = 350$~GeV. Here we have also found scenarios where
the decay into $b \, H^+ \, \chin{1}$ is possible. It turns out that this
channel is in general negligible because of kinematics. We have not
found any case where $m_{H^+} \lsim 120$~GeV if we use the mass formulae for the
Higgs bosons in the MSSM and at the same time include the experimental bound
$m_{h^0} < 70$~GeV \cite{higgsbound}.

These general features hold even if $\tanbe$ increases as can be seen in
\fig{fig:brst3tanb}. Here we have fixed $\cost = 0.7$. As expected by the
discussion above, the decay into $\sto{1} \to b \, \nutau \, \sta{1}$ gains
some importance for large $\tanbe$. Note, that for large $\tanbe$
$BR(\sto{1} \to b \, H^+ \, \chin{1})$ decreases since $m_{H^+}$ increases due
to radiative corrections. However, there are scenarios where this decay
can become important as demonstrated in \fig{fig:brst3MD}.
Here we show the branching ratios as a function of $M_D$ for $\ma = 90$~GeV,
$\tanbe = 30$ and the other parameters as in \tab{tab:brst3cospmu}.
At the lower 
\begin{minipage}{77mm}
{\setlength{\unitlength}{1mm}
\begin{picture}(77,72)
\put(-0.1,2.0){\mbox{\psfig{figure=brst3MDa.eps,height=6.3cm,width=7.5cm}}}
\put(1,66){\makebox(0,0)[bl]{{\small $BR(\sto{1} )$}}}
\put(74.5,1){\makebox(0,0)[br]{{\small $M_D$~[GeV]}}}
\end{picture}}
\figcaption{fig:brst3MD}
           {Branching ratios for three body decays of 
            $\protect \sto{1}$ as a function of $M_D$}
           {Branching ratios for $\sto{1}$ decays as a function of $M_D$
            for $\mstop{1} = 350$~GeV, $\cost = 0.7$, $\tanbe = 30$,
            $\mu = 750$~GeV,
            $M = 380$~GeV and $\ma = 90$~GeV. The other parameters are given 
           in \tab{tab:brst3cospmu}.
           The curves in a) correspond to the transitions:
           $\circ \hspace{1mm} \sto{1} \to b \, W^+ \, \chin{1}$,
           \rechtl $\sto{1} \to b \,H^+ \,  \chin{1}$, 
           $\triangle \hspace{1mm} \sto{1} \to c \chin{1}$, 
           \recht $(\sto{1} \to b \, e^+ \, \tilde{\nu}_e)$
                + $(\sto{1} \to b \, \nu_e \, \tilde{e}^+_L)$, and
           $\bullet \hspace{1mm} (\sto{1} \to b \, \tau^+ \, \tsn)$
                + $(\sto{1} \to b \, \nutau \, \sta{1})$
                + $(\sto{1} \to b \, \nutau \, \sta{2})$.
      }{4}
\end{minipage}
\hspace{4mm}
\begin{minipage}[t]{68mm}
\vspace*{-6cm}
end of the $M_D$ range we get $m_{H^+} = 114$~GeV. Moreover,
$\msbot{1}$ is approximately 
$\mstop{1} - \mw$ leading to an enhancement 
of this width. We have found that the 
relative importance of the sbottom
exchange is larger in the case of $\sto{1} \to b \, H^+ \, \chin{1}$ than in 
the case of
$\sto{1} \to b \, W^+ \, \chin{1}$. This is a consequence of the different spin
structure of the corresponding matrix elements (\eqn{eq:TfistbWchi} and
(\ref{eq:TfistbHchi})). Note, that the decrease in the 
$BR(\sto{1} \to b \, H^+ \, \chin{1})$ for
$M_D \gsim 450$~GeV is mainly due to the growth in $m_{H^+}$.
\end{minipage}

\section{The decay $\protect \sbo{1} \to c \, \protect \chim{1}$}
\label{sect:rarsbot}

In the following we assume that SUSY is realized in nature in such a way that
i) $\chin{1}$, $\chin{2}$, and $\chipm{1}$ are mainly higgsinos, 
ii) $\msbot{1} < min(\mchipm{1} + m_t, \mchin{3} + m_b)$
and iii) $\tanbe$ is small. Under 
these conditions the decay widths $\Gamma(\sbo{1} \to b \, \chin{1})$  
and $\Gamma(\sbo{1} \to b \, \chin{2})$  will be small.
Now the mixing between the squark generations and between the
the $c$-quark and the $t$-quark leads to the decay
$\sbo{1} \to c \, \chim{1}$. In the following we will assume for simplicity 
that the mixing in the squark sector is the same as in the quark sector. 
In this case the decay width
is given by:
\beq
\zerfallz{\sbo{1}}{c}{\chim{1}} \plgl
 \frac{ g^2 |K^2_{cb}| \lamh{\sbo{1}}{c}{\chipm{1}}{\einha} }
      {16 \pi  m^3_{\sbo{1}} } \no
  \plogl{-15}
     *  \left[ \Big( \kbeq{1}{ik} + \lbeq{1}{ik}
        \Big) \Big( \msbq{1} - m^2_{c} - \mchipmq{1} \Big)
  - 4 \kbe{1} \lbe{1} m_{c} \mchipm{1} \right] 
\eeq
\begin{minipage}[t]{150mm}
{\setlength{\unitlength}{1mm}
\begin{picture}(150,65)                        
\put(-4,-4){\mbox{\epsfig{figure=sbottomrar.eps,height=6.9cm,width=15.6cm}}}
\put(-1,61){{\small \bf a)}}
\put(4,60){\makebox(0,0)[bl]{{\small $BR(\sbo{1} )$}}}
\put(70.5,1){\makebox(0,0)[br]{{\small $\cosb$}}}
\put(75,60){{\small \bf b)}}
\put(80,59){\makebox(0,0)[bl]{{\small $BR(\sbo{1} )$}}}
\put(148,1){\makebox(0,0)[br]{{\small $\tanbe$}}}
\end{picture}}
\figcaption{fig:sbottomrar}
           {Branching ratios for $\protect \sbo{1} \to c \, \protect \chim{1}$}
           {Branching ratios for $\sbo{1}$ decays a) as a function of $\cosb$
            for $\tanbe = 1.5$ and b) as a function of $\tanbe$ for 
            $\cosbe = -0.9$. The other parameters are $\msbot{1} = 200$~GeV,
            $M = 500$~GeV, and $\mu = -110$~GeV.
           The curves correspond to the transitions:
           $\circ \hspace{1mm} \sbo{1} \to b \, \chin{1}$,
           \rechtl \hspace{1mm} $\sbo{1} \to b \, \chin{2}$, and
           \recht $\sbo{1} \to c \, \chim{1}$.
      }{7}
\end{minipage}

\noindent
where $K_{cb}$ is the corresponding CKM matrix element. Note, that the 
decay width is approximately proportional to $m^2_t$ if we neglect $m_c$ and
the couplings to the gaugino component of $\chim{1}$:
\beq
\zerfallz{\sbo{1}}{c}{\chim{1}} \simeq
 \frac{ g^2 \, |K^2_{cb}| \, V^2_{12} \, \cosbq \, m^2_t}
      {32 \, \pi \, \sinbeq \, \mwq \, m^3_{\sbo{1}} } 
      \Big( \msbq{1} - \mchipmq{1} \Big)^2
\label{eq:sbotrar}
\eeq

As a typical example we have taken $M = 500$~GeV, $\mu = -110$~GeV, 
$\tanbe = 1.5$, and $\msbot{1} = 200$~GeV. This corresponds to 
$\mchin{1} \simeq 110$~GeV, $\mchin{2} \simeq 126$~GeV, and
$\mchipm{1} \simeq 120$~GeV. In \fig{fig:sbottomrar}a we show the branching 
ratios as a function of $\cosb$. As expected by the approximation formula in
\eqn{eq:sbotrar} $BR(\sbo{1} \to c \, \chim{1})$ has its maximum near 
$|\cosb| = 1$ where it is of $ \sim 20\%$. The reason for this, at first glance,
large branching ratio is that $N_{11}^2$ and $N^2_{12}$ are of the same order
as $|K_{cb}|^2 m^2_t / \mwq$. The signatures will be combinations
of the following final states:
\beq
\begin{array}{ll}
\sbo{1} \to c \, \chim{1} & \to c \, l^- \, \bar{\nu}_l \, \chin{1} \\
                          & \to c \, q \, \bar{q}' \, \chin{1} \\
\sbo{1} & \to b \, \chin{1} \\
\sbo{1} \to b \, \chin{2} & \to b \, l^- \, l^+ \, \chin{1} \\
                          & \to b \, \nu_l \bar{\nu}_l \, \chin{1} \\
                          & \to b \, q \, \bar{q} \, \chin{1}
\end{array}
\eeq
with $q = u,d,c,s,b$ and $l=e,\mu,\tau$.

To get a feeling for the term 'small $\tanbe$' we show in 
\fig{fig:sbottomrar}b the branching ratios as a function of $\tanbe$ for
$\cosb = -0.9$. One sees that it rapidly decreases with the increasing of
$\tanbe$.
For example we get that $BR(\sbo{1} \to c \, \chim{1}) \simeq 0.1$ (0.02) for
$\tanbe = 3$ (10). In every case we have found that the total decay width
$\Gamma(\sbo{1})$ is of $O(10$~MeV). Therefore, we expect hadronization
to play a r\^ole in such a scenario.

%% file: conclusions.tex
\chapter{Summary}
\label{chap:con}

We have studied  the phenomenology of third generation sfermions paying
particular attention to the implications of the Yukawa couplings and to the
left-right mixing.
Analytical formulae have been given for the sfermion mixing, the production
cross sections, and for
all possible two body decay widths that can occur at tree level.

In respect to the masses and the mixing angles of the sfermions we have found 
that
a large mass splitting between sfermions of the same flavour occurs if either
the off diagonal term 
$m_f (A_f - \mu \Theta(\beta))$ and/or $|M^2_{LL} - M^2_{RR}$ is large 
($\Theta(\beta) = \cotbe$ for $f=t$ and $\tanbe$ for $f = b,\tau$). 
The size
of the mixing is determined by the ratio 
$m_f (A_f - \mu \Theta(\beta)) / (M^2_{LL} - \msferq{1})$. This can lead to a
rather large mixing for sbottoms and staus even if $\tanbe$ is small.
The production cross sections of stops, sbottoms and staus in $e^+ e^-$ 
annihilation depend on the mixing angle in a characteristic way. This dependence
is more pronounced if polarized $e^-$ are available.

Large Yukawa couplings and left-right mixing strongly influences the decays
of $\sto{1,2}$, $\sbo{1,2}$, $\sta{1,2}$, and $\tsn$. In the
case of the decays into
charginos and neutralinos there occur interferences between gaugino and
higgsino components. These interferences can be either positive or negative
depending on the parameters. In particular we have found scenarios where the
decays into the heaviest neutralino and chargino are the most important ones.

In the case that the mass splitting of the sfermions is large, decays
into gauge bosons and Higgs bosons also have large branching ratios. 
Here we have found that the following facts are important:
An extra $\lambda / m^2_V$ factor
strengthens kinematical effects in case of decays into
gauge bosons. The decay width into a $Z$ boson is enlarged if there is 
a strong mixing of the sfermions whereas in case of a $W$ boson the width is
enhanced if both sfermions are left states. Both cases can be
explained through the chirality preservation of the gauge boson couplings to 
the corresponding fermions. By contrast, Higgs bosons change the chirality of
fermions. Corresponding to this we have found that decays of sfermions
into Higgs bosons are important if one of the sfermions is a left state and the
other is a right state. The only exception is the pseudoscalar Higgs because
its couplings are independent of the mixing angles. Moreover, the decay widths
are strongly influenced by the magnitude of $A_i$ ($i=b,t,\tau$) and $\mu$,
the relative sign between these parameters, and the Higgs mixing angle.

In case of the light stop it is possible that all two body decays are forbidden
at tree level. In such a scenario three body decays will be important followed
by the one loop decay into a charm quark and the lightest neutralino. 
Additionally,
in case of the light sbottom we have found scenarios where the flavour changing
decay into a charm-quark and the lightest chargino gains some importance.
In both cases we expect hadronization effects to be important.

We have seen that stops, sbottoms, staus and the tau sneutrino have a rich
phenomenology. A precise determination of their properties will give us 
insight into the SUSY breaking parameters and hopefully also into the 
mechanism of
SUSY breaking.

%% file: appint.tex
\chapter{Formulae for the three body decay widths}
\label{appA}

In the subsequent the formulae for the three body decay widths of the light
stop are collected which have been omitted in \sect{sec:threebody}.
 
\section[Light Stop into W, Bottom, and Neutralino]
       {The Decay of the Light Stop into a W-Boson, a Bottom Quark and
         the Lightest Neutralino}

\noindent
The decay width is given by
\beq
 \Gamma(\stwbc) \plgl \no
 \plogl{-39} = \frac{\alpha^2}{16 \, \pi m^3_{\sto{1}} \sinwv}
        \eint{(\mstop{1}-m_W)^2}{(m_b + \mchin{1})^2}{s}
   \left( F_{\chip{} \chip{}} +
   F_{\chip{} t} +
   F_{\chip{} \sbo{}} +
   F_{t t} +
   F_{t \sbo{}} +
   F_{\sbo{} \sbo{}} \right) \no
\eeq
with
\beq \hspace*{-3mm}
   F_{\chip{} \chip{}} \plgl
       \sum^2_{i=1} \Big[
          ( ca_{i1} + ca_{i2} s + ca_{i3} s^2 + ca_{i4} s^3) \no
    \plogl{15} * J^0_t(\dsum{\sto{1}}{W}{b}{\chin{1}}{\chip{i}}
             ,\Gamma_{\chip{i}} \mchip{i}) \no
    \plogl{5} + ( ca_{i5} + ca_{i6} s + 2 \, ca_{i4} s^2) \no
    \plogl{15} * J^1_t(\dsum{\sto{1}}{W}{b}{\chin{1}}{\chip{i}}
             ,\Gamma_{\chip{i}} \mchip{i}) \no
    \plogl{3} + ( ca_{i7} + ca_{i4} s)
          J^2_t(\dsum{\sto{1}}{W}{b}{\chin{1}}{\chip{i}}
             ,\Gamma_{\chip{i}} \mchip{i}) \Big] \no
    \plogl{-1} +
          ( ca_{31} + ca_{32} s + ca_{33} s^2 + ca_{34} s^3) \no
    \plogl{9} * J^0_{tt}(\dsum{\sto{1}}{W}{b}{\chin{1}}{\chip{1}}
             ,\Gamma_{\chip{1}} \mchip{1}  \no
       \plogl{17} ,\dsum{\sto{1}}{W}{b}{\chin{1}}{\chip{2}}
             ,\Gamma_{\chip{2}} \mchip{2}) \no
    \plogl{-1} + ( ca_{35} + ca_{36} s + 2 \, ca_{34} s^2) \no
    \plogl{9} * J^1_{tt}(\dsum{\sto{1}}{W}{b}{\chin{1}}{\chip{1}}
             ,\Gamma_{\chip{1}} \mchip{1}  \no
       \plogl{17} ,\dsum{\sto{1}}{W}{b}{\chin{1}}{\chip{2}}
             ,\Gamma_{\chip{2}} \mchip{2}) \no
    \plogl{-1} + ( ca_{37} + ca_{34} s) \no
    \plogl{9} * J^2_{tt}(\dsum{\sto{1}}{W}{b}{\chin{1}}{\chip{1}}
             ,\Gamma_{\chip{1}} \mchip{1}  \no
       \plogl{17} ,\dsum{\sto{1}}{W}{b}{\chin{1}}{\chip{2}}
             ,\Gamma_{\chip{2}} \mchip{2})
\eeq
\beq
\hspace*{-3mm}
   F_{\chip{} t} \plgl
       \sum^2_{i=1} \Big[
          ( cb_{i1} + cb_{i2} s + cb_{i3} s^2) \no
    \plogl{10} * J^0_{tt}(\dsum{\sto{1}}{W}{b}{\chin{1}}{\chip{1}}
       ,-\Gamma_{\chip{1}} \mchip{1},m^2_t, \Gamma_t m_t) \no
    \plogl{5} + ( cb_{i4} + cb_{i5} s + cb_{i6} s^2) \no
    \plogl{10} * J^1_{tt}(\dsum{\sto{1}}{W}{b}{\chin{1}}{\chip{1}}
       ,-\Gamma_{\chip{1}} \mchip{1} ,m^2_t, \Gamma_t m_t) \no
    \plogl{5} + ( cb_{i7} + cb_{i6} s) \no
    \plogl{10} * J^2_{tt}(\dsum{\sto{1}}{W}{b}{\chin{1}}{\chip{1}}
       ,-\Gamma_{\chip{1}} \mchip{1} ,m^2_t, \Gamma_t m_t)  \Big] \no
\eeq
\beq
\hspace*{-3mm}
   F_{\chip{} \sbo{}} \plgl
       \sum^2_{k,i=1} \Big[
          ( cc_{ik1} + cc_{ik2} s + cc_{ik3} s^2 + cc_{ik4} s^3) \no
    \plogl{10} * J^0_{st}(
       \msbq{k}, \Gamma_ {\sbo{k}} \msbot{k}
       ,\dsum{\sto{1}}{W}{b}{\chin{1}}{\chip{1}}
       ,-\Gamma_{\chip{1}} \mchip{1}) \no
    \plogl{5} + ( cc_{ik5} + cc_{ik6} s + cc_{ik4} s^2) \no
    \plogl{10} * J^1_{st}(
       \msbq{k}, \Gamma_ {\sbo{k}} \msbot{k}
       ,\dsum{\sto{1}}{W}{b}{\chin{1}}{\chip{1}}
       ,-\Gamma_{\chip{1}} \mchip{1}) \Big]  \no
\eeq
\beq
\hspace*{-3mm}
   F_{t t} \plgl
          ( cd_{1} + cd_{2} s) J^0_t(m^2_t, \Gamma_t m_t) +
          ( cd_{3} + cd_{4} s) J^1_t(m^2_t, \Gamma_t m_t) \no
  \plogl{-3} +  ( cd_{5} + cd_{6} s) J^2_t(m^2_t, \Gamma_t m_t)
\eeq
\beq
\hspace*{-3mm}
   F_{t \sbo{}} \plgl \sum^2_{k=1} \Big[
   ( ce_{k1} + ce_{k2} s + ce_{k3} s^2)
   J^0_{st}(\msbq{k}, \Gamma_ {\sbo{k}} \msbot{k},m^2_t, \Gamma_t m_t) \no
  \plogl{6} + ( ce_{k4} + ce_{k5} s + ce_{k6} s^2)
   J^1_{st}(\msbq{k}, \Gamma_ {\sbo{k}} \msbot{k},m^2_t, \Gamma_t m_t)
   \Big]
\eeq
\beq
\hspace*{-3mm}
   F_{\sbo{} \sbo{}} \plgl \left\{ \sum^2_{k=1}
      \frac{(cf_{k1} + cf_{k2} s)}
        {m^2_W [(s-\msbq{k})^2 + \Gamma^2_{\sbo{k}} \msbq{k}]}  \right. \no
   \plogl{2} \left. + Re\left[
      \frac{(cf_{31} + cf_{32} s)}
           {m^2_W [(s-\msbq{1}) + i \Gamma_{\sbo{1}} \msbot{1}]
                   [(s-\msbq{2}) - i \Gamma_{\sbo{2}} \msbot{2}]}
          \right]  \right\} \no
 \plogl{-1} *\frac{\lambda(s,\mstq{1},m^2_W)\sqrt{\lambda(s,\mstq{1},m^2_W)
                  \lambda(s,\mchinq{1},m^2_b)}}{s}
\eeq
The integrals $J^{0,1,2}_{t,tt,st}$ are defined in \sect{intJ}. Their
integration range is given by
\beq
t_{max \atop min} \plgl \frac{\mstq{1} + m^2_b + m^2_W + \mchinq{1} -s}{2}
       - \frac{(\mstq{1}-m^2_W) (\mchinq{1}-m^2_b)}{2 s} \no
  \plogl{0} \pm  \frac{\sqrt{\lambda(s,\mstq{1},m^2_W)
                  \lambda(s,\mchinq{1},m^2_b)}}{2 s}
\eeq
and $s = (p_{\sto{1}} - p_W)^2$. Note, that $-\Gamma_{\chip{1}} \mchip{1}$
appears in the entries of the integrals $F_{\chip{} \sbo{j}}$ and
$F_{\chip{} t}$ because the chargino exchange is the $u$-channel in our
convention.
The coefficients are given by:
\beq
ca_{11} \plgl  6 \, \col{11} \cor{11} \Big[ (\kteq{1} + \lteq{1})
          \mchip{1} \mchin{1}  (2 \,  m^2_b + \mchinq{1} + \mwq) \no
   \plogl{19}  + 2  \, \kte{1} \lte{1} m_b \mchin{1}
          (m^2_b + \mchinq{1} + \mstq{1} + \mchipq{1} +\mwq) \Big] \no
   \plogl{-10} - \left( \lteq{1} \colq{11} + \kteq{1} \corq{11} \right)
        \Bigg[ (4 \, m^2_b + \mwq) (\mchinq{1}+m^2_b) + \mchinv{1} \no
   \plogl{3}  + \mstq{1} (2 \, \mwq + 4 \, m^2_b + \mchinq{1})
   + \left. \frac{(m^2_b+\mstq{1})
        (m^4_b + m^2_b \mstq{1} + \mchinq{1}\mstq{1})}{\mwq} \right] \no
   \plogl{-10} - \left( \kteq{1} \colq{11} + \lteq{1} \corq{11} \right)
      \mchipq{1}  \left( \mchinq{1} + \mstq{1}  + 2 \, m^2_b +
             \frac{m^4_b +m^2_b \mstq{1}}{\mwq}   \right) \no 
   \plogl{-10}  - 2 \, \kte{1} \lte{1}  \left( \colq{11} + \corq{11} \right)
          m_b \mchip{1} \no
   \plogl{0} * \left( 3 \, m^2_b + 2 \, \mchinq{1} + 3 \, \mstq{1}
                         + \frac{(\mstq{1}+m^2_b)^2}{\mwq} \right) \\
ca_{12} \plgl -12 \, \kte{1} \lte{1}  \col{11} \cor{11}  m_b \mchin{1}
             - 6 \, \left( \kteq{1} + \lteq{1} \right) \col{11} \cor{11}
              \mchip{1} \mchin{1} \no
     \plogl{-10} + 2 \, \kte{1} \lte{1} \left( \colq{11} + \corq{11} \right)
          m_b \mchip{1}
             \left( 3 + \frac{2 \,(m^2_b+\mstq{1})}{\mwq} \right) \no
     \plogl{-10} + \left( \lteq{1} \colq{11} + \kteq{1} \corq{11} \right)
            \Bigg( 6 \, m^2_b + 2 \, \mchinq{1} + 2 \, \mstq{1} + \mwq \no
     \plogl{30} + \frac{3 \, m^4_b + m^2_b \mchinq{1} + 4 \, m^2_b \mstq{1}
                   + 2 \, \mchinq{1} \mstq{1} + \mstv{1}}{\mwq} \Bigg) \no
     \plogl{-10} + \left( \kteq{1} \colq{11} + \lteq{1} \corq{11} \right)
           \mchipq{1}  \left (2 \, + \frac{m^2_b}{\mwq} \right) \\
ca_{13} \plgl - 2 \, \kte{1} \lte{1}  \left( \colq{11} + \corq{11} \right)
                \frac{m_b \mchip{1}}{\mwq} \no
     \plogl{-10}  - \left( \lteq{1} \colq{11} + \kteq{1} \corq{11} \right)
      \left( 2 \, + \frac{\mchinq{1} + 3 \, m^2_b + 2 \, \mstq{1}}{\mwq}
       \right) \\
ca_{14} \plgl \frac{\lteq{1} \colq{11} + \kteq{1} \corq{11}}{\mwq}
\eeq
\beq
ca_{15} \plgl -12 \, \kte{1} \lte{1}  \col{11} \cor{11}  m_b \mchin{1}
            - 6 \, \left( \kteq{1} + \lteq{1} \right) \col{11} \cor{11}
                \mchip{1} \mchin{1} \no
     \plogl{-10}  + 2 \, \kte{1} \lte{1} \left( \colq{11} + \corq{11} \right)
             m_b \mchip{1}
             \left( 3 \, + 2 \, \frac{m^2_b+\mstq{1}}{\mwq} \right) \no
     \plogl{-10} + \left( \lteq{1} \colq{11} + \kteq{1} \corq{11} \right)
       \Bigg( 6 \, m^2_b + 3 \, \mchinq{1} + 2 \, \mstq{1} + 2 \, \mwq \no
     \plogl{33} + \frac{2 \, m^4_b + 2 \, m^2_b \mstq{1}
                       +\mchinq{1}\mstq{1}}{\mwq} \Bigg) \no
     \plogl{-10}  + \left( \kteq{1} \colq{11} + \lteq{1} \corq{11} \right)
            \mchipq{1} \left( 1 \, + \frac{2 \, m^2_b + \mstq{1}}{\mwq}
            \right)  \\
ca_{16} \plgl -4 \, \kte{1} \lte{1} \left( \colq{11} + \corq{11} \right)
            \frac{m_b \mchip{1}}{\mwq} \no
   \plogl{-10}   - \left( \kteq{1} \colq{11} + \lteq{1} \corq{11} \right)
         \frac{\mchipq{1}}{\mwq}  \no
      \plogl{-10}  - \left( \lteq{1} \colq{11} + \kteq{1} \corq{11} \right)
        \left(4 + \frac{\mchinq{1} + 4 \, m^2_b +2 \, \mstq{1}}{\mwq}
        \right)   \\
ca_{17} \plgl - 2 \, \kte{1} \lte{1} \left( \colq{11} + \corq{11} \right)
               \frac{m_b \mchip{1}}{\mwq} \no
   \plogl{-10} - \left(\lteq{1} \colq{11} + \kteq{1} \corq{11} \right)
         \left( 2  + \frac{m^2_b}{\mwq} \right) \no
   \plogl{-10} - \left( \kteq{1} \colq{11} + \lteq{1} \corq{11} \right)
        \frac{\mchipq{1}}{\mwq}
\eeq
\beq
ca_{31} \plgl 12 \, \kzleOlzOre  m_b \mchin{1}
                \mchip{1} \mchip{2} \no
    \plogl{-10} + \, 12 \, \kelzOlzOre  m_b \mchin{1} \no
    \plogl{-5} * ( m^2_b + \mchinq{1} + \mstq{1} + \mwq) \no
    \plogl{-10} + \, 6 \, \lelzOlzOre  \mchin{1} \mchip{1}
               ( 2 \, m^2_b + \mchinq{1} + \mwq) \no
    \plogl{-10} + \, 6 \,  \kekzOlzOre \mchin{1} \mchip{2}
                 (2 \, m^2_b + \mchinq{1} + \mwq ) \no
    \plogl{-10} - \, 2 \, \lelzOleOlz \no
    \plogl{-5} * \Bigg[ (\mchinq{1}+m^2_b) \, (\mchinq{1}+\mstq{1}+\mwq+m^2_b)
             + 2 \, \mstq{1} \mwq \no
     \plogl{0} + m^2_b (3 \, m^2_b + 2 \, \mchinq{1} + 3 \, \mstq{1})
             +  \left. (m^4_b+\mstq{1} (\mchinq{1}+m^2_b))
                \frac{m^2_b + \mstq{1}}{\mwq} \right] \no
    \plogl{-10} - 2 \, \kelzOleOlz  m_b \mchip{1} \no
    \plogl{0}  * \left[ 3 \, m^2_b + 2 \, \mchinq{1} + 3 \, \mstq{1}
            + \frac{(m^2_b+\mstq{1}) \, (m^2_b+\mstq{1})}{\mwq} \right] \no
    \plogl{-10} - 2 \, \kzleOleOlz  m_b \mchip{2} \no
    \plogl{0} * \left[ 3 \, m^2_b + 2 \, \mchinq{1} + 3 \, \mstq{1}
            + \frac{(m^2_b+\mstq{1}) \, (m^2_b+\mstq{1})}{\mwq} \right] \no
    \plogl{-10} - 2 \, \kekzOleOlz  \mchip{1} \mchip{2} \no
    \plogl{0} * \left[ 2 \, m^2_b + \mchinq{1} + \mstq{1}
                + \frac{(m^2_b+\mstq{1}) m^2_b}{\mwq} \right] \\
ca_{32} \plgl - 6 \, \lelzOlzOre  \mchin{1} \mchip{1} \no
  \plogl{-10} - 12 \, \kelzOlzOre  m_b \mchin{1} \no
  \plogl{-10} - 6 \, \kekzOlzOre \mchin{1} \mchip{2} \no
  \plogl{-10} + 2 \, \lelzOleOlz \no
  \plogl{-8}  * \left[ 6 \, m^2_b + 2 \, \mchinq{1} + 2 \, \mstq{1} + \mwq
    + \frac{(2 \, m^2_b +\mchinq{1} ) \, (m^2_b+2 \,\mstq{1})
                       + m^4_b + \mstv{1}}{\mwq} \right] \no
  \plogl{-10} + 2 \, \kelzOleOlz  m_b \mchip{1}
        \frac{3 \, + 2 \, (m^2_b+\mstq{1})}{\mwq} \no
  \plogl{-10} + 2 \, \kzleOleOlz  m_b \mchip{2}
     \frac{3 \, + 2 \, (m^2_b+\mstq{1})}{\mwq} \no
  \plogl{-10} + 2 \, \kekzOleOlz  \mchip{1} \mchip{2}
               \left(2 + \frac{m^2_b}{\mwq} \right) \\
ca_{33} \plgl - 2 \, \kelzOleOlz  \frac{m_b \mchip{1} }{\mwq} \no
 \plogl{-10} - 2 \, \kzleOleOlz  \frac{m_b \mchip{2} }{\mwq} \no
 \plogl{-10} - 2 \,  \lelzOleOlz
   \left( 2 \, + \frac{\mchinq{1} + 3 \, m^2_b + 2 \, \mstq{1}}{\mwq}
   \right) \\
 ca_{34} \plgl \frac{2 \, \lelzOleOlz}{\mwq}
\eeq
\beq
ca_{35} \plgl - 6 \, \lelzOlzOre  \mchin{1} \mchip{1} \no
  \plogl{-10} - 12 \, \kelzOlzOre m_b \mchin{1} \no
  \plogl{-10} - 6 \, \kekzOlzOre \mchin{1} \mchip{2} \no
  \plogl{-10} + 2 \,\lelzOleOlz \no
  \plogl{0} * \left[ 6 \, m^2_b + 3 \, \mchinq{1}
                       + 2 \, \mstq{1} + 2 \, \mwq
      + \frac{2 \, m^4_b + \mstq{1} \,
                   (\mchinq{1}+2 \, m^2_b)}{\mwq} \right] \no
  \plogl{-10} + 2 \, \kelzOleOlz m_b \mchip{1}
         \left( 3 \, + \frac{2 \, (m^2_b+\mstq{1})}{\mwq} \right)  \no
  \plogl{-10} + 2 \, \kzleOleOlz m_b \mchip{2}
             \left( 3 \, + \frac{2 \, (m^2_b+\mstq{1})}{\mwq} \right) \no
  \plogl{-10} + 2 \, \kekzOleOlz  \mchip{1} \mchip{2}
              \left( 1 \, + \frac{2 \, m^2_b+\mstq{1}}{\mwq} \right) \\
ca_{36} \plgl - 2 \, \kekzOleOlz \frac{\mchip{1} \mchip{2}}{\mwq} \no
  \plogl{-10} - 4 \, \kelzOleOlz \frac{m_b \mchip{1}}{\mwq} \no
  \plogl{-10} - 4 \, \kzleOleOlz \frac{m_b \mchip{2}}{\mwq} \no
  \plogl{-10} - 2 \, \lelzOleOlz   \left( 4 \, +
          \frac{\mchinq{1} + 4 \, m^2_b + 2 \, \mstq{1}}{\mwq} \right) \\
ca_{37} \plgl - 2 \, \lelzOleOlz  \left( 2 + \frac{m^2_b}{\mwq} \right) \no
  \plogl{-10} - 2 \, \kelzOleOlz \frac{m_b \mchip{1}}{\mwq} \no
  \plogl{-10} - 2 \, \kzleOleOlz \frac{m_b \mchip{2}}{\mwq} \no
  \plogl{-10} - 2 \, \kekzOleOlz \frac{\mchip{1} \mchip{2}}{\mwq}
\eeq
\beq
cb_{11} \plgl - \wzw \Bigg\{ 3 \, \bcop{t}{11} \lte{1} \col{11}
             m_t \mchin{1}  (2 \, m^2_b+\mchinq{1}+\mwq)  \no
 \plogl{-10} +  6 \, \bcop{t}{11} \kte{1} \col{11}
        m_t \mchin{1} m_b \mchip{1}  + 3 \, \acop{t}{11} \kte{1} \col{11}
        m_b \mchip{1}  (\mchinq{1}-\mstq{1}) \no
  \plogl{-10} + \acop{t}{11} \lte{1} \col{11} \no
  \plogl{-4} * \left[ 2 \, (\mchinq{1}-\mstq{1}) (\mwq+\mchinq{1}+m^2_b)
   - m^2_b \mstq{1}  \left( 1 +
       \frac{\mchinq{1}-m^2_b-\mstq{1} }{\mwq} \right) \right] \no
  \plogl{-10} + \acop{t}{11} \lte{1} \cor{11} \mchin{1} \mchip{1}
           \left[ 2 \, \mwq - m^2_b
                  \left(1 \, + \frac{m^2_b}{\mwq} \right) \right] \no
  \plogl{-10} + \acop{t}{11} \kte{1} \cor{11} m_b \mchin{1}
    \left[3 \, \mwq   + \mstq{1} \left(2 \,-\frac{m^2_b}{\mwq} \right)
     - m^2_b \left(2 \, + \frac{m^2_b}{\mwq} \right) -\mchinq{1} \right] \no
  \plogl{-10} - \bcop{t}{11} \lte{1} \cor{11} m_t \mchip{1}
    \left[ \mchinq{1}  + \mstq{1} \left(1 \,+\frac{m^2_b}{\mwq} \right)
         + m^2_b \left( 2 \,+\frac{m^2_b}{\mwq} \right) \right] \no
  \plogl{-10}- \bcop{t}{11} \kte{1} \cor{11} m_b m_t \left[ 2 \, \mchinq{1} +
         (m^2_b+\mstq{1}) \left(3 \,+\frac{m^2_b+\mstq{1}}{\mwq} \right)
         \right] \Bigg\} \\
cb_{12} \plgl - \wzw \Bigg\{ \acop{t}{11} \lte{1} \col{11}
    (\mstq{1}-\mchinq{1}) \left( 2 \, -\frac{m^2_b}{\mwq} \right)
     - 3 \, \bcop{t}{11} \lte{1} \col{11} m_t \mchin{1} \no
  \plogl{-10}+ \acop{t}{11} \kte{1} \cor{11} m_b \mchin{1}
           \left(\frac{m^2_b}{\mwq}-1 \right)
        + \bcop{t}{11} \lte{1} \cor{11} m_t \mchip{1}
          \left( 2 \, + \frac{m^2_b}{\mwq} \right) \no
  \plogl{-10} + \bcop{t}{11} \kte{1} \cor{11}  m_b m_t
        \left( 3 \, + 2 \,  \frac{m^2_b+\mstq{1}}{\mwq}  \right) \Bigg\} \\
cb_{13} \plgl \wzw \bcop{t}{11} \kte{1} \cor{11} \frac{m_b m_t}{\mwq} \\
cb_{14} \plgl - \wzw \Bigg\{ 3 \, \acop{t}{11} \kte{1} \col{11} m_b \mchip{1}
               - 3 \, \bcop{t}{11} \lte{1} \col{11} m_t \mchin{1} \no
  \plogl{-12} - \acop{t}{11} \lte{1} \cor{11} \mchin{1} \mchip{1}
        \left( 1 \, - 2 \,\frac{m^2_b}{\mwq} \right)
     + \acop{t}{11} \kte{1} \cor{11} m_b \mchin{1}
        \left(1 \, + \frac{2 \, m^2_b+\mstq{1}}{\mwq} \right) \no
  \plogl{-12} + \bcop{t}{11} \lte{1} \cor{11}  m_t \mchip{1}
        \left( 1 \, + \frac{2 \, m^2_b+\mstq{1}}{\mwq} \right)
    + \bcop{t}{11} \kte{1} \cor{11} m_b m_t
        \left(3 \, + 2 \, \frac{m^2_b+\mstq{1}}{\mwq} \right) \no
  \plogl{-12} + \acop{t}{11} \lte{1} \col{11}
    \left[ 4 \, m^2_b + 2 \, \mwq  + \mstq{1}
       \left( 2 \, + \frac{\mchinq{1}-m^2_b}{\mwq} \right) \right] \Bigg\} \no
\eeq
\newpage
\beq
cb_{15} \plgl \wzw \Bigg\{ \acop{t}{11} \kte{1} \cor{11}
         \frac{m_b \mchin{1}}{\mwq}
     + \bcop{t}{11} \lte{1} \cor{11} \frac{m_t \mchip{1}}{\mwq} \no
  \plogl{-10} + \acop{t}{11} \lte{1} \col{11}
         \left( 3 \, +\frac{m^2_b+\mstq{1}+\mchinq{1}}{\mwq} \right)
   + 2 \, \bcop{t}{11} \kte{1} \cor{11} \frac{m_b m_t}{\mwq} \Bigg\} \\
cb_{16} \plgl - \frac{\wzw \acop{t}{11} \lte{1} \col{11}}{\mwq} \\
cb_{17} \plgl \wzw  \Bigg\{ 2 \, \acop{t}{11} \lte{1} \col{11}
    + \acop{t}{11} \lte{1} \cor{11} \frac{\mchin{1} \mchip{1}}{\mwq}
    + \acop{t}{11} \kte{1} \cor{11} \frac{m_b \mchin{1}}{\mwq} \no
  \plogl{-10} +\bcop{t}{11} \lte{1} \cor{11} \frac{m_t \mchip{1}}{\mwq}
    + \bcop{t}{11} \kte{1} \cor{11} \frac{m_b m_t}{\mwq} \Bigg\}
\eeq
\beq
cc_{ki1} \plgl - \wzw \cost \cosb \Bigg\{ \fbltOle  \Bigg[ m^2_b \mwq
             + m^4_b \no
  \plogl{5} - 2 \, m^2_b \mchinq{1}
             + \mchinq{1} \mstq{1} \left(\frac{\mstq{1}}{\mwq}-1 \right)
               + m^2_b \mstq{1} \left( \frac{m^2_b}{\mwq}-2 \right)
               + \frac{m^2_b \mstv{1}}{\mwq} \Bigg] \no
  \plogl{-12.5} + \hbltOle m_b \mchin{1} \left(m^2_b-\mstq{1}-2 \, \mchinq{1}
             + \mstq{1} \frac{m^2_b+\mstq{1}}{\mwq} \right) \no
  \plogl{-12.5} + \hbktOle \mchin{1} \mchip{1} \left[ m^2_b - \mwq
      - 2 \,\mchinq{1} +\mstq{1} \left( 1 + \frac{m^2_b}{\mwq} \right)
                \right] \no
  \plogl{-12.5} + \fbktOle  m_b \mchip{1} \left[ m^2_b - 2 \, \mchinq{1} +
     \mstq{1} \left(\frac{m^2_b+\mstq{1}}{\mwq}-1 \right)
     \right] \Bigg\} \no \\
cc_{ki2} \plgl  \wzw \cost \cosb \Bigg\{ \fbltOle \no
  \plogl{15} * \left[ \mchinq{1} \left(1 + 2 \frac{\mstq{1}}{\mwq} \right)
              + \frac{\mstv{1}}{\mwq}
              + \mwq + m^2_b \left(3 + \frac{m^2_b+3 \, \mstq{1}}{\mwq}
             \right) \right] \no
  \plogl{-10} + \hbltOle m_b \mchin{1}
               \left(1 + \frac{m^2_b+2 \, \mstq{1}}{\mwq} \right) \no
  \plogl{-10} + \hbktOle \mchin{1} \mchip{1}
             \left(\frac{m^2_b}{\mwq}-1 \right) \no
  \plogl{-10} + \fbktOle m_b \mchip{1}
           \left( 1 + \frac{m^2_b+2 \, \mstq{1}}{\mwq} \right) \Bigg\} 
\eeq
\newpage
\beq
cc_{ki3} \plgl - \wzw \cost \cosb \Bigg\{
              \fbktOle \frac{m_b \mchip{1}}{\mwq} \no
    \plogl{-10} + \hbltOle \frac{m_b \mchin{1}}{\mwq} \no
    \plogl{-10} +  \fbltOle \left.
          \left( 2 + \frac{2 \, m^2_b + \mchinq{1}+ 2 \, \mstq{1}}
                             {\mwq} \right) \right\} \\
cc_{ki4} \plgl  \wzw \cost \cosb \frac{\fbltOle}{\mwq} \\
cc_{ki5} \plgl - \wzw \cost \cosb
         \left(1 - \frac{\mstq{1}}{\mwq} \right) \no
  \plogl{-10} * \bigg[ \hbktOle \mchin{1} \mchip{1}
            + \fbktOle m_b \mchip{1} \no
  \plogl{-5} + \hbltOle m_b \mchin{1}  + \fbltOle m^2_b \bigg] \no \\
cc_{ki6} \plgl - \wzw \cost \cosb \Bigg\{
        \hbltOle \frac{m_b \mchin{1}}{\mwq} \no
  \plogl{-10} + \hbktOle \frac{\mchin{1} \mchip{1}}{\mwq}
              + \fbktOle \frac{m_b \mchip{1}}{\mwq} \no
  \plogl{-10} + \fbltOle
         \left( 1 + \frac{m^2_b+\mstq{1}}{\mwq} \right) \Bigg\}
\eeq
\beq
cd_{1} \plgl \acopq{t}{11} \frac{(\mchinq{1}-\mstq{1})}{2}
              \left[ 2 \, \mwq - m^2_b \left( 1 \, + \frac{m^2_b}{\mwq}
                \right) \right] \no
   \plogl{-6} - \bcopq{t}{11} \frac{m^2_t}{2}
        \left[ \mchinq{1} + \mstq{1}
            + m^2_b \left( 2 \, + \frac{m^2_b+\mstq{1}}{\mwq} \right)
        \right] \no
   \plogl{-6}+ \acop{t}{11} \bcop{t}{11} m_t \mchin{1}
         \left[2 \, \mwq - m^2_b \left( 1 \, + \frac{m^2_b}{\mwq} \right)
         \right] \\
cd_{2} \plgl \bcopq{t}{11} m^2_t
          \left( 1 \, + \frac{m^2_b}{2 \, \mwq} \right) \\
cd_{3} \plgl \acopq{t}{11}  \left( \frac{m^2_b}{2} + \mstq{1} + \mwq
        + (2 \, \mchinq{1} - \mstq{1}) \frac{m^2_b}{2 \, \mwq} \right) \no
 \plogl{-6}+ \bcopq{t}{11} m^2_t
         \frac{(\mwq + 2 \, m^2_b+\mstq{1})}{2 \, \mwq}
     - \acop{t}{11} \bcop{t}{11} m_t \mchin{1}
       \left( 1 \, - 2 \frac{m^2_b}{\mwq} \right) \\
cd_{4} \plgl - \acopq{t}{11} \left( 1 \, +\frac{m^2_b}{2 \, \mwq} \right)
             - \bcopq{t}{11} \frac{m^2_t}{2 \, \mwq} \\
cd_{5} \plgl - \acopq{t}{11}
       \left( 1 \, + \frac{\mchinq{1}}{2 \, \mwq} \right)
    - \acop{t}{11} \bcop{t}{11}  \frac{m_t \mchin{1}}{\mwq}
    - \bcopq{t}{11} \frac{m^2_t}{2 \, \mwq} \\
cd_{6} \plgl \frac{\acopq{t}{11}}{2 \, \mwq}
\eeq
\beq
ce_{11} \plgl \cost \cosb \Bigg\{
           \left( \acop{b}{11} \bcop{t}{11} m_t \mchin{1}
                + \bcop{b}{11} \acop{t}{11} m_b \mchin{1} \right) \no
  \plogl{24} * \left( m^2_b + \mstq{1} - 2 \, \mchinq{1} - \mwq
             + \frac{m^2_b \mstq{1}}{\mwq} \right) \no
  \plogl{-10} + \acop{b}{11} \acop{t}{11} \Bigg[ m^2_b (\mchinq{1}+\mstq{1})
        + 2 \, \mchinq{1} (\mstq{1}-\mchinq{1}-\mwq)
        + (\mchinq{1}-\mstq{1}) \frac{m^2_b \mstq{1}}{\mwq} \Bigg] \no
  \plogl{-10} + \bcop{b}{11} \bcop{t}{11} m_b m_t
   \left( m^2_b - 2 \, \mchinq{1} - \mstq{1}
      + \frac{\mstq{1} (m^2_b+\mstq{1})}{\mwq} \right)  \Bigg\} \\
ce_{12} \plgl \cost \cosb \Bigg\{
         \left( \acop{b}{11} \bcop{t}{11} m_t \mchin{1}
              + \bcop{b}{11} \acop{t}{11} m_b \mchin{1} \right)
         \left( 1 \, - \frac{m^2_b}{\mwq} \right) \no
 \plogl{-10} + \acop{b}{11} \acop{t}{11}
        (\mchinq{1} - \mstq{1}) \left( 2 \, - \frac{m^2_b}{\mwq} \right)
   - \bcop{b}{11} \bcop{t}{11} m_b m_t
     \left( 1 \, + \frac{m^2_b + 2 \, \mstq{1}}{\mwq} \right) \Bigg\} \\
ce_{13} \plgl \cost \cosb \bcop{b}{11} \bcop{t}{11} \frac{m_b m_t}{\mwq} \\
ce_{14} \plgl \cost \cosb \left( 1 \, - \frac{\mstq{1}}{\mwq} \right) \no
   \plogl{4} * \left( \acop{b}{11} \acop{t}{11} \mchinq{1}
                  + \acop{b}{11} \bcop{t}{11} m_t \mchin{1}
                  + \bcop{b}{11} \acop{t}{11} m_b \mchin{1}
                  + \bcop{b}{11} \bcop{t}{11} m_b m_t \right) \\
ce_{15} \plgl \cost \cosb \Bigg\{ \acop{b}{11} \acop{t}{11}
           \left( 1 \, + \frac{\mchinq{1} + \mstq{1}}{\mwq} \right)
           + \acop{b}{11} \bcop{t}{11}
                 \frac{m_t \mchin{1}}{\mwq}           \no
  \plogl{24} + \bcop{b}{11} \acop{t}{11} \frac{m_b \mchin{1}}{\mwq}
       + \cosb \bcop{b}{11} \bcop{t}{11} \frac{m_b m_t}{\mwq} \Bigg\} \\
ce_{16} \plgl - \frac{\cost \cosb \acop{b}{11} \acop{t}{11}}{\mwq}
\eeq
\beq
cf_{11} \plgl - \einha \costq \cosbq
         \left[ \left(\acopq{b}{11}+\bcopq{b}{11} \right)
                 \left( m^2_b + \mchinq{1} \right)
                + 4 \, \acop{b}{11} \bcop{b}{11} m_b \mchin{1} \right] \no \\
cf_{12} \plgl \einha \costq \cosbq \left(\acopq{b}{11}+\bcopq{b}{11} \right)
\eeq
\beq
cf_{31} \plgl \costq \cosb \sinb  \no
  \plogl{0} * \left[ \left( \acop{b}{11} \acop{b}{12}
     + \bcop{b}{11} \bcop{b}{12} \right) \left( m^2_b + \mchinq{1} \right)
     + 2 \, \left( \acop{b}{11} \bcop{b}{12}
         + \bcop{b}{11} \acop{b}{12} \right) m_b \mchin{1} \right] \\
cf_{32} \plgl - \costq \cosb \sinb
       \left( \acop{b}{11} \acop{b}{12} + \bcop{b}{11} \bcop{b}{12} \right)
\eeq

\noindent
To get the remaining coefficients one has to make the following replacements:\\
\begin{tabular}{ll}
  $ca_{1i} \to ca_{2i}$: & $\lte{1} \to \lte{2}$, $\kte{1} \to \kte{2}$,
                          $\col{11} \to \col{12}$, $\col{11} \to \col{12}$,
                          $\mchip{1} \to \mchip{2}$ \\
  $cb_{1i} \to cb_{2i}$: & $\lte{1} \to \lte{2}$, $\kte{1} \to \kte{2}$,
                          $\col{11} \to \col{12}$, $\cor{11} \to \cor{12}$,
                          $\mchip{1} \to \mchip{2}$ \\
  $cc_{11i} \to cc_{12i}$: & $\lte{1} \to \lte{2}$, $\kte{1} \to \kte{2}$,
                            $\col{11} \to \col{12}$, $\cor{11} \to \cor{12}$,
                            $\mchip{1} \to \mchip{2}$ \\
  $cc_{11i} \to cc_{21i}$: & $\acop{b}{11} \to \acop{b}{12}$,
                            $\bcop{b}{11} \to \bcop{b}{12}$,
                            $\msbot{1} \to \msbot{2}$,
                            $\cosb \to -\sinb$ \\
  $cc_{11i} \to cc_{22i}$: & $\lte{1} \to \lte{2}$, $\kte{1} \to \kte{2}$,
                            $\col{11} \to \col{12}$, $\cor{11} \to \cor{12}$,
                            $\mchip{1} \to \mchip{2}$, \\
                          & $\acop{b}{11} \to \acop{b}{12}$,
                            $\bcop{b}{11} \to \bcop{b}{12}$,
                            $\msbot{1} \to \msbot{2}$,
                          $\cosb \to -\sinb$ \\
  $ce_{1i} \to ce_{2i}$: & $\acop{b}{11} \to \acop{b}{12}$,
                          $\bcop{b}{11} \to \bcop{b}{12}$,
                          $\msbot{1} \to \msbot{2}$,
                          $\cosb \to -\sinb$ \\
  $cf_{1i} \to cf_{2i}$: & $\acop{b}{11} \to \acop{b}{12}$,
                          $\bcop{b}{11} \to \bcop{b}{12}$,
                          $\msbot{1} \to \msbot{2}$,
                          $\cosb \to -\sinb$ \\
\end{tabular}

\section[Light Stop into charged Higgs, Bottom, and Neutralino]
       {The Decay of the Light Stop into a charged Higgs Boson,
        a Bottom Quark and the Lightest Neutralino}

\noindent
The decay width is given by
\beq
 \Gamma(\sthgbc) \plgl \no
 \plogl{-43} = \frac{\alpha^2}{16 \, \pi m^3_{\sto{1}} \sinwv}
        \eint{(\mstop{1}-\mhp)^2}{(m_b + \mchin{1})^2}{s}
   \left( G_{\chip{} \chip{}} +
   G_{\chip{} t} +
   G_{\chip{} \sbo{}} +
   G_{t t} +
   G_{t \sbo{}} +
   G_{\sbo{} \sbo{}} \right) \no
\eeq
with
\beq \hspace*{-3mm}
   G_{\chip{} \chip{}} \plgl
       \sum^2_{i=1} \Big[  ( da_{i1} + da_{i2} s )
          J^0_t(\dsum{\sto{1}}{H^{\pm}}{b}{\chin{1}}{\chip{i}}
             ,\Gamma_{\chip{i}} \mchip{i}) \no
    \plogl{5} + ( da_{i3} + da_{i4} s )
       J^1_t(\dsum{\sto{1}}{H^{\pm}}{b}{\chin{1}}{\chip{i}}
             ,\Gamma_{\chip{i}} \mchip{i}) \no
    \plogl{5} + \, da_{i4} \,
          J^2_t(\dsum{\sto{1}}{H^{\pm}}{b}{\chin{1}}{\chip{i}}
             ,\Gamma_{\chip{i}} \mchip{i}) \Big] \no
    \plogl{-1} +  ( da_{31} + da_{32} s )
          J^0_{tt}(\dsum{\sto{1}}{H^{\pm}}{b}{\chin{1}}{\chip{1}}
             ,\Gamma_{\chip{1}} \mchip{1}  \no
       \plogl{33} ,\dsum{\sto{1}}{H^{\pm}}{b}{\chin{1}}{\chip{2}}
             ,\Gamma_{\chip{2}} \mchip{2}) \no
    \plogl{-1} + ( da_{33} + da_{34} s )
       J^1_{tt}(\dsum{\sto{1}}{H^{\pm}}{b}{\chin{1}}{\chip{1}}
             ,\Gamma_{\chip{1}} \mchip{1}  \no
       \plogl{33} ,\dsum{\sto{1}}{H^{\pm}}{b}{\chin{1}}{\chip{2}}
             ,\Gamma_{\chip{2}} \mchip{2}) \no
    \plogl{-1} + \, da_{34} \,
       J^2_{tt}(\dsum{\sto{1}}{H^{\pm}}{b}{\chin{1}}{\chip{1}}
             ,\Gamma_{\chip{1}} \mchip{1}  \no
       \plogl{17} ,\dsum{\sto{1}}{H^{\pm}}{b}{\chin{1}}{\chip{2}}
             ,\Gamma_{\chip{2}} \mchip{2})
\eeq
\beq
\hspace*{-3mm}
   G_{\chip{} t} \plgl
       \sum^2_{i=1} \Big[ ( db_{i1} + db_{i2} s ) \no
    \plogl{10} * J^0_{tt}(\dsum{\sto{1}}{H^{\pm}}{b}{\chin{1}}{\chip{1}}
       ,-\Gamma_{\chip{1}} \mchip{1},m^2_t, \Gamma_t m_t) \no
    \plogl{5} + ( db_{i3} + db_{i4} s ) \no
    \plogl{10} * J^1_{tt}(\dsum{\sto{1}}{H^{\pm}}{b}{\chin{1}}{\chip{1}}
       ,-\Gamma_{\chip{1}} \mchip{1} ,m^2_t, \Gamma_t m_t) \no
    \plogl{5} + \, db_{i4} \,
        J^2_{tt}(\dsum{\sto{1}}{H^{\pm}}{b}{\chin{1}}{\chip{1}}
       ,-\Gamma_{\chip{1}} \mchip{1} ,m^2_t, \Gamma_t m_t)  \Big] \no
\eeq
\beq
\hspace*{-3mm}
   G_{\chip{} \sbo{}} \plgl
       \sum^2_{k,i=1} \Big[
          ( cc_{ik1} + cc_{ik2} s ) \no
    \plogl{7} * J^0_{st}(
       \msbq{k}, \Gamma_ {\sbo{k}} \msbot{k}
       ,\dsum{\sto{1}}{H^{\pm}}{b}{\chin{1}}{\chip{1}}
       ,-\Gamma_{\chip{1}} \mchip{1}) \no
    \plogl{-3} + \, cc_{ik3} \, J^1_{st}(
       \msbq{k}, \Gamma_ {\sbo{k}} \msbot{k}
       ,\dsum{\sto{1}}{H^{\pm}}{b}{\chin{1}}{\chip{1}}
       ,-\Gamma_{\chip{1}} \mchip{1}) \Big]  \no
\eeq
\beq
\hspace*{-3mm}
   G_{t t} \plgl
          ( dd_{1} + dd_{2} s) J^0_t(m^2_t, \Gamma_t m_t) +
          ( dd_{3} + dd_{4} s) J^1_t(m^2_t, \Gamma_t m_t) \no
   \plogl{0} + \, dd_{4} \, J^2_t(m^2_t, \Gamma_t m_t) 
\eeq
\beq
\hspace*{-3mm}
   G_{t \sbo{}} \plgl \sum^2_{k=1} \Big[
   ( de_{k1} + de_{k2} s )
   J^0_{st}(\msbq{k}, \Gamma_ {\sbo{k}} \msbot{k},m^2_t, \Gamma_t m_t) \no
  \plogl{6} + \, de_{k3} \,
   J^1_{st}(\msbq{k}, \Gamma_ {\sbo{k}} \msbot{k},m^2_t, \Gamma_t m_t)
   \Big]
\eeq
\beq
\hspace*{-3mm}
G_{\sbo{} \sbo{}} \plgl \frac{\sqrt{\lambda(s,\mstq{1},\mhpq)
                         \lambda(s,\mchinq{1},m^2_b)}}{s} \no
  \plogl{-12} * \left\{ \sum^2_{k=1}
      \frac{(df_{k1} + df_{k2} s)}
           {(s-\msbq{k})^2 + \Gamma^2_{\sbo{k}} \msbq{k}}
 + Re\left[ \frac{(df_{31} + df_{32} s)}
           { (s-\msbq{1} + i \Gamma_{\sbo{1}} \msbot{1})
             (s-\msbq{2} - i \Gamma_{\sbo{2}} \msbot{2})}
          \right]  \right\} \no
\eeq
The integrals $J^{0,1,2}_{t,tt,st}$ are defined in \sect{intJ}. Their
integration range is given by
\beq
t_{max \atop min} \plgl \frac{\mstq{1} + m^2_b + \mhpq + \mchinq{1} -s}{2}
       - \frac{(\mstq{1}-\mhpq) (\mchinq{1}-m^2_b)}{2 s} \no
  \plogl{0} \pm  \frac{\sqrt{\lambda(s,\mstq{1},\mhpq)
                  \lambda(s,\mchinq{1},m^2_b)}}{2 s}
\eeq
and $s = (p_{\sto{1}} - p_{H^{\pm}})^2$.
Note, that $-\Gamma_{\chip{1}} \mchip{1}$
appears in the entries of the integrals $G_{\chip{} \sbo{j}}$ and
$G_{\chip{} t}$ because the chargino exchange is the $u$-channel in our
convention.
The coefficients are given by:
\beq
da_{11} \plgl - 4 \, \kte{1} \lte{1} \cql{11} \cqr{11} m_b \mchin{1}
       \left( m^2_b + \mchipmq{1} + \mchinq{1} + \mstq{1} + \mhpq \right) \no
  \plogl{-11} - 2 \, \cql{11} \cqr{11} \left( \kteq{1} + \lteq{1} \right)
    \mchin{1} \mchipm{1} \left( 2 \, m^2_b + \mchinq{1} + \mhpq \right)  \no
  \plogl{-11} - 2 \, \kte{1} \lte{1} \left( \cqlq{11} + \cqrq{11} \right)
     m_b \mchipm{1}  \left( m^2_b + 2 \, \mchinq{1} + \mstq{1} \right)  \no
  \plogl{-11} - \left( \kteq{1} \cqrq{11} + \lteq{1} \cqlq{11} \right) \no
   \plogl{0} * \left[ \left( m^2_b+\mchinq{1} \right)^2
    + \left( m^2_b+\mhpq \right) \left(\mchinq{1}+\mstq{1}\right) \right] \no
  \plogl{-11} - \left( \kteq{1} \cqlq{11} + \lteq{1} \cqrq{11} \right)
     \mchipmq{1}  \left(m^2_b+\mchinq{1} \right) \\
da_{12} \plgl 4 \, \kte{1} \lte{1} \cql{11} \cqr{11} m_b \mchin{1} \no
   \plogl{-11}
        + \left( \kteq{1} \cqrq{11} + \lteq{1} \cqlq{11} \right)
           \left( m^2_b+\mchinq{1} \right) \no
  \plogl{-11} + 2 \, \kte{1} \lte{1} \left( \cqlq{11} + \cqrq{11} \right)
                 m_b \mchipm{1} \no
   \plogl{-11}
        + 2 \, \cql{11} \cqr{11} \left( \kteq{1} + \lteq{1} \right)
                 \mchin{1} \mchipm{1}  \no
  \plogl{-11} + \left( \kteq{1} \cqlq{11} + \lteq{1} \cqrq{11} \right)
             \mchipmq{1}   \\
da_{13} \plgl 4 \, \kte{1} \lte{1} \cql{11} \cqr{11} m_b \mchin{1}
      + 2 \, \cql{11} \cqr{11} \left( \kteq{1} + \lteq{1} \right)
              \mchin{1} \mchipm{1} \no
  \plogl{-11} + \left( \kteq{1} \cqrq{11} + \lteq{1} \cqlq{11} \right)
         \left(2 \, m^2_b+2 \, \mchinq{1}+\mhpq+\mstq{1}\right)  \no
  \plogl{-11} + 2 \, \kte{1} \lte{1} \left( \cqlq{11} + \cqrq{11} \right)
               m_b \mchipm{1} \\
da_{14} \plgl - \kteq{1} \cqrq{11} - \lteq{1} \cqlq{11}
\eeq
\beq
da_{31} \plgl - 2  \lelzQlzQre \mchin{1} \mchipm{1}
         \left( \mchinq{1} + \mhpq + 2 \, m^2_b \right) \no
  \plogl{-11}  - 4 \, \kelzQlzQre m_b \mchin{1}
      \left( \mchinq{1} + \mhpq + m^2_b + \mstq{1} \right) \no
  \plogl{-11} - 4 \, \kzleQlzQre m_b \mchin{1} \mchipm{1} \mchipm{2} \no
  \plogl{-11} - 2 \, \kekzQlzQre \mchin{1} \mchipm{2}
      \left( \mchinq{1} + \mhpq + 2 \, m^2_b \right) \no
  \plogl{-11} - 2 \, \lelzQleQlz     \no
  \plogl{0} * \left[ \left(m^2_b+\mchinq{1}\right)^2
      + \left(\mchinq{1}+\mstq{1}\right) \left(m^2_b+\mhpq\right) \right] \no
  \plogl{-11} - 2 \, \kelzQleQlz m_b \mchipm{1}
     \left(m^2_b+\mstq{1}+2 \, \mchinq{1} \right) \no
  \plogl{-11} - 2 \, \kzleQleQlz m_b \mchipm{2}
     \left(m^2_b+\mstq{1}+2 \, \mchinq{1} \right) \no
  \plogl{-11} - 2 \, \kekzQleQlz \mchipm{1} \mchipm{2}
    \left(\mchinq{1} + m^2_b \right)  \\
da_{32} \plgl 2 \lelzQlzQre \mchin{1} \mchipm{1} \no
  \plogl{-11} + 4 \, \kelzQlzQre m_b \mchin{1} \no
  \plogl{-11} + 2 \, \kekzQlzQre \mchin{1} \mchipm{2} \no
  \plogl{-11} + 2 \, \lelzQleQlz
    \left( m^2_b+\mchinq{1} \right)  \no
  \plogl{-11} + 2 \, \kelzQleQlz m_b \mchipm{1} \no
  \plogl{-11} + 2 \, \kzleQleQlz m_b \mchipm{2} \no
  \plogl{-11} + 2 \, \kekzQleQlz \mchipm{1} \mchipm{2} \\
da_{33} \plgl 2 \lelzQlzQre \mchin{1} \mchipm{1} \no
  \plogl{-11} + 4 \, \kelzQlzQre m_b \mchin{1} \no
  \plogl{-11} + 2 \, \kekzQlzQre \mchin{1} \mchipm{2} \no
  \plogl{-11} + 2 \, \lelzQleQlz
   \left(2 \, m^2_b+2 \, \mchinq{1}+\mhpq+\mstq{1} \right) \no
  \plogl{-11} + 2 \, \kelzQleQlz m_b \mchipm{1} \no
  \plogl{-11} + 2 \, \kzleQleQlz m_b \mchipm{2} \\
da_{34} \plgl - 2 \, \lelzQleQlz
\eeq
\beq
db_{11} \plgl \frac{\wzw}{m_W} \Bigg\{
       \bcop{t}{11} \kte{1} \cql{11} \mchipm{1} m_b m_t
       \left[ \left( \mstq{1}-\mchinq{1} \right) \cotbe
            - \left( m^2_b + \mchinq{1} \right) \tanbe \right] \no
  \plogl{-11} + \bcop{t}{11} \lte{1} \cql{11} m_t
       \left[ \left( \mhpq \mstq{1} - m^2_b \mchinq{1} \right) \cotbe
            - m^2_b  \left( m^2_b + \mstq{1} + 2 \, \mchinq{1} \right)
             \tanbe \right] \no
  \plogl{-11} + \acop{t}{11} \lte{1} \cql{11} \mchin{1} \no
  \plogl{-6} * \left[ m^2_b
          \left( \mhpq+\mstq{1}-\mchinq{1} -m^2_b \right) \tanbe
          - m^2_t \left( 2 \, m^2_b +\mchinq{1} +\mhpq \right)
          \cotbe \right] \no
  \plogl{-11} + \acop{t}{11} \kte{1} \cql{11} \mchipm{1} m_b \mchin{1}
       \left[ \left(\mhpq-m^2_b \right) \tanbe
          - 2 \, m^2_t \cotbe \right] \no
  \plogl{-11} + \bcop{t}{11} \kte{1} \cqr{11} \mchin{1} m_b m_t \no
  \plogl{0} * \left[ \left(\mstq{1}+\mhpq-m^2_b -\mchinq{1} \right) \cotbe
         - \left( \mhpq+2 \, m^2_b +\mchinq{1} \right) \tanbe \right] \no
  \plogl{-11} + \bcop{t}{11} \lte{1} \cqr{11} m_t \mchin{1} \mchipm{1}
      \left[ \left(\mhpq-m^2_b \right) \cotbe
         - 2 \, m^2_b  \tanbe \right] \no
  \plogl{-11} - \acop{t}{11} \kte{1} \cqr{11} m_b \no
  \plogl{-5}    \left[ \left(m^2_b \mchinq{1} - \mstq{1} \mhpq \right) \tanbe
         + m^2_t \left( m^2_b +\mstq{1}+2 \, \mchinq{1} \right)
          \cotbe \right]  \no
  \plogl{-11} - \acop{t}{11} \lte{1} \cqr{11} \mchipm{1}
      \left[ m^2_b  \left(\mchinq{1}-\mstq{1} \right) \tanbe
         + m^2_t \left(m^2_b +\mchinq{1} \right) \cotbe \right] \Bigg\} \no \\
db_{12} \plgl \frac{\wzw}{m_W} \Bigg\{ \bcop{t}{11} \kte{1} \cql{11}
              \mchipm{1} m_b m_t \tanbe
       + \bcop{t}{11} \lte{1} \cql{11} m^2_b  m_t \tanbe \no
  \plogl{-11} + \acop{t}{11} \lte{1} \cql{11} m^2_t \mchin{1} \cotbe
              + \bcop{t}{11} \kte{1} \cqr{11} \mchin{1} m_b m_t \tanbe \no
  \plogl{-11} + \acop{t}{11} \lte{1} \cqr{11} \mchipm{1} m^2_t \cotbe
               + \acop{t}{11} \kte{1} \cqr{11} m_b m^2_t \cotbe \Bigg\} \\
db_{13} \plgl - \frac{\wzw}{m_W} \Bigg\{ \bcop{t}{11} \kte{1} \cql{11}
           \mchipm{1} m_b m_t \cotbe
    - \acop{t}{11} \lte{1} \cql{11} m^2_t \mchin{1} \cotbe \no
  \plogl{-11}  + \bcop{t}{11} \lte{1} \cql{11} m_t
     \left[ \left( \mchinq{1} +\mhpq+\mstq{1}+m^2_b \right) \cotbe
          - m^2_b  \tanbe \right] \no
  \plogl{-11}  + \acop{t}{11} \kte{1} \cql{11} \mchipm{1} m_b \mchin{1} \tanbe
    - \bcop{t}{11} \kte{1} \cqr{11} \mchin{1} m_b m_t \tanbe \no
  \plogl{-11}  + \bcop{t}{11} \lte{1} \cqr{11} \mchipm{1} m_t \mchin{1} \cotbe
   + \acop{t}{11} \lte{1} \cqr{11} m^2_b  \mchipm{1} \tanbe \no
  \plogl{-11}  + \acop{t}{11} \kte{1} \cqr{11} m_b
     \left[ \left(m^2_b +\mchinq{1} +\mhpq+\mstq{1} \right) \tanbe
          - m^2_t \cotbe \right] \Bigg\} \\
db_{14} \plgl \frac{\wzw}{m_W}
         \left( \bcop{t}{11} \lte{1} \cql{11} m_t \cotbe
               + \acop{t}{11} \kte{1} \cqr{11} m_b \tanbe \right)
\eeq
\beq
dc_{111} \plgl 2 \, C^H_{\sto{1} \sbo{1}} \Bigg\{
         \hbltQle m_b \left( m^2_b + \mstq{1} + 2 \, \mchinq{1} \right) \no
  \plogl{13} + \fbltQle \mchin{1} \left( \mchinq{1} + \mhpq + 2 \, m^2_b
             \right) \no
  \plogl{13} + \hbktQle \mchipm{1} \left( m^2_b+\mchinq{1} \right) \no
  \plogl{13} + \fbktQle 2 \, m_b \mchin{1} \mchipm{1} \Bigg\} \\
dc_{112} \plgl - 2 \, C^H_{\sto{1} \sbo{1}} \Bigg\{
     \hbltQle m_b \no
   \plogl{16}  + \fbltQle \mchin{1} \no
   \plogl{16} + \hbktQle \mchipm{1} \Bigg\} \\
dc_{113} \plgl -2 \, C^H_{\sto{1} \sbo{1}} \Bigg\{ \hbltQle m_b\no
   \plogl{16}  + \fbltQle \mchin{1} \Bigg\}
\eeq
\newpage
\beq
dd_{1} \plgl \frac{1}{2 \, \mwq} \Bigg\{
   - \left( \acopq{t}{11} m^2_t \cotbeq + \bcopq{t}{11} m^2_b \tanbeq \right)
       m^2_t \left( m^2_b + \mchinq{1} \right) \no
  \plogl{5} + \left( \acopq{t}{11} m^2_b \tanbeq
                     + \bcopq{t}{11} m^2_t \cotbeq \right)
          \left( \mchinq{1} - \mstq{1} \right)
          \left( \mhpq - m^2_b \right)  \no
  \plogl{5} + 2 \, \acop{t}{11} \bcop{t}{11} m_t \mchin{1}
      \left[ \left(\mhpq-m^2_b \right)
             \left(m^2_b \tanbeq + m^2_t \cotbeq \right)
        - 2 \, m^2_b m^2_t \right]  \no
  \plogl{5} + 2 \, \left( \acopq{t}{11} + \bcopq{t}{11} \right)
       m^2_b m^2_t \left( \mstq{1} - \mchinq{1} \right) \Bigg\}  \\
dd_{2} \plgl \frac{m^2_t}{2 \, \mwq}
 \left( \acopq{t}{11} m^2_t \cotbeq + \bcopq{t}{11} m^2_b \tanbeq \right)  \\
dd_{3} \plgl -\frac{1}{2 \, \mwq} \Bigg\{
       2 \, \left( \acopq{t}{11} + \bcopq{t}{11} \right)
         m^2_b m^2_t \no
  \plogl{15} - \left( \acopq{t}{11} m^2_b \tanbeq
                     + \bcopq{t}{11} m^2_t \cotbeq \right)
        \left(\mhpq+\mstq{1} \right) \no
  \plogl{15} + 2 \, \acop{t}{11} \bcop{t}{11} m_t \mchin{1}
   \left( m^2_b \left( 2 +\tanbeq \right) + m^2_t \cotbeq \right) \Bigg\} \\
dd_{4} \plgl - \frac{\acopq{t}{11} m^2_b \tanbeq + \bcopq{t}{11} m^2_t \cotbeq}
                {2 \, \mwq}
\eeq
\beq
de_{11} \plgl - \frac{ \wzw C^H_{\sto{1} \sbo{1}}}{\mw} \Bigg\{
       \acop{b}{11} \bcop{t}{11} m_t \mchin{1}
       \left[ \left( \mhpq - m^2_b \right) \cotbe
             - 2 \, m^2_b \tanbe \right] \no
  \plogl{21} + \bcop{b}{11} \bcop{t}{11} m_b m_t
      \left[ \left( \mstq{1} - \mchinq{1} \right) \cotbe
           - \left(m^2_b+\mchinq{1} \right) \tanbe \right] \no
  \plogl{21} - \acop{b}{11} \acop{t}{11}
       \left[ m^2_t \left( m^2_b+\mchinq{1} \right) \cotbe
            + m^2_b \left(\mchinq{1} -\mstq{1} \right) \tanbe \right] \no
  \plogl{21} - \bcop{b}{11} \acop{t}{11} m_b \mchin{1}
       \left[ 2 \, m^2_t \cotbe
             + \left( m^2_b-\mhpq \right) \tanbe \right] \Bigg\} \\
de_{12} \plgl - \frac{ \wzw C^H_{\sto{1} \sbo{1}}}{\mw} \left\{
           \acop{b}{11} \acop{t}{11} m^2_t \cotbe
           + \bcop{b}{11} \bcop{t}{11} m_b m_t \tanbe \right\} \\
de_{13} \plgl \frac{ \wzw C^H_{\sto{1} \sbo{1}}}{\mw} \bigg\{
            \acop{b}{11} \bcop{t}{11} m_t \mchin{1} \cotbe
          + \bcop{b}{11} \bcop{t}{11} m_b m_t \cotbe \no
  \plogl{17} + \, \acop{b}{11} \acop{t}{11} m^2_b \tanbe
       + \bcop{b}{11} \acop{t}{11} m_b \mchin{1} \tanbe \bigg\}
\eeq
\beq
df_{11} \plgl - (C^H_{\sto{1} \sbo{1}})^2
         \left[ \left(\acopq{b}{11}+\bcopq{b}{11} \right)
                 \left( m^2_b + \mchinq{1} \right)
                + 4 \, \acop{b}{11} \bcop{b}{11} m_b \mchin{1} \right] \\
df_{12} \plgl (C^H_{\sto{1} \sbo{1}})^2
              \left(\acopq{b}{11}+\bcopq{b}{11} \right)
\eeq
\beq
df_{31} \plgl - 2 \, C^H_{\sto{1} \sbo{1}} C^H_{\sto{1} \sbo{2}} \no
 \plogl{0} *
 \left[ \left( \acop{b}{11} \acop{b}{12}
     + \bcop{b}{11} \bcop{b}{12} \right) \left( m^2_b + \mchinq{1} \right)
     + 2 \, \left( \acop{b}{11} \bcop{b}{12}
         + \bcop{b}{11} \acop{b}{12} \right) m_b \mchin{1} \right] \\
df_{32} \plgl 2 \, C^H_{\sto{1} \sbo{1}} C^H_{\sto{1} \sbo{2}}
       \left( \acop{b}{11} \acop{b}{12} + \bcop{b}{11} \bcop{b}{12} \right)
\eeq
To get the remaining coefficients one has to make the following replacements:\\
\begin{tabular}{ll}
  $da_{1i} \to da_{2i}$: & $\lte{1} \to \lte{2}$, $\kte{1} \to \kte{2}$,
                          $\cql{11} \to \cql{12}$, $\cqr{11} \to \cqr{12}$,
                          $\mchip{1} \to \mchip{2}$ \\
  $db_{1i} \to db_{2i}$: & $\lte{1} \to \lte{2}$, $\kte{1} \to \kte{2}$,
                          $\cql{11} \to \cql{12}$, $\cqr{11} \to \cqr{12}$,
                          $\mchip{1} \to \mchip{2}$ \\
  $dc_{11i} \to dc_{12i}$: & $\lte{1} \to \lte{2}$, $\kte{1} \to \kte{2}$,
                            $\cql{11} \to \cql{12}$, $\cqr{11} \to \cqr{12}$,
                            $\mchip{1} \to \mchip{2}$ \\
  $dc_{11i} \to dc_{21i}$: & $\acop{b}{11} \to \acop{b}{12}$,
                            $\bcop{b}{11} \to \bcop{b}{12}$,
                            $\msbot{1} \to \msbot{2}$,
                            $C^H_{\sto{1} \sbo{1}} \to
                             C^H_{\sto{1} \sbo{2}}$ \\
  $dc_{11i} \to dc_{22i}$: & $\lte{1} \to \lte{2}$, $\kte{1} \to \kte{2}$,
                            $\cql{11} \to \cql{12}$, $\cqr{11} \to \cqr{12}$,
                            $\mchip{1} \to \mchip{2}$, \\
                          & $\acop{b}{11} \to \acop{b}{12}$,
                            $\bcop{b}{11} \to \bcop{b}{12}$,
                            $\msbot{1} \to \msbot{2}$,
                          $C^H_{\sto{1} \sbo{1}} \to
                           C^H_{\sto{1} \sbo{2}}$ \\
  $de_{1i} \to de_{2i}$: & $\acop{b}{11} \to \acop{b}{12}$,
                          $\bcop{b}{11} \to \bcop{b}{12}$,
                          $\msbot{1} \to \msbot{2}$,
                          $C^H_{\sto{1} \sbo{1}} \to
                           C^H_{\sto{1} \sbo{2}}$ \\
  $df_{1i} \to df_{2i}$: & $\acop{b}{11} \to \acop{b}{12}$,
                          $\bcop{b}{11} \to \bcop{b}{12}$,
                          $\msbot{1} \to \msbot{2}$,
                          $C^H_{\sto{1} \sbo{1}} \to
                           C^H_{\sto{1} \sbo{2}}$ \\
\end{tabular}

\section[Light Stop into Bottom, Slepton and Lepton]
        {The Decay of the Light Stop into a Bottom Quark, a Slepton and
         a Lepton}

Here the decay width is given by
\beq
 \Gamma(\sto{1} \to b + \tilde{l} + l') \plgl \no
 \plogl{-35} = \frac{\alpha^2}{16 \, \pi m^3_{\sto{1}} \sinwv}
        \eint{(\mstop{1}-m_b)^2}{(m_{l'} + m_{\tilde{l}})^2}{s}
        W_{l' \tilde{l}}(s)
        \sum^3_{i=1} \left( \sum^5_{j=1} c_{ij} s^{(j-4)} \right) D_i (s)
\eeq
with
\beq
 D_{1,2} (s) \plgl
  \frac{1}{(s-\mchipmq{1,2})^2 + \mchipmq{1,2} \Gamma^2_{\chipm{1,2}}} \\
 D_3 (s) \plgl Re \left(
  \frac{1}{(s-\mchipmq{1} + i \mchipm{1} \Gamma_{\chipm{1}})
            (s-\mchipmq{2} - i \mchipm{2} \Gamma_{\chipm{2}}) }  \right).
\eeq
In the case of $\sto{1} \to b + \tilde{\nu}_e + e^-$ one finds in the
limit $m_e \to 0$ that
\beq
W_{e \tilde{\nu}_e}(s) \plgl \lambda^{\einha}(s,\mstq{1},m^2_b)
                            \left(s-m^2_{\tilde{\nu}_e} \right)
\eeq
\beq
c_{11} \plgl \einha \kteq{1} V^2_{11} \mchipmq{1} m^2_{\tilde{\nu}_e}
            \left( m^2_b-\mstq{1} \right)   
\eeq
\newpage
\beq
c_{12} \plgl V^2_{11} \bigg[
      \einha \lteq{1} m^2_{\tilde{\nu}_e} \left( m^2_b - \mstq{1} \right)
    + \einha \kteq{1} \mchipmq{1}
             \left(\mstq{1}+m^2_{\tilde{\nu}_e}-m^2_b \right) \no
    \plogl{8} + 2 \, \kte{1} \lte{1} m_b \mchipm{1} m^2_{\tilde{\nu}_e}
       \bigg] \\
c_{13} \plgl V^2_{11} \left[ \einha \lteq{1}
             \left( \mstq{1} + m^2_{\tilde{\nu}_e} - m^2_b \right)
        - 2 \, \kte{1} \lte{1} m_b \mchipm{1}
        - \einha \kteq{1} \mchipmq{1} \right]                \\
c_{14} \plgl - \frac{\lteq{1} V^2_{11}}{2}
\eeq
$c_{2i}$ is obtained from $c_{1i}$ by the following replacements:
$\kte{1} \to \kte{2},\lte{1} \to \lte{2},V_{11} \to V_{12}$ and
$\mchipm{1} \to \mchipm{2}$.
\beq
c_{31} \plgl \kte{1} \kte{2} V_{11} V_{12} \mchipm{1} \mchipm{2}
            (m^2_b-\mstq{1}) m^2_{\tilde{\nu}_e}        \\
c_{32} \plgl V_{11} V_{12} \bigg[
       \lte{1} \lte{2} m^2_{\tilde{\nu}_e} \left( m^2_b - \mstq{1} \right)
       + \kte{1} \kte{2} \mchipm{1} \mchipm{2}
                   \left( \mstq{1}+m^2_{\tilde{\nu}_e}-m^2_b \right) \no
    \plogl{12} + 2 \, \kte{1} \lte{2} m_b \mchipm{1} m^2_{\tilde{\nu}_e}
        + 2 \, \kte{2} \lte{1} m_b \mchipm{2} m^2_{\tilde{\nu}_e} \bigg]\\
c_{33} \plgl V_{11} V_{12} \bigg[
        \lte{1} \lte{2} \left(\mstq{1}+m^2_{\tilde{\nu}_e}-m^2_b \right)
         - 2 \, \kte{1} \lte{2} m_b \mchipm{1}  \no
    \plogl{12}  - 2 \, \kte{2} \lte{1} m_b \mchipm{2}
         -  \kte{1} \kte{2} \mchipm{1} \mchipm{2} \bigg] \\
c_{34} \plgl - \lte{1} \lte{2} V_{11} V_{12} \\
c_{15} \plgl c_{25} = c_{35} = 0
\eeq
In the case of $\sto{1} \to b + \tsn + \tau^-$ one finds that
\beq
W_{\tau \tsn}(s) \plgl \lambda^{\einha}(s,\mstq{1},m^2_b)
                       \lambda^{\einha}(s,\mtsnq,\mtauq)
\eeq
\beq
c_{11} \plgl \einha \klqltq \mchipmq{1}
      \left( m^2_b-\mstq{1} \right) \left(\mtsn^2-\mtauq \right) \\
c_{12} \plgl \bigg[ \einha \klqktq \left(\mstq{1}-m^2_b \right)
                   \left(\mtauq - \mtsn^2 \right) \no
  \plogl{1} + \einha \klqltq \mchipmq{1}
              \left(\mstq{1}+\mtsn^2-m^2_b-\mtauq \right) \no
  \plogl{1} + 2 \, \ktlq m_{\tau} \mchipm{1}
              \left( \mstq{1}-m^2_b \right)  \no
  \plogl{1} + 2 \, \kllq m_b \mchipm{1} \left(\mtsn^2-\mtauq \right) \no
  \plogl{1} - 4 \, \kkll m_b m_{\tau} \mchipmq{1}     \bigg] 
\eeq
\newpage
\beq
c_{13} \plgl \bigg[ \einha \klqktq
               \left( \mstq{1} + \mtsn^2 - m^2_b - \mtauq \right) \no
    \plogl{1} - 2 \, \kllq m_b \mchipm{1} \no
    \plogl{1} - \einha \klqltq \mchipmq{1} \no
    \plogl{1} - 4 \, \kkll m_b m_{\tau}
       - 2 \, \ktlq \mtau \mchipm{1} \bigg]   \\
c_{14} \plgl - \einha \klqktq
\eeq
$c_{2i}$ is obtained from $c_{1i}$ by the following replacements:
$\kte{1} \to \kte{2},\lte{1} \to \lte{2},
\kcop{\nutau}{1} \to \kcop{\nutau}{2},\lcop{\nutau}{1} \to \lcop{\nutau}{2}$
and $\mchipm{1} \to \mchipm{2}$.
\beq
c_{31} \plgl \kleklzlteltz \mchipm{1} \mchipm{2}
      \left( m^2_b-\mstq{1} \right) \left( \mtsn^2-\mtauq \right) \\
c_{32} \plgl \bigg[ \kleklzktektz \left( \mstq{1}-m^2_b \right)
                   \left( \mtauq-\mtsn^2 \right) \no
 \plogl{1} +  \kleklzlteltz \mchipm{1} \mchipm{2}
             \left( \mstq{1}+\mtsn^2-m^2_b-\mtauq \right) \no
 \plogl{1} + 2 \, \klellzlteltz m_{\tau} \mchipm{1}
               \left( \mstq{1}-m^2_b \right) \no
 \plogl{1} + 2 \, \klektektzllz m_{\tau} \mchipm{2}
               \left( \mstq{1}-m^2_b \right) \no
 \plogl{1} + 2 \, \kleklzktzlte m_b \mchipm{1}
               \left( \mtsn^2-\mtauq \right) \no
 \plogl{1} + 2 \, \kleklzkteltz m_b \mchipm{2}
               \left( \mtsn^2-\mtauq \right) \no
 \plogl{1} - 4 \, \klektzllzlte m_b m_{\tau} \mchipm{1} \mchipm{2} \bigg] \\
c_{33} \plgl \bigg[ \kleklzktektz
                  \left( \mstq{1}+\mtsn^2-m^2_b-\mtauq \right)  \no
 \plogl{1} - 2 \, \kleklzktzlte m_b \mchipm{1} \no
 \plogl{1} - 2 \, \kleklzkteltz m_b \mchipm{2} \no
 \plogl{1} - \kleklzlteltz \mchipm{1} \mchipm{2} \no
 \plogl{1} - 4 \, \klektellzltz m_b m_{\tau} \no
 \plogl{1} - 2 \, \klellzlteltz m_{\tau} \mchipm{1} \no
 \plogl{1} - 2 \, \klektektzllz m_{\tau} \mchipm{2} \bigg] \\
c_{34} \plgl - \kleklzktektz \\
c_{15} \plgl c_{25} = c_{35} = 0
\eeq
In the case of $\sto{1} \to b + \sta{1} + \nutau$ one finds that
\beq
W_{\nutau \sta{1}}(s) \plgl \lambda^{\einha}(s,\mstq{1},m^2_b)
\eeq
\beq
c_{11} \plgl \einha \ltaeq{1} \lteq{1} \mchipmq{1} m^4_{\sta{1}}
            \left( \mstq{1}-m^2_b \right)  \\
c_{12} \plgl \ltaeq{1} \bigg[ \lteq{1} \mstaq{1} \mchipmq{1}
                 \left( m^2_b-\mstq{1}-\einha \mstaq{1} \right)
       + \einha \kteq{1} m^4_{\sta{1}} \left( \mstq{1}-m^2_b \right)  \no
  \plogl{10} - 2 \, \kte{1} \lte{1} m_b \mchipm{1} m^4_{\sta{1}} \bigg] \\
c_{13} \plgl \ltaeq{1} \bigg[ \einha \lteq{1} \mchipmq{1}
               \left( \mstq{1} + 2 \, \mstaq{1} - m^2_b \right)
        + 4 \, \kte{1} \lte{1} m_b \mchipm{1} \mstaq{1}      \no
  \plogl{10} + \kteq{1} \mstaq{1}
            \left( m^2_b - \mstq{1} - \einha \mstaq{1} \right) \bigg] \\
c_{14} \plgl \ltaeq{1} \bigg[ \einha \kteq{1}
         \left( 2 \, \mstaq{1} + \mstq{1} - m^2_b \right)
     - \einha \lteq{1} \mchipmq{1}
     - 2 \, \kte{1} \lte{1} m_b \mchipm{1}  \bigg] \no 
\eeq
\beq
c_{15} \plgl - \einha \ltaeq{1} \kteq{1}
\eeq
$c_{2i}$ are obtained from $c_{1i}$ by the following replacements:
$\kte{1} \to \kte{2},\lte{1} \to \lte{2},\ltae{1} \to \ltae{2}$ and
$\mchipm{1} \to \mchipm{2}$.
\beq
c_{31} \plgl \ltae{1} \ltae{2} \lte{1} \lte{2} \mchipm{1} \mchipm{2}
      m^4_{\sta{1}} \left( \mstq{1}-m^2_b \right) \\
c_{32} \plgl \ltae{1} \ltae{2} \bigg[ \kte{1} \kte{2} m^4_{\sta{1}}
         \left( \mstq{1}-m^2_b \right)
     + 2 \, \lte{1} \lte{2} \mchipm{1} \mchipm{2} \mstaq{1}
               \left( m^2_b - \einha \mstaq{1} - \mstq{1} \right)  \no
  \plogl{10} - 2 \, \lte{1} \kte{2} m_b \mchipm{1} m^4_{\sta{1}}
     - 2 \, \lte{2} \kte{1} m_b \mchipm{2} m^4_{\sta{1}}  \bigg] \\
c_{33} \plgl \ltae{1} \ltae{2} \bigg[ \lte{1} \lte{2} \mchipm{1}
        \mchipm{2} \left( 2 \, \mstaq{1} + \mstq{1}-m^2_b \right)
     + 4 \, \lte{1} \kte{2} m_b \mchipm{1} \mstaq{1}  \no
   \plogl{10} + 4 \, \lte{2} \kte{1} m_b \mchipm{2} \mstaq{1}
       - 2 \, \kte{1} \kte{2} \mstaq{1}
             \left( \mstq{1} + \einha \mstaq{1} - m^2_b \right) \bigg] \\
c_{34} \plgl \ltae{1} \ltae{2} \bigg[ \kte{1} \kte{2}
         \left( 2 \, \mstaq{1} + \mstq{1}-m^2_b \right)
     - 2 \, \lte{1} \kte{2} m_b \mchipm{1}   \no
   \plogl{10}  - 2 \, \lte{2} \kte{1} m_b \mchipm{2}
     - \lte{1} \lte{2} \mchipm{1} \mchipm{2}   \bigg]   \\
c_{35} \plgl - \ltae{1} \ltae{2} \kte{1} \kte{2}
\eeq
To get the coefficients for $\sto{1} \to b + \sta{2} + \nutau$ one has to
make the following replacements: $\ltae{i} \to \ltaz{i}$ and 
$\mstau{1} \to \mstau{2}$.
For $\sto{1} \to b + \tilde{e}_L + \nu_e$ one gets the corresponding 
coefficients
by the replacements: $\ltae{i} \to u_{1i} $ and
$\mstau{1} \to m_{\tilde{e}_L}$.

\section{Some analytical solutions of integrals}

\subsection{The integrals $J_i$}
\label{intJ}

The following integrals appear in the formulae for the three body decays:
\begin{enumerate}
 \item \beq \jintz{t}{0}{m_1}
        = \eint{t_{max}}{t_{min}}{t} \frac{1}{(t-m^2_1)^2 \, + \, \gmz{1}}
        \nonumber
        \eeq
 \item \beq \jintz{t}{1}{m_1}
        = \eint{t_{max}}{t_{min}}{t} \frac{t}{(t-m^2_1)^2 \, + \, \gmz{1}}
        \nonumber
        \eeq
 \item \beq \jintz{t}{2}{m_1}
        = \eint{t_{max}}{t_{min}}{t} \frac{t^2}{(t-m^2_1)^2 \, + \, \gmz{1}}
        \nonumber
        \eeq
 \item \beq \jintz{tt}{0}{m_1,m_2}
        = Re \eint{t_{max}}{t_{min}}{t}
        \nonumber
          \frac{1}{(t-m^2_1 \, + \,i \gme{1})(t-m^2_2 \, - \,i \gme{2})}
        \nonumber
        \eeq
 \item \beq \jintz{tt}{1}{m_1,m_2}
        = Re \eint{t_{max}}{t_{min}}{t}
          \frac{t}{(t-m^2_1 \, + \,i \gme{1})(t-m^2_2 \, - \,i \gme{2})}
        \nonumber
        \eeq
 \item \beq \jintz{tt}{2}{m_1,m_2}
        = Re \eint{t_{max}}{t_{min}}{t}
          \frac{t^2}{(t-m^2_1 \, + \,i \gme{1})(t-m^2_2 \, - \,i \gme{2})}
        \nonumber
        \eeq
 \item \beq \jintz{st}{0}{s,m_1,m_2}
        = Re \frac{1}{s-m^2_1 \, + \,i \gme{1}}
             \eint{t_{max}}{t_{min}}{t} \frac{1}{t-m^2_2 \, - \,i \gme{2}}
        \nonumber
        \eeq
 \item \beq \jintz{st}{1}{s,m_1,m_2}
        = Re \frac{1}{s-m^2_1 \, + \,i \gme{1}}
             \eint{t_{max}}{t_{min}}{t} \frac{t}{t-m^2_2 \, - \,i \gme{2}}
        \nonumber
        \eeq
\end{enumerate}

\subsection{The solution of $\jintz{t}{0}{m_1}$}

If $\Gamma_1 = 0$ then
\beq
\jintz{t}{0}{m_1} = \frac{1}{t_{min} - m^2_1}
               - \frac{1}{t_{max} - m^2_1}
\eeq
else
\beq
\jintz{t}{0}{m_1} = \frac{1}{\gme{1}} \left[
                \atan \left( \frac{t_{max} - m^2_1}{\gme{1}} \right)
               - \atan \left( \frac{t_{min} - m^2_1}{\gme{1}} \right)
               \right]
\eeq

\subsection{The solution of $\jintz{t}{1}{m_1}$}

If $\Gamma_1 = 0$ then
\beq
\jintz{t}{1}{m_1} = \frac{m^2_1}{t_{min} - m^2_1}
               - \frac{m^2_1}{t_{max} - m^2_1}
               + \log \left|
                 \frac{t_{max} - m^2_1}{t_{min} - m^2_1} \right|
\eeq
else
\beq
\jintz{t}{1}{m_1} \plgl \frac{m_1}{\Gamma_1} \left[
                \atan \left( \frac{t_{max} - m^2_1}{\gme{1}} \right)
               - \atan \left( \frac{t_{min} - m^2_1}{\gme{1}} \right)
               \right] \no
       \plogl{5} + \einhag \log \left|
                 \frac{(t_{max} - m^2_1)^2 + \gmz{1}}
                      {(t_{min} - m^2_1)^2 + \gmz{1}} \right|
\eeq

\subsection{The solution of $\jintz{t}{2}{m_1}$}

If $\Gamma_1 = 0$ then
\beq
\jintz{t}{2}{m_1} = \frac{m^4_1}{t_{min} - m^2_1}
               - \frac{m^4_1}{t_{max} - m^2_1}
               + 2 m^2_1 \log \left|
                 \frac{t_{max} - m^2_1}{t_{min} - m^2_1} \right|
               + t_{max} - t_{min}
\eeq
else
\beq
\jintz{t}{2}{m_1} \plgl \frac{m^4_1 - \gmz{1}}{\gme{1}} \left[
                \atan \left( \frac{t_{max} - m^2_1}{\gme{1}} \right)
               - \atan \left( \frac{t_{min} - m^2_1}{\gme{1}} \right)
               \right] \no
       \plogl{5} + m^2_1 \log \left|
                 \frac{(t_{max} - m^2_1)^2 + \gmz{1}}
                      {(t_{min} - m^2_1)^2 + \gmz{1}} \right|
               + t_{max} - t_{min}
\eeq

\subsection{The solution of $\jintz{tt}{0}{m_1,m_2}$}

If $\Gamma_1 = 0$ and $\Gamma_2 = 0$ then
\beq
\jintz{tt}{0}{m_1,m_2} = \frac{1}{m^2_1 - m^2_2}
          \left[
                \log \left| \frac{t_{max} - m^2_1}{t_{min} - m^2_1} \right|
               - \log \left| \frac{t_{max} - m^2_2}{t_{min} - m^2_2} \right|
          \right]
\eeq
else
\beq
\jintz{tt}{0}{m_1,m_2} \plgl  \frac{m^2_1 - m^2_2}
                             {2[(m^2_1 - m^2_2)^2 + (\gme{1}+\gme{2})^2]} \no
          \plogl{-1} *
          \left[
                \log \left| \frac{(t_{max} - m^2_1)^2 + \gmz{1}}
                                 {(t_{min} - m^2_1)^2 + \gmz{1}} \right|
               - \log \left| \frac{(t_{max} - m^2_2)^2 + \gmz{2}}
                                  {(t_{min} - m^2_2)^2 + \gmz{2}} \right|
          \right] \no
       \plogl{-5} +   \frac{\gme{1} + \gme{2}}
                          {(m^2_1 - m^2_2)^2 + (\gme{1}+\gme{2})^2} \no
          \plogl{-1} *
                  \left[ \atan \left( \frac{\gme{1}}{t_{max}-m^2_1} \right)
                       - \atan \left( \frac{\gme{1}}{t_{min}-m^2_1} \right)
                  \right. \no
          \plogl{1} \left.
                       + \atan \left( \frac{\gme{2}}{t_{max}-m^2_2} \right)
                       - \atan \left( \frac{\gme{2}}{t_{min}-m^2_2} \right)
                  \right]
\eeq

\subsection{The solution of $\jintz{tt}{1}{m_1,m_2}$}

If $\Gamma_1 = 0$ and $\Gamma_2 = 0$ then
\beq
\jintz{tt}{1}{m_1,m_2} = \frac{1}{m^2_1 - m^2_2}
          \left[ m^2_1
                \log \left| \frac{t_{max} - m^2_1}{t_{min} - m^2_1} \right|
               - m^2_2
                 \log \left| \frac{t_{max} - m^2_2}{t_{min} - m^2_2} \right|
          \right]
\eeq
else
\beq
\jintz{tt}{1}{m_1,m_2} \plgl \no
   \plogl{-12} =
         \frac{m^2_1 (m^2_1-m^2_2) + \gme{1} (\gme{1}+\gme{2})}
              {2[(m^2_1 - m^2_2)^2 + (\gme{1}+\gme{2})^2]}
          \log \left| \frac{(t_{max} - m^2_1)^2 + \gmz{1}}
                                 {(t_{min} - m^2_1)^2 + \gmz{1}} \right| \no
     \plogl{-11}
              + \frac{m^2_2 (m^2_2-m^2_1) + \gme{2} (\gme{1}+\gme{2})}
                    {2[(m^2_1 - m^2_2)^2 + (\gme{1}+\gme{2})^2]}
                \log \left| \frac{(t_{max} - m^2_2)^2 + \gmz{2}}
                                  {(t_{min} - m^2_2)^2 + \gmz{2}} \right| \no
     \plogl{-11} -   \frac{m^2_2 \gme{1} + m^2_1 \gme{2}}
                          {(m^2_1 - m^2_2)^2 + (\gme{1}+\gme{2})^2} \no
          \plogl{-1} *
                  \left[ \atan \left( \frac{\gme{1}}{t_{max}-m^2_1} \right)
                       - \atan \left( \frac{\gme{1}}{t_{min}-m^2_1} \right)
                  \right. \no
          \plogl{1} \left.
                       + \atan \left( \frac{\gme{2}}{t_{max}-m^2_2} \right)
                       - \atan \left( \frac{\gme{2}}{t_{min}-m^2_2} \right)
                  \right]
\eeq

\subsection{The solution of $\jintz{tt}{2}{m_1,m_2}$}

If $\Gamma_1 = 0$ and $\Gamma_2 = 0$ then
\beq
\jintz{tt}{2}{m_1,m_2} \plgl  t_{max} - t_{min}  \no
       \plogl{-2} + \frac{1}{m^2_1 - m^2_2}
          \left[ m^4_1
                \log \left| \frac{t_{max} - m^2_1}{t_{min} - m^2_1} \right|
               - m^4_2
                 \log \left| \frac{t_{max} - m^2_2}{t_{min} - m^2_2} \right|
          \right]
\eeq
else
\beq
\jintz{tt}{2}{m_1,m_2} \plgl  t_{max} - t_{min} \no
    \plogl{-5} + \frac{(m^4_1-\gmz{1}) (m^2_1-m^2_2)
                      + 2 m^3_1 \Gamma_1 (\gme{1}+\gme{2})}
                    {2[(m^2_1 - m^2_2)^2 + (\gme{1}+\gme{2})^2]} \no
    \plogl{0} * \log \left| \frac{(t_{max} - m^2_1)^2 + \gmz{1}}
                                 {(t_{min} - m^2_1)^2 + \gmz{1}} \right| \no
    \plogl{-5} + \frac{(m^4_2-\gmz{2}) (m^2_2-m^2_1)
                      + 2 m^3_2 \Gamma_2 (\gme{1}+\gme{2})}
                    {2[(m^2_1 - m^2_2)^2 + (\gme{1}+\gme{2})^2]} \no
    \plogl{0} * \log \left| \frac{(t_{max} - m^2_2)^2 + \gmz{2}}
                                  {(t_{min} - m^2_2)^2 + \gmz{2}} \right| \no
    \plogl{-5} -   \frac{(m^4_1 - \gmz{1}) (\gme{1} + \gme{2})
                          - 2 m^3_1 \Gamma_1 (m^2_1 - m^2_2)}
                        {(m^2_1 - m^2_2)^2 + (\gme{1}+\gme{2})^2} \no
          \plogl{0} *
                  \left[ \atan \left( \frac{\gme{1}}{t_{max}-m^2_1} \right)
                   - \atan \left( \frac{\gme{1}}{t_{min}-m^2_1} \right)
                  \right] \no
    \plogl{-5} -   \frac{(m^4_2 - \gmz{2}) (\gme{1} + \gme{2})
                          - 2 m^3_2 \Gamma_2 (m^2_2 - m^2_1)}
                        {(m^2_1 - m^2_2)^2 + (\gme{1}+\gme{2})^2} \no
          \plogl{0} * \left[
                        \atan \left( \frac{\gme{2}}{t_{max}-m^2_2} \right)
                       - \atan \left( \frac{\gme{2}}{t_{min}-m^2_2} \right)
                  \right]
\eeq

\subsection{The solution of $\jintz{st}{0}{s,m_1,m_2}$}

If $\Gamma_1 = 0$ and $\Gamma_2 = 0$ then
\beq
\jintz{st}{0}{s,m_1,m_2} = \frac{1}{s - m^2_1}
                \log \left| \frac{t_{max} - m^2_2}{t_{min} - m^2_2} \right|
\eeq
else
\beq
\jintz{st}{0}{s,m_1,m_2} \plgl 
               \frac{s-m^2_1}{2[(s - m^2_1)^2 + \gmz{1}]}
                \log \left| \frac{(t_{max} - m^2_2)^2 + \gmz{2}}
                                 {(t_{min} - m^2_2)^2 + \gmz{2}} \right| \no
    \plogl{-5} - \frac{\gme{1}}{(s - m^2_1)^2 + \gmz{1}} \no
    \plogl{0} * \left[ \atan \left( \frac{\gme{2}}{t_{max}-m^2_2} \right)
                   - \atan \left( \frac{\gme{2}}{t_{min}-m^2_2} \right)
                  \right] \no
\eeq

\subsection{The solution of $\jintz{st}{1}{s,m_1,m_2}$}

If $\Gamma_1 = 0$ and $\Gamma_2 = 0$ then
\beq
\jintz{st}{1}{s,m_1,m_2} = \frac{1}{s - m^2_1}
         \left( t_{max} - t_{min}
                + m^2_2
                 \log \left| \frac{t_{max} - m^2_2}{t_{min} - m^2_2} \right|
         \right)
\eeq
else
\beq
\jintz{st}{1}{s,m_1,m_2} \plgl
               \frac{s-m^2_1}{(s - m^2_1)^2 + \gmz{1}}
                 \left( t_{max} - t_{min} \right) \no
    \plogl{-5} + \frac{m^2_2 (s-m^2_1) + \gme{1} \gme{2}}
                      {2[(s - m^2_1)^2 + \gmz{1}]}
                \log \left| \frac{(t_{max} - m^2_2)^2 + \gmz{2}}
                                 {(t_{min} - m^2_2)^2 + \gmz{2}} \right| \no
    \plogl{-5} - \frac{\gme{2} (s - m^2_1) - m^2_2 \gme{1}}
                      {(s - m^2_1)^2 + \gmz{1}} \no
    \plogl{0} * \left[ \atan \left( \frac{\gme{2}}{t_{max}-m^2_2} \right)
                   - \atan \left( \frac{\gme{2}}{t_{min}-m^2_2} \right)
                  \right] \no
\eeq

%% file: determin.tex
\chapter{Calculation of sfermion parameters}
\label{appB}

Here we present the formulae for the computation of the soft
SUSY parameters from sfermion masses and mixing angles.
The following assumptions are made:
\begin{itemize}
 \item $\mu$ and $\tanbe$ are known from other experiments (Higgs, chargino
       and neutralino sector)
 \item there is hardly any mixing between the generations.
\end{itemize}
From the inversion of the
eigenvalue-problem for the squark-masses one obtains:
\beq
\label{eq:determq1}
 \mqq &=& - \mzq \coszbe ( \einha - \zwdr \sinwq ) - m^2_t \no
 &&\hspace{2cm}         + \einha \left( \mstq{1} + \mstq{2} +
                   ( \mstq{1} - \mstq{2})  \coszt  \right) \\
       &=&  \mzq \coszbe ( \einha - \eindr \sinwq )
                              - m^2_b  \no
 &&        + \einha \left( \msbq{1} + \msbq{2} +
                   ( \msbq{1} - \msbq{2})  \coszb  \right) \\
\label{eq:determq2}
  \muq &=& - \mzq \coszbe \zwdr \sinwq  - m^2_t \no
 &&\hspace{2cm}        + \einha \left( \mstq{1} + \mstq{2} +
                   ( \mstq{2} - \mstq{1})  \coszt  \right) \\
  \mdq &=& - \mzq \coszbe \eindr \sinwq  - m^2_b  \no
 &&\hspace{2cm}        + \einha \left( \msbq{1} + \msbq{2} +
                   ( \msbq{2} - \msbq{1})  \coszb  \right) \\
 A_t &=& \mu \cotbe + \frac{ (\mstq{1} - \mstq{2}) \sinzt }
                        {2 m_t} \\
 A_b &=& \mu \tanbe + \frac{ (\msbq{1} - \msbq{2}) \sinzb }
                        {2 m_b }
\eeq
As one can see, there are two equations to determine $\mqq$. Therefore,
one gets at tree-level the condition:
\beq
\hspace*{-7mm}
\mwq \coszbe = \mstq{1} \costq + \mstq{2} \sintq - m^2_t
             -  \msbq{1} \cosbq - \msbq{2} \sinbq + m^2_b
\eeq
This condition implies that one of the six quantities
$\mstop{1}$, $\mstop{2}$, $\costq$, $\msbot{1}$, $\msbot{2}$, $\cosbq$
can be predicted by the other five, if the above assumptions are fulfilled.
However, the above relations are only valid at tree-level.
For a precise calculation one needs also the inclusion of one-loop
corrections \cite{Helmut97}.

For the sleptons the corresponding results are:
\beq
 M^2_{\tilde L} &=& \mtsnq - \einha \mzq \coszbe \\
       &=&  \mzq \coszbe ( \einha - \sinwq )
                              - m^2_{\tau}  \no
 &&        + \einha \left( \mstaq{1} + \mstaq{2} +
                   ( \mstaq{1} - \mstaq{2})  \coszta  \right) \\
\label{eq:determl1}
 M^2_{\tilde E} &=& - \mzq \coszbe \sinwq  - m^2_{\tau}  \no
 &&\hspace{2cm}        + \einha \left( \mstaq{1} + \mstaq{2} +
                   ( \mstaq{2} - \mstaq{1})  \coszta  \right) \\
 A_b &=& \mu \tanbe + \frac{ (\mstaq{1} - \mstaq{2}) \sinzta }
                        {2 m_{\tau} }
\eeq
Again, there are two equations for $M^2_{\tilde L}$ giving the following
condition:
\beq
\label{eq:determl2}
\mwq \coszbe = \mtsnq
             -  \mstaq{1} \costaq - \mstaq{2} \sintaq + m^2_{\tau}
\eeq
Therefore, one of the four quantities
$m^2_{\nutau}$,  $\mstau{1}$, $\mstau{2}$, $\costaq$
can be predicted by the other three under the above assumptions.

%% file: wirk.tex
\chapter{Production cross sections}
\label{appC}

\section{Tree Level}

The reaction $\ee \sfer{i} \asfer{j}$ proceeds via $\gamma$ and $Z^{0}$ 
exchange (see \fig{fig:prodsfer}).
The tree level cross section at a center-of-mass energy of $\sqrt{s}$ 
is given by (no summation over i and j):
\beq
\sigma^{tree}(e^+ e^- \to \sfer{i} \asfer{j})
= c_{ij}\left[ e_f^2 \delta_{ij} - T_{\gamma Z} e_f a_{ij} \delta_{ij}
+ T_{ZZ} a_{ij}^2 \right] \, ,
\label{sigtree}
\eeq
with
\beq
c_{ij} & = & \frac{\pi N_C \alpha^2}{3 \, s}\lambda^{3/2}_{ij} \, ,
          \label{cij}\\
T_{\gamma Z} & = &  \frac{(L_e + R_e)}{\sinwq \coswq}
          \frac{s(s-m_Z^2)}{(s-m_Z^2)^2+ \Gamma^2_Z m_Z^2}\, ,
          \label{TgZ}\\
T_{ZZ} & = & \frac{(L_e^2+R_e^2)}{2 \, \sinwv \coswv }
          \frac{s^2}{(s-m_Z^2)^2+ \Gamma^2_Z m_Z^2}
          \label{TZZ}\, .
\eeq
Here $N_C$ is a colour factor which is 3 for squarks and 1 for sleptons;
$\lambda_{ij}= (1-\mu_i^2 - \mu_j^2)^2 - 4 \mu_i^2 \mu_j^2,
\mu^2_{i,j}= \msferq{i,j}/s$;
$e_f$ is the charge of the sfermions ($e_t = 2/3, e_b = - 1/3, e_{\tau} = -1$) 
in units of $e ( = \sqrt{4 \pi \alpha})$. $L_e$
($R_e$) is the coupling of the left-handed (right-handed)
electron to the
$Z$ boson: $L_e = -\einha + \sinwq$ ($R_e = \sinwq$), 
and $a_{ij}$ are the corresponding couplings $Z \sfer{i} \asfer{j}$:
\beq
a_{11} \plgl 4(I_f^{3L} \cosfq  - \sinwq e_f) \ , \no
a_{22} \plgl 4(I_f^{3L} \sinfq - \sinwq e_f) \ , \no
a_{12} \plgl a_{21}= -2 I_q^{3L} \sinfq \, ,
\label{acoup}
\eeq
where $I_f^{3L}$ is the third component of the weak isospin of the fermion $f$.

\begin{minipage}{142mm}
{\setlength{\unitlength}{1mm}
\begin{picture}(142,55)
\put(8,4){\mbox{\psfig{figure=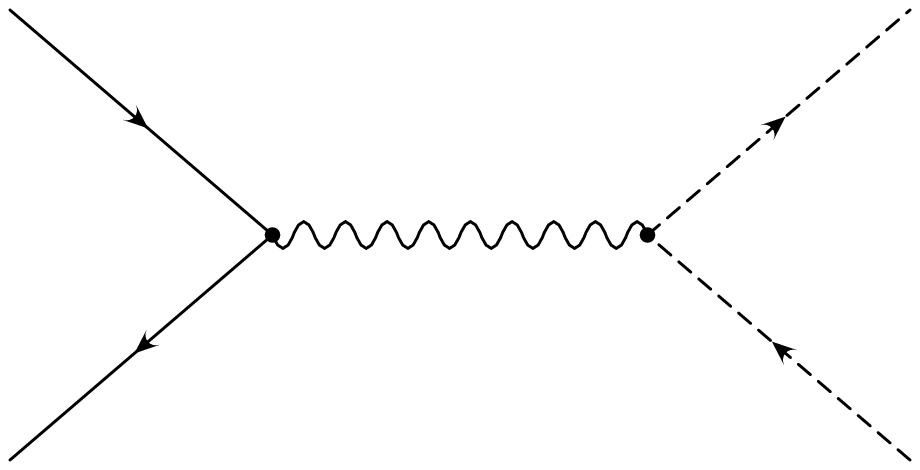,height=4.6cm}}}
\put(7,51){\makebox(0,0)[r]{$e^-$}}
\put(7,5){\makebox(0,0)[r]{$e^+$}}
\put(55,31){\makebox(0,0)[b]{$\gamma, Z^0$}}
\put(101,51){\makebox(0,0)[l]{$\sfer{i}$}}
\put(101,5){\makebox(0,0)[l]{$\asfer{j}$}}
\end{picture}}
\figcaption{fig:prodsfer}
           {Feynman diagrams for the process $\protect\ee \protect\sfer{i}
            \protect\asfer{j}$ ($f \neq e, \nu_e$).}
           {Feynman diagrams for the process $\ee \sfer{i} \asfer{j}$
            ($f \neq e, \nu_e$).}{7}
\end{minipage}

$\sigma^{tree}$ shows the typical $\lambda^{3/2}$ suppression.
The interference of the $\gamma$ and $Z^{0}$ exchange 
contributions leads to a characteristic minimum of the cross section 
at 
\beq
  \cosfq{\big |}_{min} = 
    \frac{Q_{f}}{I^{3}_{f}}\sinwq \,
    \left[ 1 + (1-\frac{\mzq}{s})\coswq
           \frac{L_{e}+R_{e}}{L_{e}^{2}+R_{e}^{2}} \,\right]
\eeq
in case of $\sfer{1} \asfer{1}$ production. In the case of $\sfer{2} \asfer{2}$ 
production $\cosfq$ has to be replaced by $\sinfq$.
The angular distribution has the familiar $\sin^2\vartheta$ 
shape, with $\vartheta$ the scattering angle:
\beq
\frac{{\rm d}\,\sigma^{tree}}{{\rm d} \cos\vartheta} = \dreivi
                     \,\sin^2\vartheta\, \sigma^{tree}.
\eeq
These formulae for the production cross sections are valid for the case of
unpolarized electrons. In the case of left-polarized electrons one has to make
the substitutions $(L_e + R_e) \to 2 \, L_e$ and 
$(L^2_e + R^2_e) \to 2 \, L^2_e$ and in case of right-polarized electrons:
$(L_e + R_e) \to 2 \, R_e$ and $(L^2_e + R^2_e) \to 2 \, R^2_e$.
The cross section for a certain $e^-$ beam polarization 
$\xi \in [-1,1]$, where $\xi = (-1,0,1) \equiv$ (left-polarized, unpolarized,
right-polarized) $e^-$ beam, has the form
\beq
\sigma(\xi) & =& (\Theta(\xi) - \xi) \, \sigma_L
               + (\Theta(-\xi) + \xi) \, \sigma_R \no
& & \mathrm{with} \hspace{1cm}
\Theta(\xi) = \left\{ \begin{array}{cc}
                     1 & \xi > 0 \\
                     1/2 & \xi = 0 \\
                     0 & \xi < 0 
                      \end{array} \right.
\eeq
                  

\section{SUSY-QCD Corrections}

The supersymmetric QCD corrected cross section in $O(\alpha_s)$ 
can be written as \cite{Eberl96}:
\beq
\sigma^{QCD} = \sigma^{tree} + \delta \sigma^{g} + \delta \sigma^{\tilde g}
+ \delta \sigma^{\tilde q} \, .
 \label{sig1}
\eeq
$\delta \sigma^{g}$ gives the standard gluonic correction,
$\delta \sigma^{\glu}$ is the correction 
due to the gluino exchange, and
$\delta \sigma^{\squ{}}$ is the correction due to squark exchange.
$\delta \sigma^{\squ{}}$ is 0
in the renormalization prescription used.

$\delta \sigma^{g}$ is given by
\beq
\delta \sigma^{g} = \sigma^{tree} \left[
\frac{4 \, \alpha_s}{3 \, \pi} \Delta_{ij}\right]
 \label{dsigg}
\eeq
with
\beq
\hspace*{-5mm}
\Delta_{ij} &=& \log (\mu_i \mu_j) +2
+ \frac{2+\mu_i^2+\mu_j^2}{\lambda_{ij}^{1/2}} \log \lambda_0
+ \frac{1+2 \mu_i^2}{\lambda_{ij}^{1/2}}  \log \lambda_1 +
\frac{1+2\mu_j^2}{\lambda_{ij}^{1/2}} \log \lambda_2 \no
&+&
\frac{1- \mu_i^2-\mu_j^2}{\lambda_{ij}^{1/2}}
 \log \frac{1-\mu_i^2 -\mu_j^2+\lambda_{ij}^{1/2}}
{1-\mu_i^2 -\mu_j^2 -\lambda_{ij}^{1/2}} \no
&+&
\left[\frac{(1-\mu_i^2-\mu_j^2)}{\lambda_{ij}^{1/2}} \log\lambda_0 -1
\right] \log \frac{\lambda_{ij}^2}{\mu_i^2 \mu_j^2} \no
&+&  \frac{4}{\lambda_{ij}^{3/2}}
 \left[ \frac{1}{4} \lambda_{ij}^{1/2} (1+\mu_i^2+ \mu_j^2) + \mu_i^2
\log \lambda_2+ \mu_j^2 \log \lambda_1+ \mu_i^2 \mu_j^2 \log \lambda_0 \right]
\no
&+&\frac{1-\mu_i^2-\mu_j^2}{\lambda_{ij}^{1/2}}\Bigg[\frac{2\pi^2}{3}
+ 2 \Li(1-\lambda_0^2)+\Li(\lambda_1^2)- \Li(1-\lambda_1^2)+
\Li(\lambda_2^2) \no
\plogl{21} - \Li(1-\lambda_2^2) + 2 \log^2 \lambda_0 -\log\lambda_{ij}
\log\lambda_0\Bigg]\, ,
\label{delgluon}
\eeq
\beq
\lambda_0= \frac{1}{2\mu_i \mu_j}(1-\mu_i^2 - \mu_j^2 +\lambda_{ij}^{1/2})
\ \ \ \ , \ \
\lambda_{1,2}= \frac{1}{2\mu_{j,i}} (1 \mp \mu_i^2 \pm \mu_j^2 -\lambda_{ij}^{
1/2})\, ,
\label{lam}
\eeq
and $\Li(x) = - \int_0^1 \log (1-xt)/ t\, \rm{d}t$.
In \eqn{delgluon} soft and hard gluon radiation is included.

The gluino correction can be written as:
\beq
\hspace*{-2mm}
\delta \sigma^{\glu} = c_{ij}\left[ 2 e_q \Delta (e_q)_{ij}^{(\glu)} \delta_{ij} -
 T_{\gamma Z} (e_q \delta_{ij}\Delta a_{ij}^{(\glu)} +
\Delta (e_q)_{ij}^{(\glu)} a_{ij})
+ 2 T_{ZZ} a_{ij} \Delta a_{ij}^{(\glu)} \right]
 \label{dsiga}
\eeq
Here $\Delta a_{ij}^{(\glu)}$ and $\Delta a_{ij}^{(\glu)}$ are given by:
\beq 
 \hspace*{-5mm}
 \Delta a_{ij}^{(\glu)} &=& \frac{2 \, \alpha_s}{3 \, \pi}
    \Bigg\{ 2 \, \mglu m_q v_q (S^{\squ{}})_{ij} 
         \left( 2 \, C^+_{ij} + C^0_{ij} \right) \no 
 \plogl{-5} + v_q \delta_{ij} \left[
     \left( 2 \, \mgluq + 2 \, m_q^2 + \msquq{i} + \msquq{j} \right) C^+_{ij}
     + 2 \, \mgluq \, C^0_{ij} + B^0(s,m^2_q,m^2_q) \right] \no
 \plogl{-5} + a_q (A^{\squ{}})_{ij} \bigg[
        \left( 2 \, \mgluq - 2 \, m_q^2 + \msquq{i} + \msquq{j} \right) C^+_{ij}
        \no
 \plogl{12} + \left(\msquq{i} - \msquq{j} \right) C^-_{ij}
            +2 \, \mgluq \, C^0_{ij} + B^0(s,m^2_q,m^2_q) \bigg] \no  
 \plogl{-5} - \frac{a_{ij}}{2} \bigg[
      B_0(\msquq{i}, \mgluq,m_q^2)  +  B_0(\msquq{j}, \mgluq,m_q^2) \no
 \plogl{4}  + (\msquq{i}-m_q^2-\mgluq -2 m_q \mglu (-1)^i \sinzq) 
      B_0'(\msquq{i}, \mgluq,m_q^2) \no
 \plogl{4} 
    + (\msquq{j}-m_q^2-\mgluq -2 m_q \mglu (-1)^j \sinzq) 
      B_0'(\msquq{j}, \mgluq,m_q^2) \bigg] \no
 \plogl{-5} - 
      \frac{2 \, \mglu m_q}{\msquq{1} - \msquq{2}} \delta_{ij} 
       \bigg[ B^0(\msquq{i},\mgluq,m_q^2)
            \left( (-1)^{i+1} 2 a_{ii'} \coszq  - a_{i'i'} \sinzq \right) \no
 \plogl{23}
      + B^0(\msquq{i'},\mgluq,m_q^2) a_{ii} \sinzq \bigg]
   \Bigg\} \, , \, i' \neq i \, , \, j' \neq j \, ,
\label{22}
\eeq
and, if $i=j$,
\beq
 \hspace*{-5mm}
 \Delta {(e_q)}_{ij}^{(\glu)} &=&
   \frac{2 \, e_q \alpha_s}{3 \, \pi} \Bigg\{ 
   \left(2 \, \mgluq+ 2 \, m_q^2+ \msquq{i} + \msquq{j}
   + 4 \, \mglu m_q (S^{\squ{}})_{ii} \right) C^+_{ij} \no
 \plogl{0} + 
    2 \, \left(\mgluq + \mglu m_q (S^{\squ{}})_{ii} \right) C^0_{ij} 
     + B^0(s,m^2_q,m^2_q) - B_0(\msquq{i}, \mgluq,m_q^2)   \no
  \plogl{0}
  - (\msquq{i}-m_q^2-\mgluq -2 m_q \mglu (-1)^i \sinzq) 
      B_0'(\msquq{i}, \mgluq,m_q^2)  \Bigg\} \, ,
 \label{eqvertex1}
\eeq
and, if $i \neq j$,
\beq
 \hspace*{-5mm}
 \Delta {(e_q)}_{ij}^{(\glu)} &=&
   \frac{2 \, e_q \alpha_s}{3 \, \pi} \Bigg\{
    2 \, \mglu m_q (S^{\squ{}})_{ij}
   \left( 2 C^+_{ij}+ C^0_{ij} \right) \no
 \plogl{0} + \frac{2 \, \mglu m_q \coszq}{\msquq{1} - \msquq{2}}
   \left( B_0(\msquq{2}, \mgluq,m_q^2)  -  B_0(\msquq{1}, \mgluq,m_q^2)
   \right) \Bigg\} \, ,
 \label{eqvertex2}
\eeq
where $\delta_{ij}$ is the identity matrix, $v_q = 2 I_q^{3L} - 4 \sinwq e_q$,
$a_q = 2 I_q^{3L}$,  
\beq
 A^{\squ{}} = \left( \begin{array}{cc} \coszq & -\sinzq \\
                                   - \sinzq & -\coszq
            \end{array} \right) &,&
S^{\squ{}} = \left( \begin{array}{cc} -\sinzq & -\coszq \\
                                      -\coszq & \sinzq
             \end{array} \right)\, .
\eeq
The functions $C^\pm_{ij}$ are defined by
\begin{equation}
C^+ = \frac{C^1+C^2}{2} \qquad , \qquad C^- = \frac{C^1-C^2}{2} \, .
\end{equation}
The arguments of all C--functions are $(\msquq{i}, s, \msquq{j},
\mgluq, m_q^2, m_q^2)$. $B^0, C^0, C^1$, and $C^2$ are the usual two--
and three--point functions \cite{Passarino79}. Here we use the conventions
of \cite{Denner93}:
\beq
B^0(k^2, m_1^2, m_2^2)&=&
\int\frac{d^Dq}{i\pi^2}\frac{1}{(q^2-m_1^2)((q+k)^2-m_2^2)}\,, \nonumber\\
\left[C^0, k^\mu C^1 - \bar{k}^\mu C^2\right]&=&
\int\frac{d^Dq}{i\pi^2}\frac{[1, q^\mu]}{(q^2-\mgluq)((q+k)^2-m_q^2)
((q-\bar{k})^2-m_q^2)} \nonumber\, .
\eeq

\section{Initial-State Radiation}

The effect of initial-state radiation is most easily included by introducing
an energy-dependent $e^+ e^-$ luminosity function $L_{ee}$. For our purpose
it is sufficient to use the $O(\alpha)$ expression, which reads after 
summing up  the leading logarithms 
\cite{Drees90,Peskin89} (Here we use the notation of \cite{Drees90}):
\beq
\sigma^{total} =
       \left( 1 + \alpha(s) \left( \frac{\pi}{3} - \frac{1}{2 \, \pi}
                \right) \right)
    \eint{1}{x_{min}}{x}
               L_{ee}(x) \, \sigma^{QCD}(x \, s)
\eeq
with (Obviously there are no QCD corrections in case of sleptons)
\beq
x_{min} =  \frac{(\msfer{i} + \msfer{j})^2}{s},
\eeq
\beq
L_{ee}(x) = \beta_{em} (1-x)^{\beta_{em}-1}
               \left( 1 + \dreivi \beta_{em} \right)
             -\einha \beta_{em} \left( 1 + x \right)
\eeq
and 
\beq
\beta_{em} = \frac{2 \alpha_{em}}{\pi} 
             \left( \log\left(\frac{s}{m^2_e}\right) - 1 \right).
\eeq
For a discussion on 
$O(\alpha^2)$ corrections in various schemes see for instance \cite{Kuraev85}.

%% file: leben.tex
\addcontentsline{toc}{chapter}{\protect\numberline{Zusammenfassung}}

\chapter*{Zusammenfassung}
\markright{ZUSAMMENFASSUNG}

Die Vorhersagen des Standardmodells der Elementarteilchenphysik stimmen mit 
einer
erstaunlichen Genauigkeit mit den experimentellen Daten \"uberein. Trotz dieses
gro{\ss}artigen Erfolges mu{\ss} dieses Modell als Grenzfall 
einer \"ubergeordneten
Theorie angesehen werden. Supersymmetrie gilt als einer der aussichtsreichsten
Erweiterungen des Standardmodells. Im Rahmen dieser neuen Theorie wird die
Poincar\'e-Algebra um fermionische, antikommutierende Generatoren erweitert, 
die zu einer Symmetrie f\"uhren, die fermionische und bosonische Freiheitsgrade
ineinander \"uberf\"uhrt. Im Rahmen des Minimalen Supersymmetrischen Modells
wird dieses Konzept auf eine ph\"anomenologische Grundlage gestellt.

Diese grundlegende Ver\"anderung der Symmetrie hat eine wesentliche
Erweiterung des Teilchengehaltes zur Folge: Zu den Leptonen und Quarks kommen
die so\-ge\-nann\-ten Sleptonen und Squarks als Partner hinzu. 
Der Higgssektor mu{\ss}
um ein Dup\-lett erweitert werden. Dies impliziert die Existenz von f\"unf 
physikalischen Higgs-Bosonen.
Die supersymmetrischen
Partner der Eich- und Higgs-Bosonen treten als Mischzust\"ande auf, die 
Charginos und Neutralinos genannt werden. Da man bis heute keine 
supersymmtrischen Teilchen gefunden hat, mu{\ss} Supersymmetrie 
gebrochen sein. Dieser Brechung wird durch die Einf\"uhrung sogenannter
,,soft SUSY breaking'' Terme Rechnung getragen. Diese werden unterteilt in
die Massenterme f\"ur die Gauginos und die Sfermionen, trilineare Kopplungen 
zwischen Higgs-Bosonen und Sfermionen und einem bilinearen Term im Higgs-Sektor.

Es hat sich experimentell gezeigt, da{\ss} das Tau-Lepton, das Bottom-Quark 
und im besonderen das Top-Quark eine wesentlich gr\"o{\ss}ere Masse haben als
die entsprechenden Fermionen der ersten beiden Generationen. Diesem Umstand
wird in den Modellen durch entsprechend gro{\ss}e Yukawakopplungen
Rechnung getragen. Diese Yukawakopplungen beeinflussen auch wesentlich die
Ph\"anomenologie der ent\-sprechen\-den supersymmetrischen Partner. 
In der vorliegenden Arbeit werden systematisch die Zerfallsbreiten s\"amtlicher 
Zwei\-k\"orper\-zer\-f\"alle auf Tree-Level berechnet. Im Falle des leichten
Stops, einem der beiden Partner des Top-Quarks, ist es dar\"uber 
hinaus notwendig, die Zerfallsbreiten der wichtigsten Dreik\"orper\-zerf\"alle 
zu berechnen. Damit ist die Grundlage zur Erarbeitung der Ph\"anomenologie 
dieser Teilchen an laufenden und zuk\"unftigen Beschleunigern gegeben.
Je nach Massenbereich sind dies der LEP-Beschleuniger (CERN), Tevatron
(Fermilab), LHC (CERN) und ein zuk\"unftiger Leptonbeschleuniger mit
einem Energiebereich von 300~GeV bis 2~TeV.

Im Folgenden bezeichnen wir die supersymmetrischen Partner des Top-Quarks, des
Bottom-Quarks, des Tau-Leptons und des Tau-Neutrinos als Stops, Sbottoms,
Staus und Tau-Sneutrino.
Die Yukawakopplungen in den Massenmatrizen dieser Teilchen haben zur Folge, 
da{\ss} die elektroschwachen Eigenzust\"ande mischen. Die physikalischen
Teilchen werden durch die Massen und den Mischungswinkel charakterisiert.
Die Produktion dieser Teilchen bei $e^+ e^-$-Beschleunigern h\"angt in einer
charakteristischen Weise vom Mischungswinkel ab. Diese Abh\"angigkeit ist
wesentlich ausgepr\"agter, falls polarisierte Elektronen zur Verf\"ugung 
stehen.

Die gro{\ss}en Yukawakopplungen und die Mischung der elektroschwachen
Eigenzust\"ande haben zur Folge, da{\ss} sich die Zerf\"alle dieser Teilchen
wesentlich von den Zerf\"allen der entsprechenden Teilchen der ersten beiden
Generationen unterscheiden. Im Falle der Zerf\"alle in Charginos und Neutralinos
treten Interferenzen zwischen den Gaugino- und Higgsinokomponenten dieser 
Teilchen auf, die abh\"angig von den Parametern positiv oder negativ sein
k\"onnen. Im Besonderen gibt es Szenarien, in denen aufgrund dieser 
Interferenzen der Zerfall in das schwerste Chargino bzw. in das schwerste
Neutralino am wichtigsten ist.

Ein wesentlicher Aspekt ist die Tatsache, da{\ss} aufgrund der Mischung und
der Massenaufspaltung Zerf\"alle in Eich- und Higgsbosonen m\"oglich sind.
Es zeigt sich, da{\ss} diese Zerf\"alle dominieren k\"onnen, besonders im
Fall der Stops und Sbottoms. Folgende Punkte haben sich als wichtig 
erwiesen: Kinematische Effekte sind im Falle der Zerf\"alle in Eichbosonen
besonders ausgepr\"agt. Die Zerfallsbreite in ein $Z$-Boson ist umso
gr\"o{\ss}er je gr\"o{\ss}er die Mischung ist. Der Zerfall in ein  $W$-Boson 
wird
umso wichtiger  je gr\"o{\ss}er die Links-Komponente der Sfermionen ist. 
Beide Tatsachen lassen sich dadurch verstehen, da{\ss} die analogen Kopplungen
der Eichbosonen an die entsprechenden Fermionen deren Chiralit\"at 
erhalten. Im Gegensatz dazu \"andern die Kopplungen der Higgsbosonen
an die Fermionen deren Chiralit\"at. Entsprechend dazu stellt sich heraus, 
da{\ss} die Zerf\"alle in Higgsbosonen besonders wichtig sind falls
das eine Sfermion \"uberwiegend ein Linkszustand und das andere Sfermion
\"uberwiegend ein Rechtszustand ist. Die einzige Ausnahme ist hier das
pseudoskalare Higgs, da dessen Kopplungen unabh\"angig von der Mischung im
Sfermionsektor ist. 

Im Falle des leichten Stops kann nun der Fall auftreten, da{\ss} s\"amtliche
Zweik\"orper\-zerf\"alle auf Tree-Level kinematisch verboten sind. In solch
einem Szenario sind die Dreik\"orper\-zerf\"alle dominant gefolgt von dem
,,Flavour Changing'' Zerfall in ein Charm-Quark und das leichteste Neutralino.
Auch im Falle des leichten Sbottoms haben wir Szenarien gefunden, in denen
der ,,Flavour Changing'' Zerfall in ein Charm-Quark und das leichteste 
Chargino wichtig sein kann. In beiden F\"allen sind die totalen Zerfallsbreiten
so klein, da{\ss} h\"ochstwahrscheinlich Hadronisation dieser Teilchen 
wichtig sein wird.

Es hat sich herausgestellt, da{\ss} die Ph\"anomenologie der Stops, Sbottoms,
Staus und des Tau-Sneutrinos sehr reichhaltig ist. Eine genaue Bestimmung
ihrer Eigenschaften wird uns Aufschlu{\ss} \"uber die ,,SUSY breaking'' 
Parameter geben und dar\"uber hinaus hoffentlich auch Einsicht in den
Mechanismus der Supersymmetriebrechung.

\newpage

\addcontentsline{toc}{chapter}{\protect\numberline{Lebenslauf}}
\thispagestyle{empty}

\noindent \underline{{\large \bf Lebenslauf}} \hfill \\
\begin{tabbing}
	   aaaaaaaaaaaaa \= aaaaaaaaaaaaaaaaa \=       \kill
\underline{Personalien:}         \> Name:         \> Werner Porod \\
				 \> Geburtsdatum: \> 10. August 1967 \\
				 \> Geburtsort:   \> Waidhofen an der Ybbs \\
				 \> Eltern:       \> Stefanie Porod, Hausfrau \\
				 \>               \> Klaus Porod,
						     Justizwachbeamter \\ \\
\underline{Bildungsgang:} \> 4 Jahre \= Volksschule in Feldkirch-Tisis \\
			  \> 2 Jahre \> Hauptschule in Feldkirch-Gisingen \\
			  \> 6 Jahre \> Gymnasium in Feldkirch-Levis \\
			  \> 7 Semester \hspace{2mm} \= Physik-Diplomstudium
						  an der TU-Graz \\
			  \> 10 Semester \> Physik-Diplomstudium
						 an der Universit"at Wien \\
			  \> Abschlu\3 des Diplomstudiums mit
					 Auszeichnung am 4.8.1993 \\
			  \> 9 Semester \> Physik-Doktoratstudium
				      an der Universit"at Wien \\ \\
\underline{Zivildienst:} \> abgeleistet vom 1.10.1996 bis 31.8.1997 in der 
                            Pfarre St.~Cyril \\ \> und Method \\ \\

\underline{Mitwirkung an folgenden Lehrveranstaltungen:} \\
	    \> Tutorium zum Proseminar zur theoretischen Physik \\
         \> \hspace{5mm} \= f"ur Lehramtskanidaten 1 im WS 1992/93 und 1993/94\\
	    \>  Tutorium zum Proseminar zur theoretischen Physik \\
	    \>   \> f"ur Lehramtskanidaten 2 im SS 1993 und 1994\\ \\

\underline{Auslandsaufenthalte:} \\
      \> Summerschool on high energy physics and cosmology,\\
      \> International
	 Center for Theoretical Physics in Triest, \\
      \> 13.6.94-29.7.94 \\ \\
      \> 27. Herbstschule f\"ur Hochenergiephysik, \\
      \> Maria Laach, 5.9.1995-15.9.1995 \\ \\
\end{tabbing}